\documentclass[superscriptaddress,aps,prd,showpacs,preprint]{revtex4-1}
\usepackage{lingmacros}
\usepackage{tree-dvips}
\usepackage{graphicx}
\usepackage{subfigure}
\usepackage{amsmath}
\usepackage{amssymb}
\usepackage{latexsym}
\usepackage{color}
\usepackage{fullpage}
\usepackage{commath}
\usepackage{array}
\usepackage{dcolumn}
\usepackage[latin1]{inputenc}
\usepackage{graphicx, psfrag}
\usepackage{float}
\usepackage{tensor}
\usepackage{amsfonts}
\usepackage{hyperref}
\usepackage{pgfplots}
\usepackage{epstopdf}
\usepackage{booktabs}
\usepackage{comment}

\begin{document}

%\title{On the null geodesics and the optical appearance of Einstein-Maxwell-dilaton black holes}

\title{Remarks on the light ring images and the optical appearance of hairy black holes in Einstein-Maxwell-dilaton gravity}

\author{Chatchai Promsiri} \email{chatchaipromsiri@gmail.com} 
\affiliation{Quantum Computing and Information Research Centre (QX), Faculty of Science, King Mongkut's University of Technology Thonburi,  Bangkok 10140, Thailand}

\author{Weerawit Horinouchi} 
\email{wee.hori@gmail.com}
\affiliation{Department of Physics, Faculty of Science, Chulalongkorn University, Bangkok 10330, Thailand}

\author{Ekapong Hirunsirisawat}
\email{ ekapong.hir@kmutt.ac.th}
\affiliation{Quantum Computing and Information Research Centre (QX), Faculty of Science, King Mongkut's University of Technology Thonburi, Bangkok 10140, Thailand}
\affiliation{Learning Institute, King Mongkut's University of Technology Thonburi, Bangkok 10140, Thailand}

\begin{abstract}
The behaviors of null geodesics in the spherical symmetric black holes in Einstein-Maxwell-dilaton (EMD) theory with coupling function $f(\Phi)=e^{-2\alpha \Phi}$ are meticulously analyzed. 
We investigate the effects of coupling constant $\alpha$ on the effective potential of photon trajectories within three ranges, namely $0<\alpha <1$, $\alpha =1$ and $\alpha >1$.
We find that the thicknesses of lensing and photon rings are smaller at larger $\alpha$ and fixed electric charge in the unit of mass $q$, whereas they are larger at fixed $\alpha$ and larger $q$. 
This behavior can be described by using the angular Lyapunov exponent $\gamma$ in the vicinity of the critical curve. 
Remarkably, the behaviors of photon trajectories are found to be more interesting when $\alpha>1$. Namely, the radius of the black hole shadow $R_\text{s}$ becomes to be smaller than the photon sphere radius $r_\text{ph}$ when $\alpha > 1$ and $q>q^*$. 
Moreover, $R_\text{s}$ goes to zero as $q$ saturates the extremal limit, beyond which the photon orbit becomes absent.  
%This tends to be consistent with the result of Wilczek.
Furthermore, we construct the optical appearance of black holes surrounded by optically and geometrically thin accretion disk with three cases of Gralla-Lupsasca-Marrone (GLM) emission profile.  
Our results indicate that the observed flux originating from the lensing and photon rings exhibits suppression as $\alpha$ increases, while it undergoes amplification with the increasing parameter $q$.

\end{abstract}

\maketitle

\section{Introduction} \label{sec:Introduction}
%Review black hole's discovery from the GW and Image of M87 and photon ring/sphere shadow and the effects of the accretion disk on them
Interestingly, a black hole (BH), as predicted by general relativity (GR), is an immensely gravitating compact object, characterized by its highly curved spacetime that allows nothing inside its event horizon to escape. 
Despite this, advancements in recent years have demonstrated that the existence of BHs can be substantiated through the observation of certain signals directly emitted from the effects of BH spacetime. 
Notably, significant progress was achieved in 2016 when the observation of gravitational waves from the GW150914 event by the LIGO and VIRGO collaborations, supposed to be originated from the merger of binary BHs. 
This provided compelling evidence for the existence of BHs ~\cite{LIGOScientific:2016aoc}. 
However, studying the detailed features of BH through optical astronomical observation poses considerable challenges.
Recently, Event Horizon Telescope (EHT) collaboration exhibited images of two supermassive BHs, i.e., M87*~\cite{ETH1,*ETH2,*ETH3,*ETH4,*ETH5,*ETH6} at the center of the giant elliptical galaxy M87 in the Virgo cluster and SgrA*~\cite{EventHorizonTelescope:2022wkp, *EventHorizonTelescope:2022apq, *EventHorizonTelescope:2022wok, *EventHorizonTelescope:2022exc, *EventHorizonTelescope:2022urf, *EventHorizonTelescope:2022xqj} at the center of Milky Way galaxy.

When considering photons emitted from luminous objects surrounding a BH and traversing in the vicinity of a BH, their trajectories undergo bending due to the highly curved spacetime of the hole, leading to the BH image appearing to a far distant observer as the BH shadow region terminating by light rings.
For a Schwarzschild BH, the critical impact parameter value, representing the radius of photon ring in the BH image, is given by $b_c=3\sqrt{3}M$ and the corresponding radius of the photon sphere around the BH is $r_\text{ph}=3M$, as investigated in \cite{1959RSPSA.249180D, *Synge:1966okc, *ohanian1987black}. 
The size and shape of the BH shadow depend on the spacetime geometry in the vicinity of the event horizon. 
For instance, the shadow of a non-rotating BH takes the form of a circular shape with an apparent radius of  $b_c$.  
In the case of a Reissner-Nordstr\"om BH, the apparent radius of the shadow region decreases as the electric charge increases~\cite{Zakharov:2005ek, *Zakharov:2014lqa}. Moreover, Bardeen~\cite{bardeen1970kerr, *Bardeen:1973tla} initially studied the shadow of a rotating BH, revealing its D-shaped appearance of light ring, which depends on the angular momentum parameter.
Interestingly, Hioki and Maeda introduced a considerable difference in the appearance of BH shadow to distinguish nonzero-spin BHs with naked singularity from common Kerr BHs with event horizons~\cite{Hioki:2009na}. 
%The optical appearance of a nonzero spin naked singularity has found to have two parts, namely, it consists of an open dark arc and a dark spot when an observer is above the equatorial plane, whereas it consists of a dark point with \fixme{a line that is connected to the dark arc when a far distant observer is on the equatorial plane.} Since the neighborhood of the arc \fixme{darken due to gravitational redshift}, the shadow may appear as a dark lunate. 
This holds the potential to utilize observational data to examine whether an astronomical object is a BH candidate with naked singularity or not. 
There are several further investigations of the optical appearance in many BH spacetime  backgrounds and alternative gravitational theories, see \cite{Takahashi:2005hy, *Tsukamoto:2017fxq, *Abdujabbarov:2012bn, *Mandal:2022oma, *PhysRevD.97.104062, *Khodadi:2020jij, *Wei:2013kza, *Das:2019sty, *Papnoi:2021rvw, *Panpanich:2019mll, *Hendi:2022qgi, *Khodadi:2020gns, *Zhu:2019ura} for example.

Classically, the size of a black hole is characterized by its event horizon. 
However, the radius of this horizon can be challenging to directly observe, either locally or from asymptotic infinity.
In the realm of astrophysical observation, a significant focus lies in the examination of the photon sphere and its associated shadow radius. 
The study of how they are determined by the black hole mass, charges and other hairs, is crucial for measuring the size of black holes since photons moving in an unstable orbit, ensuring their eventual escape from the photon sphere and thereby rendering it observable.
A comparison between the radius of shadow and the radius of the photon sphere, namely the ratio $R_s/r_\text{ph}$ is made in $d$-dimensional quintessential BHs \cite{Belhaj:2020rdb}. 
It is found that $R_s/r_\text{ph}\approx 1$ in the case of $d>6$, while $R_s/r_\text{ph}>1$ for $d\leq 6$.
Remarkably, by using the weak energy condition and the trace of energy-momentum tensor condition, Ref.~\cite{Hod:2013jhd, Cvetic:2016bxi} shows a bound between $R_s$ and $r_\text{ph}$.
Recently, Ref.~\cite{Lu:2019zxb, Ma:2019ybz, Chakraborty:2021dmu} has further proposed an inequality between $r_+, r_\text{ph}$ and $R_s$ to the case of more general BHs. 
Thus, by comparing the values of three radii parameters, namely event horizon $r_+$, photon sphere $r_\text{ph}$ and shadow $R_s$ can used to characterized the nature of BH solutions in the theories of gravity.  

To examine the optical appearance of astronomical BHs, it is typically assumed that there exist distant localized sources emitting light into the vicinity near black hole's event horizon, leading to optical features such as shadow, lensing ring and photon ring, etc. 
Recent astronomical observations have shown that many realistic BHs may be surrounded by accretion matter~\cite{Abramowicz:2011xu}, giving rise to the detailed features of the black hole's image.
In other words, different models of accretion matter surrounding the same BH spacetime can cause different images of black hole's shadow and light rings. 
For instance, the first visualization of a BH surrounded by a luminous accretion disk was presented in \cite{luminet1979image}.
In Refs. \cite{Falcke:1999pj, *Narayan2019xty, *Heydari-Fard:2022jdu, *Heydari-Fard:2023ent}, the spherical accretion has been taken into account to consider the shadow of BH. 
The total number of photon orbits around BH, denoted by $n=\phi/2\pi$ where $\phi$ is angular distance of orbiting photons, is an important variable in considering the luminosity of detailed features of the BH image.
%classifying the null geodesics and luminosity of detailed features of the BH image.
Recently, the work by Gralla, Holz, and Wald (GHW) \cite{Gralla:2019xty} has proposed a more convenient way to investigate the image of a BH with accretion matter.
They suggest that the image formed by three types of photon trajectories, namely, the direct emission, the lensing ring, and the photon ring which correspond to $\displaystyle n<\frac{3}{4}$, $\displaystyle \frac{3}{4}<n<\frac{5}{4}$ and $\displaystyle n>\frac{5}{4}$, respectively.
According to GHW classification, the studies on optical appearance of a BH shadow in using different models of accretion matter have attracted more attention leading to the explosion of subsequent works, see \cite{Zeng2020arx, *Zeng:2022fdm, *Zeng2020qun, *Zeng:2021dlj, *Wang:2022yvi, *PhysRevD.104.044049, *Gan:2021pwu, *Uniyal:2022vdu, *Wang:2023vcv, *Peng:2020wun, *Peng:2021osd, *PhysRevD.105.084057, *Guerrero:2022msp, *Huang:2023yqd} for example.

Some clues have suggested that our current understanding of gravitational physics remains incomplete. For example, there are conundrums of dark matter and dark energy stem from astrophysical observations, specifically concerning the unusually high velocities of stars at galaxies' edges and the surprisingly accelerated expansion of the late-time universe, respectively.
Moreover, there are more clues from the considerations in BH information.  
Namely, the unitarity of BH information could be violated due to the Hawking evaporation process of BHs, so-called the information loss paradox~\cite{PhysRevD.14.2460, *Page:1993wv, *Page:2013dx, *Mathur:2009hf, *RevModPhys.88.015002, *Raju:2020smc}, for some current developments see~\cite{RevModPhys.93.035002} and references therein.  
This led to the considerations for exploring quantum descriptions of gravity. 

One of the promising candidates for quantum gravity is 
string theory, whose several BH solutions have a variety of interesting aspects to explore.
Upon examining the low energy regime of string theory, it becomes apparent that the dilaton field can exhibit a non-minimal coupling to the Maxwell field, characterized by a nonzero coupling constant $\alpha$, in the Einstein-Hilbert action. Consequently, solutions to the Einstein field equation are expected to incorporate a varying dilaton field. Notably, investigations into the static, spherically symmetric, charged BH solutions in this context of Einstein-Maxwell-dilaton (EMD) gravity have been studied in~\cite{Gibbons:1987ps, *Garfinkle:1990qj}. 
Interestingly, this theory allows an exception of the no-hair theorem (NHT)~\cite{gravitation}, namely there exist BHs with scalar hair degrees of freedom additional to classical three parameters, i.e., mass, angular momentum, and electric charge.
Thus, EMD BHs are one of the counterexamples of the NHT. 
In recent times, notable interest has been directed towards a dynamic mechanism that entails the non-minimal coupling of a scalar field to certain source terms. This intriguing phenomenon gives rise to the appearance of hairy BH solutions, referred to as spontaneous scalarization~\cite{PhysRevLett.120.131103, *PhysRevLett.120.131104, *PhysRevLett.121.101102, *Fernandes:2019rez, *Guo:2021zed, *PhysRevD.108.024015}. Consequently, this compelling avenue of research involves investigating optical properties associated with these hairy BHs. 
Such endeavors hold the potential to offer valuable insights into the validity of the NHT or impose constraints on their parameters through forthcoming astronomical observations.

In the realm of BH phenomena in the EMD theory, diverse investigations have been conducted concerning BHs with dilaton hairs. 
Previous studies have focused on scalar and fermionic fields' perturbations around BHs~\cite{Fernando:2003wc}, null geodesics of massless particles~\cite{PhysRevD.85.024033, PhysRevD.105.124009} and timelike geodesics of massive particles~\cite{Maki:1992up, Olivares:2013jza, Pradhan:2012id, Blaga:2014spa} in the vicinity of BHs, and the gravitational lensing phenomenon~\cite{Bhadra:2003zs, Mukherjee:2006ru, Ghosh:2010uw}.
There are further investigations on an extension of this static and spherical symmetric solution to rotating dilaton BHs in \cite{Shiraishi:1992np,*Horne:1992zy,*Sen:1992ua}, as well as on their implications for BH astrophysics in \cite{Younsi:2016azx, *Konoplya:2021slg, *Badia:2022phg, *Heydari-Fard:2020ugv}.
Remarkably, it was shown in \cite{holzhey1992black} that there exist infinite potential
barriers of scalar fields forming around extremal black holes when $\alpha > 1$. This
led to the interpretation that extremal black holes can behave like elementary particles. 

In this present work, our primary focus lies in the thorough examination of null geodesics encompassing BHs within the EMD theory. 
Notably, during the course of this research, a publication has emerged~\cite{PhysRevD.105.124009}, which emphasizes the impacts of the change of dilaton coupling $\alpha$ at fixed $q$ and the change of $q$ at fixed $\alpha$ on the sizes of photon ring radius enclosing the shadow image in the range of $\alpha \leq1$. 
In contrast, our investigation centers on the behaviors of the $q$ and  $\alpha$ parameters on shadow and photon rings, as well as an optical appearance with the presence of thin accretion disks, in three ranges of $\alpha$, namely $0<\alpha<1$, $\alpha=1$, and $\alpha >1$.  We find that notable results of the BH images arise when $\alpha \geq 1$. The optical features we uncover through this study bear potential significance in distinguishing relevant astronomical observations and enhancing our understanding of the underlying phenomena.

This paper is organized as follows: In Section~\ref{section 2}, an overview of the BH solution within the framework of the EMD theory is presented, accompanied by an examination of its behavioral dependence on parameters $\alpha$ and $q$. 
Subsequently, Section~\ref{section 3} delves into the derivation of null geodesics encircling the EMD BH, employing the Hamilton-Jacobi approach. 
Notably, this section extensively discusses the circular photon orbit and the GHW classification for null geodesics of the BH in the EMD theory.
Furthermore, we also derive angular Lyapunov exponent associated with unstable bound geodesic of photons moving in the charged dilaton BH background to elucidate more about physical implications of the exponentially narrowing behavior of photon subrings near the edge of BH shadow.
Section~\ref{section 4} is dedicated to the investigation of images of BH shadow. 
Intriguing outcomes concerning the BH images are observed, particularly in the case of $\alpha > 1$. 
Section~\ref{section 5} focuses on examining the transfer functions to explore the optical characteristics of the BH image in EMD gravity. 
% This analysis of the EMD BH's optical characteristics is conducted by considering the light emitted from thin accretion disks, illuminated in accordance with the Gralla-Lupsasca-Marrone (GLM) model, across three distinct scenarios [38].
 Our analysis on these optical features is conducted by considering the light emitted from thin accretion disks, in accordance with the Gralla-Lupsasca-Marrone (GLM) model across three distinct scenarios~\cite{PhysRevD.102.124004}.
%various models of illuminating thin accretion disks into the vicinity of the EMD BH. 
Finally, the paper culminates in Section~~\ref{section 6}, where additional discussion and concluding remarks are provided, summarizing the key findings and implications of the study.

\section{Black holes in the EMD gravity} \label{section 2}

The action of Einstein-Maxwell-dilaton (EMD) gravity in the Einstein frame is given as follows 
\begin{eqnarray}
S_\text{EMD}=\int d^{4}x\sqrt{-g}\left( \mathcal{R} - 2g^{\mu \nu} \nabla_\mu \Phi \nabla_\nu \Phi - f(\Phi)F_{\mu \nu}F^{\mu \nu}  \right), \label{action}
\end{eqnarray}
where $\mathcal{R}$ is the Ricci scalar, $F_{\mu \nu}=\partial_\mu A_\nu - \partial_\nu A_\mu$ is the electromagnetic field strength tensor of Maxwell gauge field $A_\mu$, and $\Phi$ is the dilaton field. 
%Note that, when $\alpha =0$, the action reduces to Einstein-Maxwell theory with real scalar and Maxwell fields decoupled, and hence the static black hole solution is the usual Reissner-Nordst\"om solution. 
%For $\alpha =1$, this action is the low energy limit of string theory where the black hole solution is known as GMGHS black hole [Gibbon, Horowitz]. When $\alpha =\sqrt{3}$, the action represents the five-dimensional Kaluza-Klein gravity and the black hole solution emerges from the dimensional reduction of five spacetime dimensions. 
In this work, we are interested in the coupling function between Maxwell and dilaton fields as $f(\Phi)=e^{-2\alpha \Phi}$, where a constant $\alpha$ denotes the dilaton coupling constant.
One can see that $\Phi$ is minimally coupled to gravity but non-minimally coupled to the gauge field.
In asymptotically flat spacetime, the action \eqref{action} with $f(\Phi)=e^{-2\alpha \Phi}$ admits a static spherically symmetric solution with an arbitrary value of $\alpha$, which is given by \cite{Gibbons:1987ps, Garfinkle:1990qj},
\begin{equation}
    ds^2 = -g(r)dt^2+\frac{dr^2}{g(r)}+R^2(r)\left( d\theta^2 + \sin^2\theta d\phi^2 \right), \label{EMD BH solution}  
\end{equation}
where
\begin{eqnarray}
g(r) &=& \left( 1 - \frac{r_+}{r} \right)\left( 1 - \frac{r_-}{r} \right)^{\frac{1-\alpha^2}{1+\alpha^2}}, \label{g} \\
R(r) &=& r\left( 1 - \frac{r_-}{r} \right)^{\frac{\alpha^2}{1+\alpha^2}}. \label{R}
\end{eqnarray}
Note that $r_+$ and $r_-$ represent the outer and inner horizon radius of BH, respectively, which can be written as functions of the ADM mass $M$ and  electric charge $Q$ of the BH, as well as $\alpha$.
The electric and dilaton fields are given by
\begin{eqnarray}
F_{tr} = \frac{e^{2\alpha \Phi}Q}{R^2(r)}, \ \ \ \ \ \Phi (r) = \frac{\alpha}{1+\alpha^2}\ln \left( 1-\frac{r_-}{r} \right). \label{field profile}
\end{eqnarray}
$M$ and $Q$ can be written in terms of  $r_+$ and $r_-$ as
\begin{eqnarray}
M = \frac{1}{2}\left[ r_+ + \left(\frac{1-\alpha^2}{1+\alpha^2}\right)r_- \right], \ \ \ \ \ Q^2 = \frac{r_+ r_-}{1+\alpha^2}.
\end{eqnarray}
For later convenience, we adopt the convention $M=1$, thus allowing variables with dimensions of length and energy to be readily expressed in unit of the BH mass $M$.  Accordingly, the BH charge and the event horizon can be written as 
\begin{eqnarray}
    q=\frac{Q}{M}=Q, \ \ \ \text{and} \ \ \  r_\pm = \frac{r_\pm}{M}.
\end{eqnarray}
It is important to note that we will occasionally reintroduce the parameter $M$ into certain formulas, serving the purpose of both reminder and discussion. However, please keep in mind that all quantities of length and energy dimensions will be expressed in the unit of $M$ from now on.
With nonzero value of $\alpha$, the  spacetime metric becomes singular when $r_+$ and $r_-$ satisfy the relations 
\begin{eqnarray}
r_+ &=&  1 + \sqrt{1 - q^2(1-\alpha^2)}, \label{outer} \\
r_- &=& \frac{(1+\alpha^2)\left[ 1 - \sqrt{1 - q^2(1-\alpha^2)} \right] }{(1-\alpha^2)}. \label{inner}
\end{eqnarray}
It should be remarked here that  $r_+>r_-$ if $q^2 < 1+\alpha^2$.  
When the charge $q$ saturates the extremal limit,
\begin{eqnarray}
q_\text{ext}= \sqrt{1+\alpha^2}. \label{extremal}
\end{eqnarray}
The radii of these two horizons coincide at $r_+=r_-=1+\alpha^2$.
Obviously, Schwarzschild BH with  horizon radius $r_+=2$ can obtain at $q=0$ limit.
For scalar-free solution $\alpha =0$, we obtain the usual Reissner-Nordstr\"om BH with $r_\pm = 1\pm \sqrt{1-q^2}$.
Thus the extremal black hole can be achieved when $q=1$.

\begin{figure}[H]
    \centering
    \includegraphics[width = 5.cm]{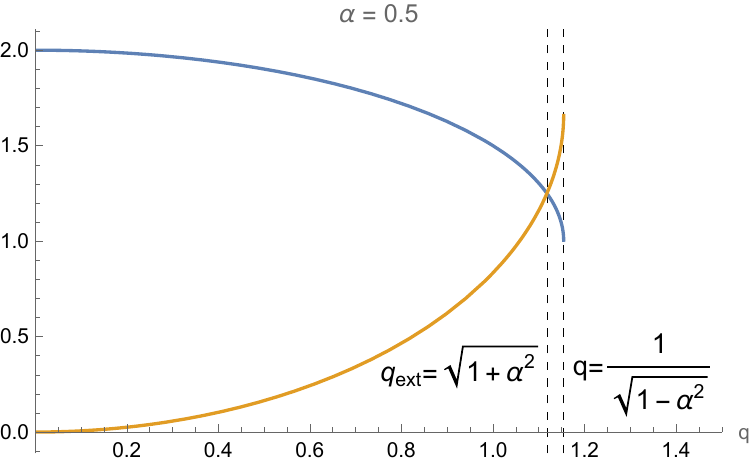}\hfill
    \includegraphics[width = 5.cm]{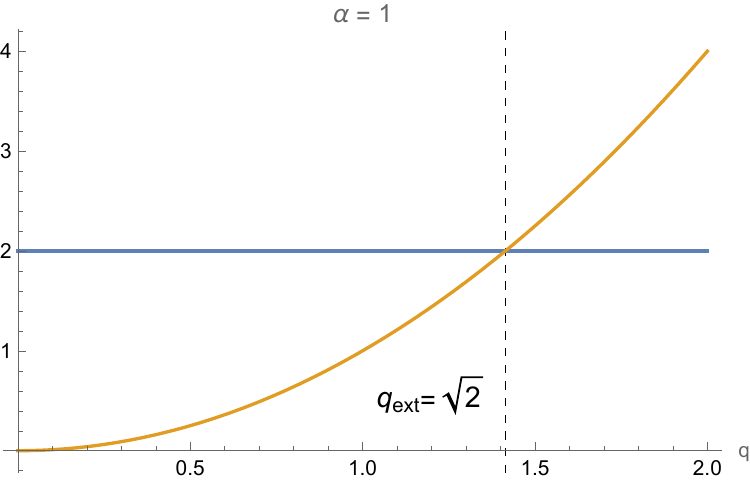}\hfill
    \includegraphics[width = 5.cm]{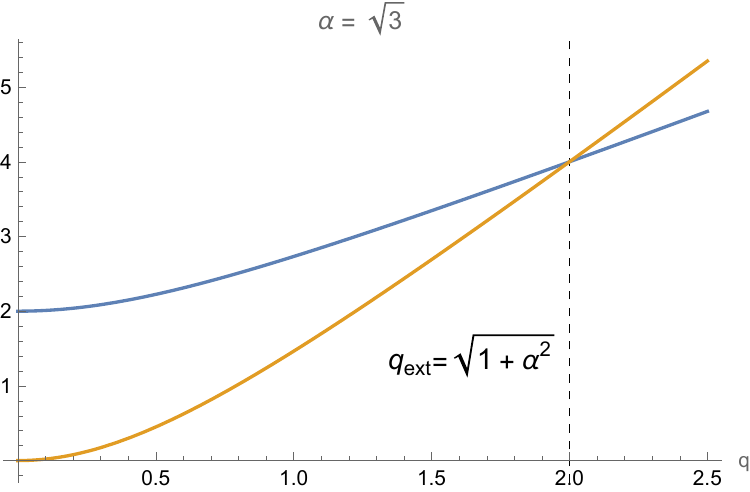}
    \caption{The plots of  horizon radii $r_+$ (solid blue curve) and $r_-$ (solid orange curve) versus $q$ are shown. From the left to the right panels, we fix the dilaton coupling constant $\alpha =0.5, 1$ and $\sqrt{3}$, respectively.} 
    \label{fig: Horizon radius vs q}
\end{figure}

Spacetime near BHs in the EMD gravity are different depending on three ranges of $\alpha$, namely $0<\alpha <1$, $\alpha =1$ and $\alpha >1$.
In Fig.~\ref{fig: Horizon radius vs q}, we show the behavior of the horizon radii $r_\pm$ as a function of $q$. 
In the left panel, we consider $0<\alpha <1$, the results indicate that  $r_+$ ($r_-$) decreases (increases) as $q$ is larger and these two radii coincide at $q_\text{ext}$.
Interestingly, $r_{-}$ becomes larger than $r_{+}$ when
$\displaystyle q_\text{ext}<q<\frac{1}{\sqrt{1-\alpha^2}}$. Hence, $r_{-}$ and $r_{+}$ turn out to be outer and inner horizons, respectively. 
%the inner horizon $r_-$ becomes larger than the outer horizon $r_+$ when $\displaystyle q > q_\text{ext}$ to some extent, i.e., not more than $\displaystyle \frac{1}{\sqrt{1-\alpha^2}}$. 
However, when $\displaystyle q>\frac{1}{\sqrt{1-\alpha^2}}$, both $r_+$ and $r_-$ encounter the changes from real number to complex one, implying that these two horizons do not exist.
One can see from that the inner horizon $r_-$ is not well defined if $\alpha =1$.
As shown in the middle panel, by taking the limit $\alpha \rightarrow 1$, the inner horizon $\displaystyle r_- = q^2$ while the outer horizon is $r_+ =2$ for all values of $q$.   
In other words, $r_+$ does not depend on $q$, while $r_-$ increases as $q$ is larger. 
Hence, these two radii are degenerate at the extremal limit where $q_\text{ext}=\sqrt{2}$.
Unlike the first case, these two radii are real number for all values of $q$.
Substituting these $r_\pm$ and setting $\alpha =1$ into Eqs.\eqref{g} and \eqref{R}, the metric takes the form that 
corresponds to the metric of the Gibbon-Maeda-Garfinkle-Horowitz-Strominger (GMGHS) BH solution.
For $\alpha >1$, we find that both horizon radii increase as $q$ is larger as shown in the right panel.  
In a similar way as in the case of $\alpha =1$, $r_+$ and $r_-$ become degenerate at  $q_\text{ext}$ and always exist for any values of $q$.  
In this plot, we have used $\alpha =\sqrt{3}$ corresponding to the Kaluza-Klein theory~\cite{PhysRevD.46.1340}, where, by using Eq.~\eqref{extremal},  $q_\text{ext}$ is simply 2.
Moreover, $r_-$ becomes larger than $r_+$ at  $q>q_\text{ext}$.
Interestingly, one can see that dilatonic BHs exhibit distinct characteristics due to different ranges of the coupling constant $\alpha$.
It is important to note that we can consider $\alpha$ with only the non-negative values, since the action in \eqref{action} is invariant under $\alpha \rightarrow -\alpha$ and $\Phi \rightarrow -\Phi$ and hence the hairy black hole solution for negative dilaton couplings can be provided by changing the sign of the scalar field $\Phi$~\cite{Astefanesei:2019pfq,Khalil:2018aaj}.

To illustrate the spacetime structure for the charged dilaton BH, we investigate the Kretschmann scalar $\mathcal{K}$ of the metric in Eq.~\eqref{EMD BH solution}, which can be expressed as follows:
\begin{eqnarray}
    \mathcal{K}=\frac{R^4 \left(g''\right)^2+4 g^2 \left(2  \left(R R''\right)^2+\left(R'\right)^4\right)-8 g R' \left(R'- g' R^2 R''\right)+4\left( g' R R'\right)^2+4}{R^4}.
\end{eqnarray}
The geometry has a curvature singularity at $r=r_-$ where $R(r_-)$ becomes vanish with any coupling $\alpha$.
Thus, spacetime singularity is always cloaked by $r_+$ for nonextremal BH.
In contrast to the extremal RN BH where the Kretschmann scalar remains regular at the horizon, the extremal charged dilaton BH exhibits a naked singularity at the horizon. 
In some sense, this singularity is, in principle, accessible to distant observers. 
It is noteworthy that our examination will focus on the null geodesics and optical characteristics of the charged dilaton BH within the parameter range of $q\leq q_\text{ext}$ where $r_+$ consistently surpasses or equals $r_-$.

\section{Null geodesics in EMD black hole geometry} \label{section 3}

To analyze the geodesics of test particles in the spacetime of Eq.~\eqref{EMD BH solution}, we use the Lagrangian and Hamilton-Jacobi equation by following the approach of Chandrasekhar's textbook \cite{Chandrasekhar:1985kt}.   
The equations of motion can be derived from the following Lagrangian
\begin{eqnarray}
2\mathcal{L} &=& g_{\mu \nu}\dot{x}^\mu \dot{x}^\nu \nonumber \\
&=& -g(r)\dot{t}^2 + \frac{\dot{r}^2}{g(r)} + R^2(r)\left( \dot{\theta}^2 + \sin^2\theta \dot{\phi}^2 \right), \label{Lagrangian}
\end{eqnarray}
where $\lambda$ is an affine parameter along the geodesic $x^\mu (\lambda)$ and $\displaystyle \dot{x}^\mu = \frac{d x^\mu}{d\lambda}$. 
The corresponding canonical momenta of particles moving in the gravitational field can be calculated via the standard relation, i.e., $\displaystyle p_\mu = \frac{\partial \mathcal{L}}{\partial \dot{x}^\mu} = g_{\mu \nu}\dot{x}^\nu$ as
\begin{eqnarray}
p_t &=& -g(r)\dot{t}, \\
p_r &=& \frac{\dot{r}}{g(r)}, \\
p_\theta &=& R^2(r)\dot{\theta}, \\
p_\phi &=& R^2(r)\sin^2\theta \dot{\phi}. \label{gm}
\end{eqnarray}
Since the metric in \eqref{EMD BH solution} has two Killing vectors associated with time translation and angular rotation, so there are two conserved quantities along the geodesics, namely the energy $E$ and the angular momentum $L$.
Thus, we have
\begin{eqnarray}
p_t &=& g(r)\dot{t} = E, \label{E} \\
p_\phi &=& R^2(r)\sin^2\theta \dot{\phi} = L. \label{L}
\end{eqnarray}
For simplicity, we will use the Hamilton-Jacobi method to analyze the geodesics in $r$ and $\theta$ coordinates since it gives the equations of motion in the first derivative concerning $\lambda$.
The Hamilton-Jacobi equation can be obtained from the differentiation of the action $S$ with respect to $\lambda$, such that
\begin{eqnarray}
\frac{\partial S}{\partial \lambda} = -\frac{1}{2}g^{\mu \nu}\frac{\partial S}{\partial x^{\mu}}\frac{\partial S}{\partial x^{\nu}} = -\frac{1}{2}g^{\mu \nu}p_\mu p_\nu , \label{Hamilton-Jacobi}
\end{eqnarray}
where we have defined the canonical momenta $p_\mu$ with the relation $\displaystyle p_\mu =\frac{\partial S}{\partial x^\mu}$.
Here, $S$ is the Jacobi action which can be separated as
\begin{eqnarray}
S=\frac{1}{2}\delta \lambda -Et+L\phi +S_r(r)+S_\theta (\theta), \label{Jacobi action}
\end{eqnarray}
where $S_r$ and $S_\theta$ are functions of $r$ and $\theta$, respectively. Moreover, $\delta$ is the parameter that is defined to specify the action to be of the form for either timelike or null geodesics.   
%The particle's mass, energy, and angular momentum are labeled as $m$, $E$ and $L$, respectively. 
Here, $\delta =1$ and $0$ for timelike and null geodesics, respectively. 
Substituting Eq.~\eqref{Jacobi action} into Eq.~\eqref{Hamilton-Jacobi}, we have
\begin{eqnarray}
\delta R^2(r)=\frac{R^2(r)E^2}{g(r)}-R^2(r)g(r)\left( \frac{\partial S_r}{\partial r} \right)^2-\left( \frac{\partial S_\theta}{\partial \theta} \right)^2-L^2\csc^2\theta .
\end{eqnarray}
One can use the identity, $\csc^2\theta =1+\cot^2\theta$, to rewrite the above equation as
\begin{eqnarray}
\delta R^2(r) + R^2(r)g(r)\left( \frac{\partial S_r}{\partial r} \right)^2 - \frac{R^2(r)}{g(r)}E^2 + L^2 = -\left( \frac{\partial S_\theta}{\partial \theta} \right)^2 - L^2\cot^2\theta . \label{separation.var}
\end{eqnarray}
Using the method of the separation of variables, we set both the left-hand side and the right-hand side of this equation to be the constant $-\mathcal{Q}$, which can be called the Carter constant~\cite{PhysRev.174.1559}.
In this way, we obtain
\begin{eqnarray}
R^4(r)g^2(r)\left( \frac{\partial S_r}{\partial r} \right)^2&=&\mathcal{R}(r), \label{r coordinate} \\
\left( \frac{\partial S_\theta}{\partial \theta} \right)^2&=&\Theta (\theta) , \label{theta coordinate}
\end{eqnarray}
where $\mathcal{R}(r)$ and $\Theta (\theta)$ are defined as the follows:
\begin{eqnarray}
\mathcal{R}(r)&=&R^4(r)E^2-R^2(r)g(r)\left(\delta R^2(r)+L^2+\mathcal{Q}\right), \\
\Theta (\theta)&=&\mathcal{Q}-L^2\cot^2\theta .
\end{eqnarray}
Then, using the relations $\displaystyle p_r =\frac{\partial S_r}{\partial r}$ and $\displaystyle p_\theta =\frac{\partial S_\theta}{\partial \theta}$, the Eqs.~\eqref{r coordinate} and \eqref{theta coordinate} becomes
\begin{eqnarray}
R^2(r)\left( \frac{dr}{d\lambda} \right)&=&\pm \sqrt{\mathcal{R}(r)}, \label{r dot} \\
R^2(r)\left( \frac{d\theta}{d\lambda} \right)&=&\pm \sqrt{\Theta (\theta)}. \label{theta dot}
\end{eqnarray}
The  $\pm$ choices depend on whether velocity is moving in a positive or negative in both the radial and angular directions.
Eqs.~\eqref{E}, \eqref{L}, \eqref{r dot} and \eqref{theta dot} are the complete set of first integrals that governs the timelike and null geodesics in the spherically symmetric spacetime.

\subsection{Circular photon orbit}

Eq.~\eqref{r dot} describing the radial motion of particles around the EMD BHs can be rewritten in the form
\begin{eqnarray}
\left( \frac{d r}{d \lambda} \right)^2+\bar{V}_\text{eff}(r) =E^2. \label{new r dot}
\end{eqnarray}
So, one can define the effective potential as
\begin{eqnarray}
\bar{V}_\text{eff} (r) = \frac{g(r)}{R^2(r)}\left(\delta R^2(r)+L^2+\mathcal{Q} \right). \label{Veff prime}
\end{eqnarray}

\begin{figure}[H]
    \centering
    \includegraphics[width = 4.7cm]{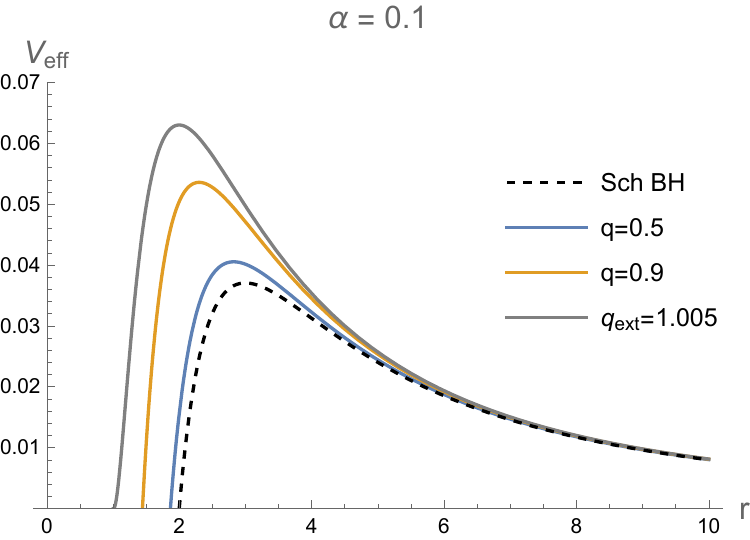}\hspace{1cm}
    \includegraphics[width = 4.7cm]{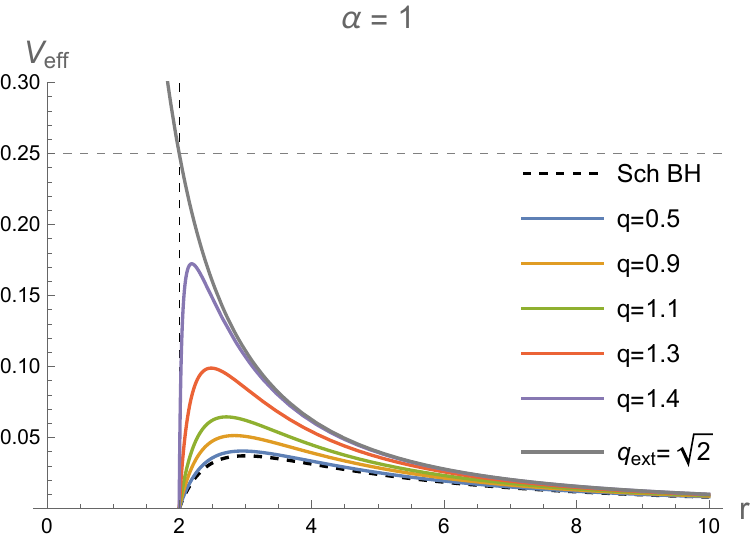}\hspace{1cm}
    \includegraphics[width = 4.7cm]{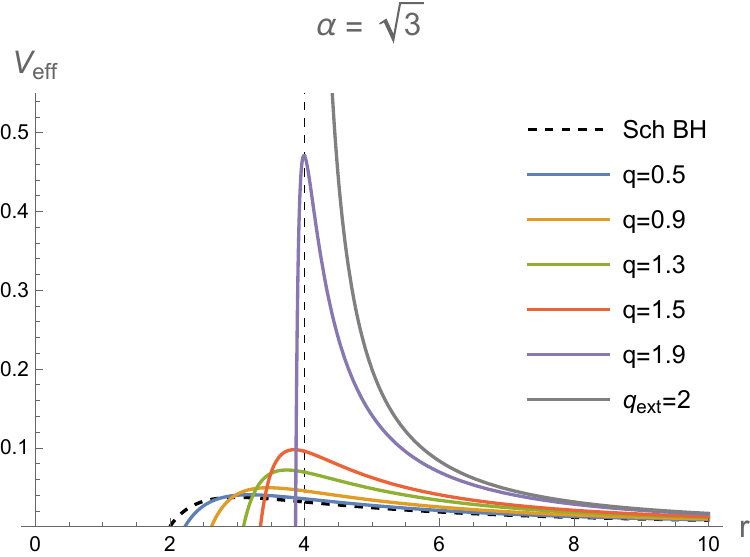}
    \caption{Different behaviors of the  effective potentials $V_\text{eff}$ versus $r$ of photons moving in the EMD BH geometry for the cases of  $0<\alpha<1$, $\alpha=1$, and $\alpha>1$ are shown here. We choose the value of $\alpha$ to be 0.1, 1 and $\sqrt{3}$ , as the example of each individual case. } 
    \label{fig: Veff vs r}
\end{figure}
\noindent 
However, with a suitable choice of coordinates, the motion of a test particle in the static and spherically symmetric spacetime is limited to a plane, yielding $p^\theta =0$.
One can choose the particles to move in the plane of $\displaystyle \theta = \frac{\pi}{2}$ and  $\dot{\theta}=0$, without loss of generality.  Accordingly, the right-hand side of Eq.~\eqref{separation.var} becomes vanished, implying that  the Carter constant $\mathcal{Q}=0$. 

For massless particles, i.e., $\delta =0$, the radial equation of motion Eq.~\eqref{new r dot} becomes 
\begin{eqnarray}
\left( \frac{d r}{d \Tilde{\lambda}} \right)^2 + V_\text{eff}(r) &=& \frac{1}{b^2}, \label{Hamiltonian}
\end{eqnarray}
where  the effective potential $V_\text{eff}(r)=\bar{V}_\text{eff}(r)\big|_{\delta=0,\mathcal{Q}=0}$ and  $\displaystyle b = \frac{|L|}{E}$,  the impact parameter for the ray of photons coming from infinity.
Note that we have defined a new affine parameter $\Tilde{\lambda}$ related to the previous one by $\Tilde{\lambda} = |L| \lambda$.
The effective potential in the equatorial plane is given by
\begin{eqnarray}
V_\text{eff}(r) = \frac{g(r)}{R^2(r)} = \frac{1}{r^2}\left( 1 - \frac{r_+}{r} \right)\left( 1 - \frac{r_-}{r} \right)^{\frac{1-3\alpha^2}{1+\alpha^2}}.
\end{eqnarray}
We plot the  effective potential $V_\text{eff}\equiv M^2V_\text{eff}$ of the photon as the function of $\displaystyle r\equiv \frac{r}{M}$ with different values of $\alpha$ and $q$ in Fig.~\ref{fig: Veff vs r}.
One can see that there are three different characteristics of the effective potential for $0<\alpha <1$, $\alpha =1$ and $\alpha >1$.
In the left panel, we fix $\alpha =0.1$, the behavior of effective potential is qualitatively similar to the case of Reissner-Nordstr\"om BH because the maxima of effective potentials have higher values and the horizon radii, where $V_\text{eff}=0$, have lower values as $q$ increases.
For $\alpha =1$, the peaks of the effective potentials have higher values when $q$ increases, as in the first case. However, $V_\text{eff}=0$ at $r=r_+=2$ for all values of $q$ in the $q<q_\text{ext}$ region, rather than depending of $q$.
Interestingly, when $q$ approaches the extremal limit, namely, $q_\text{ext}=\sqrt{2}$, the local maximum of $V_\text{eff}$ outside the horizon disappears since the potential profile changes to one that blows up to infinity as $r\rightarrow 0$.  Particularly, the effective potential at the horizon is equal to $0.25$ instead of vanishing,  as it does in the case of $q<q_\text{ext}$. 
See the middle panel of the Fig.~\ref{fig: Veff vs r}.
The effective potential profile for $\alpha >1$ is shown in the right panel.
In this plot, choosing $\alpha =\sqrt{3}$, it is clear that with increasing $q$ both the peaks of effective potentials and horizon radii have higher values. 
Remarkably, the value of effective potential becomes growing without bound at the horizon when $q=q_\text{ext}=2$, so there is no maximum point of the effective potential outside the horizon. This is in the same fashion as the results discussed in \cite{holzhey1992black}, which consider the effective potential for massless scalar field at $\alpha > 1$ and $q=q_\text{ext}$.

\begin{figure}[H]
    \centering
    \includegraphics[width = 4.7cm]{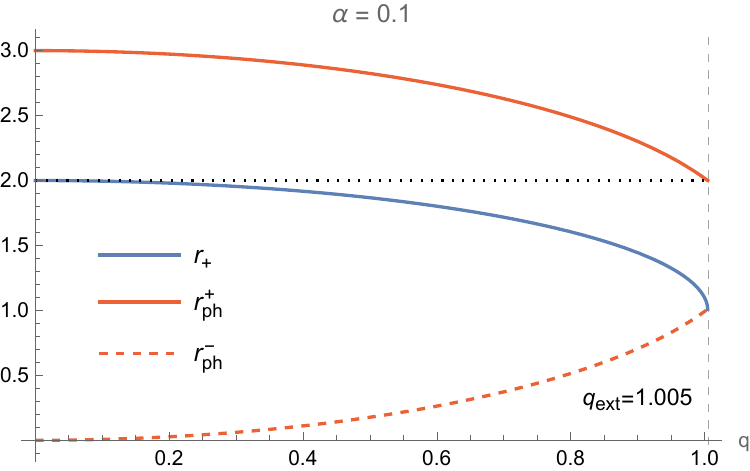}\hspace{1cm}
    \includegraphics[width = 4.7cm]{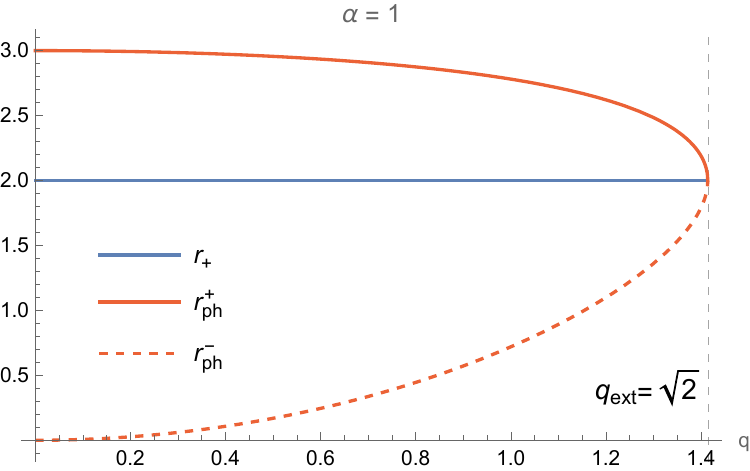}\hspace{1cm}
    \includegraphics[width = 4.7cm]{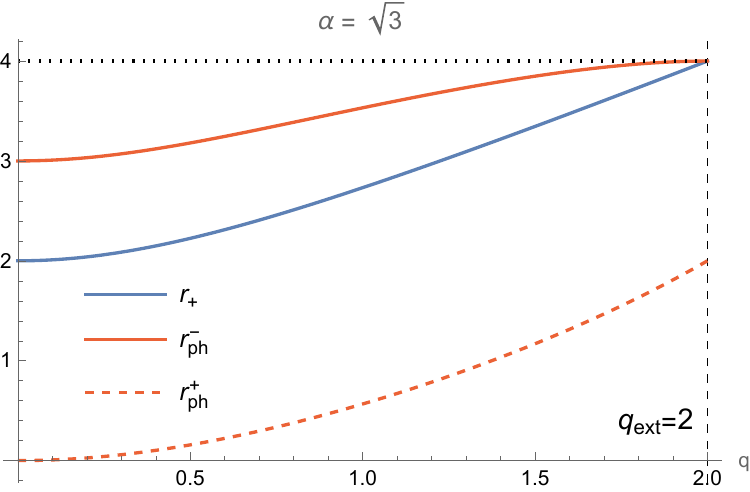}
    \caption{Different behaviors of two solutions of  photon sphere radius $r^\pm _\text{ph}$, compare with the  event horizon $r_+$, for $0<\alpha<1$, $\alpha=1$, and $\alpha>1$.  From the left to the right panels, we fixe $\alpha =0.1, 1$ and $\sqrt{3}$, respectively} 
    \label{fig: rph vs q}
\end{figure}
When the effective potential reaches its peak, it becomes possible for photons to move in circular orbits.
Nevertheless, these circular orbits are unstable because even a small change in $b$ will cause the photon to deviate from the peak and move away.
The position of the circular photon orbits can be determined by two following conditions:
\begin{eqnarray}
\partial_r V_\text{eff}(r_\text{ph}) = 0 \ \ \ \text{and} \ \ \
 V_\text{eff}(r_\text{ph}) = \frac{1}{b^2_c},
 \label{condition}
\end{eqnarray}
where $r_\text{ph}$ is the radius of the photon sphere and $b_c$ is its corresponding impact parameter. 
The first condition gives,
\begin{eqnarray}
\frac{R^2(r)g'(r)-2g(r)R(r)R'(r)}{R^4(r)} = 0. \label{first condition}
\end{eqnarray}
Note that at the event horizon where $g(r_+)=0$, the second term vanishes, while the first term is generally nonzero.
This means that the location of $r_+$ and $r_\text{ph}$ are not coincide. 
Solving Eq.~\eqref{first condition}, the photon sphere radius $r_\text{ph}$ can be expressed as \cite{PhysRevD.105.124009}
\begin{eqnarray}
r^{\pm}_\text{ph} &=& \frac{1}{2(1-\alpha^2)}\left[ 3-2\alpha^2-\alpha^2 \sqrt{1-q^2(1-\alpha^2)} \right. \nonumber \\
&& \left. \pm \sqrt{\left( 3-2\alpha^2-\alpha^2 \sqrt{1-q^2(1-\alpha^2)} \right)^2 - 8q^2(1-\alpha^2)^2} \right]. \label{rph}
\end{eqnarray}
For the GMGHS BH, we have \cite{PhysRevD.85.024033}
\begin{eqnarray}
r^\pm_\text{ph} = \frac{1}{4}\left( 6+q^2\pm \sqrt{q^4-20q^2+36} \right). \label{rph for alpha = 1}
\end{eqnarray}
From Eq.~\eqref{rph}, there are two solutions of photon sphere radius. 
For $0<\alpha <1$ ($\alpha >1$), we find that $r^{+}_\text{ph}>r_+$ ($r^{+}_\text{ph}<r_+$) while $r^{-}_\text{ph}<r_+$ ($r^{-}_\text{ph}>r_+$).
Since the radius of photon sphere should lie outside the event horizon, so $r^{+}_\text{ph}$ and $r^{-}_\text{ph}$ represent the radius of photon orbit for EMD BH with $0<\alpha <1$ and $\alpha >1$, respectively.
For GMGHS BH, Eq.~\eqref{rph for alpha = 1} indicates that $r^+_\text{ph}\geq 2$ and $r^-_\text{ph}\leq 2$.
Thus $r^+_\text{ph}$ is the physical radius of the unstable photon orbit near the GMGHS BH.

In Fig.~\ref{fig: rph vs q}, we plot the  horizon radius (solid blue curve), the physical  photon sphere radius (solid red curve) and the unphysical  photon sphere radius (dashed red curve) as a function of $q$.
There are different behaviors of circular photon orbits with respect to $q$ in the EMD BH spacetime which correspond to three kinds of effective potential as follows.
By considering $0<\alpha <1$ in the left panel, our result demonstrates that the radius of photon orbit decreases as $q$ is larger. 
When $q$ approaches the extremal limit, $r^+_\text{ph}$ in Eq.~\eqref{rph} converges to $2$.
In the middle panel, we fix $\alpha =1$, and the radius of circular photon orbit decreases as $q$ is larger in the same way as in the first case.
However, we find that three parameters are degenerate, namely $r^+_\text{ph}=r^-_\text{ph}=r_+=2$ at the extremal limit. 
In this case, the function $R(r_\text{ph})=0$, and hence the condition in Eq.~\eqref{first condition} does not satisfied. 
Lastly, the right panel shows the behavior of photon sphere radius as a function of $q$ with $\alpha >1$, where $r^-_\text{ph}$ becomes the physical radius of circular photon orbit instead of $r^+_\text{ph}$.
Contrary to the first two cases, the result shows that $r^-_\text{ph}$ increases as $q$ increases.
For extremal BH, we find that $r^-_\text{ph}$ equal to the horizon of extremal BH, i.e., $r^-_\text{ph}=1+\alpha^2$.
Therefore, the denominator in Eq.~\eqref{first condition} becomes vanished, as in the $\alpha =1$ case.
However, these two characteristics are different, as we will discuss these interesting issues later.

To obtain the critical impact parameter $b_c$, we solve the second condition in Eq.~\eqref{condition} as
\begin{eqnarray}
b^{\pm}_c &=& \frac{R(r^{\pm}_\text{ph})}{\sqrt{g(r^{\pm}_\text{ph})}} = \frac{r^{\pm}_\text{ph}}{\sqrt{\left( 1- \frac{r_+}{r^{\pm}_\text{ph}} \right)\left( 1- \frac{r_-}{r^{\pm}_\text{ph}} \right)^{\frac{1-3\alpha^2}{1+\alpha^2}}}}. \label{bc}
\end{eqnarray}
The solutions $b^+_c$ and $b^-_c$ correspond to critical impact parameter with $0<\alpha <1$ and $\alpha >1$, respectively.
For GMGHS BH, we obtain
\begin{eqnarray}
b_c=r_\text{ph}\sqrt{\frac{r_\text{ph}-r_-}{r_\text{ph}-r_+}},
\end{eqnarray}
where $r_\text{ph}$ is the positive real root in Eq.~\eqref{rph for alpha = 1}.
We show some results of the  horizon radii $r_\pm$, physical  photon sphere radius $r_\text{ph}$,  critical impact parameter $b_c$ and  innermost stable circular orbit $r_\text{ISCO}$ for different values of $q$ with $\alpha =0.1, 1$ and $\sqrt{3}$ in Tables \ref{table: alpha 0.1}, \ref{table: alpha 1} and \ref{table: alpha root 3}, respectively.
Further details regarding the calculation of $r_\text{ISCO}$ for dilatonic BH with parameters $q$ and dilaton coupling $\alpha$ can be found in Appendix~\ref{App A}.
Intriguingly, there is no real solution for $r$ satisfying Eq.~\eqref{risco}. This indicates that the ISCO does not exist in the case of extremal BHs for $\alpha >1$.
\begin{table}[!htbp]
\centering
\caption{The values of $r_+$, $r_-$, $r_\text{ph}$, $b_c$ and $r_\text{ISCO}$ of the EMD BH with different values of $q$ and fix $\alpha =0.1$} 
\begin{tabular}{c c c c c c c}
\hline
\multicolumn{7}{c}{$\alpha = 0.1$} \\
\cline{1-7}
\hspace{.3 cm}$q$\hspace{.3 cm} &\hspace{.3 cm}$q=0.1$\hspace{.3 cm} &\hspace{.3 cm} $q=0.3$ \hspace{.3 cm} 
&\hspace{.3 cm} $q=0.5$ \hspace{.3 cm}
&\hspace{.3 cm} $q=0.7$ \hspace{.3 cm}
&\hspace{.3 cm} $q=0.9$ \hspace{.3 cm}
&\hspace{.3 cm} $q_\text{ext}=1.005$ \hspace{.3 cm} \\
\hline
\hspace{.3 cm}$r_+$\hspace{.3 cm}   & $1.9950$  & $1.9544$ & $1.8675$ & $1.7176$ & $1.4451$ & $1.0087$ \\
\hspace{.3 cm}$r_-$\hspace{.3 cm}   & $0.0051$  & $0.0465$ & $0.1352$ & $0.2881$ & $0.5661$ & $1.0114$ \\
\hspace{.3 cm}$r_\text{ph}$\hspace{.3 cm}  & $2.9934$  & $2.9392$ & $2.8243$ & $2.6303$ & $2.3018$ & 1.9999 \\
\hspace{.3 cm}$b_c$\hspace{.3 cm}  & $5.1875$  & $5.1168$ & $4.9679$ & $4.7209$ & $4.3207$ & $3.9845$ \\
\hspace{.3 cm}$r_{\text{isco}}$\hspace{.3 cm}  & $5.9850$  & $5.8326$ & $5.6081$ & $5.1887$ & $4.5221$ & $3.9865$ \\
\hline
\end{tabular}
\label{table: alpha 0.1}
\end{table}

\begin{table}[!htbp]
\centering
\caption{The values of  $r_+$, $r_-$, $r_\text{ph}$, $b_c$ and $r_\text{ISCO}$ of the EMD BH with different values of $q$ and fix $\alpha =1$ (GMGHS BH).}
\begin{tabular}{c c c c c c c}
\hline
\multicolumn{7}{c}{$\alpha = 1$} \\
\cline{1-7}
\hspace{.3 cm}$q$\hspace{.3 cm} &\hspace{.3 cm}$q=0.5$\hspace{.3 cm} &\hspace{.3 cm} $q=0.9$ \hspace{.3 cm} 
&\hspace{.3 cm} $q=1.1$ \hspace{.3 cm}
&\hspace{.3 cm} $q=1.3$ \hspace{.3 cm}
&\hspace{.3 cm} $q=1.4$ \hspace{.3 cm}
&\hspace{.3 cm} $q_\text{ext}=\sqrt{2}$ \hspace{.3 cm} \\
\hline
\hspace{.3 cm}$r_+$\hspace{.3 cm}   & $2.0000$  & $2.0000$ & $2.0000$ & $2.0000$ & $2.0000$ & $2.0000$ \\
\hspace{.3 cm}$r_-$\hspace{.3 cm}   & $0.2500$  & $0.8100$ & $1.2100$ & $1.6900$ & $1.9600$ & $2.0000$ \\
\hspace{.3 cm}$r_\text{ph}$\hspace{.3 cm}  & $2.9558$  & $2.8332$ & $2.7130$ & $2.4846$ & $2.1903$ & $2.0000$ \\
\hspace{.3 cm}$b_c$\hspace{.3 cm}  & $4.9732$  & $4.4149$ & $3.9389$ & $3.1816$ & $2.4095$ & $2.0000$ \\
\hspace{.3 cm}$r_{\text{isco}}$\hspace{.3 cm}  & $5.7426$  & $5.0970$ & $4.5442$ & $3.6514$ & $2.6902$ & $2.0000$ \\
\hline
\end{tabular}
\label{table: alpha 1}
\end{table}

\begin{table}[!htbp]
\centering
\caption{The values of $r_+$, $r_-$, $r_\text{ph}$, $b_c$ and $r_\text{ISCO}$ of the EMD BH with different values of $q$ and fix $\alpha =\sqrt{3}$.} 
\begin{tabular}{c c c c c c c}
\hline
\multicolumn{7}{c}{$\alpha = \sqrt{3}$} \\
\cline{1-7}
\hspace{.3 cm}$q$\hspace{.3 cm} &\hspace{.3 cm}$q=0.5$\hspace{.3 cm} &\hspace{.3 cm} $q=0.9$ \hspace{.3 cm} 
&\hspace{.3 cm} $q=1.3$ \hspace{.3 cm}
&\hspace{.3 cm} $q=1.5$ \hspace{.3 cm}
&\hspace{.3 cm} $q=1.9$ \hspace{.3 cm}
&\hspace{.3 cm} $q_\text{ext}=2.0$ \hspace{.3 cm} \\
\hline
\hspace{.3 cm}$r_+$\hspace{.3 cm}   & $2.2247$  & $2.6186$ & $3.0928$ & $3.3452$ & $3.8671$ & $4$ \\
\hspace{.3 cm}$r_-$\hspace{.3 cm}   & $0.4495$  & $1.2373$ & $2.1857$ & $2.6904$ & $3.7341$ & $4$ \\
\hspace{.3 cm}$r_\text{ph}$\hspace{.3 cm}  & $3.1799$  & $3.4597$ & $3.7341$ & $3.8485$ & $3.9919$ & $-$ \\
\hspace{.3 cm}$b_c$\hspace{.3 cm}  & $4.9819$  & $4.5075$ & $3.7365$ & $3.2024$ & $1.4577$ & $-$ \\
\hspace{.3 cm}$r_{\text{isco}}$\hspace{.3 cm}  & $5.9724$  & $5.7945$ & $5.3665$ & $5.0446$ & $4.2157$ & $-$ \\
\hline
\end{tabular}
\label{table: alpha root 3}
\end{table}

\subsection{The GHW classification of null geodesics in the black hole spacetime}

%Since the behavior of effective potential for photons that moving in the EMD black hole geometry is different for $0<\alpha <1$, $\alpha =1$ and $\alpha >1$, so in this section we will investigate the corresponding null geodesics dynamically in these three regime of $\alpha$.
In this section, we will focus on the characterization of  the photon's dynamical motions near the EMD BH.
In general, to analyze a motion of a test particle in the gravitational field, we are interested in a particle trajectory in the $r - \phi$ plane rather than $r(\lambda)$. 
One can rewrite the radial equation of motion $r$ in terms of $\phi$ via Eqs.~\eqref{L} and \eqref{Hamiltonian} as follows
\begin{eqnarray}
\left( \frac{d r}{d \phi} \right)^2 = \left(\frac{dr/d\Tilde{\lambda}}{d\phi /d\Tilde{\lambda}}\right)^2= \frac{R^4(r)}{b^2}-R^2(r)g(r).
\end{eqnarray}
It is convenient to use the variable $\displaystyle u=\frac{1}{r}$, so the above equation takes the form
\begin{eqnarray}
\left( \frac{du}{d\phi} \right)^2 = P(u), \label{1st}
\end{eqnarray}
where 
\begin{eqnarray}
P(u)=\frac{(1 - r_- u)^{\frac{4\alpha^2}{1+\alpha^2}}}{b^2} - (1-r_+ u)(1 - r_- u)u^2.
\end{eqnarray}
By differentiating \eqref{1st} with respect to $\phi$, we have
\begin{eqnarray}
\frac{d^2 u}{d \phi^2} + u = \frac{3}{2}(r_+ + r_-)u^2 - 2r_+r_- u^3 - \frac{2\alpha^2 r_- }{b^2(1+\alpha^2)(1 - r_- u)^{\frac{1-3\alpha^2}{1+\alpha^2}}}. \label{geodesics}
\end{eqnarray}
As mentioned in Section~\ref{sec:Introduction}, astrophysical BHs are always accompanied by a substantial amount of accretion matter in motion around them.
To consider the effects of accretion matter on the optical appearance of BHs in later, it is helpful to define the total number $n$ of photon orbits as 
\begin{eqnarray}
n=\frac{\phi}{2\pi},
\end{eqnarray}
where $\phi$ denotes the total deflection angle. 
%Note that $n$ can be determined by the impact parameter $b$ of the incoming photon, namely, $n=n(b)$. 
Following \cite{Gralla:2019xty}, the range of $n$ can be divided into three categories, as follow:
\begin{enumerate}
    \item Direct emission: $\displaystyle n<\frac{3}{4}$; the light rays intersect the equatorial plane at most one time.
    \item Lensing ring: $\displaystyle \frac{3}{4}<n<\frac{5}{4}$; the light rays intersect the equatorial plane at least twice times.
    \item Photon ring: $\displaystyle n>\frac{5}{4}$; the light rays intersect the equatorial plane at least three times.
\end{enumerate}
Assuming the initial condition of the photon trajectory with $r\rightarrow \infty$ and $\phi \rightarrow 0$ and using  $\displaystyle \frac{du}{d\phi}=\frac{1}{b}$, one can numerically solve the equation of motion, Eq.~\eqref{geodesics}, to obtain the total deflection angle of a photon trajectory before passing through the event horizon or scatter away back to infinity.
Thus, we can find $n$ as a function of $b$.

Before reporting our results in detail, we will briefly discuss the shape of the $n(b)$ graphs which can describe different behaviors of photon trajectories for different values of $b$. 
The number $n=0$ if $b=0$ since photons travel radially towards the BH in a straight line.
With nonzero impact parameter $b < b_c$, the incoming photons approach the BH along a curved path before falling into the hole, so the function $n$ will increase as $b$ is larger.  Remarkably, it turns out to diverge at the critical impact parameter $b_c$, resulting from that photons can revolve an infinite number of times around the BH. 
On the other hand, when $b>b_c$, the function $n$ will decrease as $b$ increases since the photon is far away from the BH. 
Thus, it has a small bending by the gravitational field and later it will scatter back to the distant asymptotic region.

In the BH spacetime in EMD gravity, there are three different kinds of effective potential depending on $\alpha$, i.e., $0<\alpha <1$, $\alpha =1$ and $\alpha >1$.
We will explore the direct emission, lensing ring, and photon ring regions in the dilatonic BH spacetime for different values of $q$ at fixed values of coupling constant $\alpha =0.1$, $\alpha =1$ and $\alpha =\sqrt{3}$ , as representatives of each individual range of $\alpha$.

\begin{figure}[H]
    \centering
    \subfigure[]{\includegraphics[width = 4.5cm]{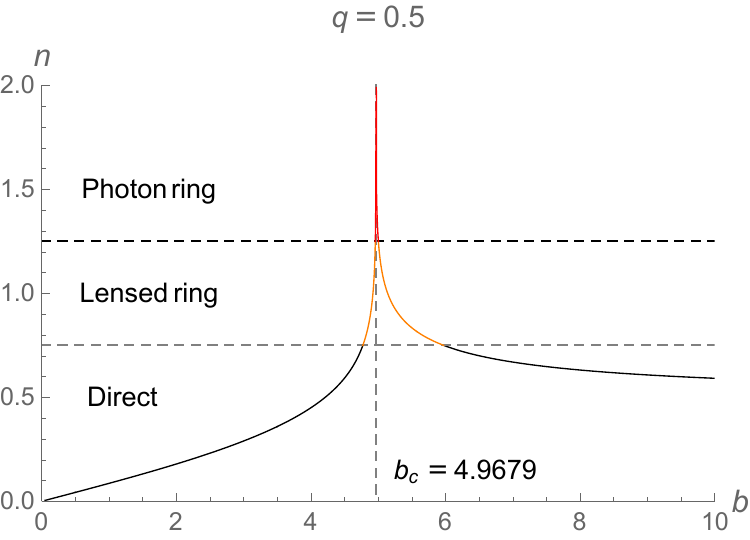}}
     \hspace{1 cm}
    \subfigure[]{\includegraphics[width = 4.5cm]{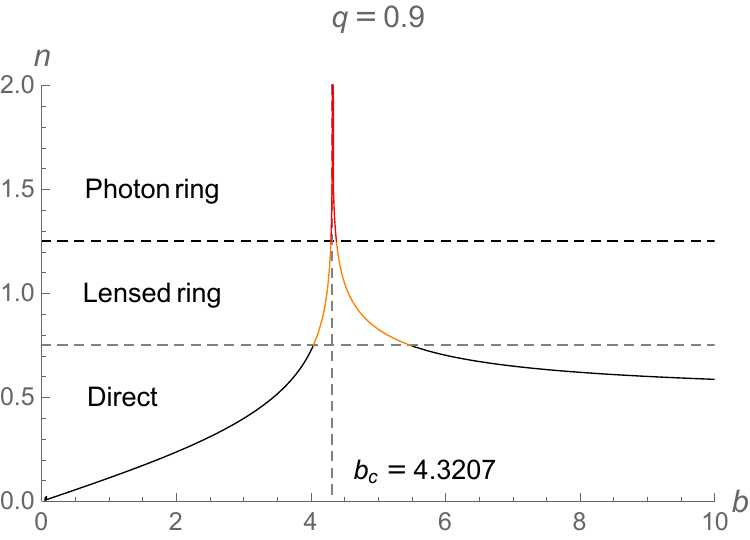}}
    \hspace{1 cm}
    \subfigure[]{\includegraphics[width = 4.5cm]{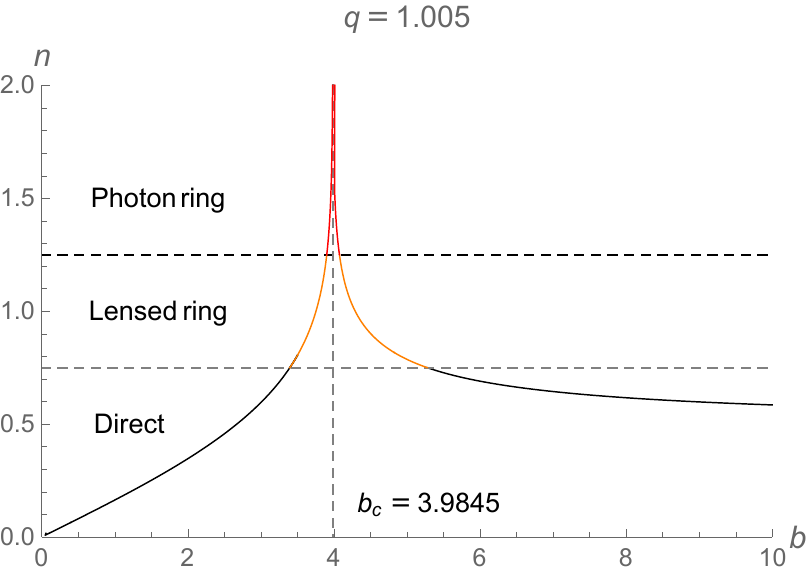}} \\
    %\vskip 0.3 cm
    \subfigure[]{\includegraphics[width = 4.5cm]{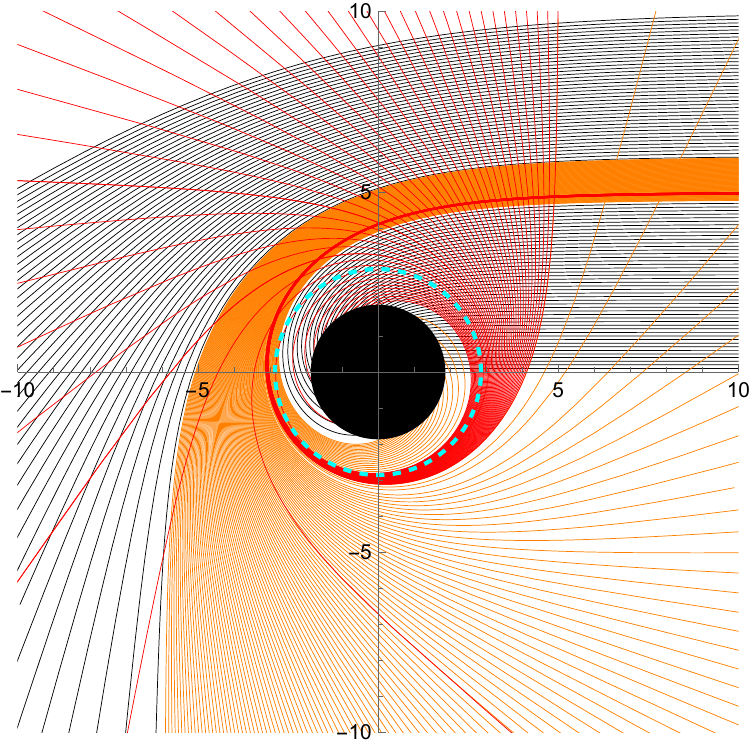}}
    \hspace{1 cm}
    \subfigure[]{\includegraphics[width = 4.5cm]{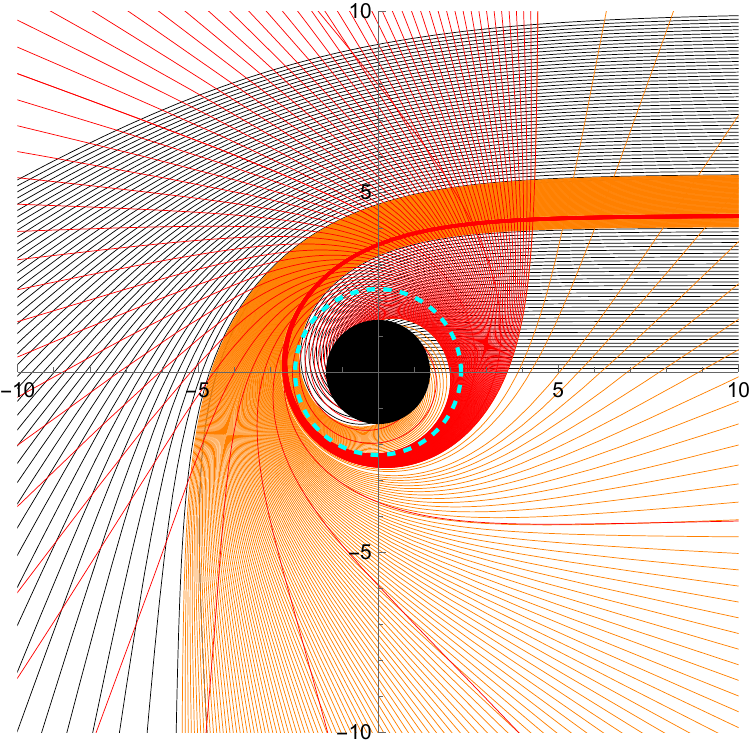}}
    \hspace{1 cm}
    \subfigure[]{\includegraphics[width = 4.5cm]{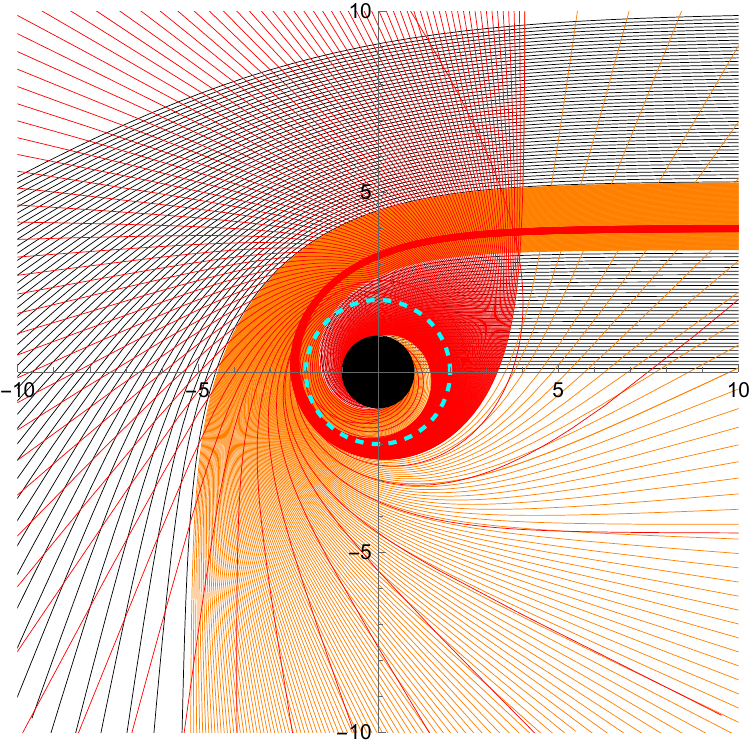}}
    \caption{The plots of the number of photon orbits $n$ versus the  impact parameter $b$ at fixed $\alpha =0.1$ for three different values of $q=0.5$ (a), $0.9$ (b) and $1.005$ (extremal BH) (c), are shown.  The photon trajectories around BH with direct emission (black), lensing ring (orange) and photon ring (red) are also  shown at fixed $\alpha =0.1$ for three different values of $q=0.5$ (d), $0.9$ (e) and $1.005$ (extremal BH) (f). }
    \label{fig: n vs b alpha=0.1}
\end{figure}

For $\alpha =0.1$, the graphs of $n$  versus $b$ with three different values of $q$ as $0.5$, $0.9$ and $1.005$ (extremal BH) are plotted, as shown in Fig.~\ref{fig: n vs b alpha=0.1} (a), (b) and (c), respectively. 
The direct emission, lensing ring and photon ring regions are represented in black, orange and red curves, respectively.
The intervals of $b$ corresponding to these regions are listed in Table \ref{tab: alpha 0.1}.
The results indicate that with increasing $q$, the position of the peak in the graph $n(b)$ will shift to the left, this means that the critical impact parameter and the radius of unstable photon orbit decreases as $q$ is larger, for numerical values see in Table \ref{table: alpha 0.1}.
According to Table \ref{tab: alpha 0.1}, the widths of the lensing ring and photon ring increase as the values of $q$ increases. 
In other words, the brightness regions become broader as $q$ becomes larger.
The photon trajectories near dilatonic BH in Euclidean polar coordinates $(r, \phi )$ are shown in Fig.~\ref{fig: n vs b alpha=0.1}, namely, (d), (e) and (f), with corresponding to $n(b)$ profile.
In these plots, the spacing in $b$ is $0.1$, $0.01$ and $0.001$ in the direct, lensing and photon ring bands, respectively. 
In the figure, the region inside the event horizon of the black hole appears as a black disk. It is surrounded by a dashed cyan curve which represents the circular photon orbit.

\begin{figure}[H]
    \centering
    \subfigure[]{\includegraphics[width = 4.5cm]{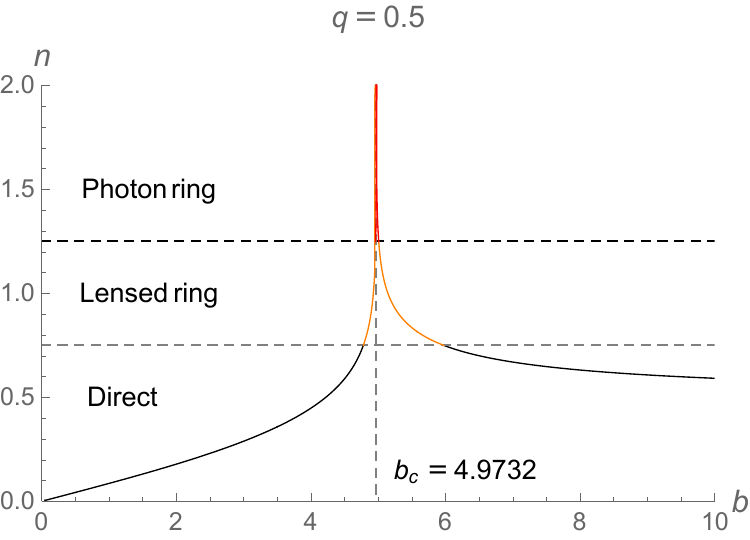} 
    \hspace{1 cm}}
    \subfigure[]{\includegraphics[width = 4.5cm]{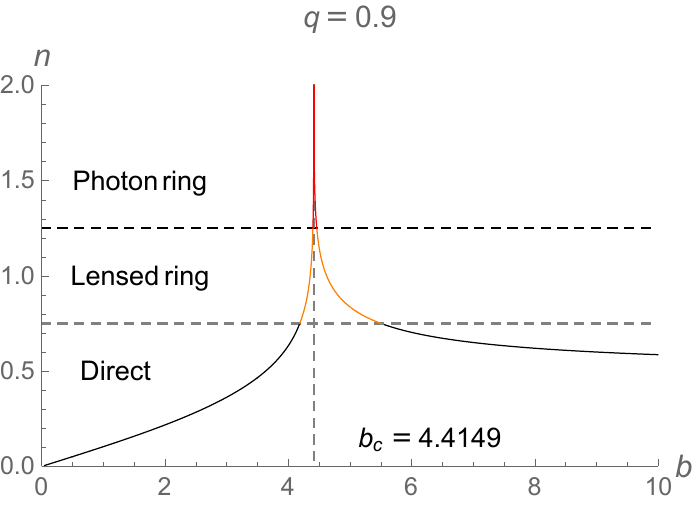}}
    \hspace{1 cm}
    \subfigure[]{\includegraphics[width = 4.5cm]{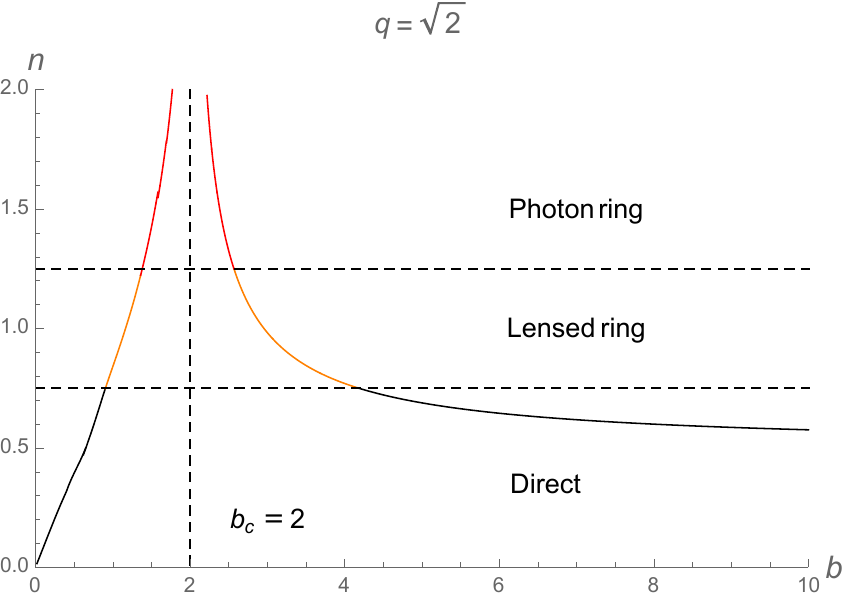}}\\
    %\vskip 0.3 cm
    \subfigure[]{\includegraphics[width = 4.5cm]{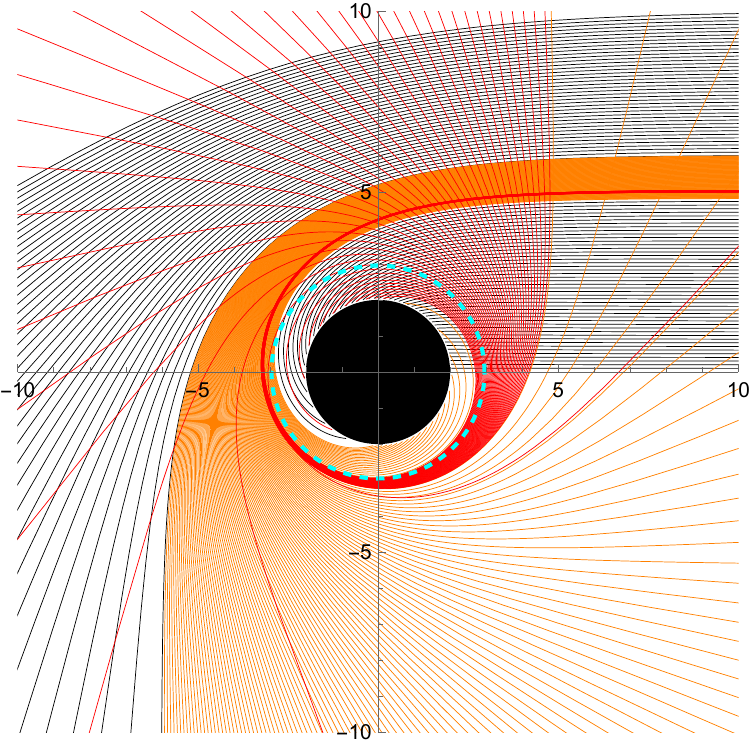}}
    \hspace{1 cm}
    \subfigure[]{\includegraphics[width = 4.5cm]{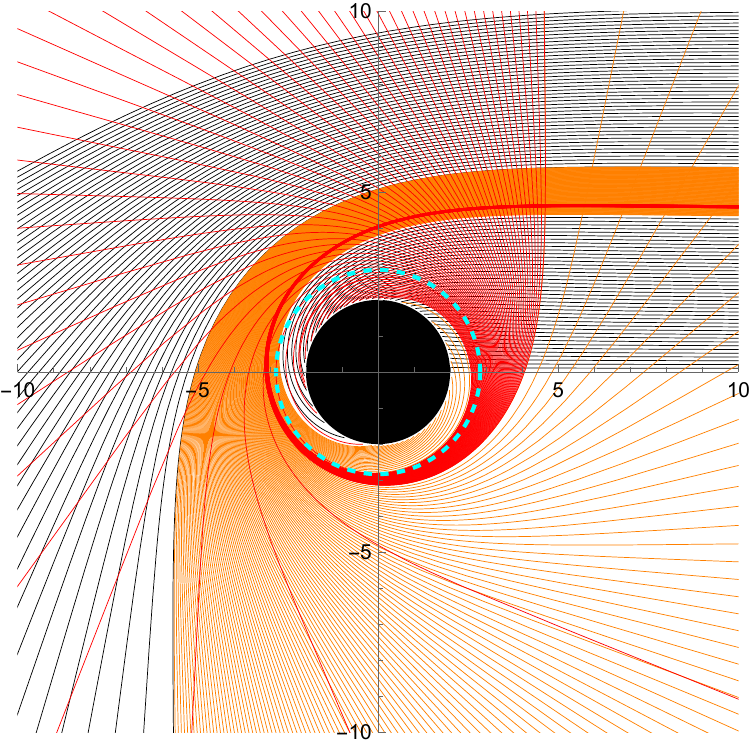}}
    \hspace{1 cm}
    \subfigure[]{\includegraphics[width = 4.5cm]{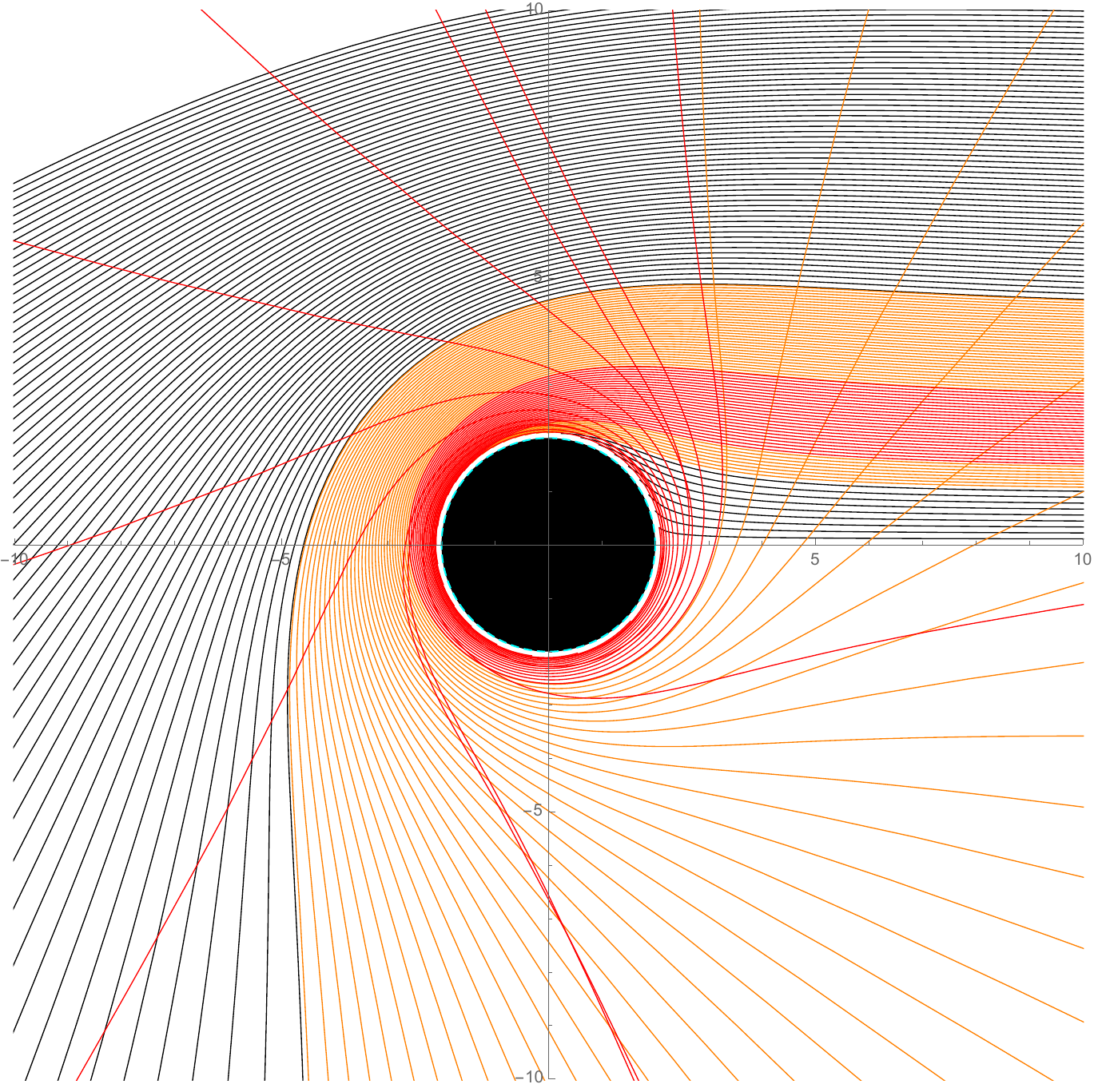}}
    \caption{The plots of the number of photon orbits $n$ versus the impact parameter $b$ at fixed $\alpha =1$ for three different values of $q=0.5$ (a), $0.9$ (b) and $\sqrt{2}$ (extremal BH) (c), are shown.  The photon trajectories around BH with direct emission (black), lensing ring (orange) and photon ring (red) are also shown at fixed $\alpha =1$ for three different values of $q=0.5$ (d), $0.9$ (e) and $\sqrt{2}$ (extremal BH) (f).}
    %\label{fig: n vs b alpha=0.1}
    \label{fig: n vs b alpha=1}
\end{figure}

For $\alpha =1$ or GMGHS BH, we find that the position of peak in the graph $n$ versus $b$ shifts to the left and the width of lensing ring and photon ring are enlarged with the increased value of $q$.
The graphs $n(b)$ and its corresponding photon trajectories with varying $q=0.5$, $0.9$ and $\sqrt{2}$ (extremal BH) are presented in (a), (b) and (c) in Fig.~\ref{fig: n vs b alpha=1}, respectively.
For trajectories of photons near the extremal BH in figure (f), the spacing in $b$ is 0.1 for direct and 0.05 for both lensing and photon rings regions.
In Table \ref{tab: alpha=1}, we list the regions of $b$ of the direct emission, lensing ring and photon ring for different values of $q$.

As mentioned in the previous subsection that the peak of effective potential outside the GMGHS BH disappears when $q$ approaches the extremal limit.
Consequently, the two conditions of circular photon orbit in Eq.~\eqref{condition} cannot be satisfied.
However, the graph of $n(b)$ with $q_\text{ext}=\sqrt{2}$ shows that the incoming photon with $1.3609<b<2.5683$ will intersect the equatorial plane at least three times and $n$ diverges at $b=2$.
According to the GHW classification, it makes sense to argue that this range of $b$ represents photon ring region with the critical  $b_c=2$, even though Eq.~\eqref{condition} is not satisfied.

\begin{figure}[H]
    \centering
    \subfigure[]{\includegraphics[width = 4.5cm]{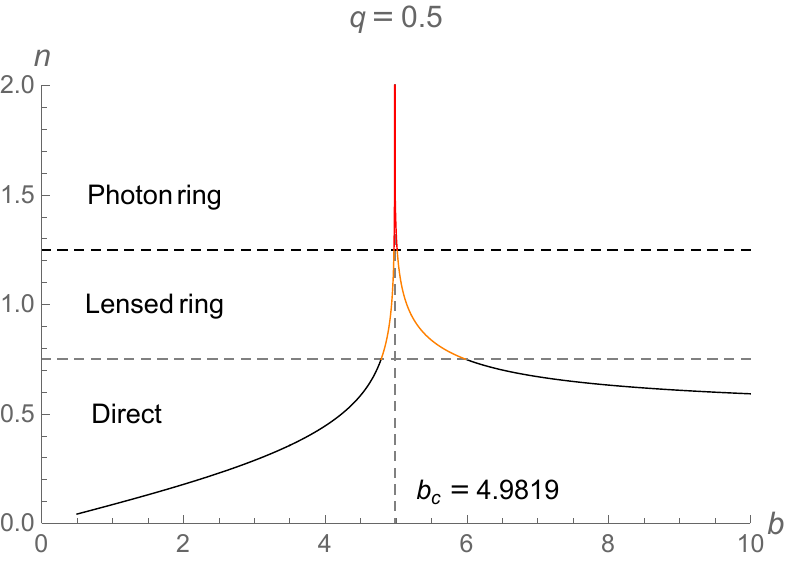}} 
    \hspace{1 cm}
    \subfigure[]{\includegraphics[width = 4.5cm]{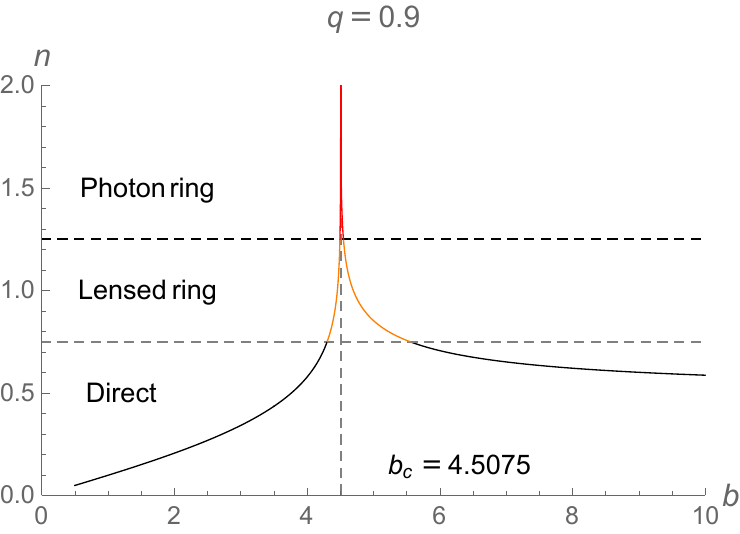}}
    \hspace{1 cm}
    \subfigure[]{\includegraphics[width = 4.5cm]{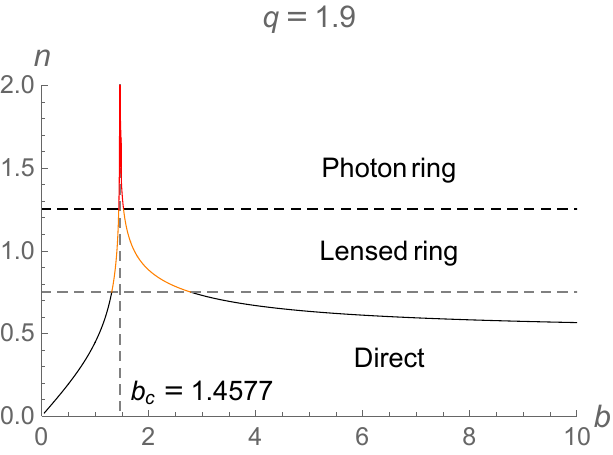}}\\
    %\vskip 1cm
    \subfigure[]{\includegraphics[width = 4.5cm]{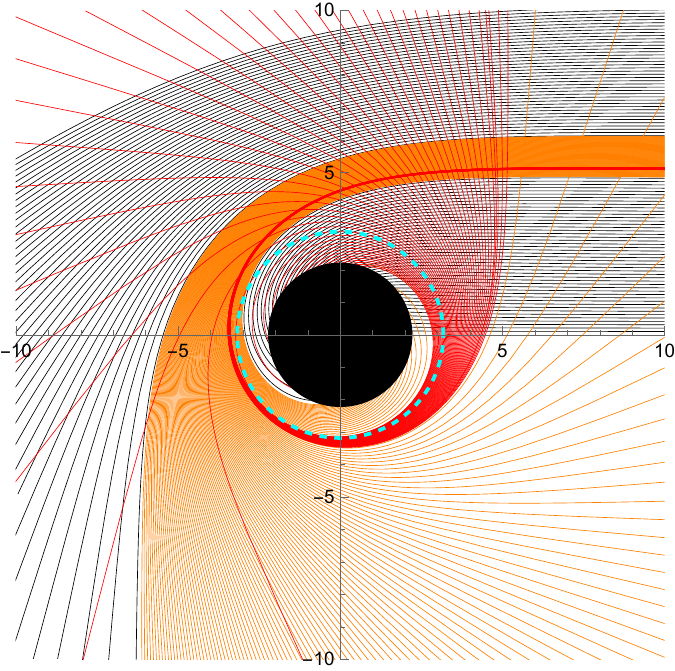}}
    \hspace{1 cm}
    \subfigure[]{\includegraphics[width = 4.5cm]{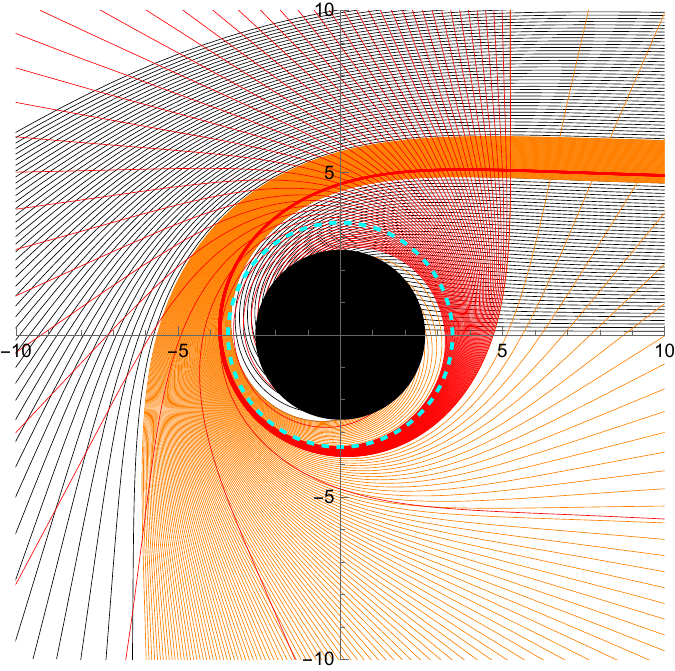}}
    \hspace{1 cm}
    \subfigure[]{\includegraphics[width = 4.5cm]{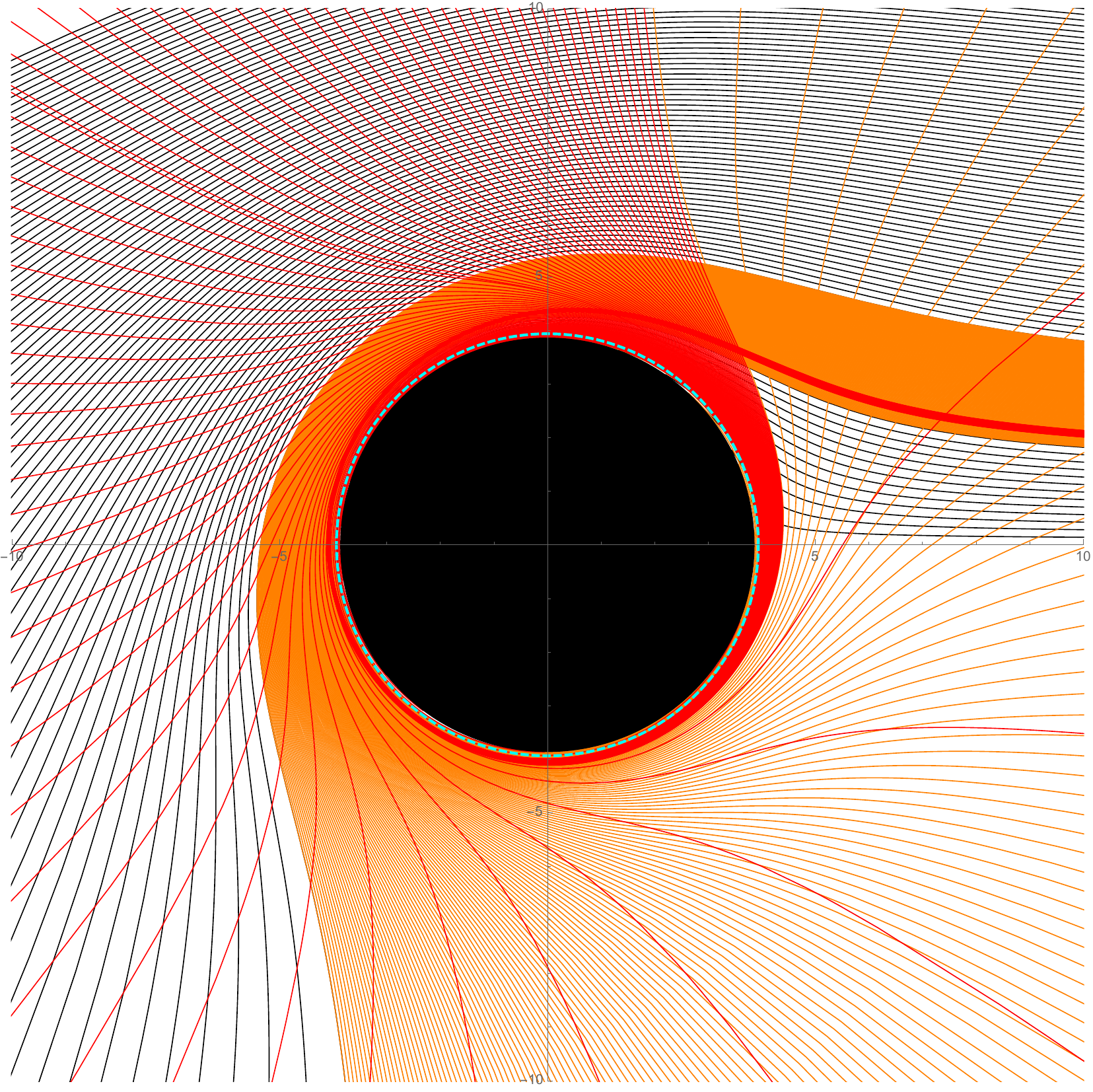}}
    \caption{The plots of the number of photon orbits $n$ versus the  impact parameter $b$ at fixed $\alpha =\sqrt{3}$ for three different values of $q=0.5$ (a), $0.9$ (b) and $1.9$ (c), are shown.  The photon trajectories around BH with direct emission (black), lensing ring (orange) and photon ring (red) are also shown at fixed $\alpha =\sqrt{3}$ for three different values of $q=0.5$ (d), $0.9$ (e) and $1.9$ (f). }
    \label{fig: n vs b alpha=root3}
\end{figure}

For $\alpha =\sqrt{3}$ with $q=0.5, 0.9$ and $1.9$ (near extremal limit), Table \ref{table: alpha root 3} shows that $r_\text{ph}$ increases and $b_c$ decreases when $q$ becomes larger.  
Thus, the peak of graph $n$ versus $b$ will shift to the left while the radius of circular dashed cyan curve which represents $r_\text{ph}$ will increase with increasing $q$, as shown in Fig.~\ref{fig: n vs b alpha=root3}.  
Surprisingly, the photon sphere radius becomes larger than the critical impact parameter when $q\geq 1.301$.
The parameter interval of $b$ for direct emission, lensing ring and photon ring are listed in Table \ref{tab: alpha=root3}.
The width of lensing ring and photon ring is enlarged with increased value of $q$ as in the $0<\alpha <1$ and $\alpha =1$ cases.  
Remarkably, for dilatonic BHs with coupling constant of the range  $\alpha >1$ with $q=q_\text{ext}$, we find that there is no photon ring region since incoming photons moving toward the BH always scatter back to the asymptotic region of spacetime due to infinitely high barrier potential just outside the horizon as shown in the right panel in Fig.~\ref{fig: Veff vs r}.
Because of this, despite $r_\text{ph}$ coincides with $r_+$ at the extremal limit, i.e.,  $r_\text{ph}=r_+ = 4$ at $q=q_\text{ext}$, as shown in  Fig.~\ref{fig: rph vs q}, the photon sphere cannot be said to exist in this case.

One can see the effect of dilaton coupling $\alpha$ on photon trajectories by considering the graphs $n$ versus $b$ with fixed values of $q=0.5$ and $0.9$ in the panels (a), (b), (d) and (e) of Fig~\ref{fig: n vs b alpha=0.1}, \ref{fig: n vs b alpha=1} and \ref{fig: n vs b alpha=root3}, where $\alpha$ has the values of $0.1, 1$ and $\sqrt{3}$, respectively. 
As $\alpha$ increases, we find that both $b_c$ and $r_\text{ph}$ becomes larger corresponding to that the peak of $n(b)$ shifts to the right, for numerical values see Tables \ref{table: alpha 0.1}, \ref{table: alpha 1} and \ref{table: alpha root 3}. 
In addition, the thickness of lensing and photon rings are smaller at larger coupling constant $\alpha$ and fixed $q$.

\begin{comment}
\begin{figure}[H]
    \centering
    \includegraphics[width = 7.cm]{FigGamma_q.pdf}
    \caption{Behavior of the angle-dependent Lyapunov exponent $\gamma$ as a function of $q$ with different values of $\alpha$.} 
    \label{fig: gamma vs q}
\end{figure}
\end{comment}
To elucidate how a larger value of $q$ ($\alpha$) with fixed $\alpha$ ($q$) results in a wider (narrower) range of photon ring image, we linearize Eq.~\eqref{Hamiltonian} around $r=r_\text{ph}$.
Thus, we can obtain the equation of motion of $\delta r=r-r_\text{ph}$ in the form
\begin{eqnarray}
    \frac{d~\delta r}{d\Tilde{\lambda}}&=&\sqrt{\frac{E^2}{L^2}-V_\text{eff}(r_\text{ph}+\delta r)} \nonumber \\
    &\approx &\sqrt{-\frac{1}{2}V_\text{eff}^{\prime \prime}(r_\text{ph})}~\delta r, \label{delta_r}
\end{eqnarray}
where we have used $\displaystyle \frac{E^2}{L^2}-V_\text{eff}(r_\text{ph})=V_\text{eff}^\prime (r_\text{ph})=0$. 
Applying Eq.~\eqref{L} into Eq.~\eqref{delta_r} leads to the equation of motion of $\delta r$ with respect to $\phi$ in the form 
\begin{eqnarray}
    \pi \frac{d~\delta r}{d\phi}=\pi R^2(r_\text{ph})\sqrt{-\frac{1}{2}V_\text{eff}^{\prime \prime}(r_\text{ph})}~\delta r.
\end{eqnarray}
Defining the \textit{angular Lyapunov exponent} $\gamma$~\cite{Johnson:2019ljv, Broderick:2023jfl, 2003GReGr..35.1909T, Gralla:2019drh}
\begin{eqnarray}
    \gamma \equiv \pi R^2(r_\text{ph})\sqrt{-\frac{1}{2}V_\text{eff}^{\prime \prime}(r_\text{ph})},
\end{eqnarray}
the solution of the above equation can be written as
\begin{eqnarray}
    \delta r(\phi)= \delta r_0e^{\gamma \phi /\pi}, \label{perturb}
\end{eqnarray}
where $\delta r_0$ is the initial deviation of a geodesic from the circular critical orbit. The deviation in radius of the geodesic grows by a factor of $e^\gamma$ for every half-orbit around the BH, causing photons to ultimately move either towards the BH or towards the observer at infinity. Remarkably, a photon starting with the geodesic closer to $r_\text{ph}$ can complete more half-orbits $m$, denoted as $\displaystyle \frac{\phi}{\pi}$, around the BH before reaching the observer at infinity. This results in the thinner subring of order $m$th in the photon subring structure~\cite{Johnson:2019ljv}. %the observed photon ring image arises from photons moving nearly photon sphere radius.
%In this way, the edge of shadow is actually not a single photon ring.
%Instead, the BH image contains an infinite sequence of discrete photon rings which we called them as \textit{subrings} .
%The $m$th photon ring corresponds to photon orbiting $m/2$ loops before reaching the observer's screen

\begin{figure}[H]
    \centering
    \includegraphics[width = 7.cm]{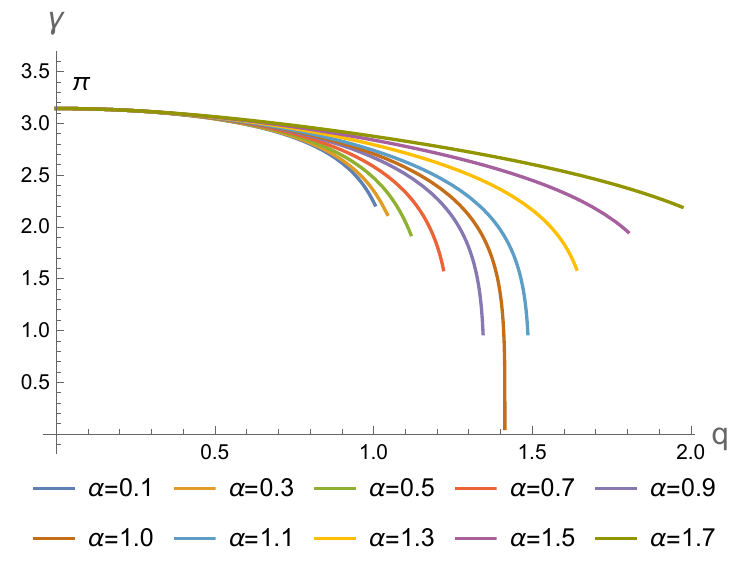}
    \caption{Behavior of the angle-dependent Lyapunov exponent $\gamma$ as a function of $q$ with different values of $\alpha$.} 
    \label{fig: gamma vs q}
\end{figure}

To obtain the width of photon subring on the observer's screen, we
linearize Eq.~\eqref{bc} about $r_\text{ph}$ to approximate $b$ of the geodesic that differ from the bound geodesic, or critical curve, by $\delta r_0$ as following
\begin{eqnarray}
    b^2 &=& \frac{1}{V_\text{eff}(r_\text{ph}+\delta r_0)} \nonumber \\
    &\approx& b_c^2+\frac{1}{2}\left[ \frac{R^2(r)}{g(r)}\right]_{r_\text{ph}}^{\prime \prime}\delta r_0^2. \label{app b}
\end{eqnarray}
Note that the Taylor series expansion of this radial equation of motion is still well defined until $\delta r$ reaches $\delta r_\text{max} \approx 1$ (or $\delta r_\text{max} \approx M$ since we measure the length in unit of mass), at which geodesics can become well separated from the photon sphere and go to infinity.
The number of half-orbits $m$ before the breakdown of perturbative expansion, as mentioned above, can be derived from Eq.~\eqref{perturb} as 
\begin{eqnarray}
    m \approx \frac{2}{\gamma}\ln \left( \frac{\delta r_\text{max}}{\delta r_0} \right). \label{m}
\end{eqnarray}
Recall that the radius of the BH shadow $R_s=b_c$, the $m$th-order photon subring denoted as $R_m$ can be identified with the impact parameter $b$ associated with $\delta r_\text{max}$. 
Using this identification with Eq.~\eqref{app b} and Eq.~\eqref{m}, we obtain the width of photon subring
\begin{eqnarray}
    R_m-R_s \approx \frac{1}{4}\frac{g(r_\text{ph})}{R^2(r_\text{ph})}\left[ \frac{R^2(r)}{g(r)}\right]_{r_\text{ph}}^{\prime \prime}\delta r_\text{max}^2e^{-m\gamma}, \label{width}
\end{eqnarray}
which indicates that the width is narrower by a factor $e^{-m\gamma}$.
As shown in \cite{Johnson:2019ljv, Bisnovatyi-Kogan:2022ujt}, the subring flux ratio between two consecutive orders does not depend on the source brightness profile but is always suppressed by the factor
\begin{eqnarray}
    \frac{F_{m+1}}{F_m}\approx e^{-\gamma}. \label{flux}
\end{eqnarray}
One can see that the width and ratio of subring flux are determined by the exponent $\gamma$ as shown in Eqs.~\eqref{width} and \eqref{flux}, respectively. 
To explain the effects of $q$ and $\alpha$ on the black hole's images,
we display the behavior of $\gamma$ with different values of parameter $q$ and $\alpha$ in Fig~\ref{fig: gamma vs q}.
In the limit of Schwarzschild BH where $q=0$ and $\alpha =0$, the angular Lyapunov exponent has a maximum value, i.e., $\gamma =\pi$. 
When $\alpha$ is fixed, it is evident that $\gamma$ decreases as $q$ increases, while $\gamma$ is larger as $\alpha$ increases with a fixed $q$.
The width and relative flux of subrings tend to be larger for smaller $\gamma$, which depends on the spacetime with different $q$ and $\alpha$. 
In other words, the width of the photon subring is wider when $q$ is larger and constant $\alpha$ ($\gamma$ decreases), while the width is narrower when $\alpha$ is larger and constant $q$ ($\gamma$ increases).

\begin{table}[!htbp]
\centering
\caption{The region of direct emission, lensing ring and photon ring for different values of $q$ with fixed coupling constant $\alpha =0.1$.}
\begin{tabular}{l c c c}
\hline
\multicolumn{4}{c}{$\alpha =0.1$} \\
\cline{1-4}
Charge/Mass           & $q=0.5$ & $q=0.9$         & $q_\text{ext}=1.005$ \\
\hline
Direct emission  & $b<4.7730$  & $b<4.0371$      & $b<3.3957$   \\
                 & $b>5.9745$            & $b>5.4824$       & $b>5.2773$   \\
Lensing ring     & \hspace{0.5cm}$4.7730<b<4.9580$\hspace{0.5cm}      & \hspace{0.5cm}$4.0371<b<4.2979$\hspace{0.5cm} & \hspace{0.5cm}$3.3957<b<3.9019$\hspace{0.5cm} \\
                 & $5.0040<b<5.9745$      & $4.3806<b<5.4824$ & $4.0759<b<5.27773$ \\
Photon ring      & $4.9580<b<5.0040$      & $4.2979<b<4.3806$ & $3.9019<b<4.0759$ \\
\hline
\end{tabular}
\label{tab: alpha 0.1}
\end{table}

\begin{table}[!htbp]
\centering
\caption{The region of direct emission, lensing ring and photon ring for different values of $q$ with fixed coupling constant $\alpha =1$.}
\begin{tabular}{l c c c}
\hline
\multicolumn{4}{c}{$\alpha =1$} \\
\cline{1-4}
Charge/Mass & $q=0.5$        & $q=0.9$    & $q_\text{ext}=\sqrt{2}$ \\
\hline
Direct emission  & $b<4.7801$   & $b<4.1863$     & $b<0.9041$   \\
                 & $b>5.9763$            &   $b>5.5086$     & $b>4.1827$   \\
Lensing ring     & \hspace{0.5cm}$4.7801<b<4.9634$\hspace{0.5cm}      & \hspace{0.5cm}$4.1863<b<4.3999$\hspace{0.5cm}  & \hspace{0.5cm}$0.9041<b<1.3609$\hspace{0.5cm} \\
                 & $5.0088<b<5.9763$      & $4.4629<b<5.5086$ & $2.5683<b<4.1827$ \\
Photon ring      & $4.9634<b<5.0088$      & $4.3999<b<4.4629$ & $1.3609<b<2.5683$ \\
\hline
\end{tabular}
\label{tab: alpha=1}
\end{table}

\begin{table}[!htbp]
\centering
\caption{The region of direct emission, lensing ring and photon ring for different values of $q$ with fixed coupling constant $\alpha =\sqrt{3}$.}
\begin{tabular}{l c c c}
\hline
\multicolumn{4}{c}{$\alpha =\sqrt{3}$} \\
\cline{1-4}
Charge/Mass & $q=0.5$        & $q=0.9$    & $q=1.9$ \\
\hline
Direct emission  & $b<4.7913$ & $b<4.3033$  & $b<1.3098$ \\
                 & $b>5.9799$ & $b>5.5514$  & $b>2.7838$  \\
Lensing ring     & \hspace{0.5cm}$4.7913<b<4.9724$\hspace{0.5cm}  & \hspace{0.5cm}$4.3033<b<4.4956$\hspace{0.5cm}  & \hspace{0.5cm}$1.3098<b<1.4409$\hspace{0.5cm} \\
                 & $5.0169<b<5.9799$ & $4.5488<b<5.5514$ & $1.5309<b<2.7838$ \\
Photon ring      & $4.9724<b<5.0169$ & $4.4956<b<4.5488$ & $1.4409<b<1.5309$ \\
\hline
\end{tabular}
\label{tab: alpha=root3}
\end{table}

\section{Black Hole Shadows in the EMD gravity} \label{section 4}
Considering the profiles of effective potential at different values of $\alpha$ and $q$ as shown in Fig,~\ref{fig: Veff vs r}, the trajectories of null geodesics are in general of three types, namely falling, scattering and unstable circular motion. An observer located at a far distance from the BH is able to detect only the latter two types scattered away from the BH.
Specifically, the unstable photon circular orbit, associated with the maximum of the effective potential, serves as a boundary that separates the scattering trajectories from the falling trajectories.  As a BH image, this ``critical curve" around BH appears as a photon ring enclosing the dark region, which is the so-called BH shadow.

In this section, we study the effect of $q$ and $\alpha$ on the shadow of spherical symmetric BH in the EMD gravity.
The configuration of the light source and the motion of the observer significantly influence the shape and size of the black hole's shadow \cite{Chang:2020miq, *Chang:2020lmg}. 
In our study, we position the selected light sources at a substantial distance from the black hole to ensure that incoming light rays are parallel.
We choose the static observer where only the $t$-component of four-velocity in nonzero, i.e., $U^\mu _\text{st}=(1, 0, 0, 0)$.
To identify the shadow region of BH on the image plane, it is more convenient to define the celestial coordinates $X$ and $Y$ as \cite{Vazquez:2003zm, *PhysRevD.90.024073}
\begin{eqnarray}
X &=& -\lim_{r_*\rightarrow \infty}\left( r^2_*\sin \theta_{O}\frac{d\phi}{dr}\right), \label{alpha}\\
Y &=& \lim_{r_*\rightarrow \infty}\left( r^2_*\frac{d\theta}{dr}\right), \label{beta}
\end{eqnarray}
where $r_*$ is the distance of an observer from the BH and $\theta_{O}$ is the angular coordinate of the observer representing the inclination angle between the line of sight to the observer and the normal to a celestial plane.  
Note that we use units with $G=c=1$, so the coordinates $(X, Y)$ that describe the image will have the units of black hole's mass $M$.
Restricting the study to the equatorial plane with $\displaystyle \theta_O=\frac{\pi}{2}$, the radius of the BH shadow is provided by
\begin{eqnarray}
R_s = \sqrt{X^2+Y^2}. \label{def BH shadow}
\end{eqnarray}
To calculate the celestial coordinates $X$ and $Y$, we used Eqs.~\eqref{L}, \eqref{r dot} and \eqref{theta dot} to obtain
\begin{eqnarray}
\frac{d\phi}{dr} &=& \frac{L\csc^2\theta}{R^2(r)\displaystyle \sqrt{E^2-\frac{g(r)}{R^2(r)}(L^2+\mathcal{Q})}}, \\
\frac{d\theta}{dr} &=& \frac{\sqrt{\mathcal{Q}-L^2\cot^2\theta}}{R^2(r)\displaystyle \sqrt{E^2-\frac{g(r)}{R^2(r)}(L^2+\mathcal{Q})}}.
\end{eqnarray}
Substituting the above expressions into the Eqs.~\eqref{alpha} and \eqref{beta} with $\displaystyle \theta_O=\frac{\pi}{2}$, we obtain
$X = \displaystyle -\frac{L}{E}$ and $\displaystyle Y = \frac{\sqrt{\mathcal{Q}}}{E}$.
For unstable photon orbits, the radial equation of motion in Eq.~\eqref{new r dot} becomes $\bar{V}_\text{eff}(r_\text{ph})=E^2$, then
\begin{eqnarray}
    \bar{V}_\text{eff}(r_\text{ph})\big |_{\delta =0}=\frac{g(r_\text{ph})}{R^2(r_\text{ph})}\left( L^2+\mathcal{Q} \right)&=&E^2, \nonumber \\
    \frac{g(r_\text{ph})}{R^2(r_\text{ph})}&=&\frac{E^2}{ L^2+\mathcal{Q}}.
\end{eqnarray}
By using Eqs.~\eqref{bc} and \eqref{def BH shadow}, one can show that
the apparent radius of shadow equal to the critical impact parameter of critical curve,
\begin{eqnarray}
R_s=b_c.
\end{eqnarray}
For Schwarzschild BH, the apparent shadow radius $R_s=3\sqrt{3}$ and the bound orbits of photon occur at $r_\text{ph}=3$, where we measure the shadow radius in unit of $M$ such that $\displaystyle R_s=\frac{R_s}{M}$.
Note that the solar radius is much larger than $3M_\odot\ \approx 4.5$ km, precluding the possibility of a circular photon orbit around the Sun \cite{2003gieg.book.H, *Gott:2018ocn}.
Consequently, the presence of the shadow and photon ring serves as compelling evidence for the existence of BHs and other compact objects in the universe \cite{Perlick:2021aok, *Cardoso:2019rvt}.

\begin{figure}[H]
    \centering
    \includegraphics[scale=0.5]{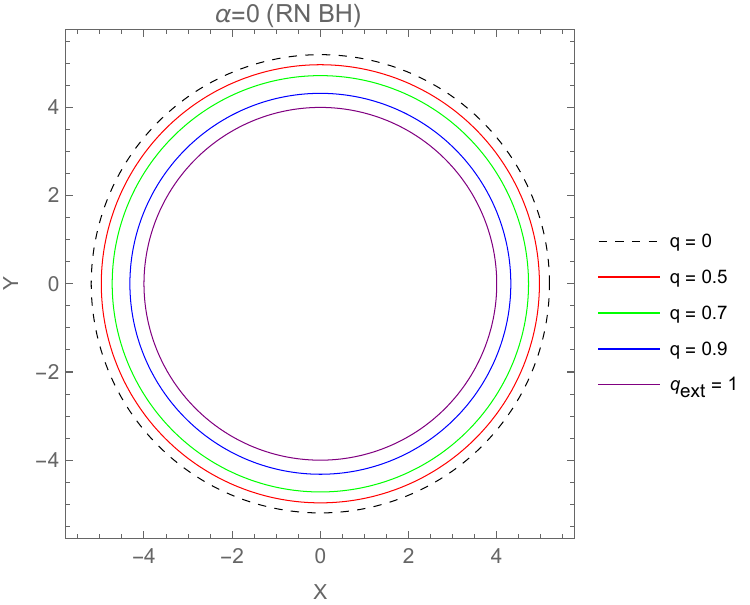}\hspace{1cm}
    \includegraphics[scale=0.53]{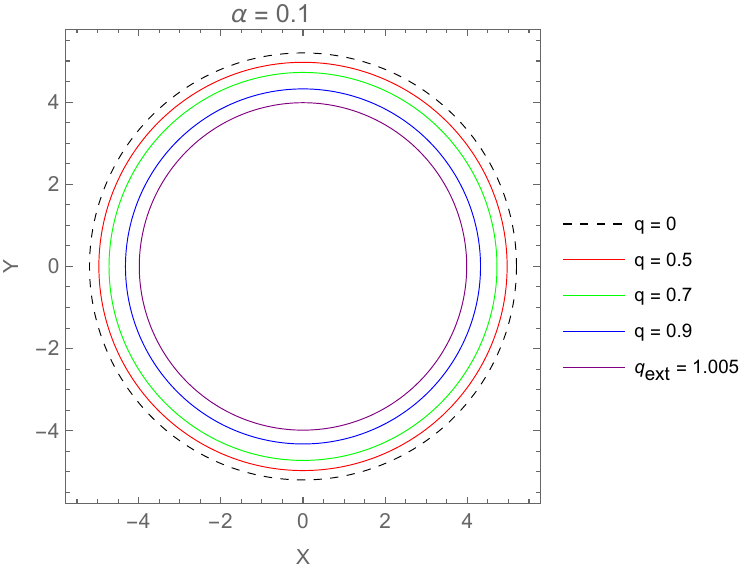} \\
    \includegraphics[scale=0.53]{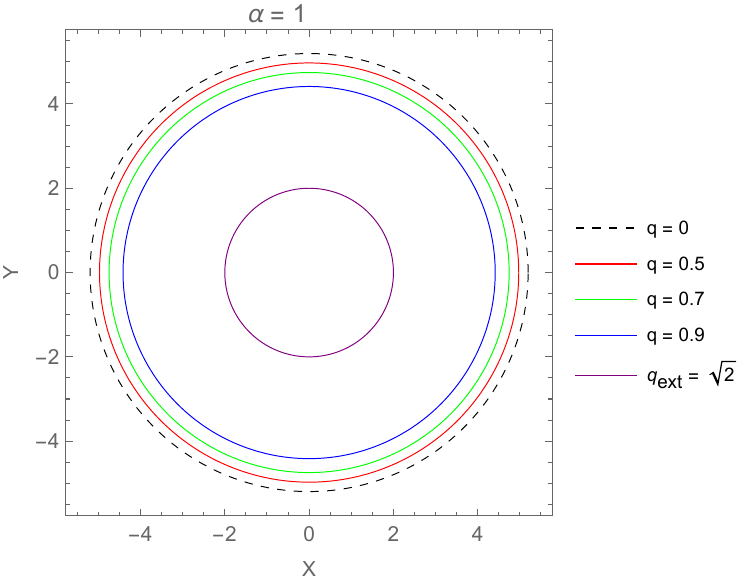}\hspace{1cm} 
    \includegraphics[scale=0.53]{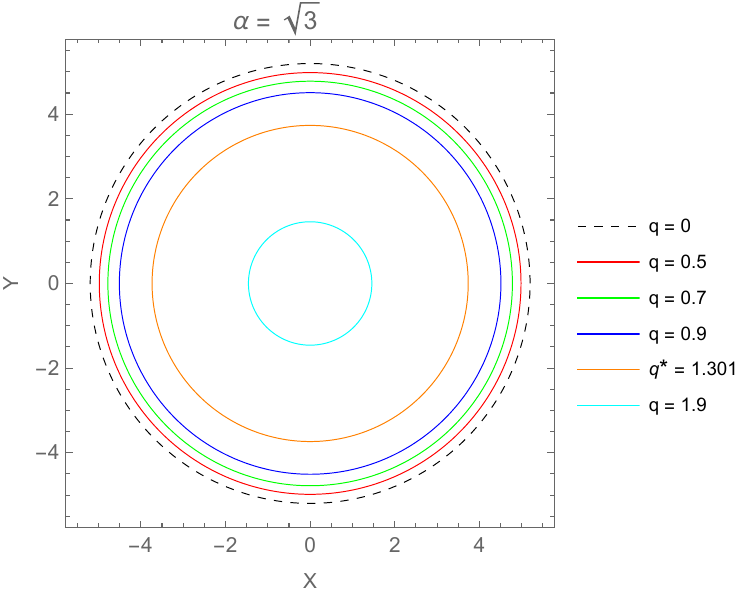}
    \caption{The shadow region of BH in the EMD gravity for diferent values of $q$ with $\alpha =0 \ \text{(RN BH)}, 0.1, 1$ and $\sqrt{3}$. In each panel, the dashed-black curve represents the boundary of shadow for Schwarzschild BH, where we use $\displaystyle X=\frac{X}{M}$ and $\displaystyle Y=\frac{Y}{M}$}. 
    \label{fig:Shadow}
\end{figure}

\begin{figure}[H]
    \centering
    \includegraphics[width=4.5cm]{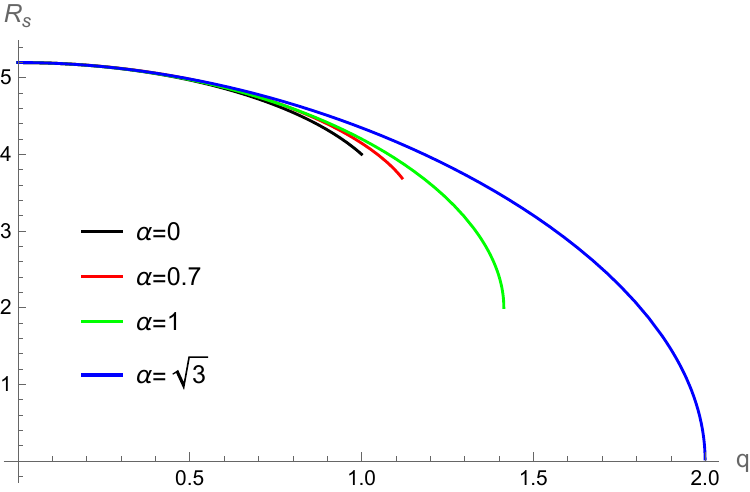}\hspace{1cm}
    \includegraphics[width=4.5cm]{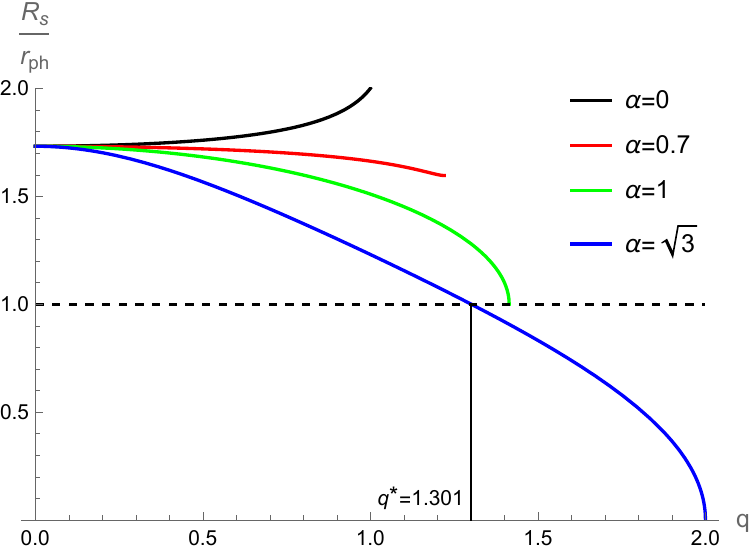}\hspace{1cm}
    \includegraphics[width=4.2cm]{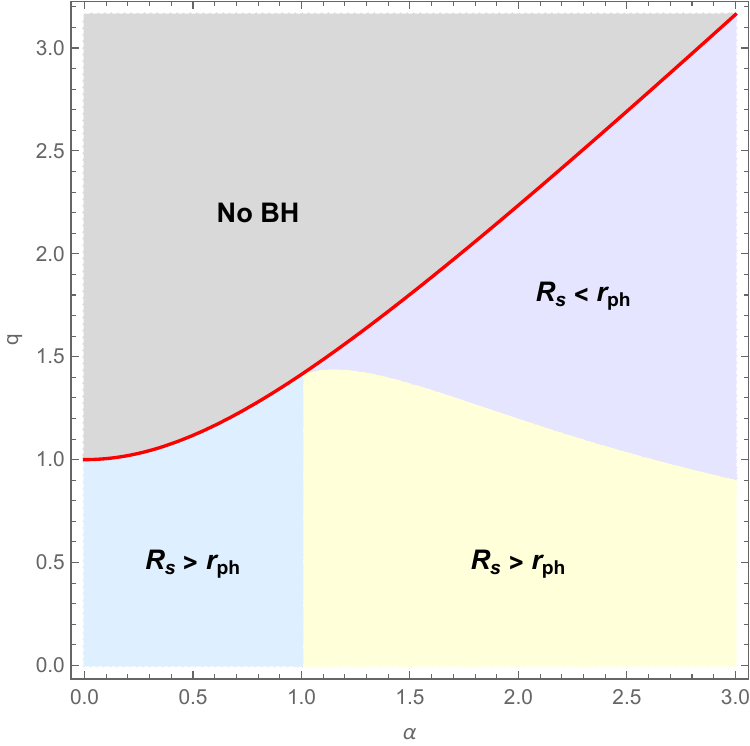}
    \caption{\textit{Left} Plot the shadow radius $R_s$ against $q$ for $\alpha =0$ (black), $0.7$ (red), $1$ (green) and $\sqrt{3}$ (blue). \textit{Middle} The plots of $R_s/r_\text{ph}$ as a function of $q$ at different values of $\alpha$, where the dashed line represents $R_s=r_\text{ph}$. 
    \textit{Right} The region plot for comparing $R_s$ and $r_\text{ph}$ in the $q-\alpha$  plane is shown. While $R_s$ is always larger than $r_\text{ph}$ for $0<\alpha <1$ (blue), it is interesting to point out that, when $\alpha >1$, there is a curve $q^*(\alpha)$ separating the $R_s>r_\text{ph}$ (yellow) and $R_s<r_\text{ph}$ (purple) regions.  In other words, at a fixed value of $\alpha$ for $\alpha>1$, $R_s>r_\text{ph}$ only at $q$ less than the maximum value which is a point on the curve, beyond which $R_s<r_\text{ph}$.   Note that the BH solution does not exist in the gray region, where the red line corresponds to extremal BHs.} 
    \label{fig: Shadow radius comparing}
\end{figure}

In Fig~\ref{fig:Shadow}, the variation in the size of dilatonic BH shadow boundary for different values of $q$ is shown graphically with $\alpha =0.1, 1$ and $\sqrt{3}$, representing three ranges of the dilaton coupling values, namely $0<\alpha <1$, $\alpha =1$ and $\alpha >1$, respectively. Moreover, we compare these cases with one of $\alpha =0$ (RN BH), which is also shown in the figure.
We find that the dark zone in the BH image is smaller as $q$ increases and it has a minimum size when $q=q_\text{ext}$ for any non-vanishing $\alpha$.
The effects of $q$ and $\alpha$ on the sizes of $R_s$ are shown in the left panel of Fig~\ref{fig: Shadow radius comparing}.
It can be observed that, for a fixed value of $q$, the increasing value of $\alpha$ results in a larger value of $R_s$.
In the middle panel in Fig~\ref{fig: Shadow radius comparing}, we compare the sizes of the shadow radius and the photon sphere radius by considering the ratio $R_s/r_\text{ph}$ as a function of $q$ for static and distant observers in asymptotically flat spacetime. 
It has been observed that $R_s/r_\text{ph}$ is larger as $q$ increases in the case of RN BH while it becomes smaller with increasing $q$ for EMD BH.
For a given value of $q$, the presence of $\alpha$ leads to a smaller value of $R_s/r_\text{ph}$. 
In the case of $\alpha =1$, when $q=\sqrt{2}$ (extremal BH), three radii, namely event horizon, shadow radius, and photon ring are degenerate. 
Interestingly, for $\alpha =\sqrt{3}$, although $R_s$ is larger than $r_\text{ph}$ for small values of $q$ as usual, the photon sphere radius is surprisingly larger as $q$ increases and these two radii thus coincide at $q=q^*=1.301$, beyond which $R_s<r_\text{ph}$. 
When $q$ saturates the extremal limit, the results indicate that $R_s\rightarrow 0$ while $r_\text{ph}$ becomes absent. 
These anomalous behaviors exhibited by photon trajectories for $\alpha > 1$ are linked with the fact that the effective potential as a function of $r$ for $\alpha>1$ has a steeper negative slope outside the event horizon as $q$ increases, as shown in Fig.~\ref{fig: Veff vs r}.
Noteworthy, one may use the ratio $R_s/r_\text{ph}$ as an indicator to distinguish the charged dilaton BH from the RN BH. 

To make the above discussion more general, we present the region plot for comparing the values of $R_s$ and $r_\text{ph}$ in the $q-\alpha$ plane in the right panel in Fig~\ref{fig: Shadow radius comparing}. 
The blue and yellow regions correspond to the cases of $R_s>r_\text{ph}$ within the ranges of $0<\alpha<1$ and $\alpha>1$, respectively. 
The purple region corresponds to the region of $R_s<r_\text{ph}$ for $\alpha >1$. 
The boundary between the yellow and purple regions can be represented by the curve $q^*(\alpha)$ at which $R_s=r_\text{ph}$.
We find that $q^*(\alpha)$ is smaller as $\alpha$ increases, and $q^*=1.301$ at $\alpha=\sqrt{3}$ as shown in the vertical line of the middle panel in Fig.~\ref{fig: Shadow radius comparing}.
The red line represents the extremal BHs where an area of shadow within the critical curve has a minimum value for $0<\alpha \leq1$ and it turns out to vanish for $\alpha >1$.
Remarkably, the BH shadow becomes absent in the latter case since the falling trajectories and the bound orbits of photons do not exist due to the infinitely high barrier of effective potential just outside the horizon.
Note that, the existence and nonexistence of photon spheres play an important role in the observation of astrophysical compact objects, for further details about these interesting issues see~\cite{Claudel:2000yi, *Cunha:2017qtt, *Cunha:2018acu, *PhysRevD.102.064039}.

It is important to emphasize our discovery that $R_s$ can be less than $r_\text{ph}$ when $\alpha >1$ and $q>q^*$ for the EMD black hole in the present work is interesting in its own right. While light bending around typical black hole spacetime exhibits the behaviors that $R_s$ is larger than $r_\text{ph}$, the photons from the critical orbits in this case seem to have remarkable trajectories due to the peculiar spacetime geometry of the EMD black hole in this range of parameters. 
One might wonder whether these results contradict the focusing theorem for null geodesic congruences. As demonstrated in Appendix~\ref{Appendix B}, considering the null geodesics moving in the spacetime of the EMD black hole with an arbitrary value of \(\alpha\), there is no violation of the focusing theorem or the null energy condition (NEC) as long as \(q\) does not exceed the extremal limit.
This indicates the EMD black hole in this range of parameters may attract more attention to the community to study further in many aspects.

\begin{comment}
\begin{figure}[H]
    \centering
    \includegraphics[width=5cm]{FigShadowRadius.pdf}
    \caption{Plot the shadow radius $R_s$ against $q$ for $\alpha =0$ (black), $0.7$ (red), $1$ (green) and $\sqrt{3}$ (blue).} 
    \label{fig:yyy}
\end{figure}
\end{comment}

\begin{figure}[H]
    \centering
    {\includegraphics[width=5cm]{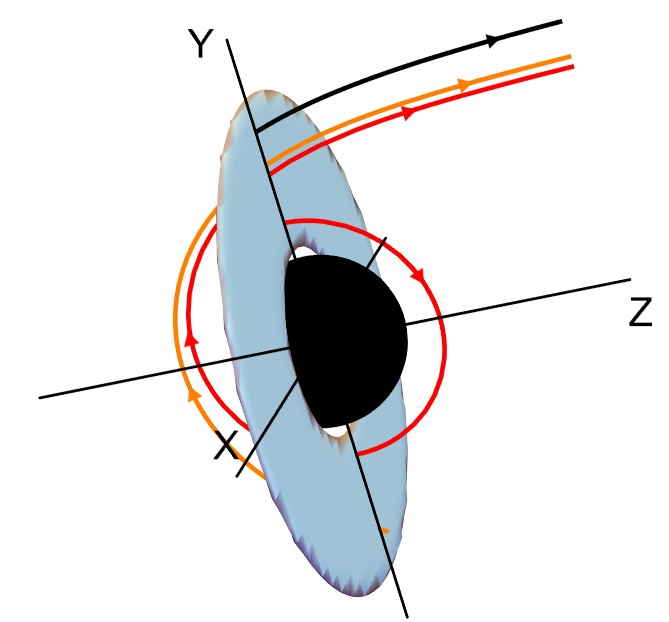}
    \hspace{1 cm}
    \includegraphics[width=5cm]{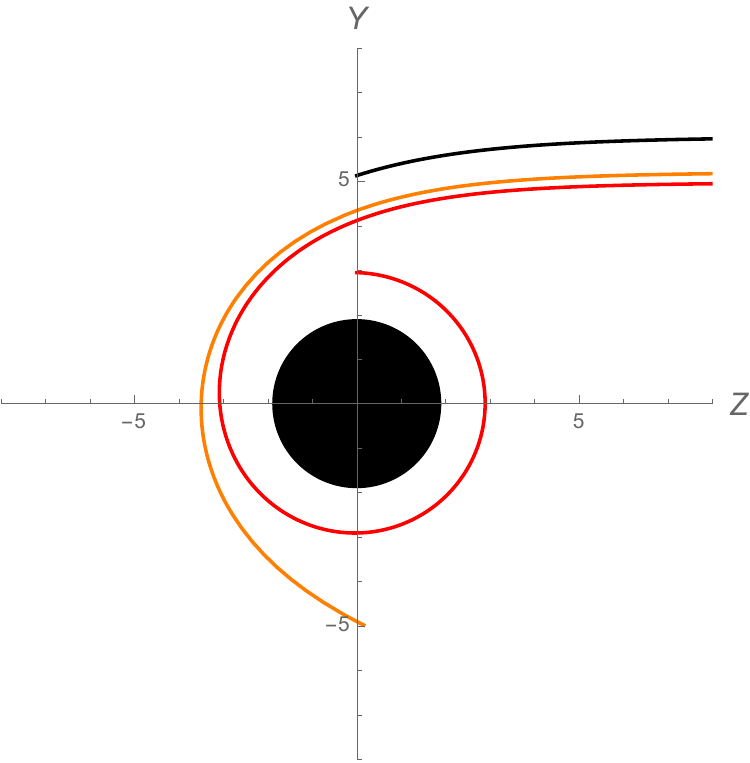}}
    \caption{ 
    The model depicts a geometrically and optically thin accretion disk, illustrated as a light blue solid disk, situated in the $XY$ equatorial plane of the black hole (represented by a solid black sphere), with a distant observer positioned along the $Z$ direction (at the north pole). This model in three dimensions (left) and in the $YZ$-plane (right) are shown here with null geodesics. In this study, we explore three distinct cases of GLM emission profiles, each characterized by an intensity profile that peaks at different positions of the inner edge of the disk.  Namely, these are $r_\text{ISCO}$, $r_\text{ph}$, and $r_+$ for \textit{Case I, II}, and \textit{III}, respectively. 
    It is crucial to understand that the disk extends spatially to infinity, although the light blue disk is depicted in the figure extending to a certain radius for illustrative purposes. Moreover, the paths of photons emanating from the direct, lensing, and photon ring regions are delineated by black, orange, and red curves, respectively.} 
    \label{fig: Accretion disk}
\end{figure}

\section{Transfer Functions and Image of Black Hole} \label{section 5}

In this section, we investigate the effect of an optically and geometrically thin accretion disk on the BH's image.
Under the assumption that the BH is solely illuminated by a thin accretion disk situated in the equatorial plane, neglecting the impact of photons from other sources, and with a distant observer positioned in the direction perpendicular to the plane as shown in Fig.~\ref{fig: Accretion disk}
The photons emitted from the thin accretion disk will undergo bending due to the curved spacetime in the vicinity of the BH before reaching the screen of a distant observer.
According to the ray-tracing procedure, these photon trajectories can be traced backward from the observer's screen to intersect the disk and pick up the brightness from the disk emission.
Since photons traced back from the regions of lensing and photon rings  can move in such a way that they intersect the accretion disk more than once,
the resulting observed intensity by the observer can be expressed as a collection of intensities from each intersection.
Additionally, the radial position $r_m(b)$ of the $m$th intersection with the disk plays an important role to computes the brightness acquired by photons from the disk. 
This is due to the fact that the emission profiles of the disk depend on the radial distance from BH.
The radial position $r_m(b)$ is commonly referred to as the ``transfer function".
We first focus on the examination of the transfer function for dilatonic BHs characterized by parameters $q$ and dilaton coupling $\alpha$. 
Subsequently, we proceed to unveil the appearing images of BHs in the EMD model through an exploration of the total observed intensity with three models of accretion disks.

\subsection{Transfer function}

\begin{figure}[H]
\centering
\subfigure[]{\includegraphics[scale=0.3]{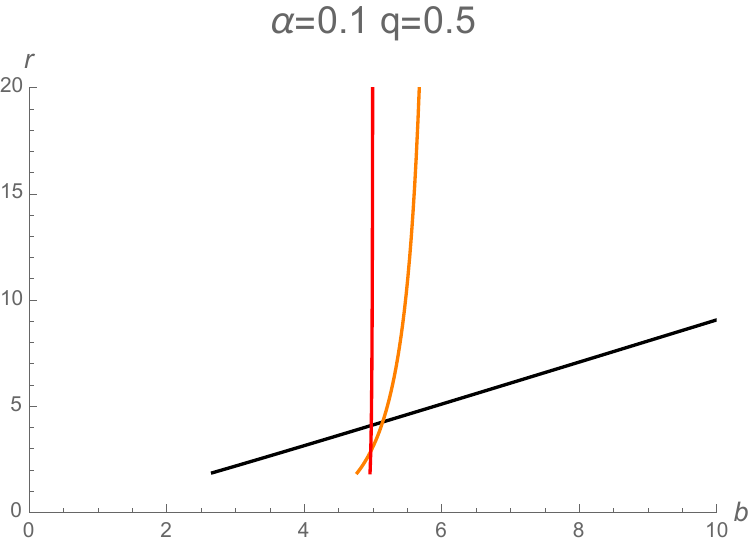}}\hspace{1cm}
\subfigure[]{\includegraphics[scale=0.3]{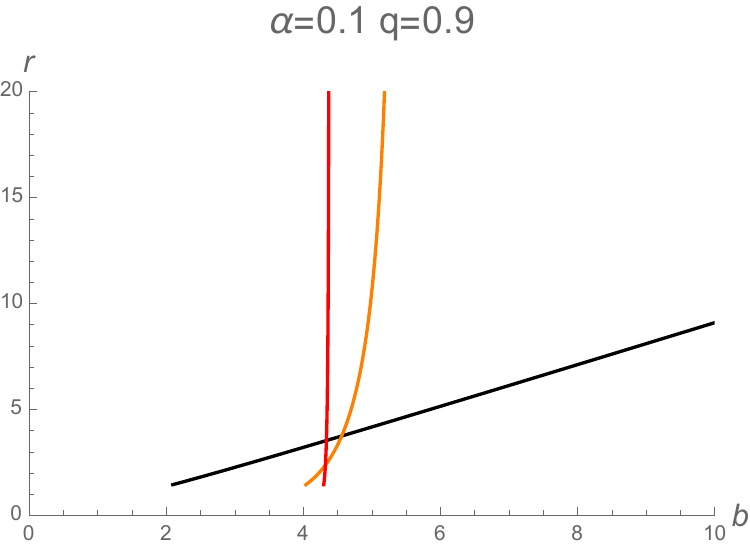}}\hspace{1cm}
\subfigure[]{\includegraphics[scale=0.3]{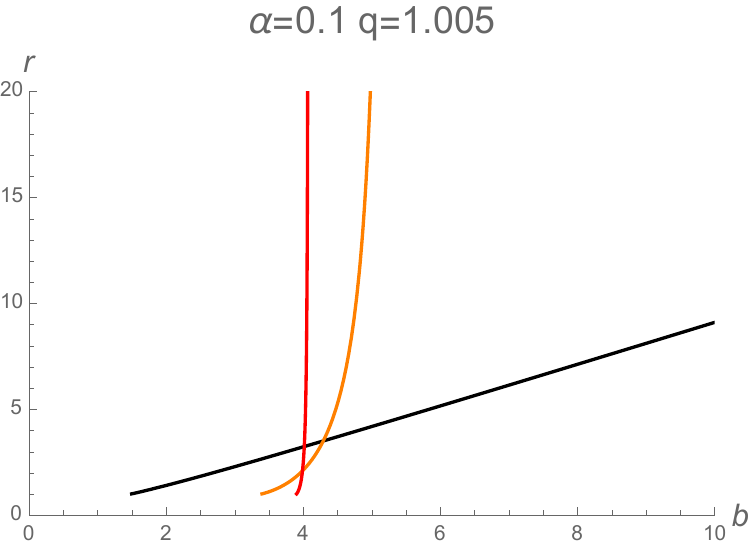}}\\
%\vskip 0.5cm
\subfigure[]{\includegraphics[scale=0.3]{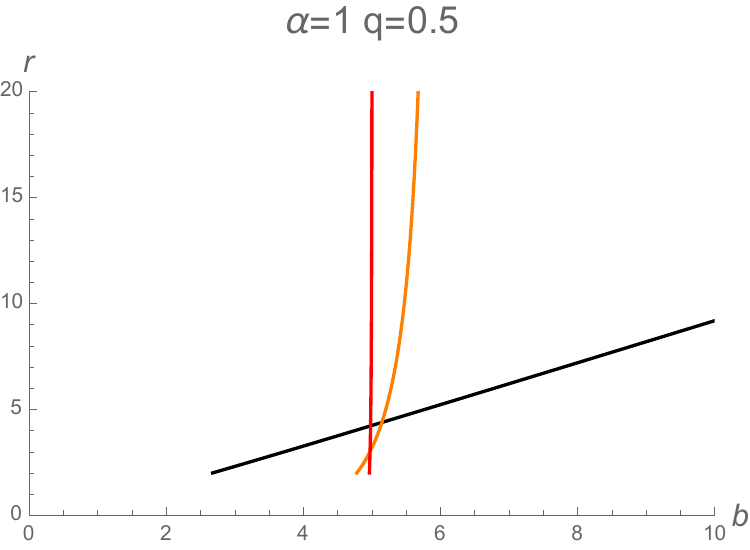}}\hspace{1cm}
\subfigure[]{\includegraphics[scale=0.3]{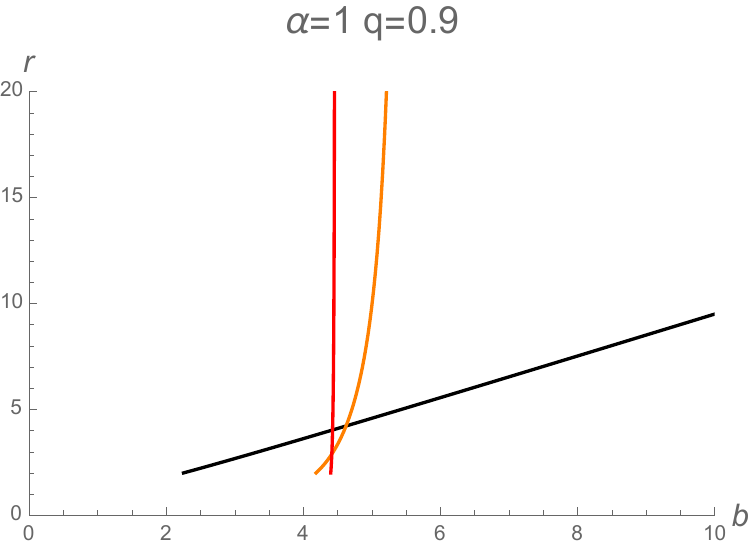}}\hspace{1cm}
\subfigure[]{\includegraphics[scale=0.3]{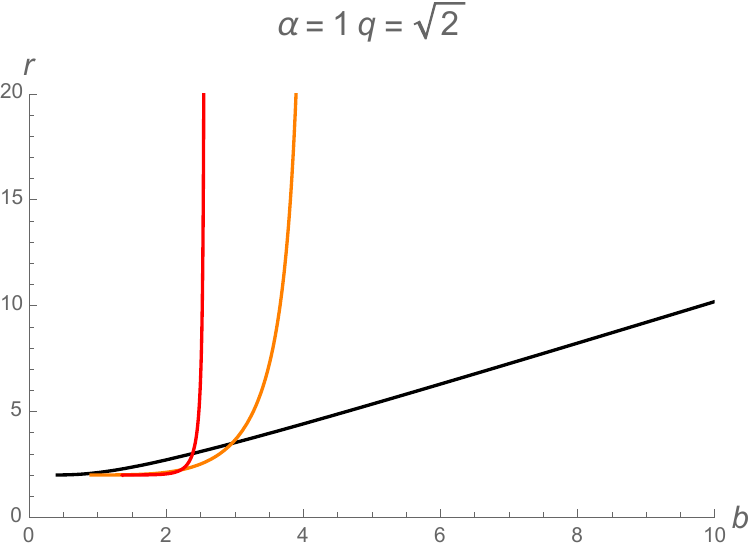}}\\
\subfigure[]{\includegraphics[scale=0.3]{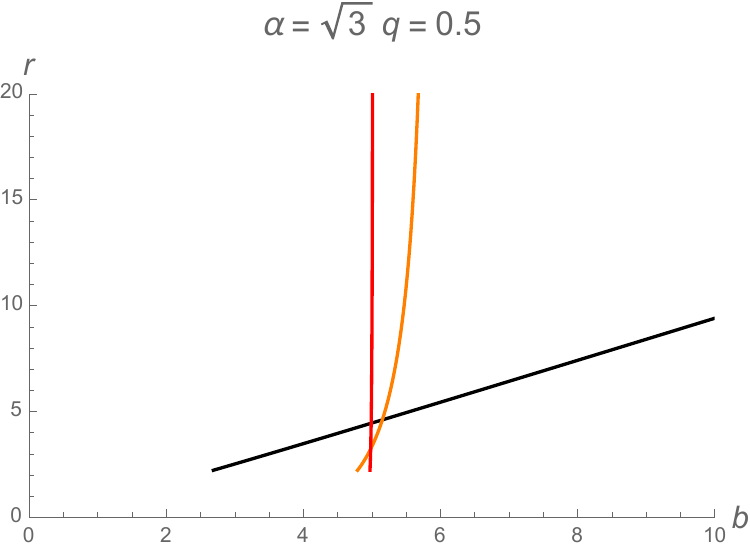}}\hspace{1cm}
\subfigure[]{\includegraphics[scale=0.3]{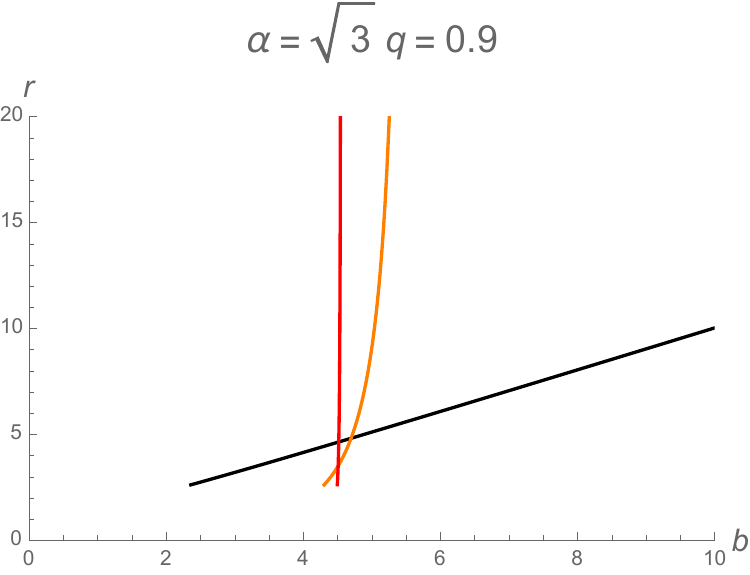}}\hspace{1cm}
\subfigure[]{\includegraphics[scale=0.3]{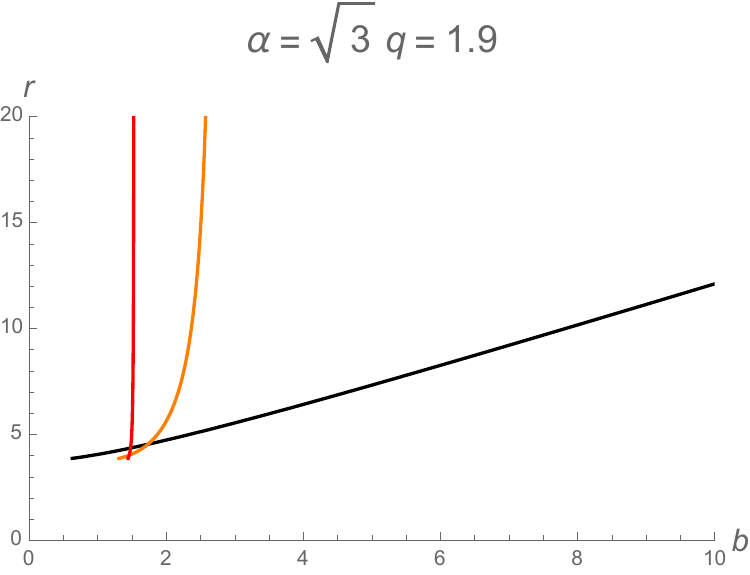}}\hspace{1cm}\\

\caption{The first three transfer functions, $r_1(b)$, $r_2(b)$ and $r_3(b)$, of dilatonic BH are plotted with black, orange, and red curves, respectively.  The panels in the first, second and third rows exhibit the plots of these three transfer functions at different values of $q$ for $\alpha=0.1, 1$ and $\sqrt{3}$, respectively.}
\label{fig:Transfer fn}
\end{figure}

We know that local frequencies of emitted photons $\nu$ and received photons $\nu'$ by a distant observer are different due to the gravitational redshift, 
\begin{eqnarray}
\displaystyle \frac{\nu'}{\nu}=\sqrt{g(r)}, \label{redshift}
\end{eqnarray}
where $g(r)$ has been defined in Eq.~\eqref{g} . 
Remarkably, $\displaystyle \frac{I_{\nu}}{\nu^3}$ is conserved along the ray, such that we have
\begin{eqnarray}
\frac{I^\text{obs}_{\nu'}}{\nu'^3} = \frac{I^\text{em}_\nu}{\nu^3}, \label{conserved}  
\end{eqnarray}
where $I^\text{em}_\nu$ and $I^\text{obs}_{\nu'}$ are the emitted specific intensity and the observed specific intensity, respectively. 
Using Eqs.~\eqref{redshift} and \eqref{conserved}, the observed specific intensity can be written as
\begin{eqnarray}
I^\text{obs}_{\nu'}(r) = g^{3/2}(r)I^\text{em}_\nu (r). 
%\ \ \ g(r)=\left( 1 - \frac{r_+}{r} \right)\left( 1 - \frac{r_-}{r} \right)^{\frac{1-\alpha^2}{1+\alpha^2}}.   
\end{eqnarray}
In order to obtain the total observed intensity $I_\text{obs}(r)$, we integrate $I^\text{obs}_{\nu'}(r)$  with respect to $\nu'$ as
\begin{eqnarray}
I_\text{obs}(r)&=&\int I^\text{obs}_{\nu'}(r) d\nu', \nonumber \\
&=& g^2(r)\int I^\text{em}_\nu (r)d\nu, \nonumber \\
&=& g^2(r)I_\text{em}(r),
\end{eqnarray}
where $\displaystyle I_\text{em}(r)=\int I^\text{em}_\nu (r)d\nu$, representing the total emission intensity. 
The above equation implies that $I_{\text{obs}}(r)$ scales as $g^2(r)$. 
Since some photons can intersect the disk more than once and it will pick up extra brightness from the disk emission, the observed intensity is a sum of the intensities from each intersection.  Hence, we can write 
\begin{eqnarray}
I_\text{obs}(b) = \sum_{m} g^2(r)I_\text{em}(r)\bigg |_{r=r_m(b)},
\end{eqnarray}
where $r_m(b)$ is a transfer function, which gives information about the radial position of the $m$th intersection with the disk plane.
In Fig~\ref{fig:Transfer fn}, we show the first three transfer functions against $b$ for different values of $q$ and $\alpha$, namely
 $r_1(b)$ (black),  $r_2(b)$ (orange) and  $r_3(b)$ (red), obtained by numerically solving Eq.~\eqref{geodesics} for $u(\phi, b)$ intersecting the disk plane at $\displaystyle \phi=\frac{\pi}{2}, \frac{3\pi}{2}$ and $\displaystyle \frac{5\pi}{2}$ with respect to $b$, respectively.  Note that the $\mathtt{ParametricNDSolve}$ function in $Mathematica$ has been applied.
Generally, we focus our attention on the first three transfer functions since the contributions from higher $m$th intersections tend to be less significant in the total luminosity.

Note that the value of transfer functions can be evaluated at a given impact parameter $b$ and its slope, represented as $\displaystyle \frac{dr}{db}$ is referred to the demagnification factor when mapping the size of rays on the intersection plane into that on the image~\cite{Gralla:2019xty}.
In the context of optical images, a demagnification factor greater than 1 denotes a decrease in size, while this factor less than 1 indicates an increase in size.
As discussed in \cite{Gralla:2019xty}, the first $r_1(b)$ function originates from direct, lensing ring and photon ring emission.
Since its demagnification factor, expressed by the slope, is almost $1$, the direct image profile is just the redshifted source profile.
The second $r_2(b)$ function originates from the emission of photons from the lensing and photon rings. 
In this particular case, the steep slope indicates that the contribution of emission from this region to the total photon flux will be much smaller compared to that of the direct image.
Finally, the third $r_3(b)$ function corresponds only to the photon ring.
The slope of the third transfer function in this range of $b$ approaches infinite.
In other words, the observer will see an extremely demagnified image that contributes negligibly to the total photons flux.
Our results on the transfer functions of dilatonic BH, as illustrated in Fig.~\ref{fig:Transfer fn}, gives us information about the optical appearance of the BH image as follows: 
\begin{itemize}
    \item For fixed dilaton coupling parameter $\alpha$ where $q$ is varied, the results suggest that the demagnification factor decreases with increasing $q$. As depicted in the panels (c), (f), and (i) of Fig~\ref{fig:Transfer fn}, the slope of the second transfer function is found to be nearly equal to or even smaller than that of the first transfer function within certain values of $b$ when $q$ is large and near the extremal limit. Thus, the thickness of lensing and photon rings appear larger on the image with increasing $q$. 
    \item Conversely, if $q$ remains fixed while $\alpha$ is varied, we find that the demagnification factor of lensing and photon rings are greater as $\alpha$ increases. Consider the panels (b), (e) and (h) of Fig.~\ref{fig:Transfer fn} for example. This implies that the observed total photons flux from emission is suppressed by the dilaton hair, while these two rings appear as sharper circles on the image.
\end{itemize}

\newpage

\begin{figure}[H]
\centering
\includegraphics[width=4.7cm]{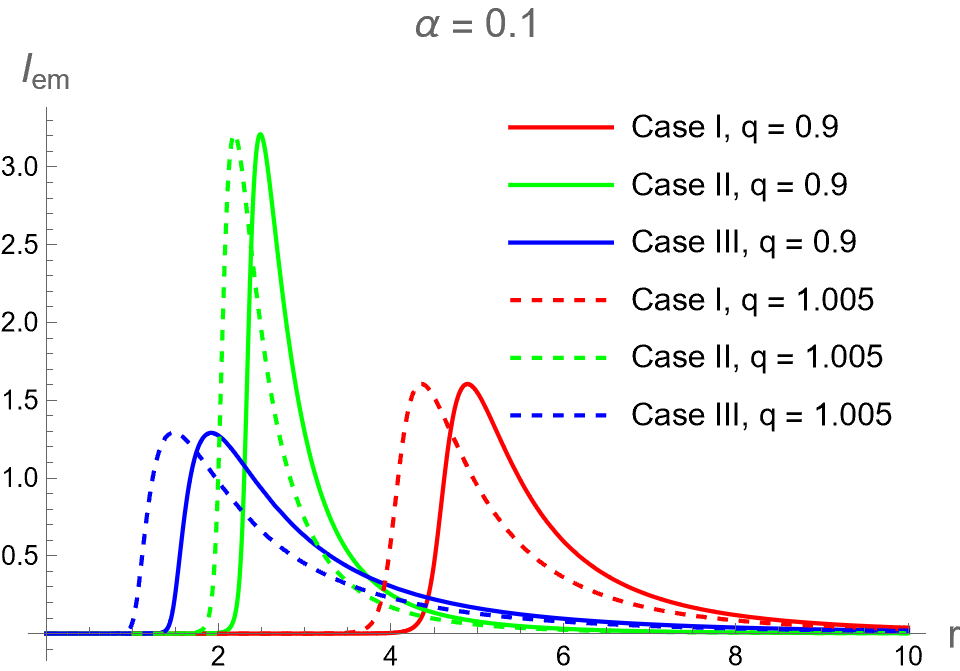}\hspace{1cm}
\includegraphics[width=4.7cm]{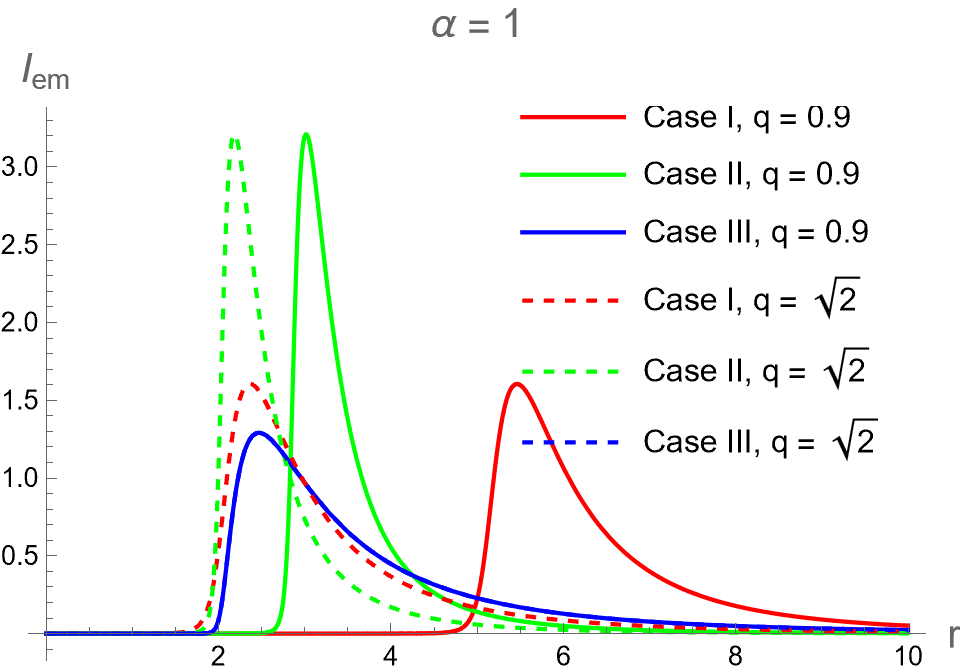}\hspace{1cm}
\includegraphics[width=4.7cm]{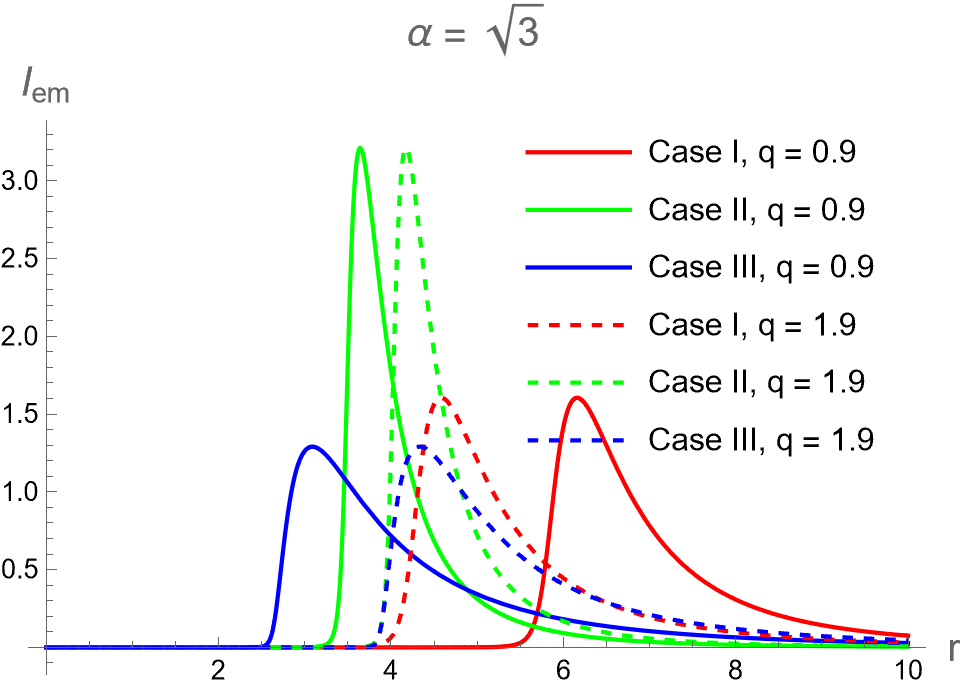}
\caption{The GLM emission profiles of thin accretion disk in \textit{Case I}, \textit{Case II} and \textit{Case III} around the charged dilaton BHs with $\alpha=0.1$ (left panel), $1$ (middle panel), $\sqrt{3}$ (right panel) are plotted. In each panel, the profiles of each case with $q=0.9$ (solid) and extremal charge limit $q = q_\text{ext}$ (dashed) are shown.}
\label{fig: GLM profiles}
\end{figure}

\subsection{Image of charged dilaton black hole surrounded by thin accretion disk}

We choose to use the GLM model of thin accretion disks \cite{PhysRevD.102.124004} as the light source to simulate images of charged dilaton BH.
The GLM disk model has been demonstrated to closely align with the observational predictions derived from general relativistic magnetohydrodynamics (GRMHD) simulations of astrophysical accretion disks~\cite{Vincent:2022fwj}. 
Recent applications of the GLM model have extended to investigations of observed images in various BH solutions \cite{DeMartino:2023ovj,Gao:2023mjb,macedo2024optical} and compact objects \cite{Olmo:2023lil,Rosa:2023hfm,Rosa:2023qcv}. 
The intensity profile in the GLM model is described by 
\begin{eqnarray}
    I_\text{em}(r;\gamma ,\alpha , \beta)=\frac{\text{exp}\left[ -\frac{1}{2}\left(\gamma +\text{arcsinh}\left( \frac{r-\alpha}{\beta}\right) \right)^2\right]}{\sqrt{(r-\alpha)^2+\beta^2}},
\end{eqnarray}
where the shape of $I_\text{em}(r;\gamma,\alpha,\beta)$ is characterized by the parameters $\gamma$, $\alpha$ and $\beta$.  
This model's intensity profile extends from a region of low intensity, starting at infinity and increasing to a peak around a specific radial position, which represents the innermost radius. Beyond this point, the intensity suppresses rapidly.   Namely, $\gamma$ influences the rate at which the intensity profile increases from infinity to its peak, $\alpha$ is responsible for shifting the radial position of the emission peak, and $\beta$ determines the overall dilation or expansion of the intensity profile.
Motivated by some astrophysical reasons as suggested in \cite{Rosa:2023hfm,Gao:2023mjb}, we consider the GLM profile in three cases, in which the parameters $\gamma$, $\alpha$, and $\beta$ are determined as follows: 
\begin{itemize}
    \item \textit{Case I}: $\gamma =-2, \alpha =r_\text{ISCO}$ and $\beta =M/4$. The emission profile has its peak near $r_\text{ISCO}$, corresponding to the absence of a stable orbit of massive particles below this radius. The calculational details of $r_\text{ISCO}$ can be seen in Appendix~\ref{App A}.
    \item \textit{Case II}: $\gamma =-2, \alpha =r_\text{ph}$ and $\beta =M/8$. The emission profile has its peak near the radial position of photon sphere $r_\text{ph}$. 
    \item \textit{Case III}: $\gamma =-3, \alpha =r_+$ and $\beta =M/8$. The emission profile has its peak near the radial position of event horizon $r_+$. 
\end{itemize}
The emission intensity profiles $I_\text{em}(r)$ of these three cases of the GLM model of thin accretion disks surrounding the EMD BHs with $\alpha =0.1, 1$ and $\sqrt{3}$ are illustrated in the left, middle and right panels of Fig.~\ref{fig: GLM profiles}, respectively.
Note that, in each case, we consider the profiles with $q=0.9$ (solid) and extremal charge limit $q=q_\text{ext}$ (dashed).

\begin{figure}[H]
\centering
\subfigure[]{\includegraphics[scale=0.3]{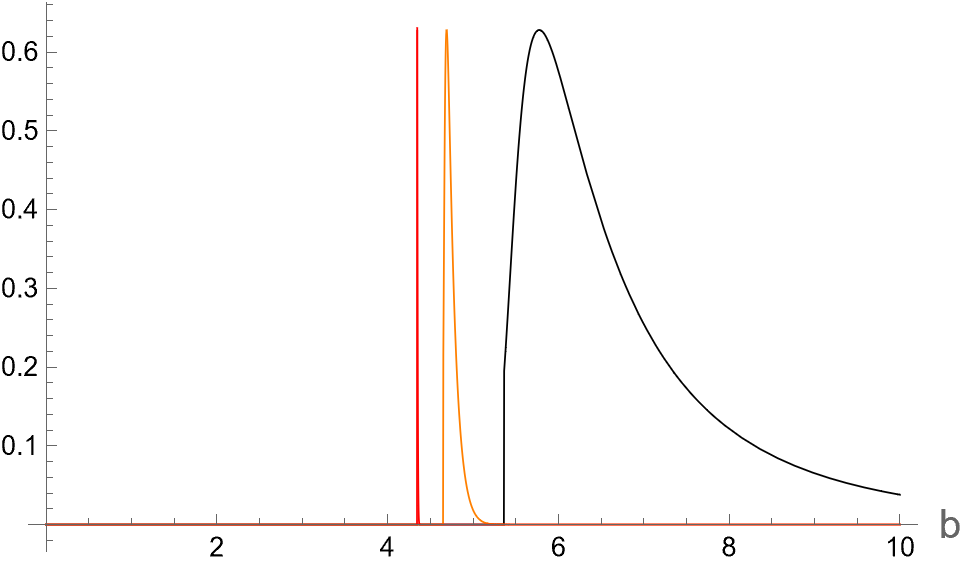}}\hfill
\subfigure[]{\includegraphics[scale=0.3]{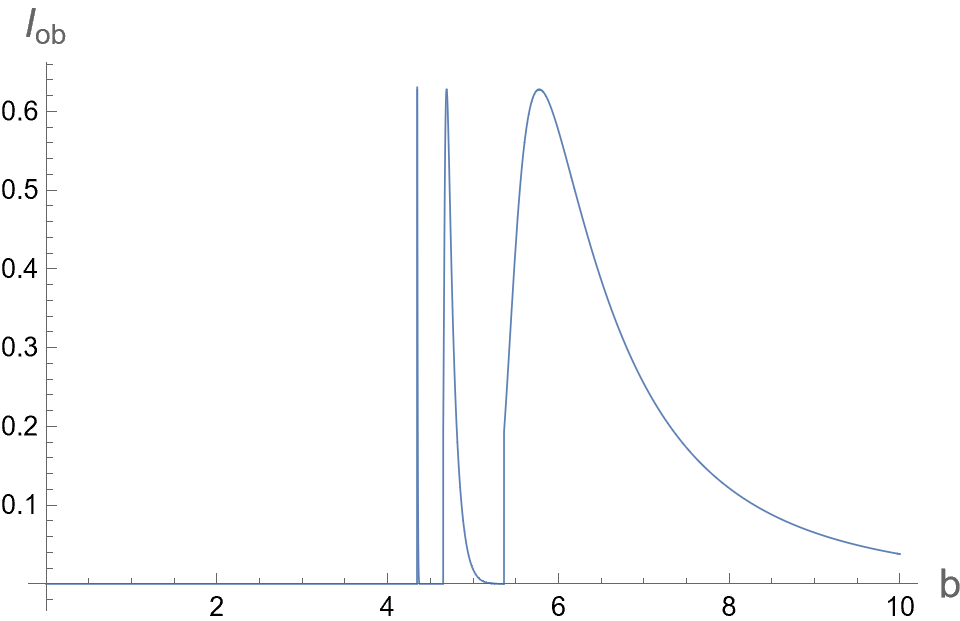}}\hfill
\subfigure[]{\includegraphics[scale=0.31]{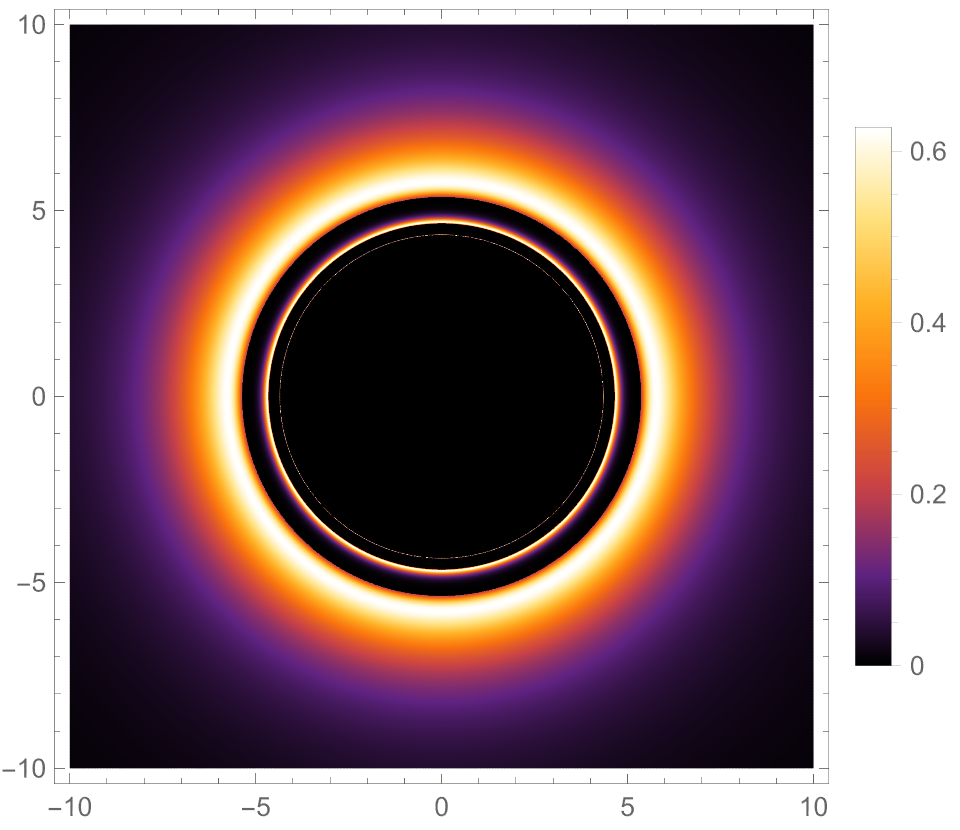}}\hfill
\subfigure[]{\includegraphics[scale=0.25]{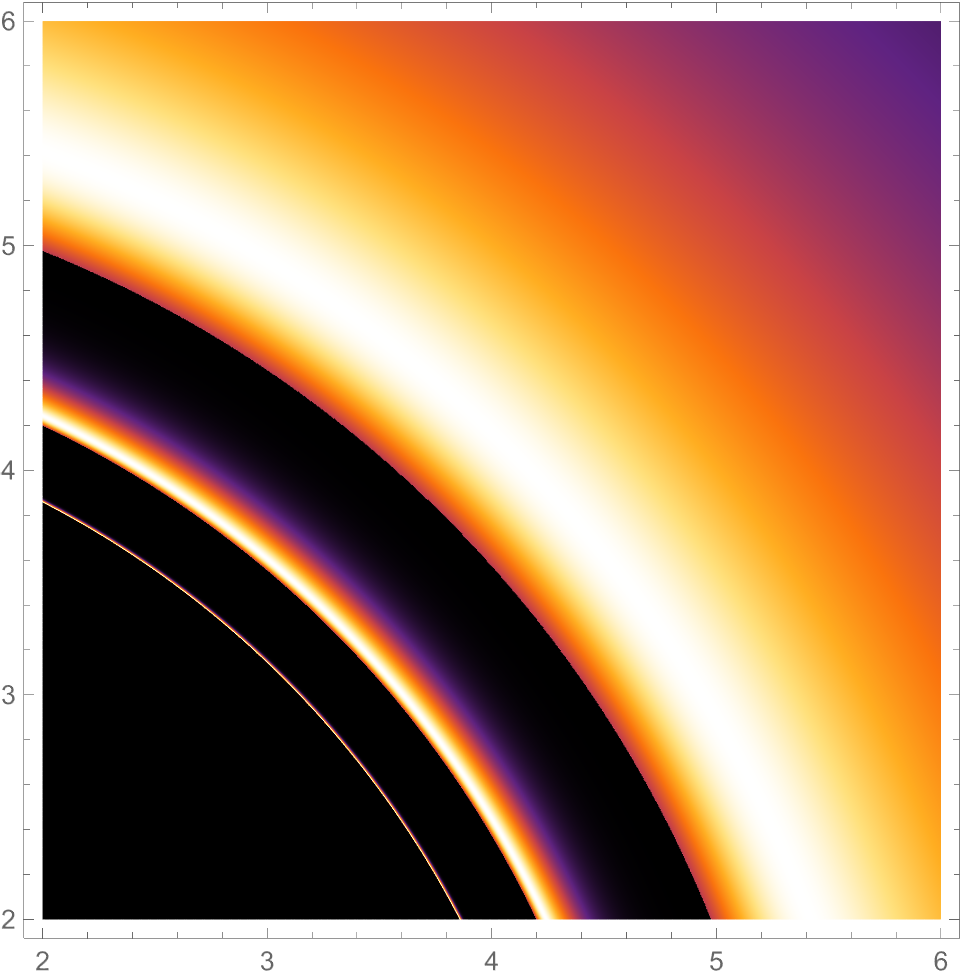}} \\
%\vskip 0.5cm
\subfigure[]{\includegraphics[scale=0.3]{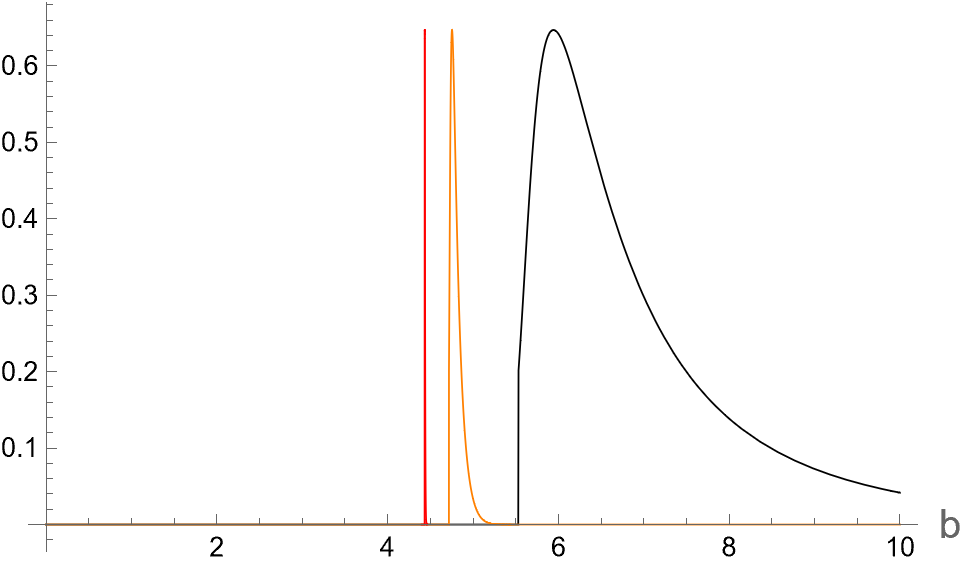}}\hfill
\subfigure[]{\includegraphics[scale=0.3]{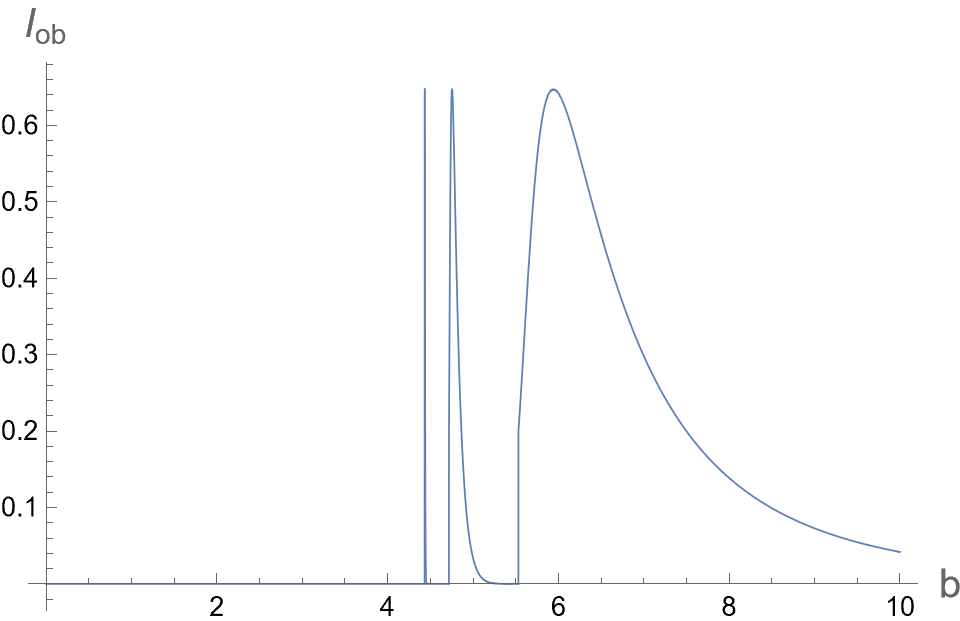}}\hfill
\subfigure[]{\includegraphics[scale=0.31]{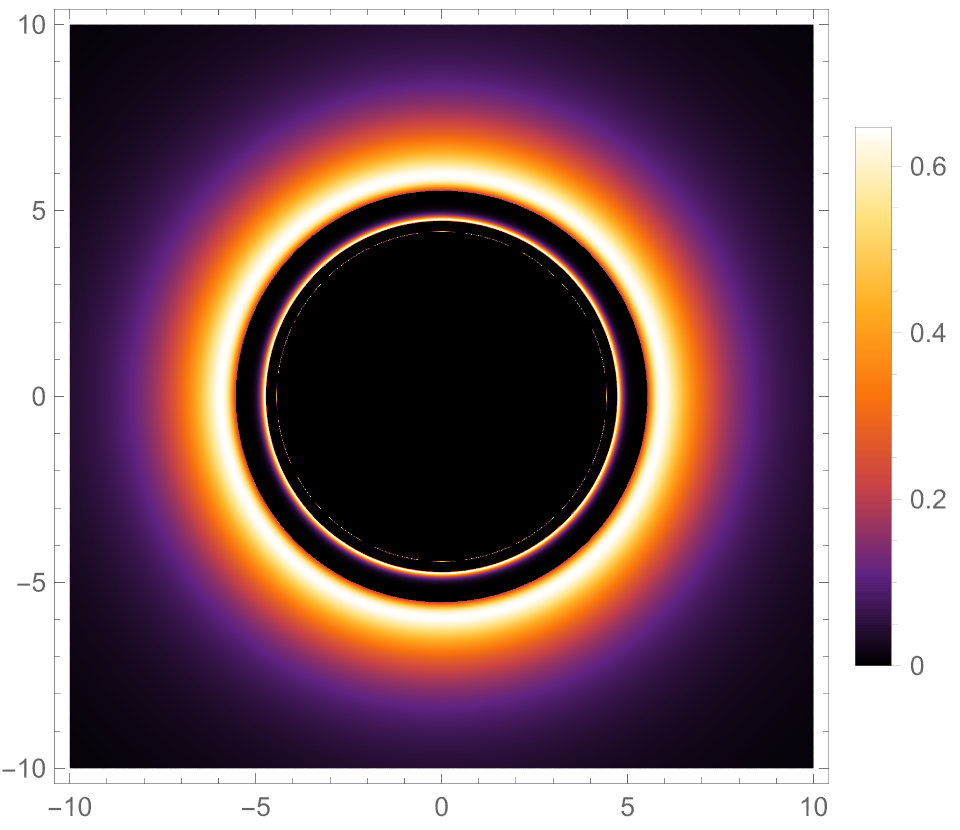}}\hfill
\subfigure[]{\includegraphics[scale=0.25]{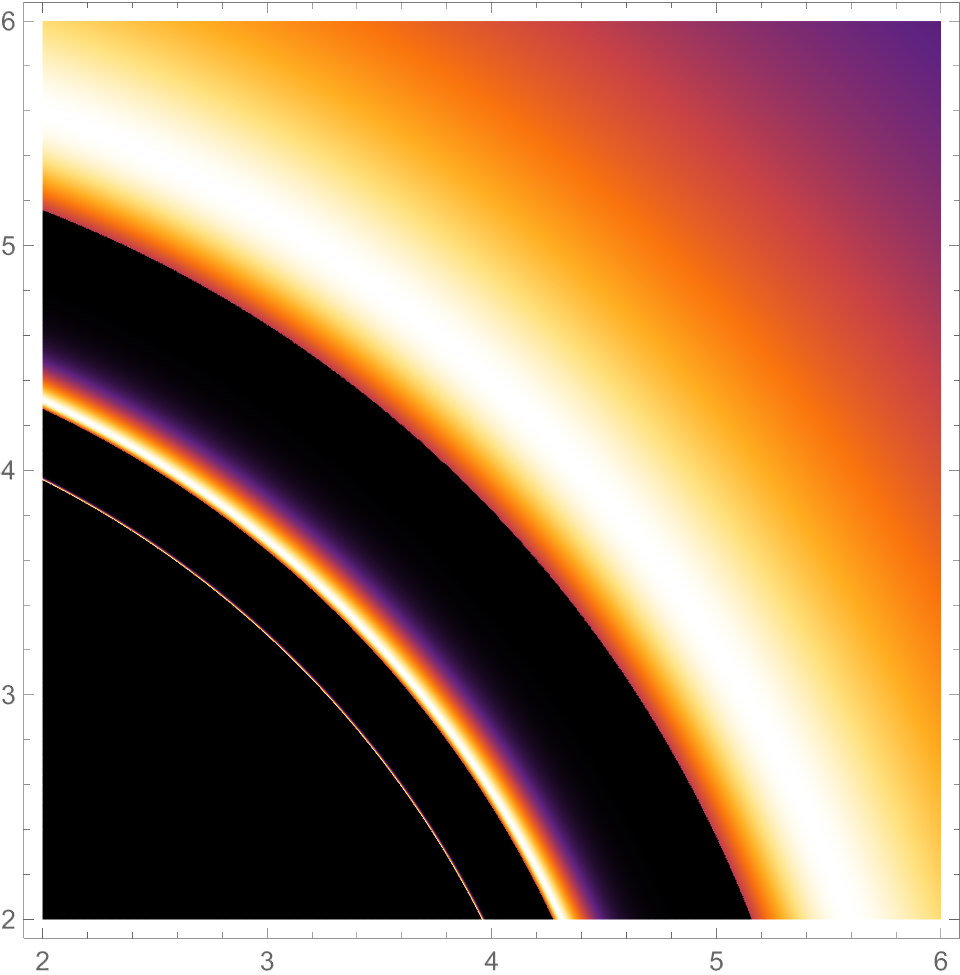}} \\
%\vskip 0.5cm
\subfigure[]{\includegraphics[scale=0.3]{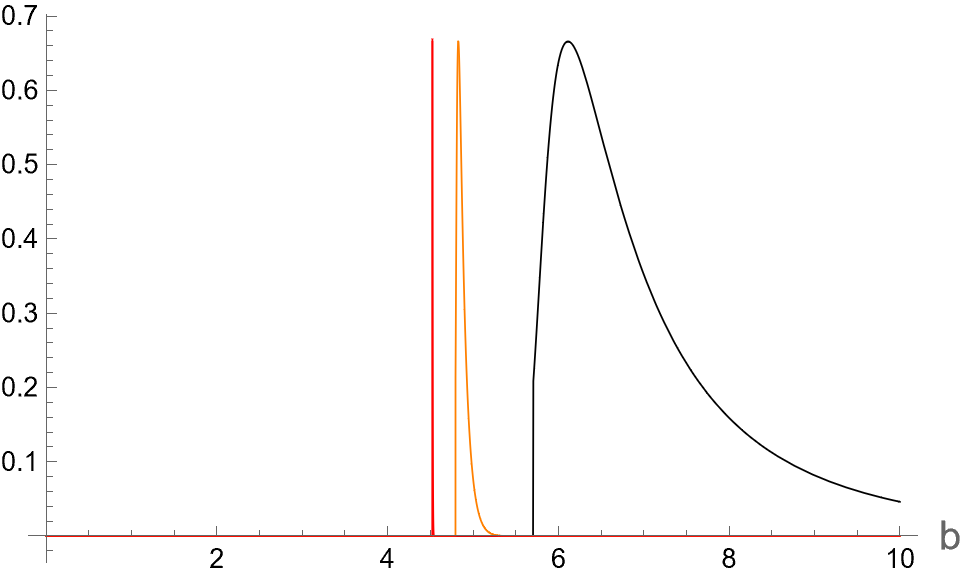}}\hfill
\subfigure[]{\includegraphics[scale=0.3]{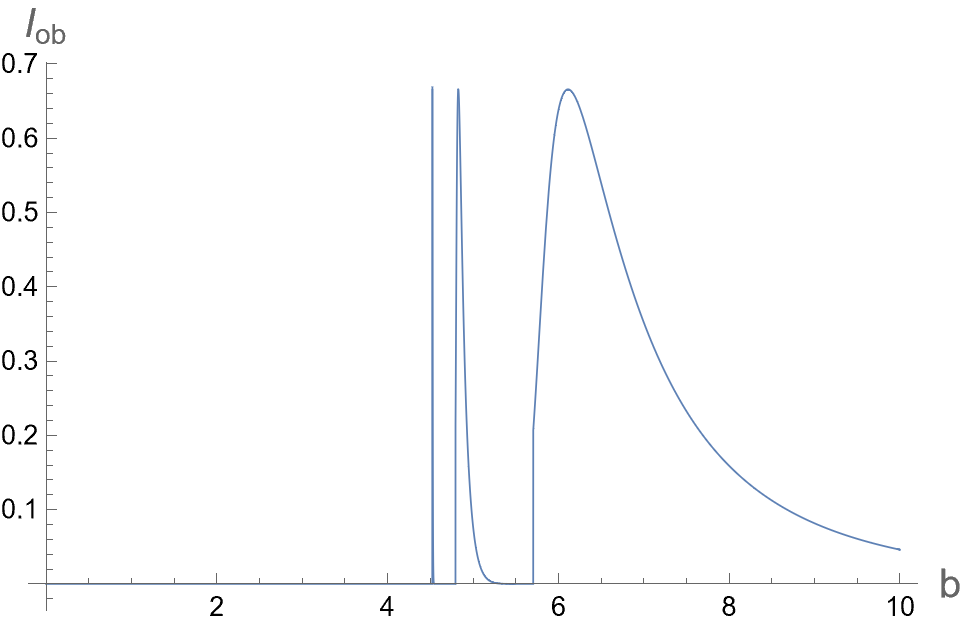}}\hfill
\subfigure[]{\includegraphics[scale=0.31]{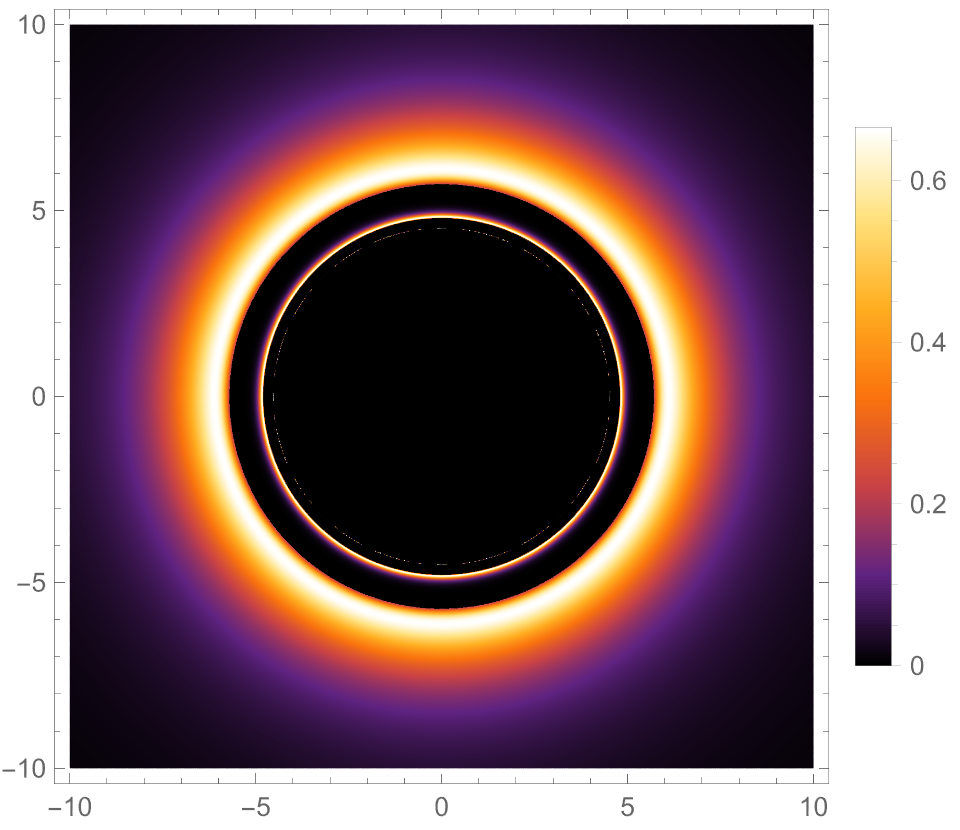}}\hfill
\subfigure[]{\includegraphics[scale=0.25]{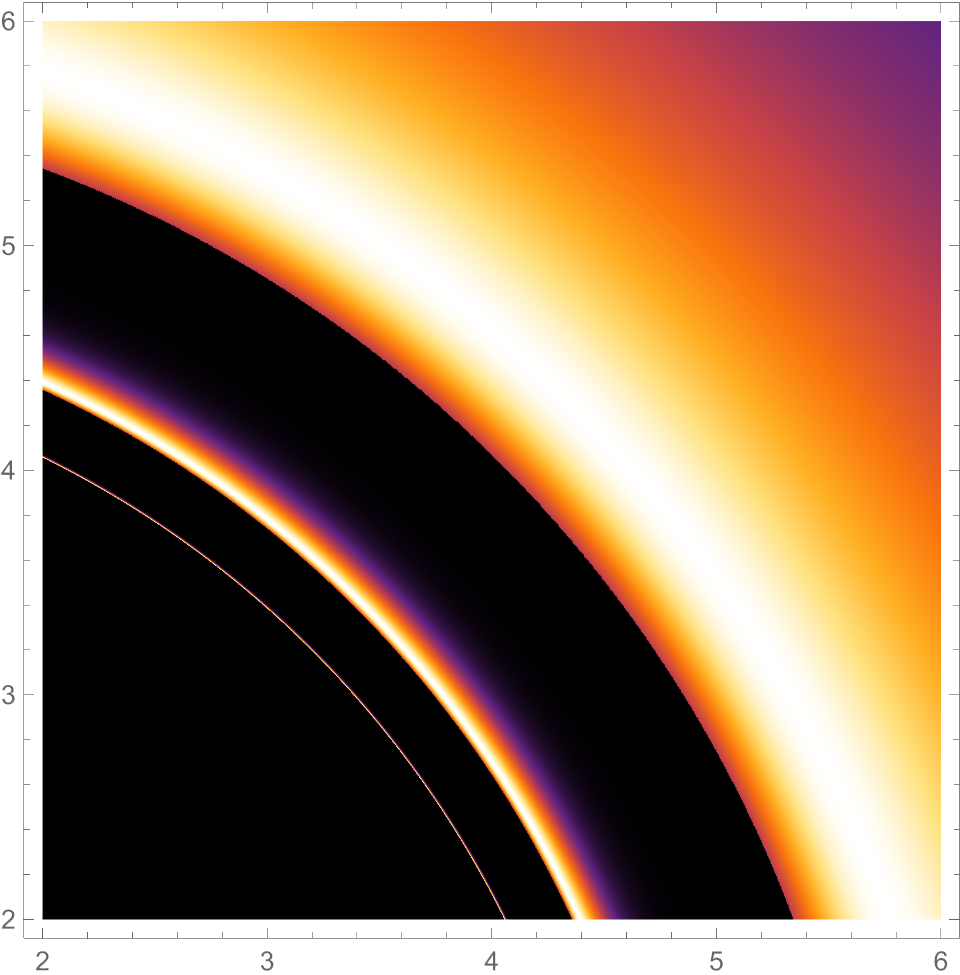}}
\caption{Observational appearance for the \textit{Case I} of GLM profile near the EMD BH with dilaton coupling $\alpha =0.1$ (upper row), $1$ (second row) and $\sqrt{3}$ (third row).
\textit{First column} represents separately the observed intensities of $m=1, 2$ and $3$ in black, orange and red curves, respectively.
\textit{Second column}, we plot the total observed intensities $I_\text{ob}$ against induced impact parameter $b$.
\textit{Third column} is the density plots of $I_\text{ob}$.
\textit{Fourth column} represents zoom in of the third column.
Note that we fix $q=0.9$ in all figures.}
\label{fig: GLM image case 1}
\end{figure}

\begin{figure}[H]
\centering
\subfigure[]{\includegraphics[scale=0.3]{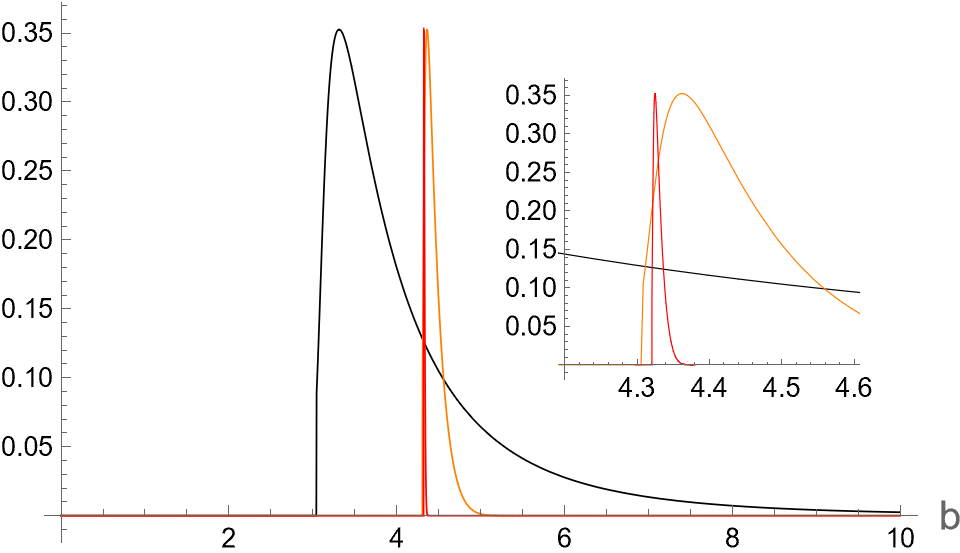}}\hfill
\subfigure[]{\includegraphics[scale=0.3]{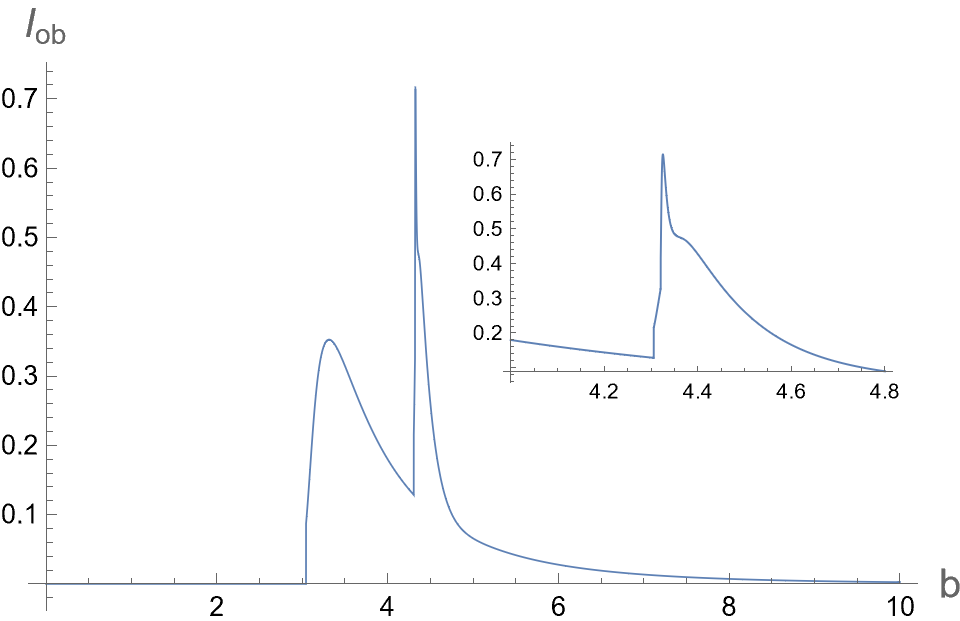}}\hfill
\subfigure[]{\includegraphics[scale=0.31]{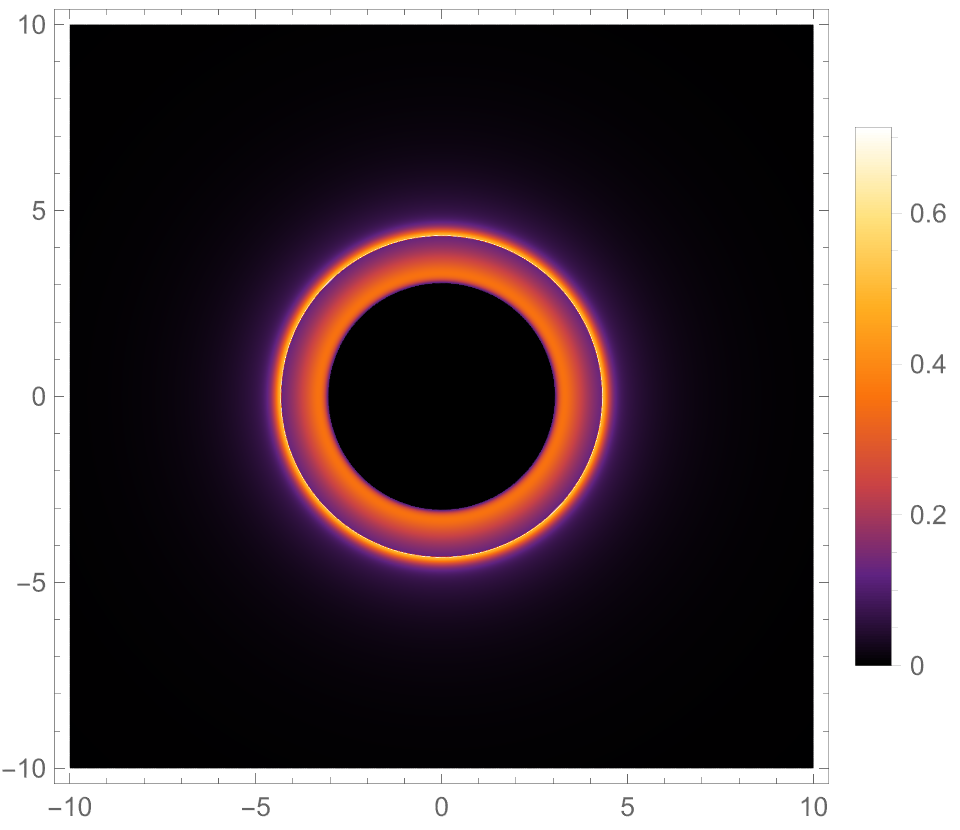}}\hfill
\subfigure[]{\includegraphics[scale=0.25]{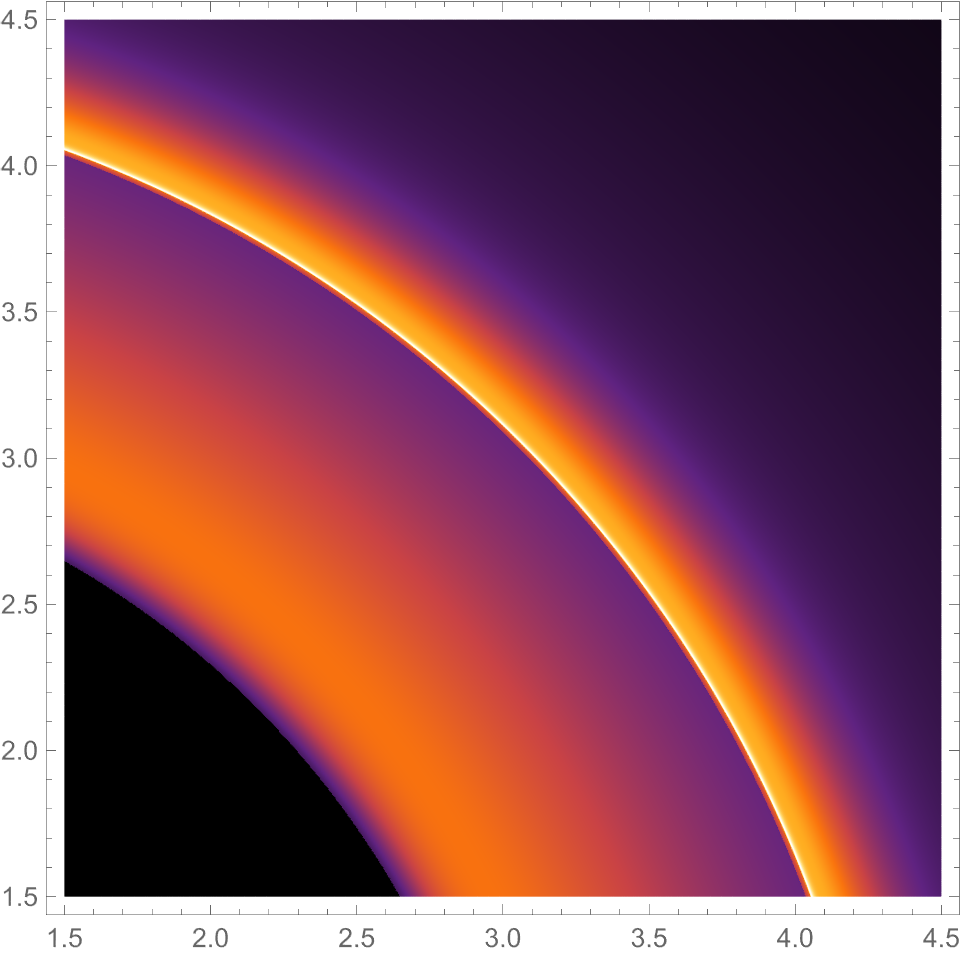}} \\
%\vskip 0.5cm
\subfigure[]{\includegraphics[scale=0.3]{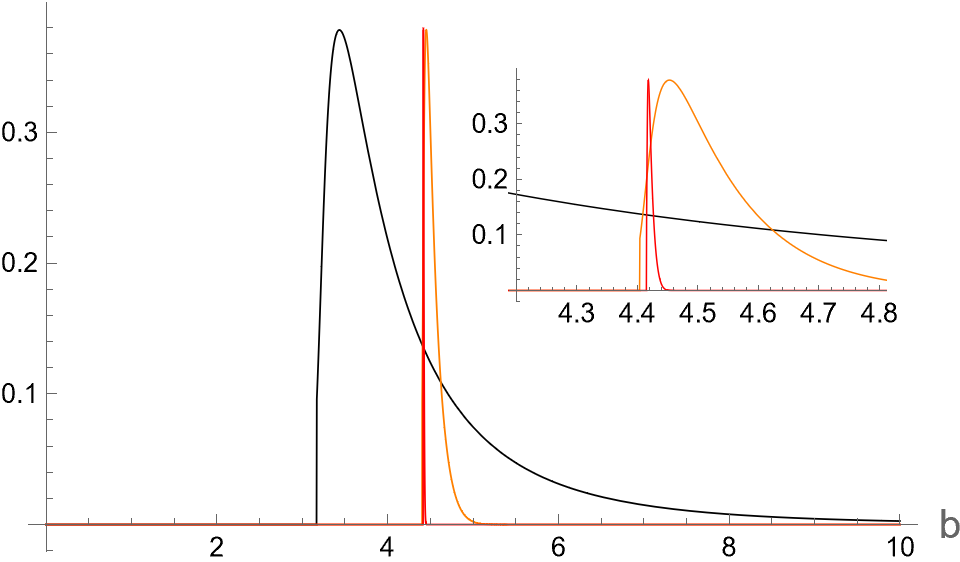}}\hfill
\subfigure[]{\includegraphics[scale=0.3]{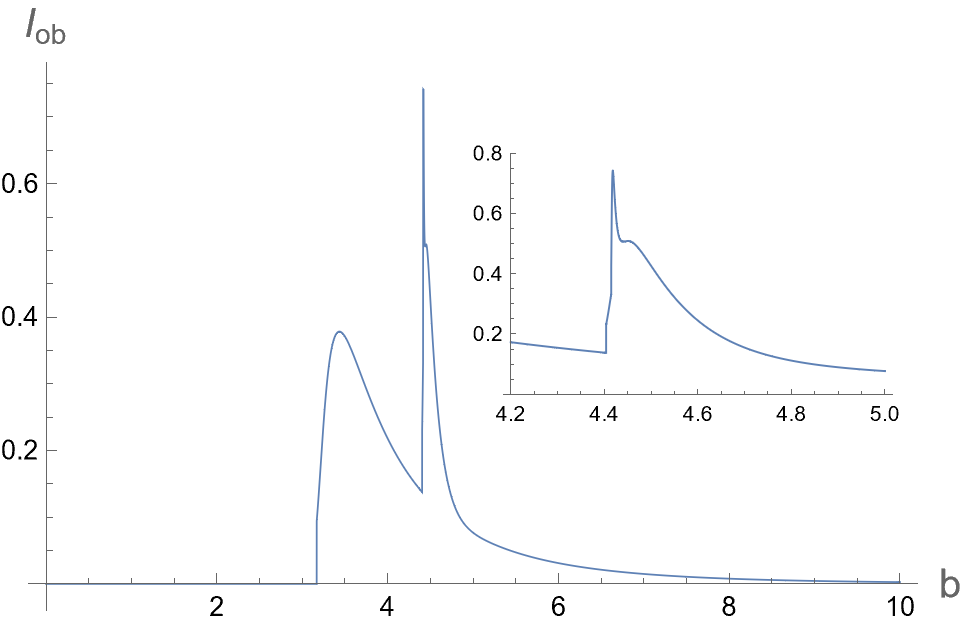}}\hfill
\subfigure[]{\includegraphics[scale=0.31]{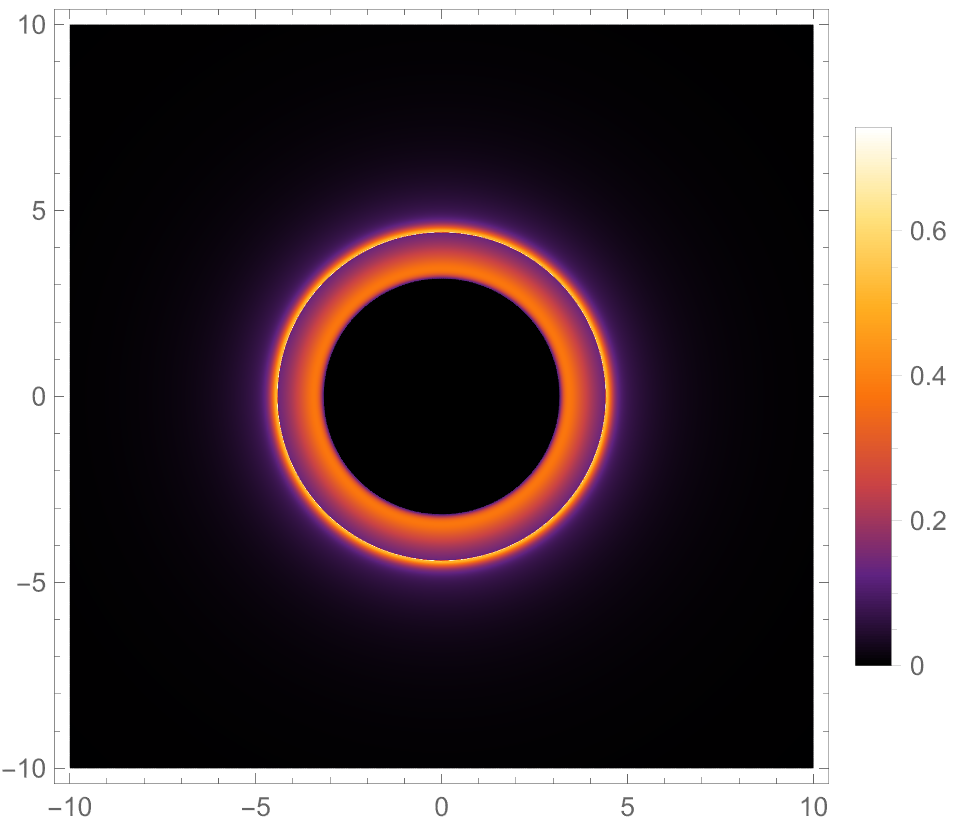}}\hfill
\subfigure[]{\includegraphics[scale=0.25]{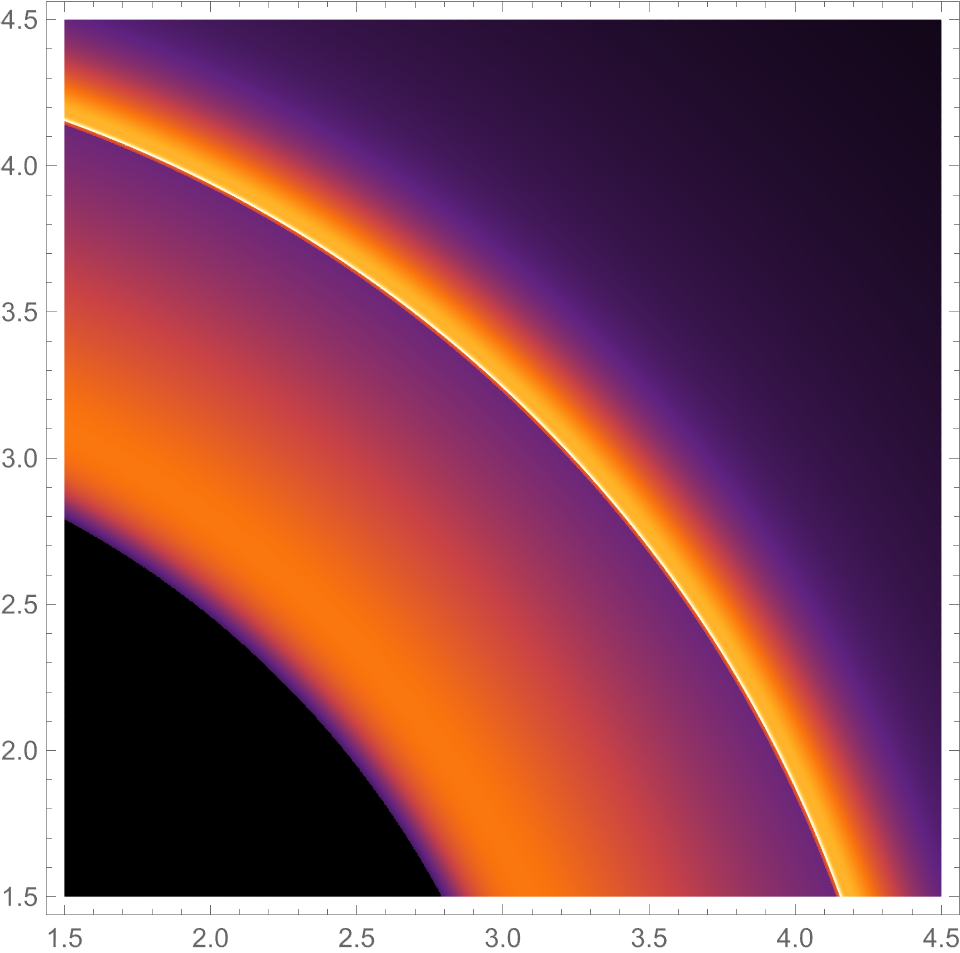}} \\
%\vskip 0.5cm
\subfigure[]{\includegraphics[scale=0.3]{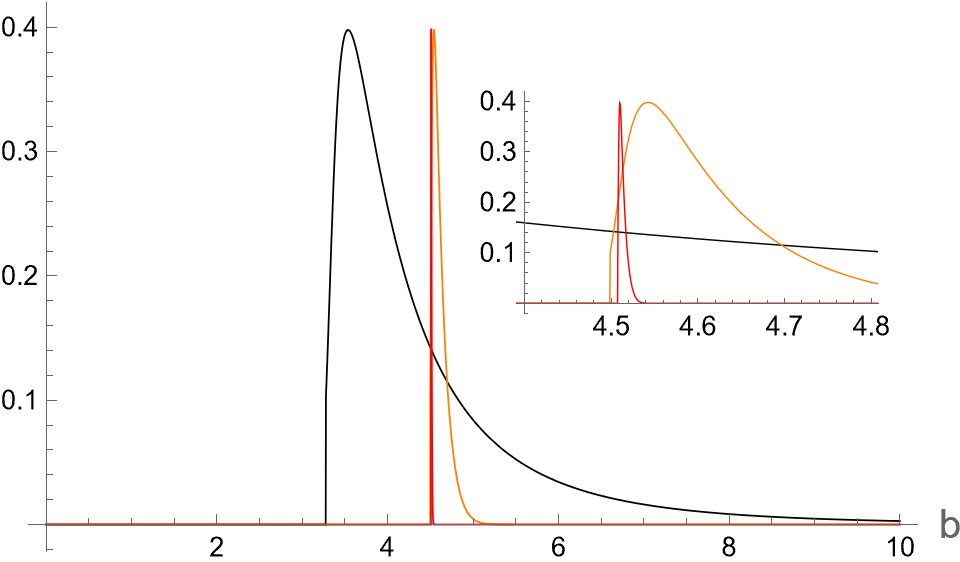}}\hfill
\subfigure[]{\includegraphics[scale=0.3]{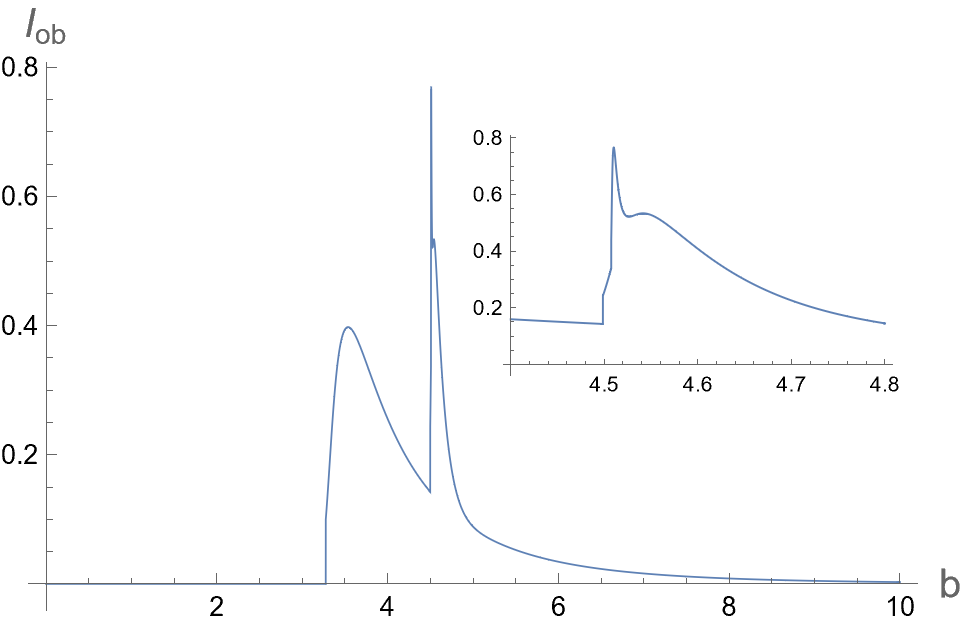}}\hfill
\subfigure[]{\includegraphics[scale=0.31]{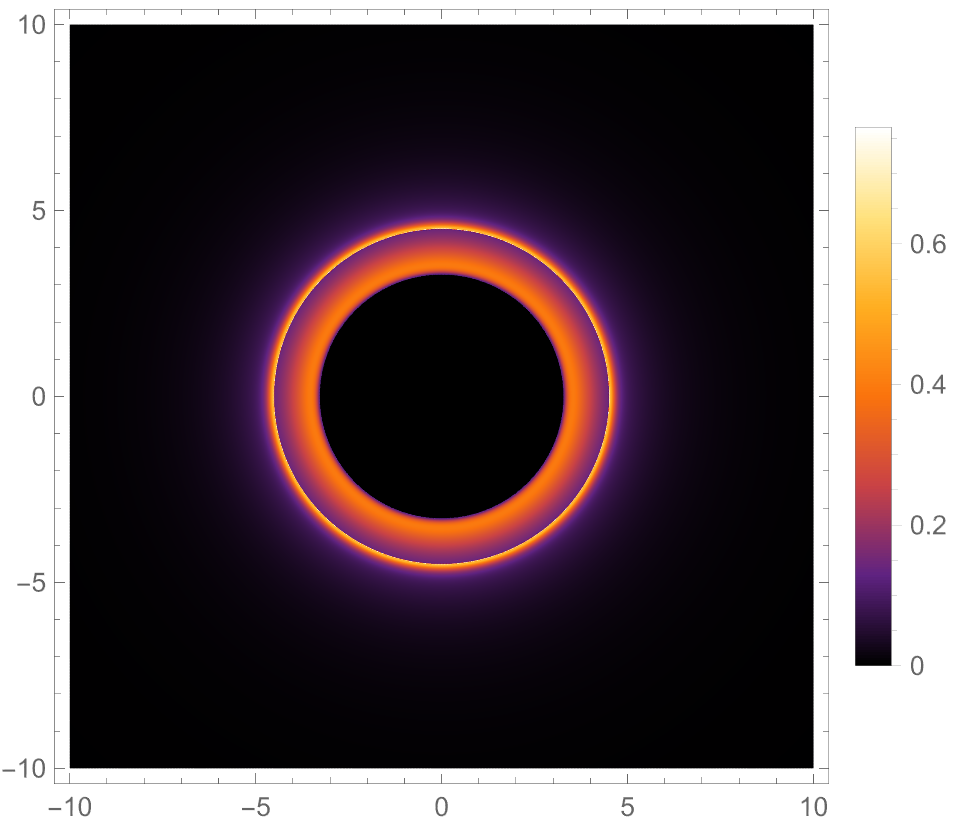}}\hfill
\subfigure[]{\includegraphics[scale=0.25]{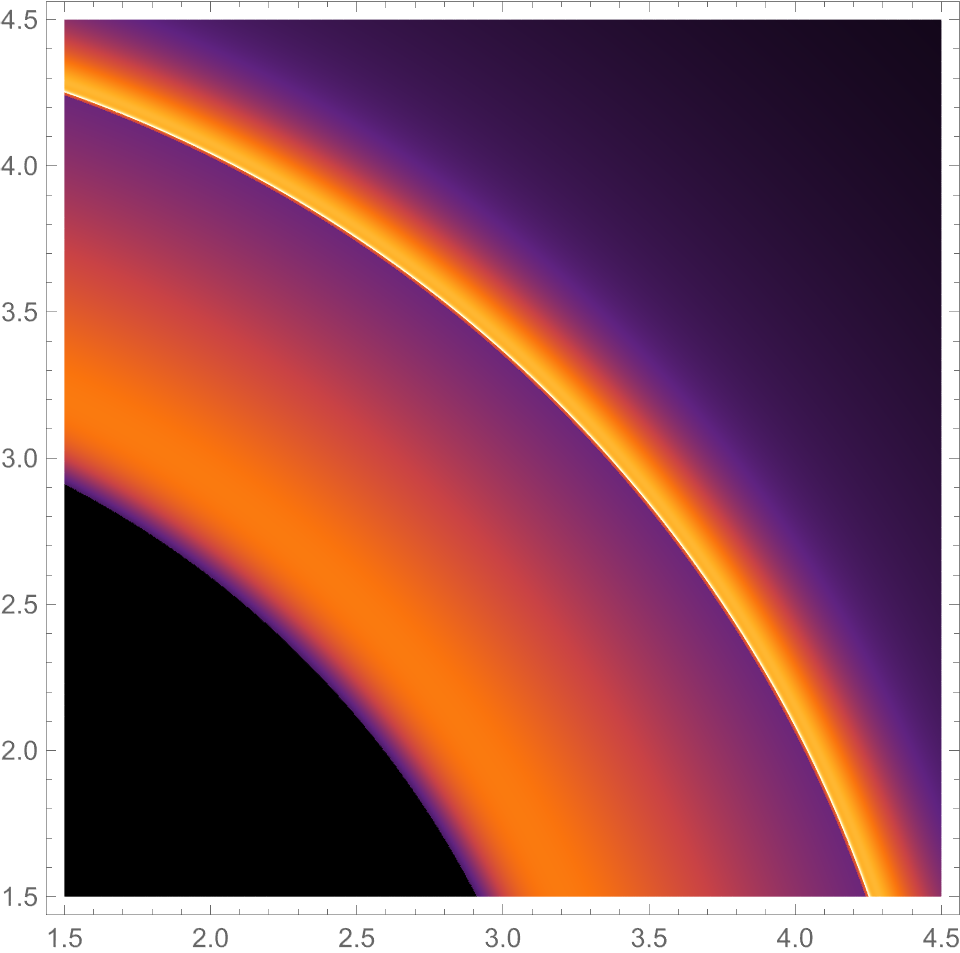}}
\caption{Observational appearance for the \textit{Case II} of GLM profile near the EMD BH with dilaton coupling $\alpha =0.1$ (upper row), $1$ (second row) and $\sqrt{3}$ (third row).
\textit{First column} represents separately the observed intensities of $m=1, 2$ and $3$ in black, orange and red curves, respectively.
\textit{Second column}, we plot the total observed intensities $I_\text{ob}$ against induced impact parameter $b$.
\textit{Third column} is the density plots of $I_\text{ob}$.
\textit{Fourth column} represents zoom in of the third column.
Note that we fix $q=0.9$ in all figures.}
\label{fig: GLM image case 2}
\end{figure}

\begin{figure}[H]
\centering
\subfigure[]{\includegraphics[scale=0.3]{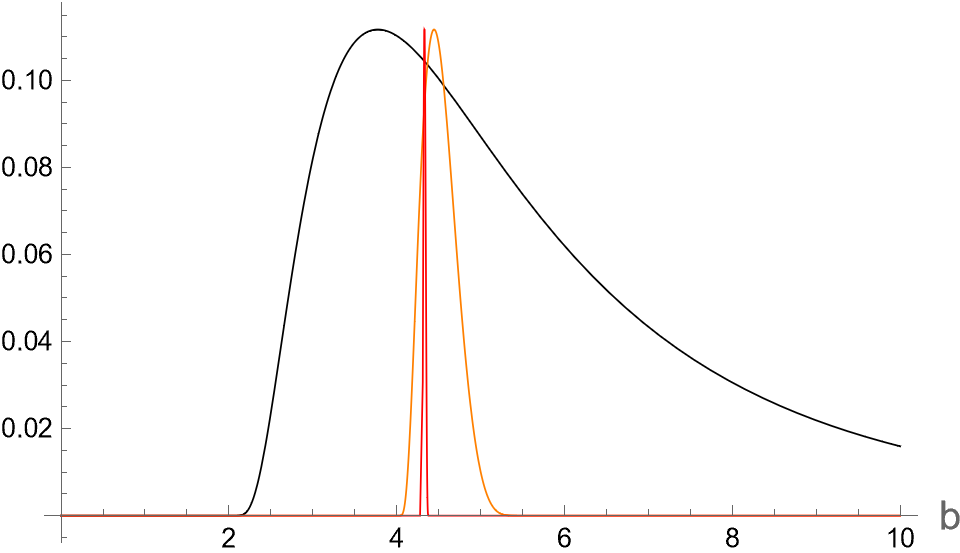}}\hfill
\subfigure[]{\includegraphics[scale=0.3]{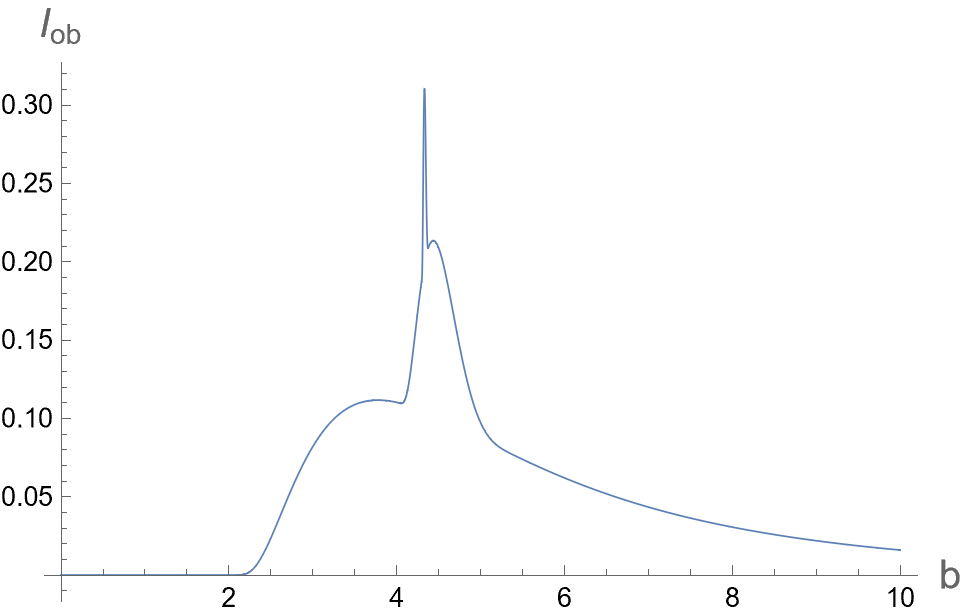}}\hfill
\subfigure[]{\includegraphics[scale=0.31]{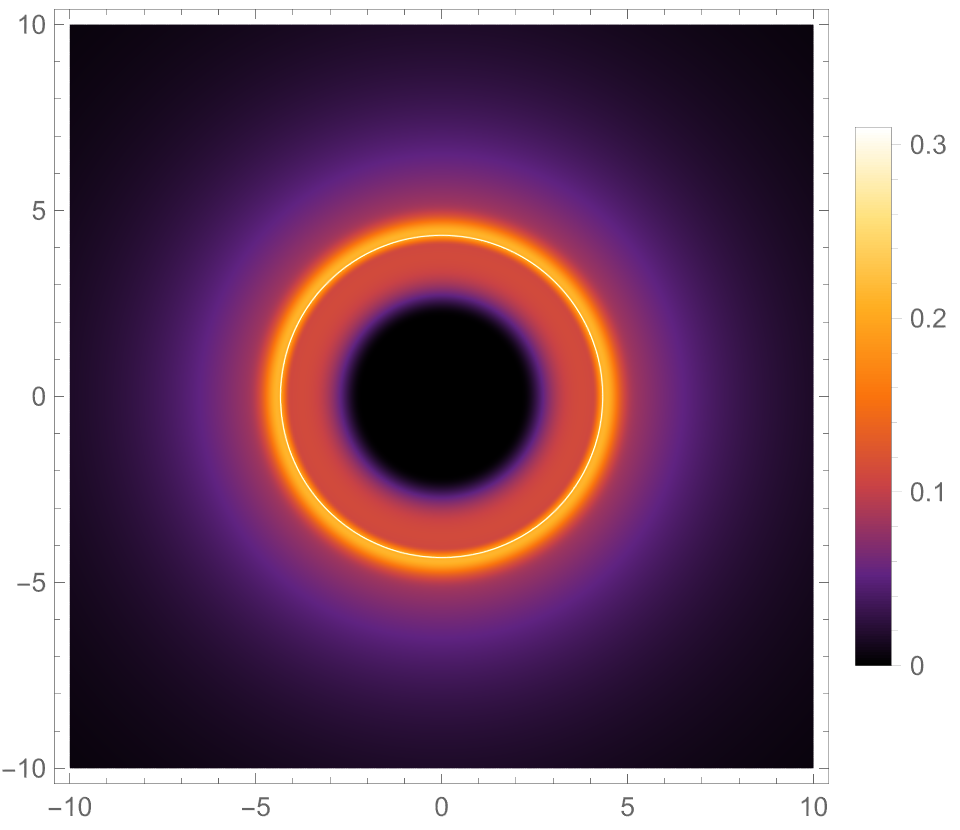}}\hfill
\subfigure[]{\includegraphics[scale=0.25]{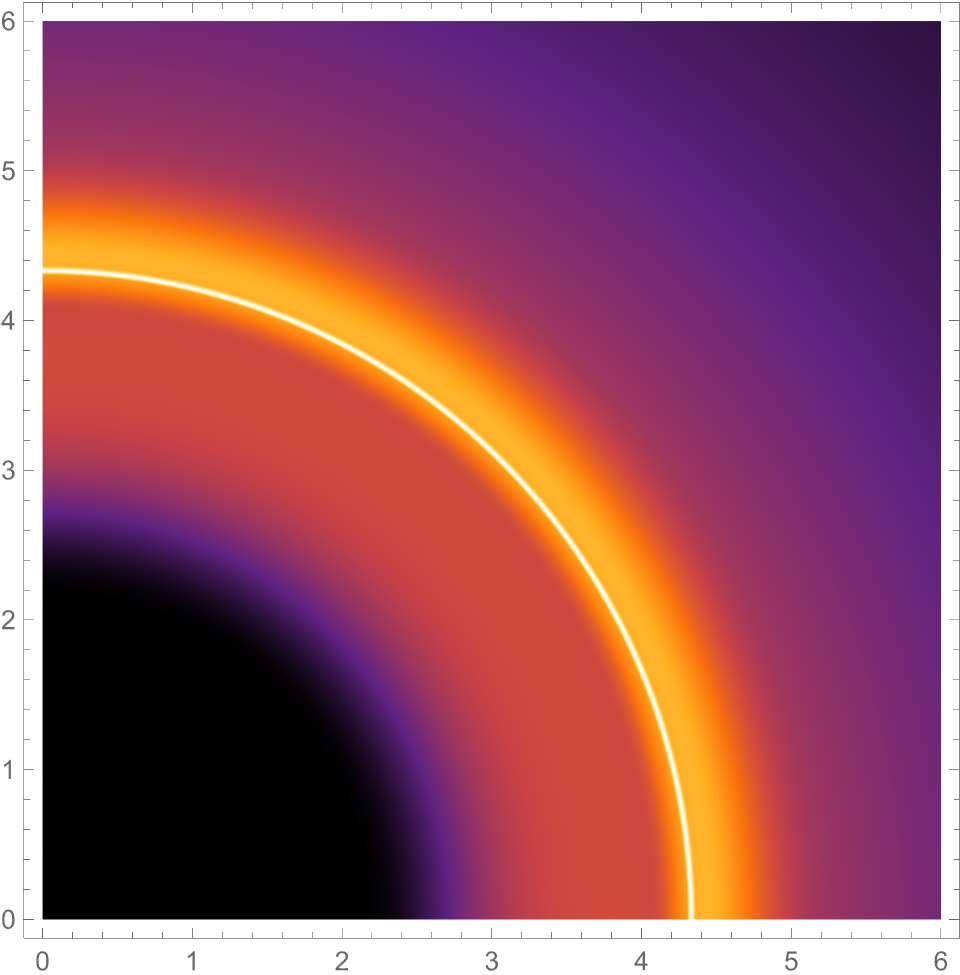}} \\
%\vskip 0.5cm
\subfigure[]{\includegraphics[scale=0.3]{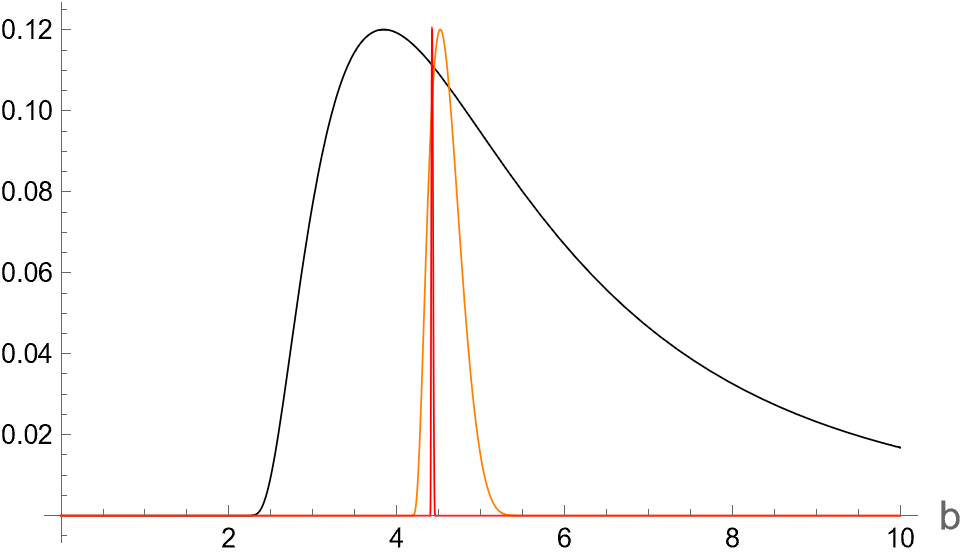}}\hfill
\subfigure[]{\includegraphics[scale=0.3]{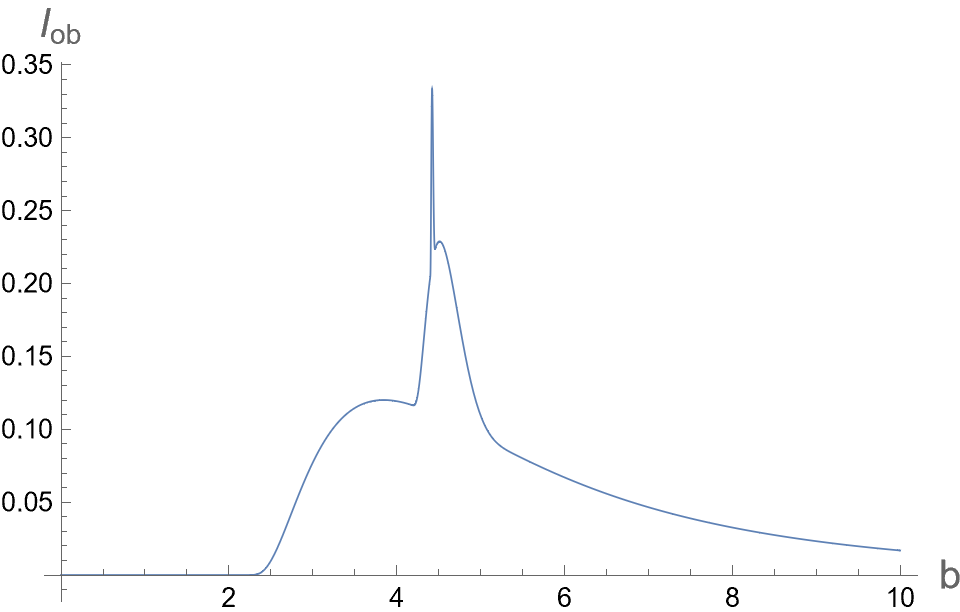}}\hfill
\subfigure[]{\includegraphics[scale=0.31]{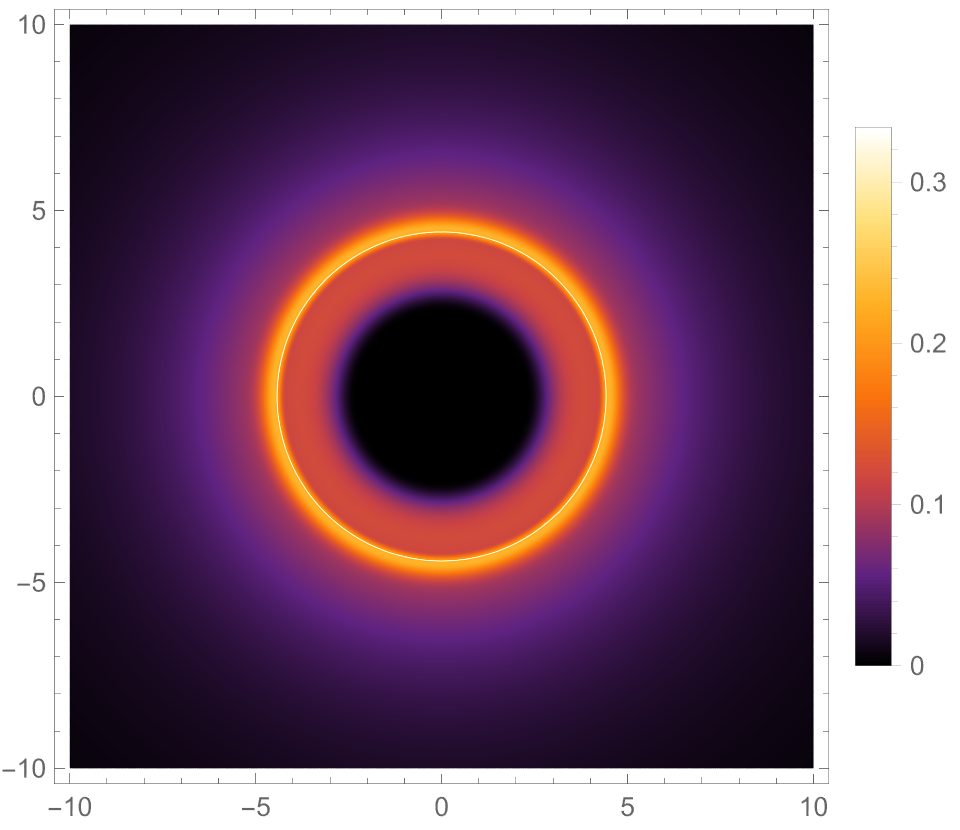}}\hfill
\subfigure[]{\includegraphics[scale=0.25]{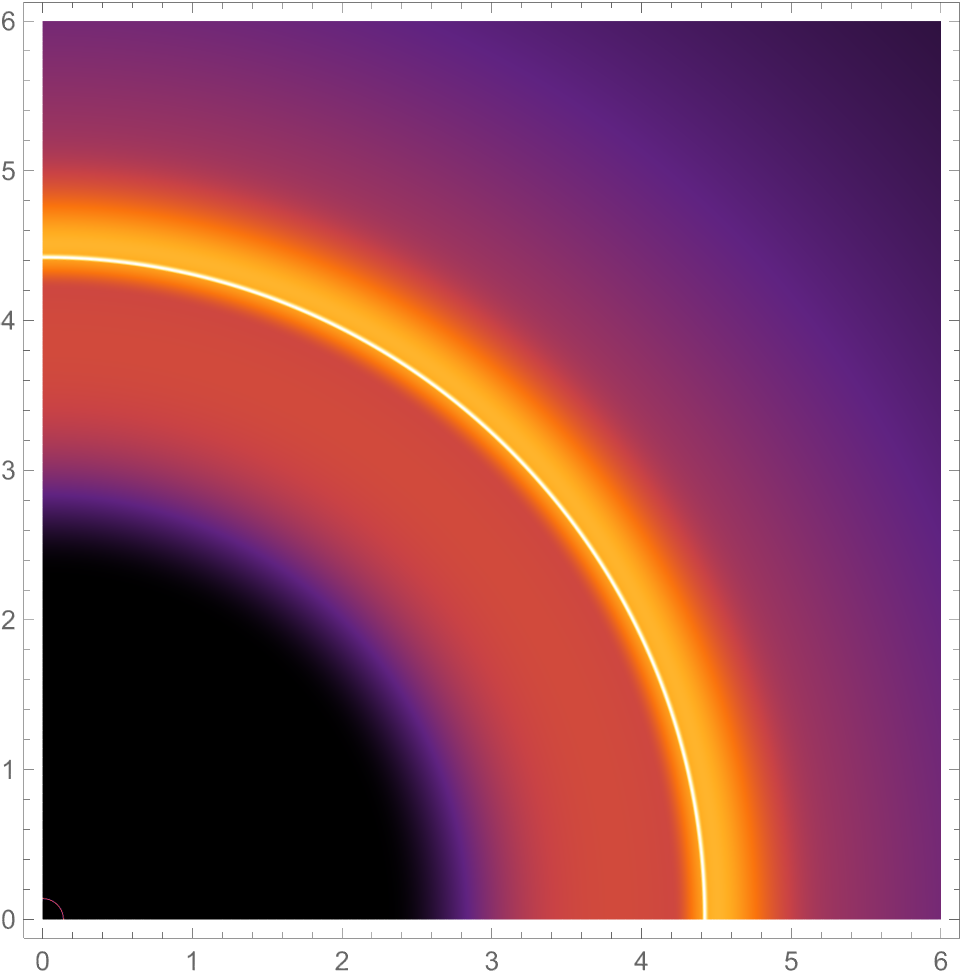}} \\
%\vskip 0.5cm
\subfigure[]{\includegraphics[scale=0.3]{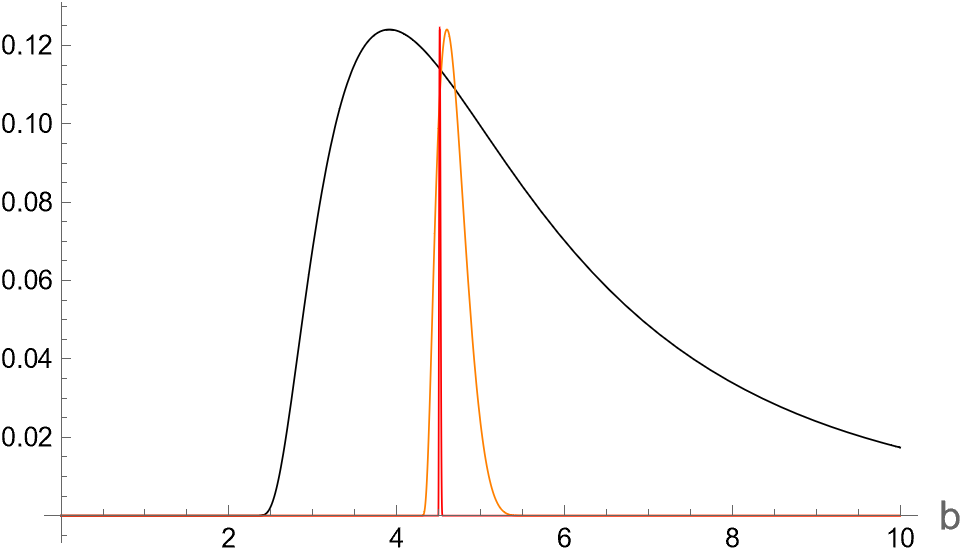}}\hfill
\subfigure[]{\includegraphics[scale=0.3]{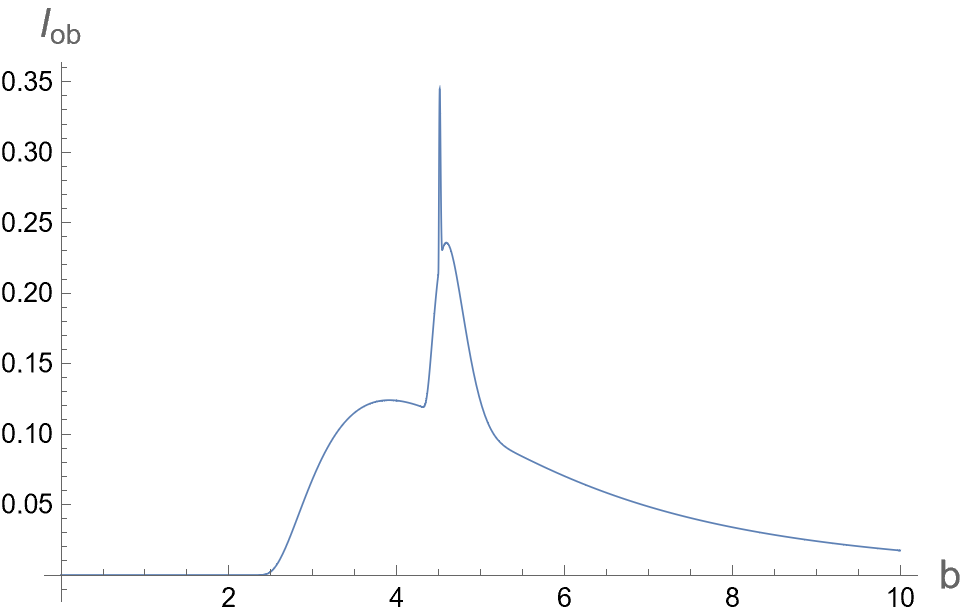}}\hfill
\subfigure[]{\includegraphics[scale=0.31]{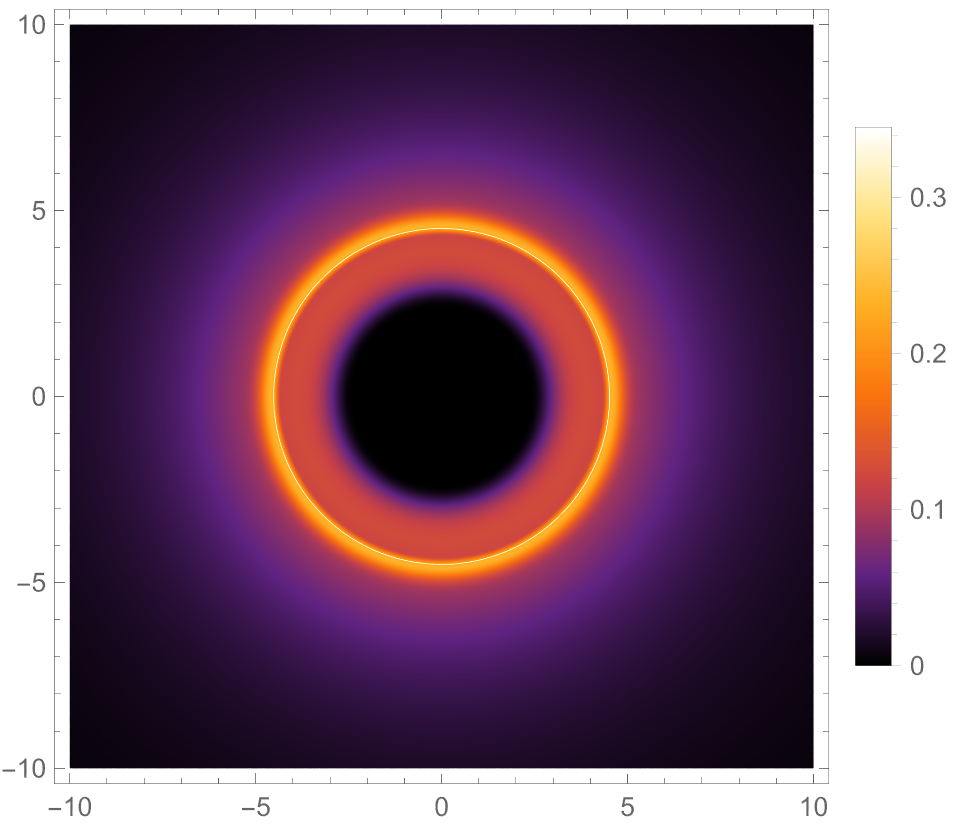}}\hfill
\subfigure[]{\includegraphics[scale=0.25]{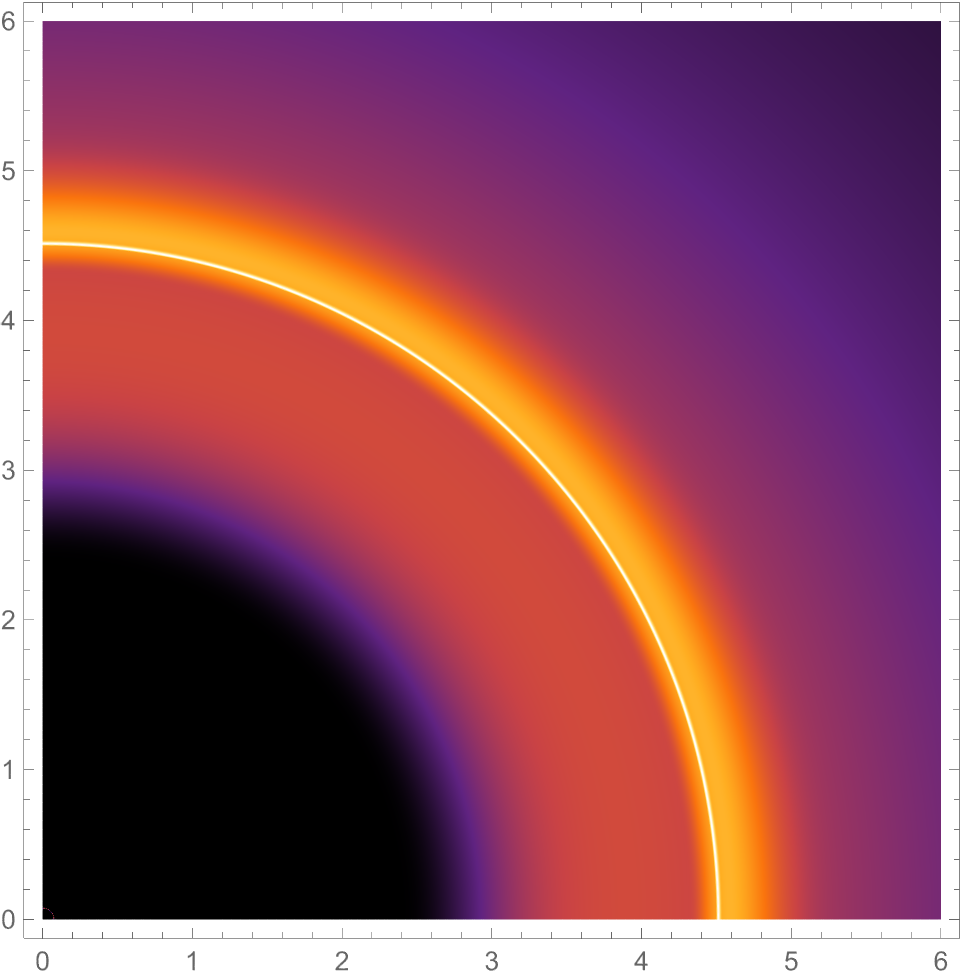}}
\caption{Observational appearance for the \textit{Case III} of GLM profile near the EMD BH with dilaton coupling $\alpha =0.1$ (upper row), $1$ (second row) and $\sqrt{3}$ (third row).
\textit{First column} represents separately the observed intensities of $m=1, 2$ and $3$ in black, orange and red curves, respectively.
\textit{Second column}, we plot the total observed intensities $I_\text{ob}$ against induced impact parameter $b$.
\textit{Third column} is the density plots of $I_\text{ob}$.
\textit{Fourth column} represents zoom in of the third column.
Note that we fix $q=0.9$ in all figures.}
\label{fig: GLM image case 3}
\end{figure}

\begin{figure}[H]
\centering
\subfigure[]{\includegraphics[scale=0.3]{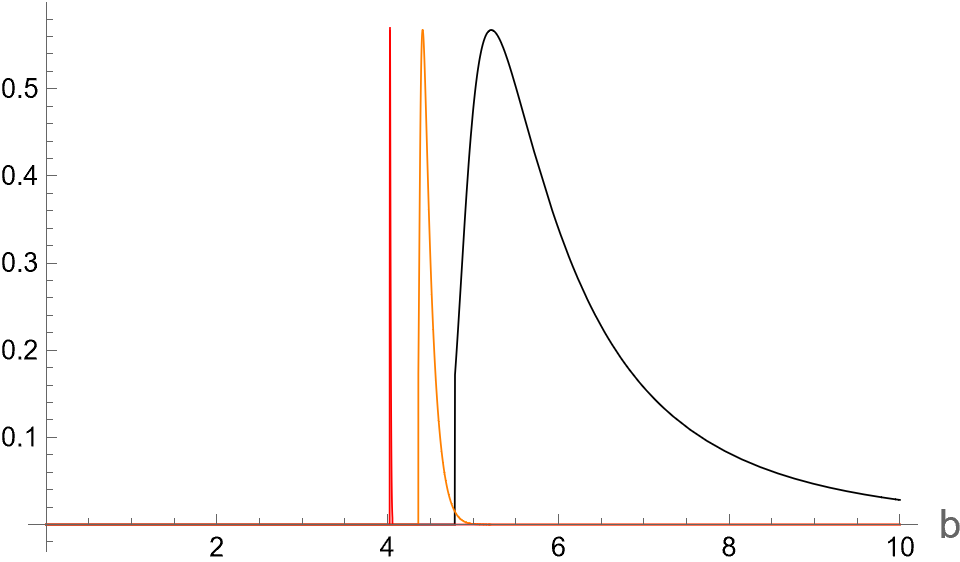}}\hfill
\subfigure[]{\includegraphics[scale=0.3]{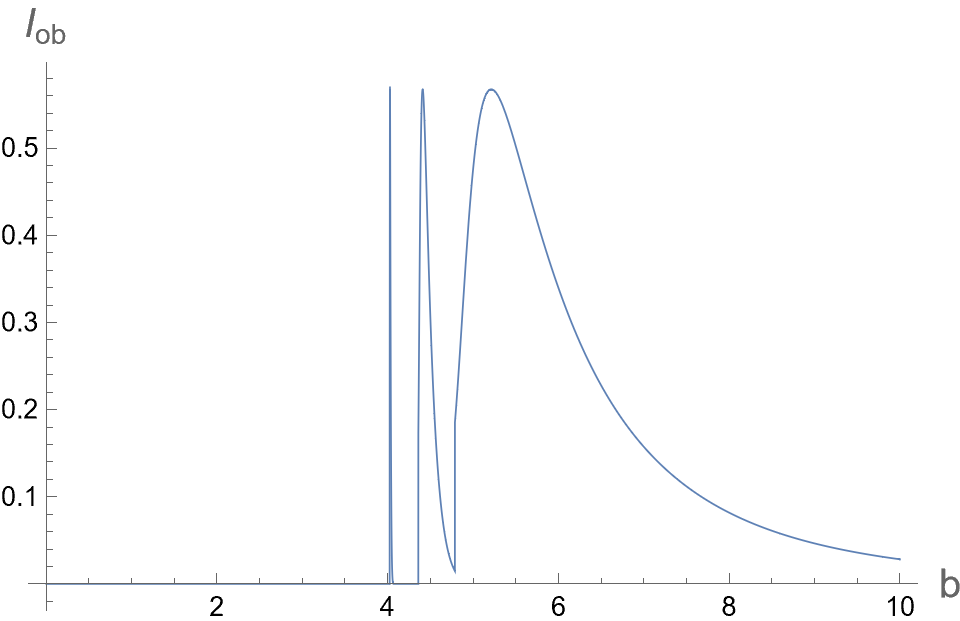}}\hfill
\subfigure[]{\includegraphics[scale=0.31]{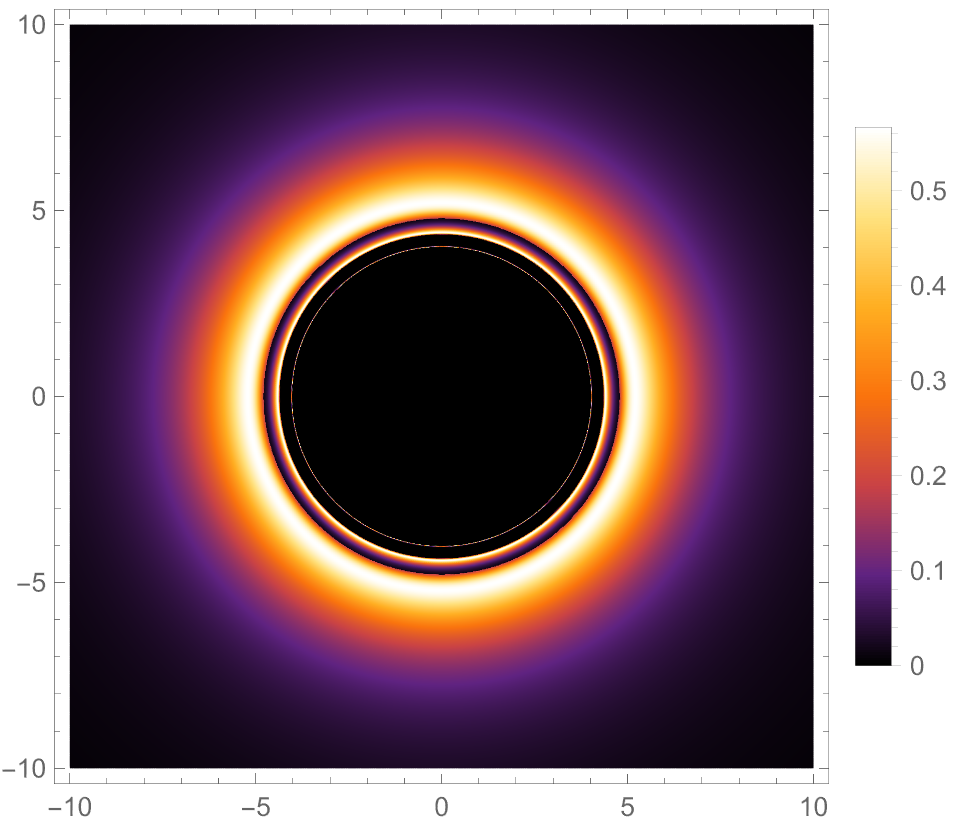}}\hfill
\subfigure[]{\includegraphics[scale=0.25]{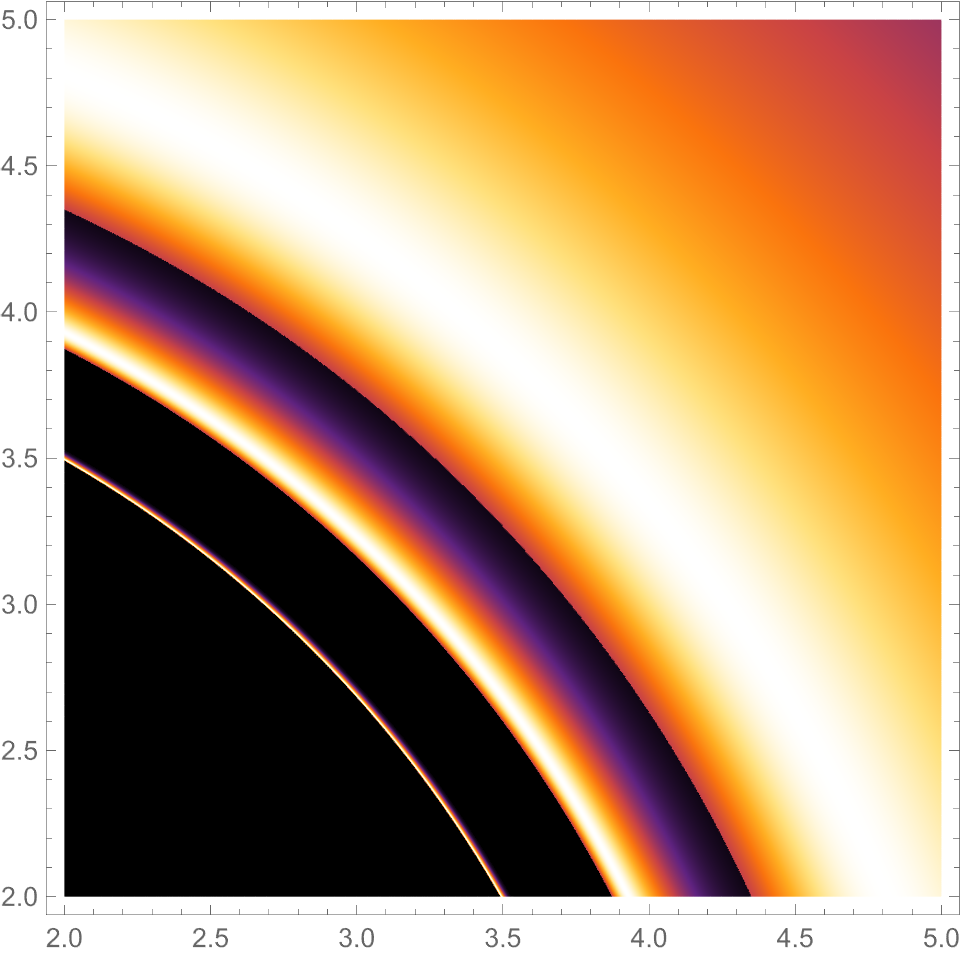}} \\
%\vskip 0.5cm
\subfigure[]{\includegraphics[scale=0.3]{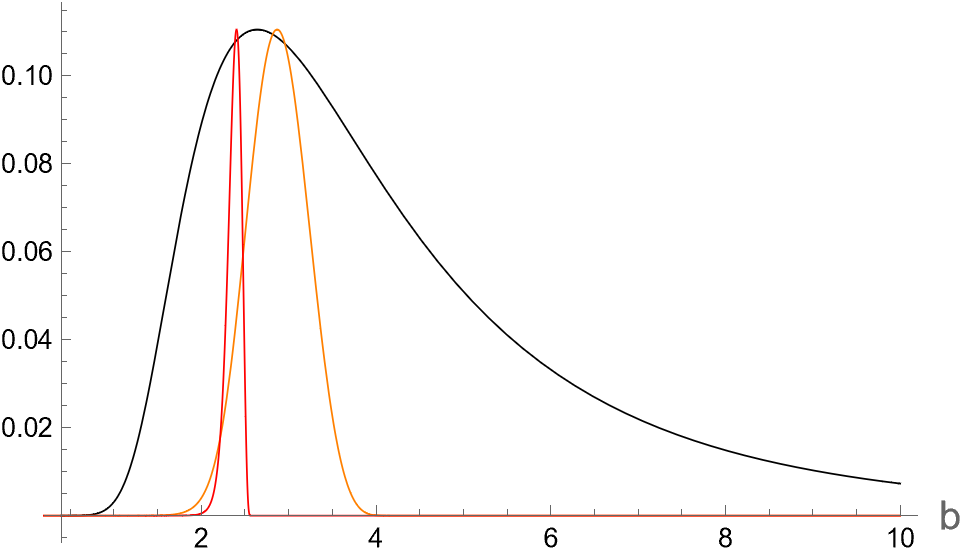}}\hfill
\subfigure[]{\includegraphics[scale=0.3]{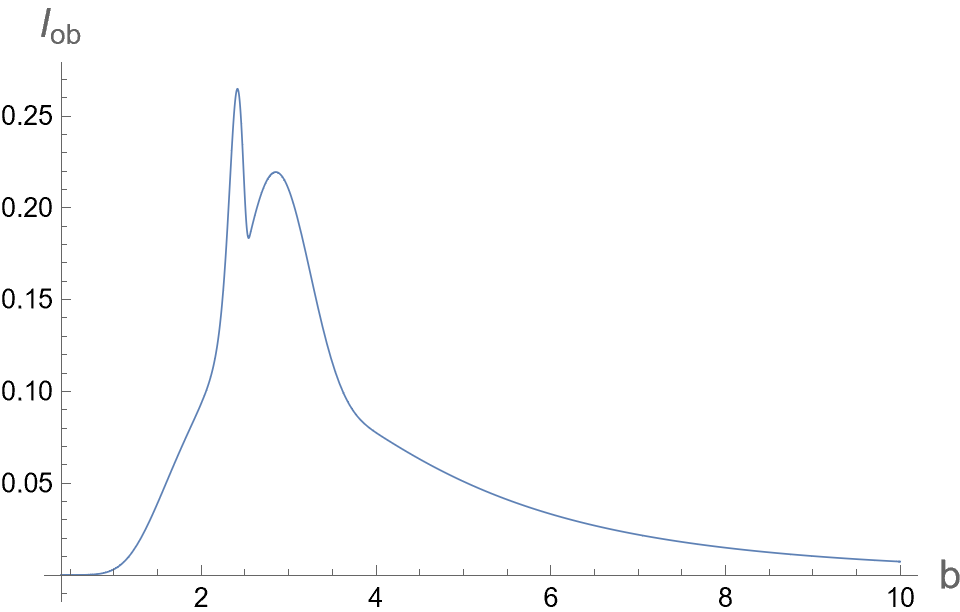}}\hfill
\subfigure[]{\includegraphics[scale=0.31]{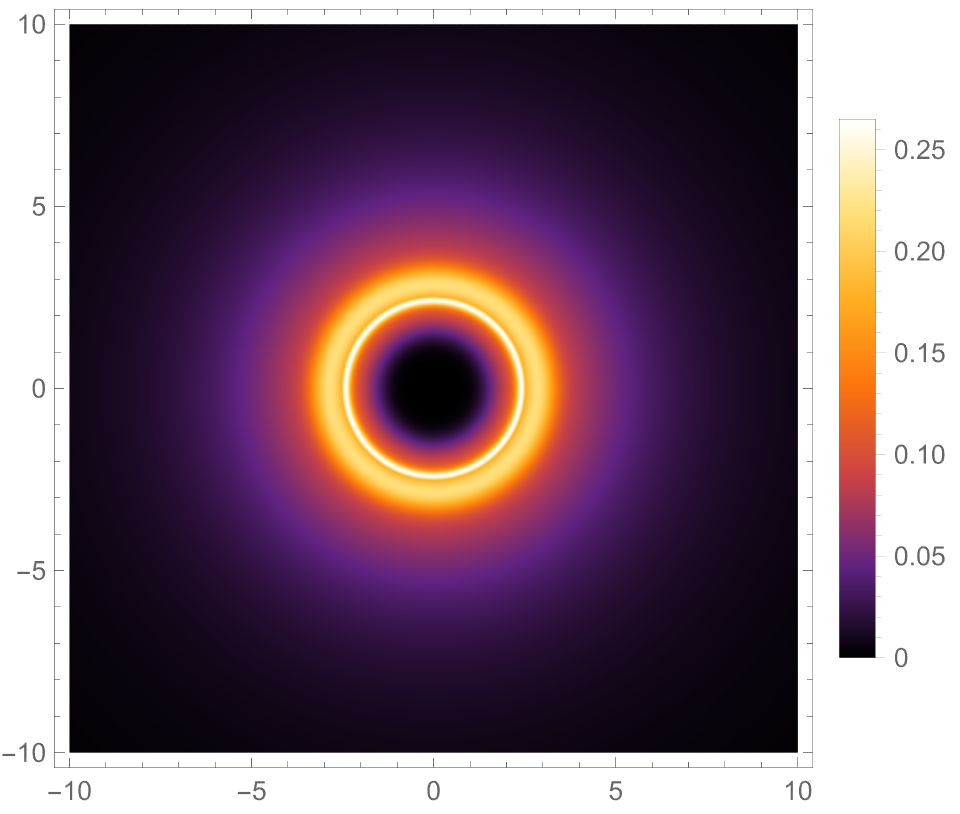}}\hfill
\subfigure[]{\includegraphics[scale=0.25]{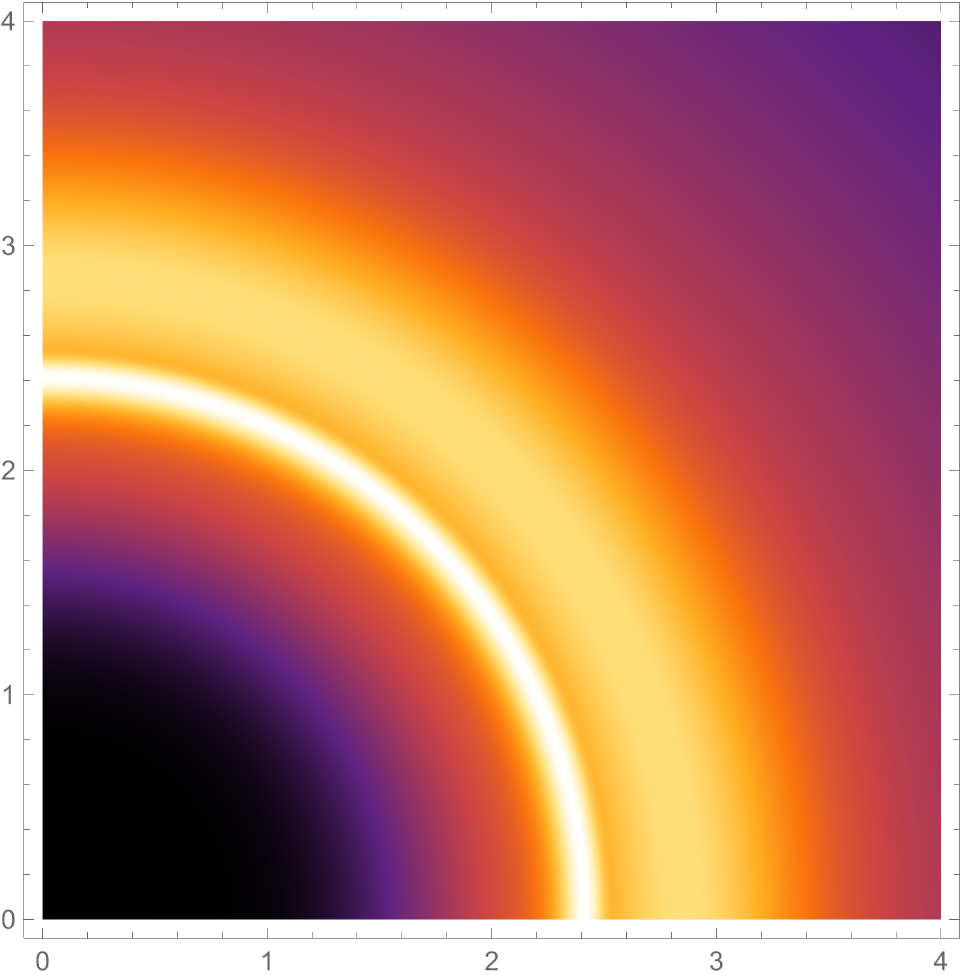}} \\
%\vskip 0.5cm
\subfigure[]{\includegraphics[scale=0.3]{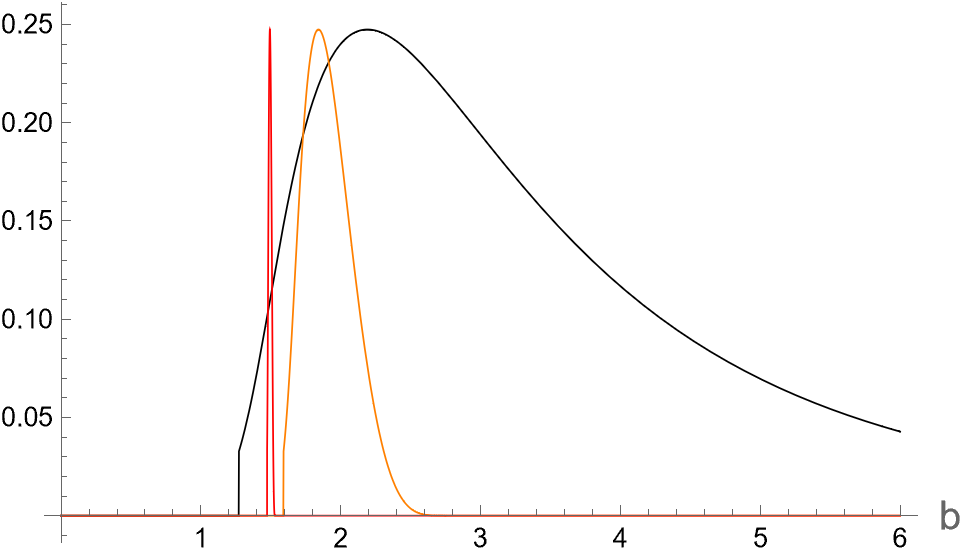}}\hfill
\subfigure[]{\includegraphics[scale=0.3]{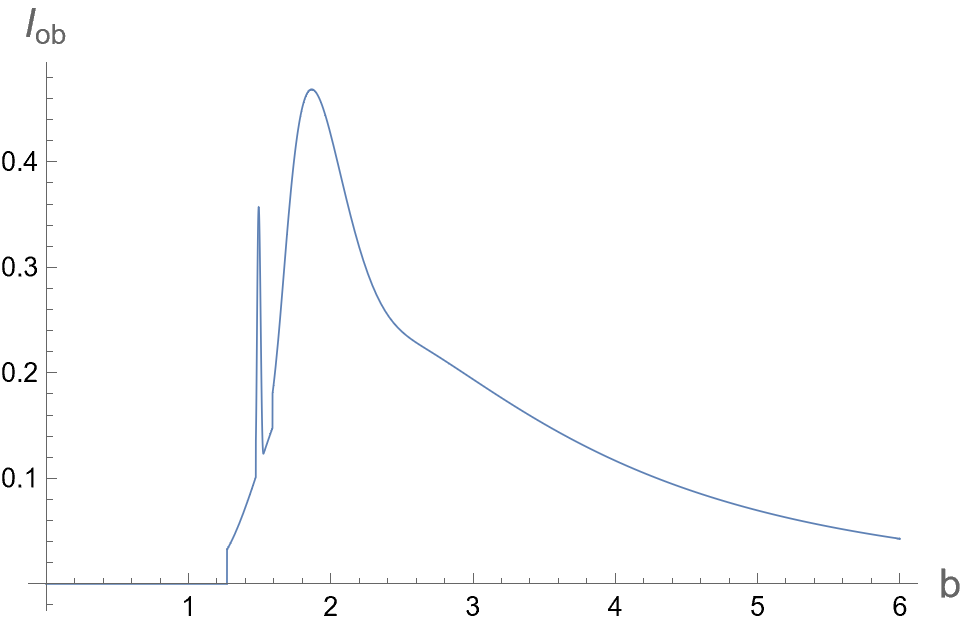}}\hfill
\subfigure[]{\includegraphics[scale=0.31]{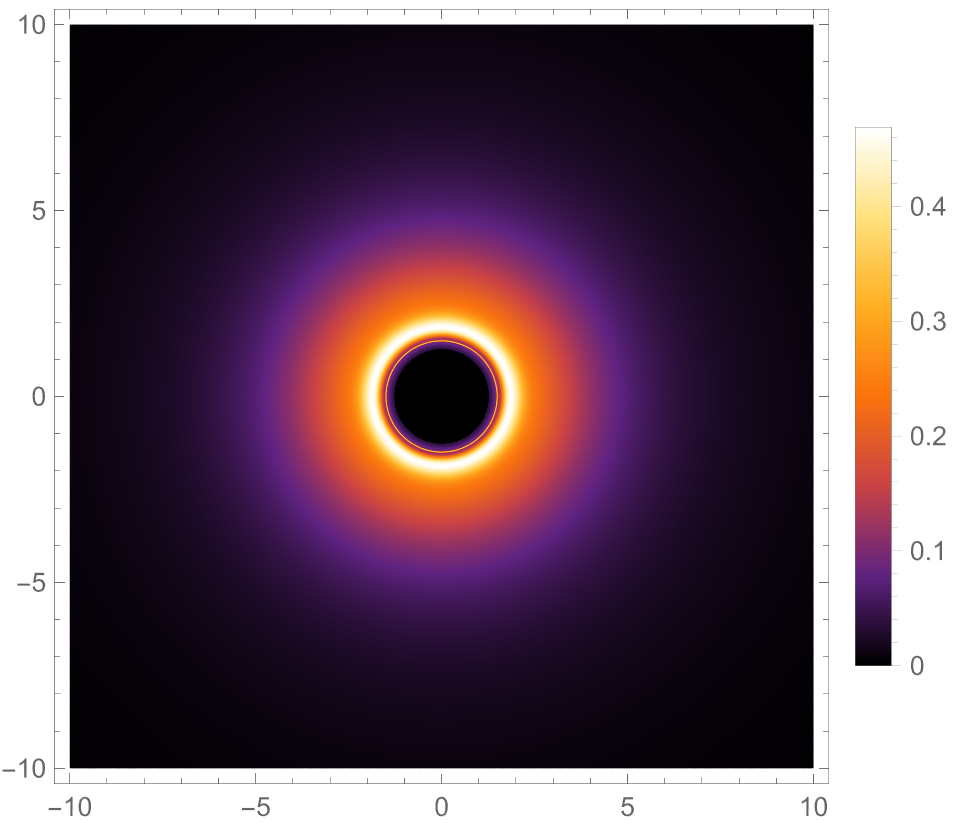}}\hfill
\subfigure[]{\includegraphics[scale=0.25]{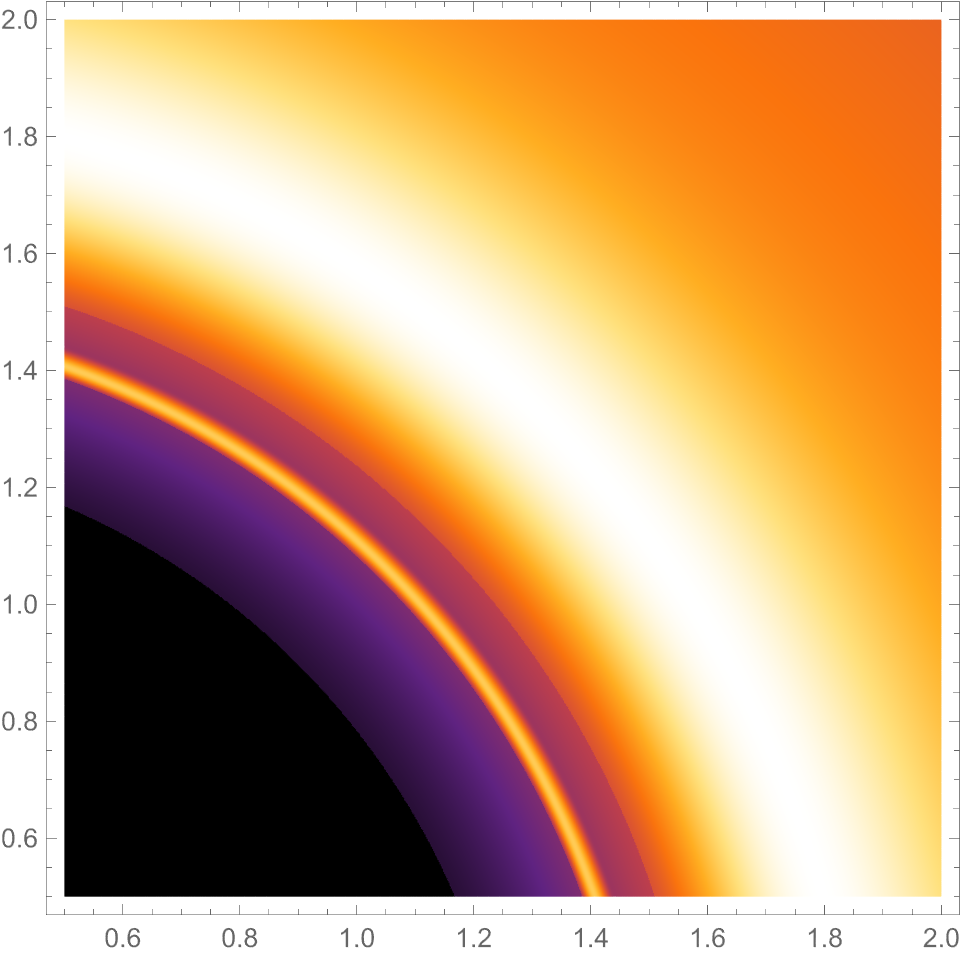}}
\caption{Observational appearance of the \textit{Case I} in the GLM profile for the extremal EMD BHs, i.e., $q_\text{ext}=1.005$ (upper row), $q_\text{ext}=\sqrt{2}$ (second row) and $q=1.9$ (third row) which corresponding to $\alpha  =0.1, 1$ and $\sqrt{3}$, respectively.
Note that we consider near extremal BH in $\alpha =\sqrt{3}$ case.
\textit{First column} represents separately the observed intensities of $m=1, 2$ and $3$ in black, orange and red curves, respectively.
\textit{Second column}, we plot the total observed intensities $I_\text{ob}$ against induced impact parameter $b$.
\textit{Third column} is the density plots of $I_\text{ob}$.
\textit{Fourth column} represents zoom in of the third column.}
\label{fig: GLM image ext case 1}
\end{figure}

\begin{figure}[H]
\centering
\subfigure[]{\includegraphics[scale=0.3]{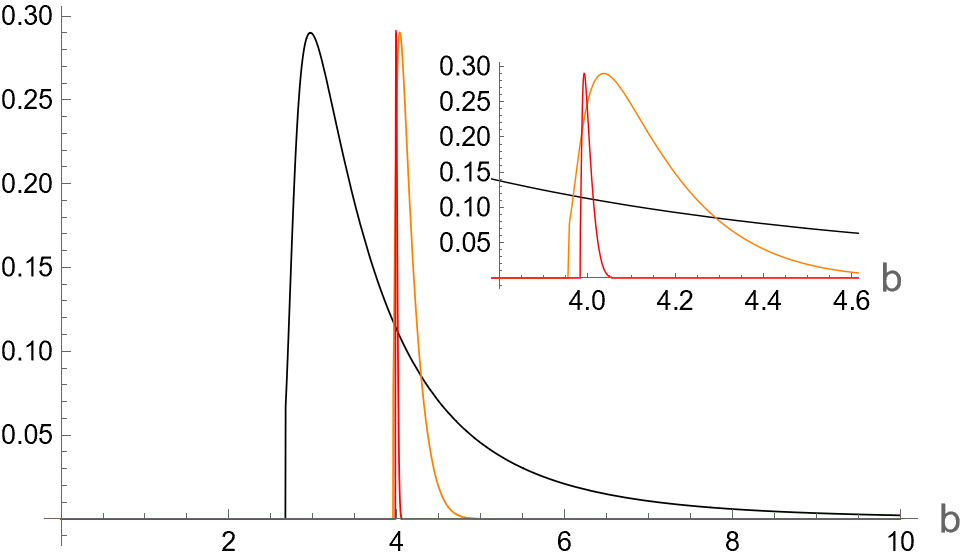}}\hfill
\subfigure[]{\includegraphics[scale=0.3]{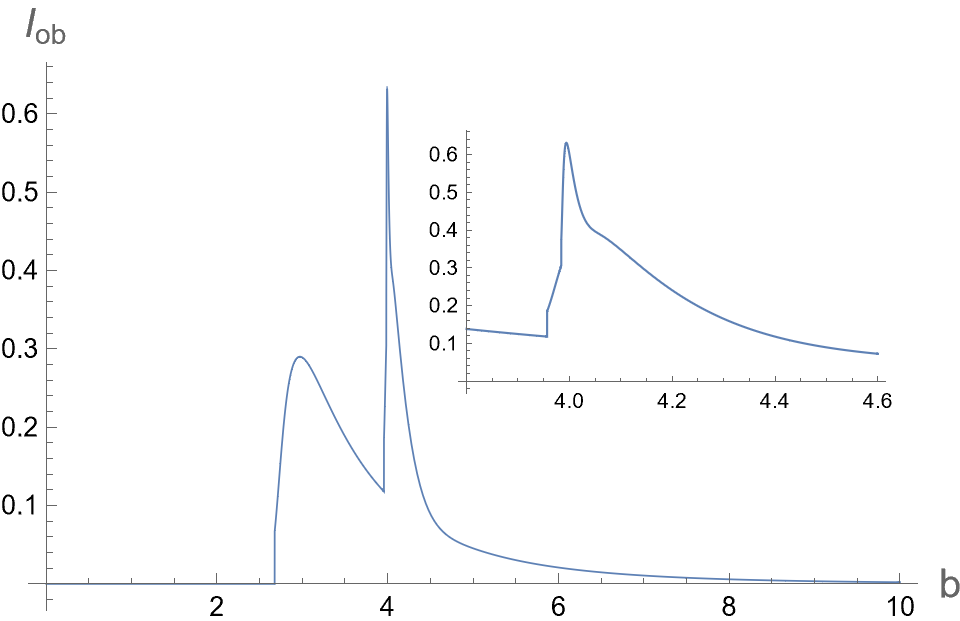}}\hfill
\subfigure[]{\includegraphics[scale=0.31]{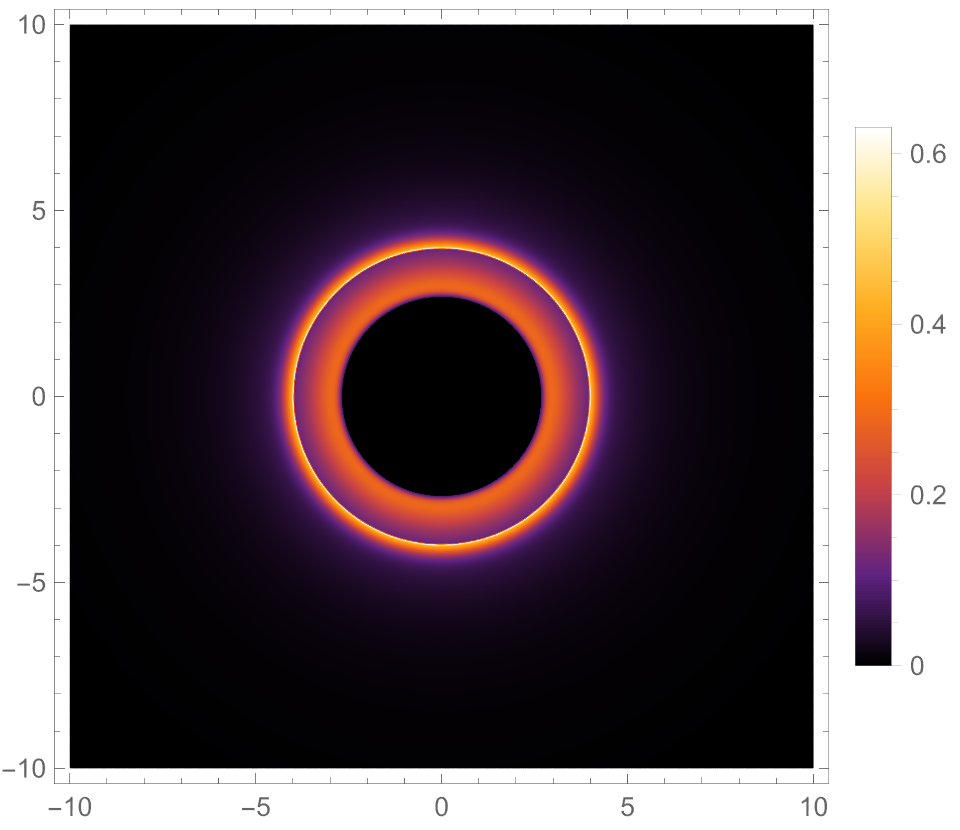}}\hfill
\subfigure[]{\includegraphics[scale=0.25]{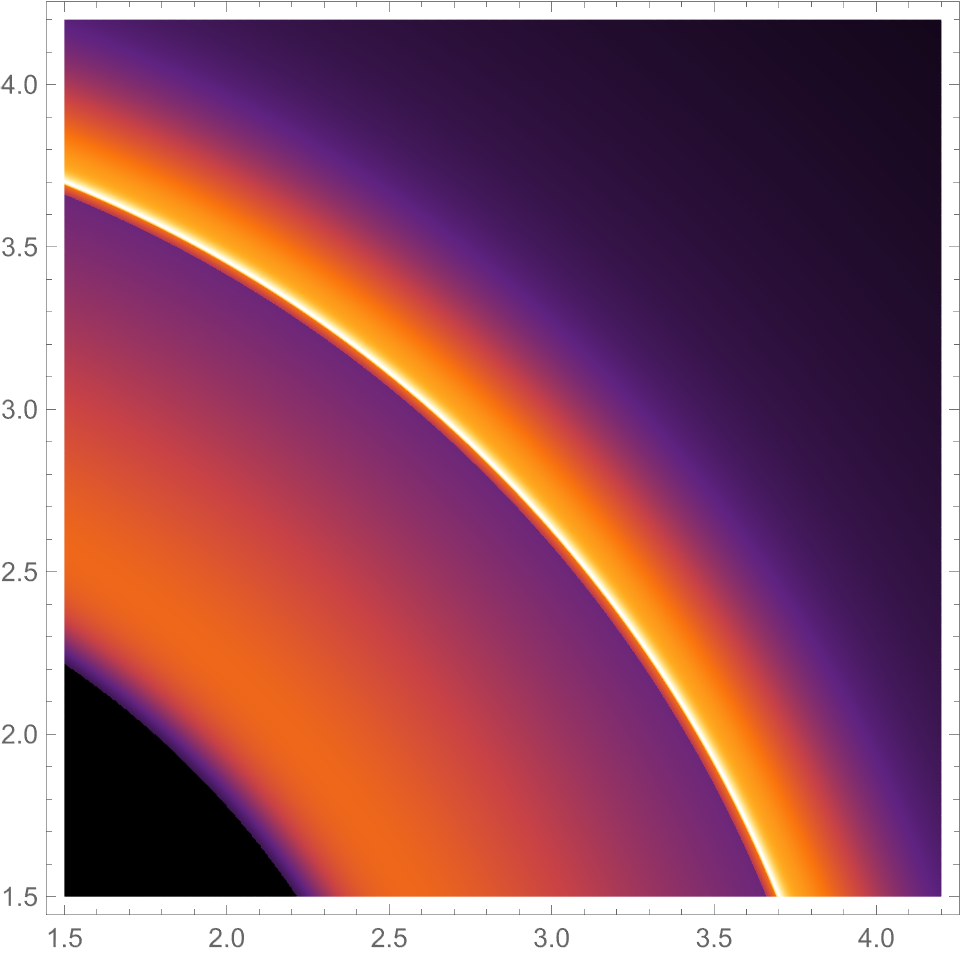}} \\
%\vskip 0.5cm
\subfigure[]{\includegraphics[scale=0.3]{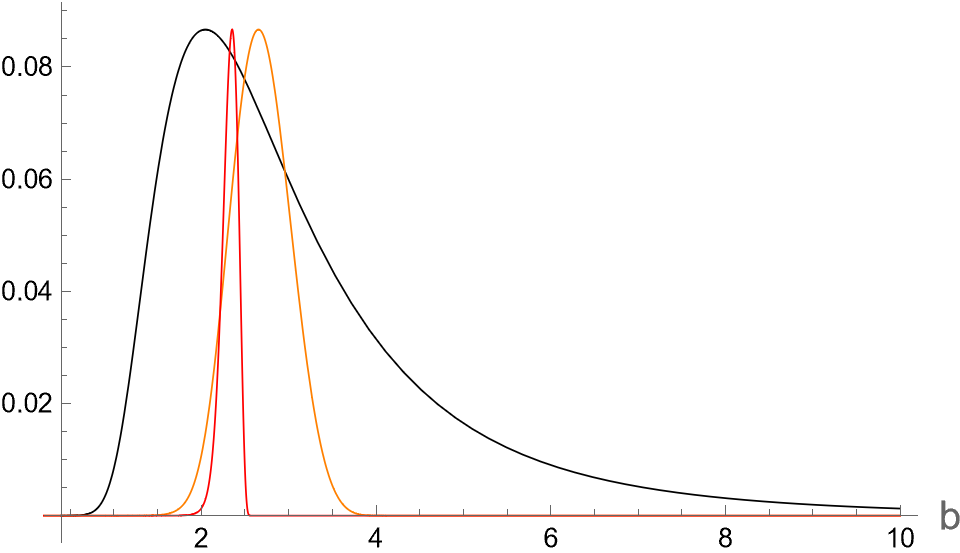}}\hfill
\subfigure[]{\includegraphics[scale=0.3]{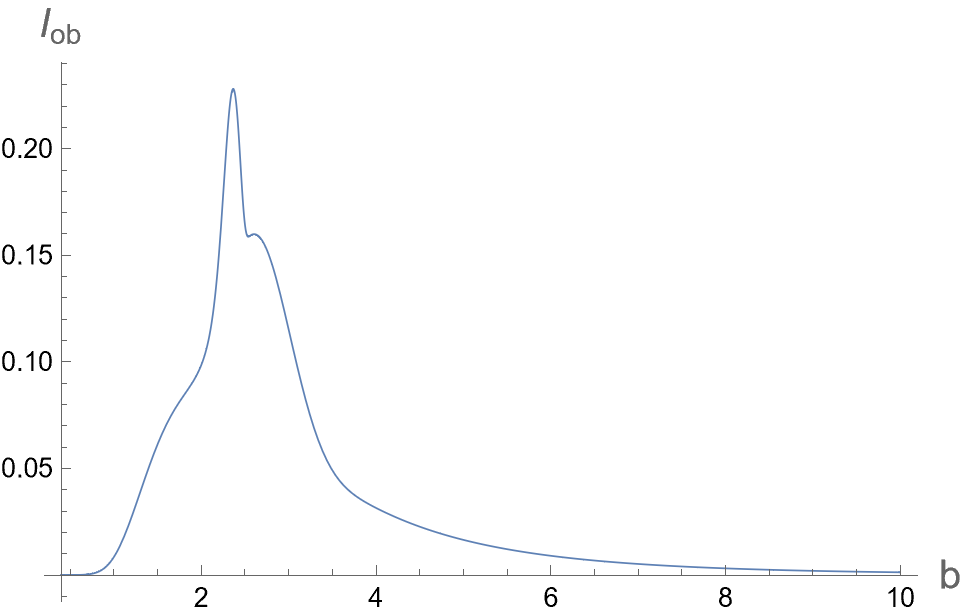}}\hfill
\subfigure[]{\includegraphics[scale=0.31]{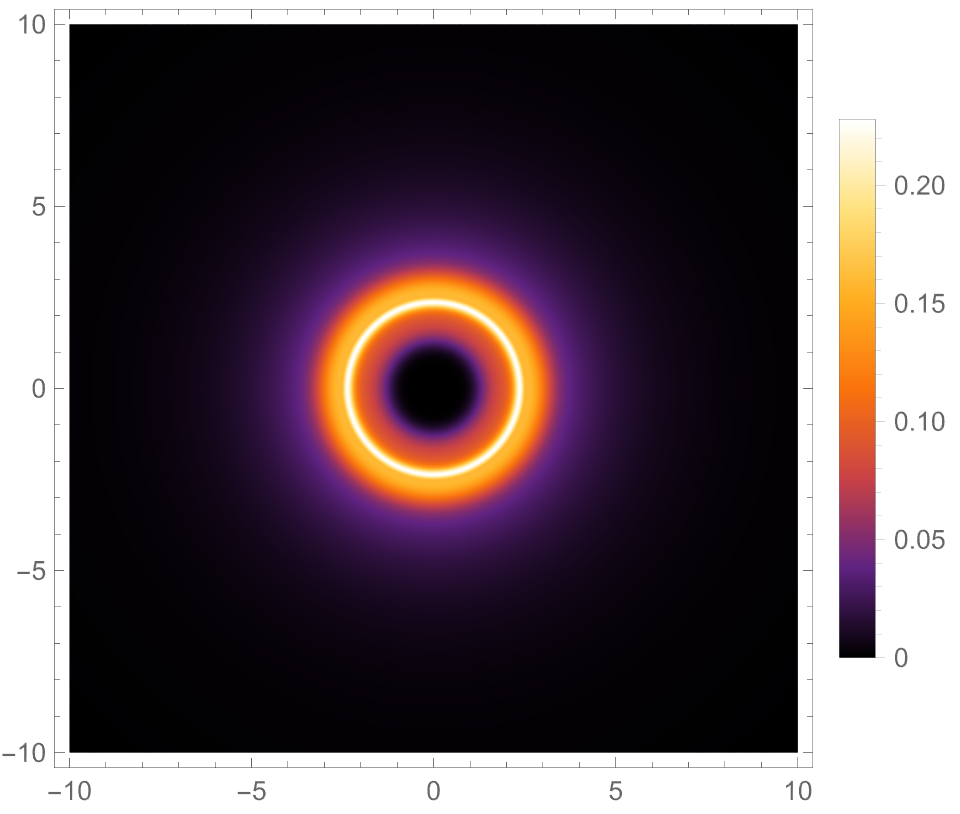}}\hfill
\subfigure[]{\includegraphics[scale=0.25]{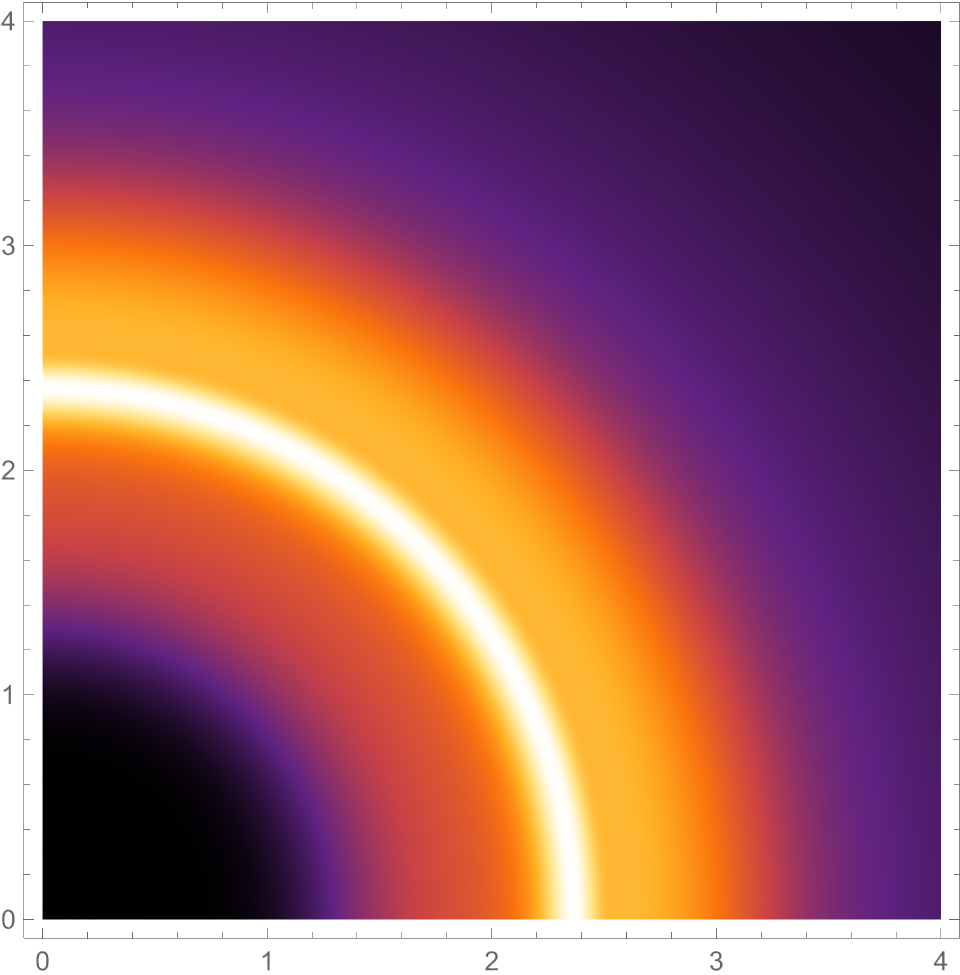}} \\
%\vskip 0.5cm
\subfigure[]{\includegraphics[scale=0.3]{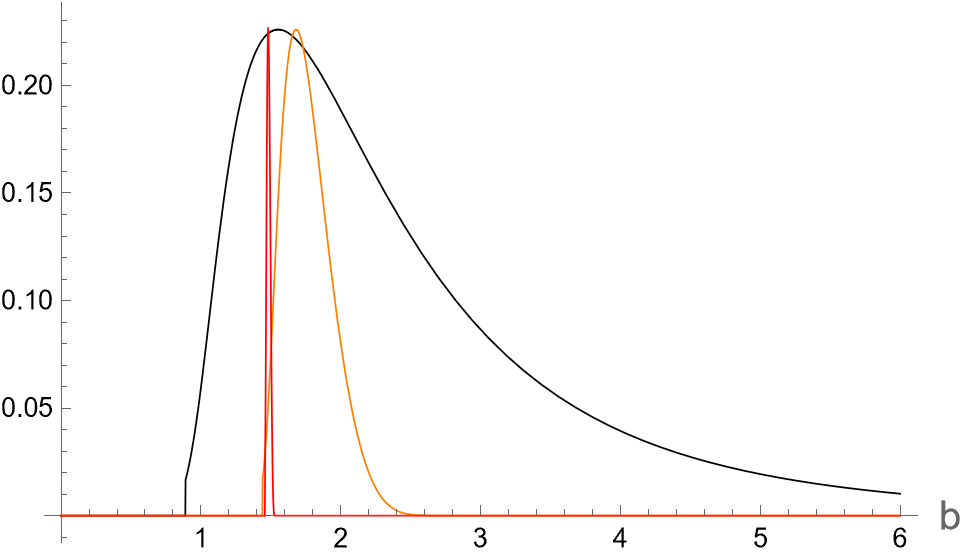}}\hfill
\subfigure[]{\includegraphics[scale=0.3]{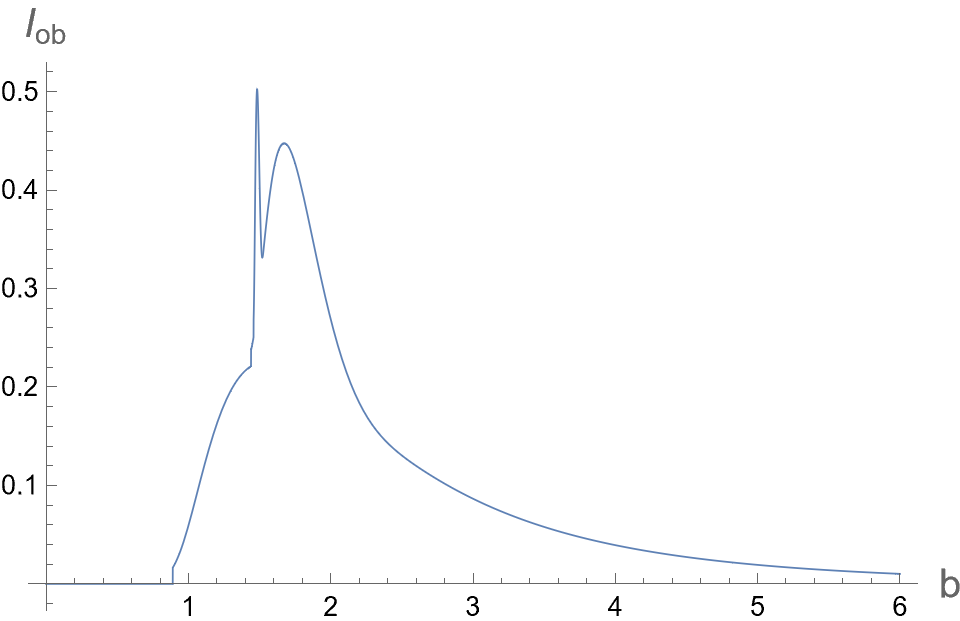}}\hfill
\subfigure[]{\includegraphics[scale=0.31]{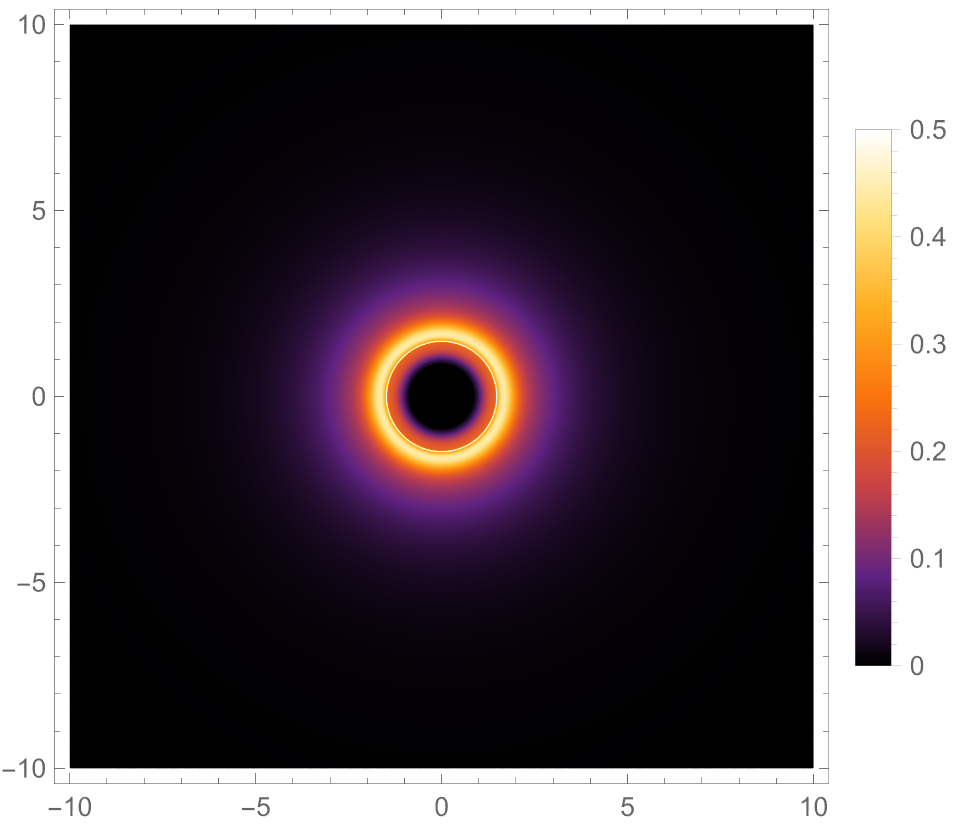}}\hfill
\subfigure[]{\includegraphics[scale=0.25]{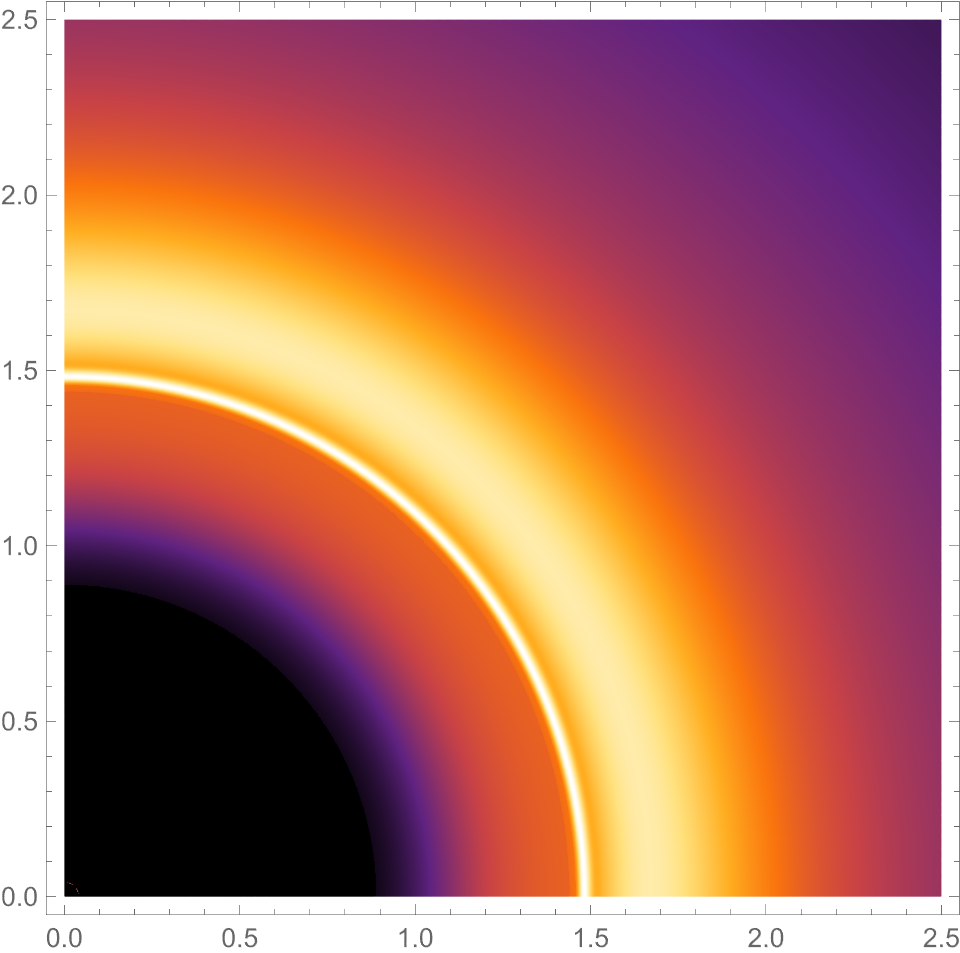}}
\caption{Observational appearance of the \textit{Case II} in the GLM profile for the extremal EMD BHs, i.e., $q_\text{ext}=1.005$ (upper row), $q_\text{ext}=\sqrt{2}$ (second row) and $q=1.9$ (third row) which corresponding to $\alpha  =0.1, 1$ and $\sqrt{3}$, respectively.
Note that we consider near extremal BH in $\alpha =\sqrt{3}$ case.
\textit{First column} represents separately the observed intensities of $m=1, 2$ and $3$ in black, orange and red curves, respectively.
\textit{Second column}, we plot the total observed intensities $I_\text{ob}$ against induced impact parameter $b$.
\textit{Third column} is the density plots of $I_\text{ob}$.
\textit{Fourth column} represents zoom in of the third column.}
\label{fig: GLM image ext case 2}
\end{figure}

\begin{figure}[H]
\centering
\subfigure[]{\includegraphics[scale=0.3]{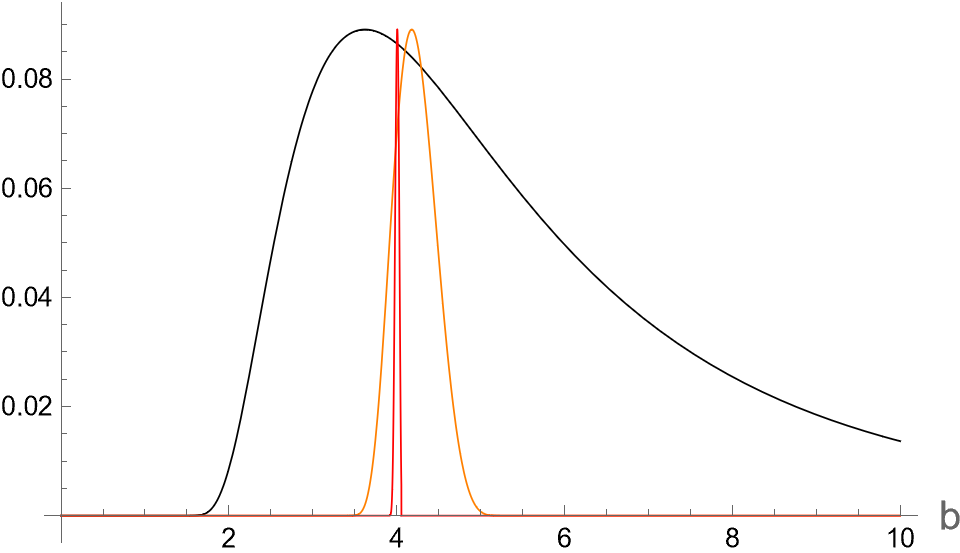}}\hfill
\subfigure[]{\includegraphics[scale=0.3]{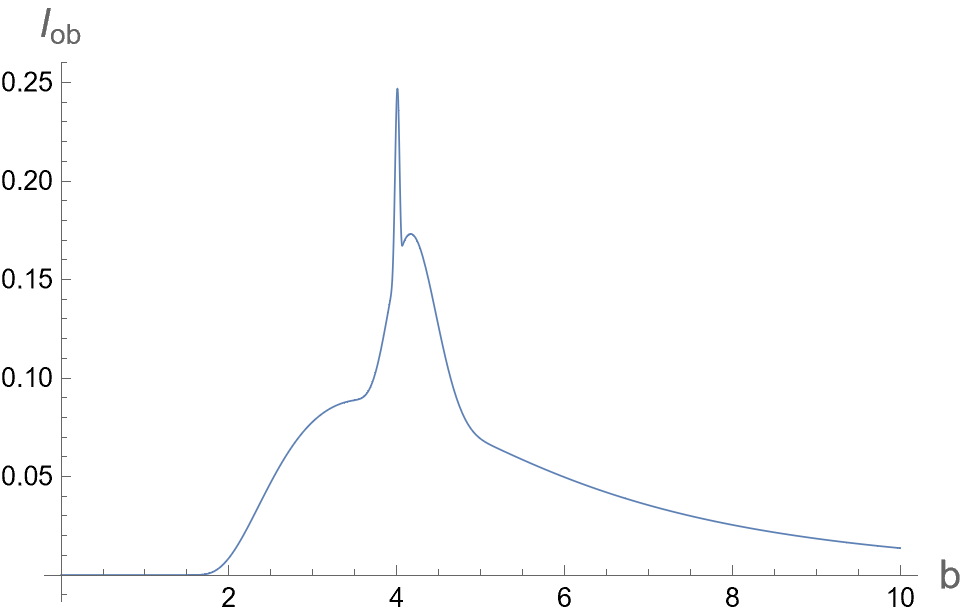}}\hfill
\subfigure[]{\includegraphics[scale=0.31]{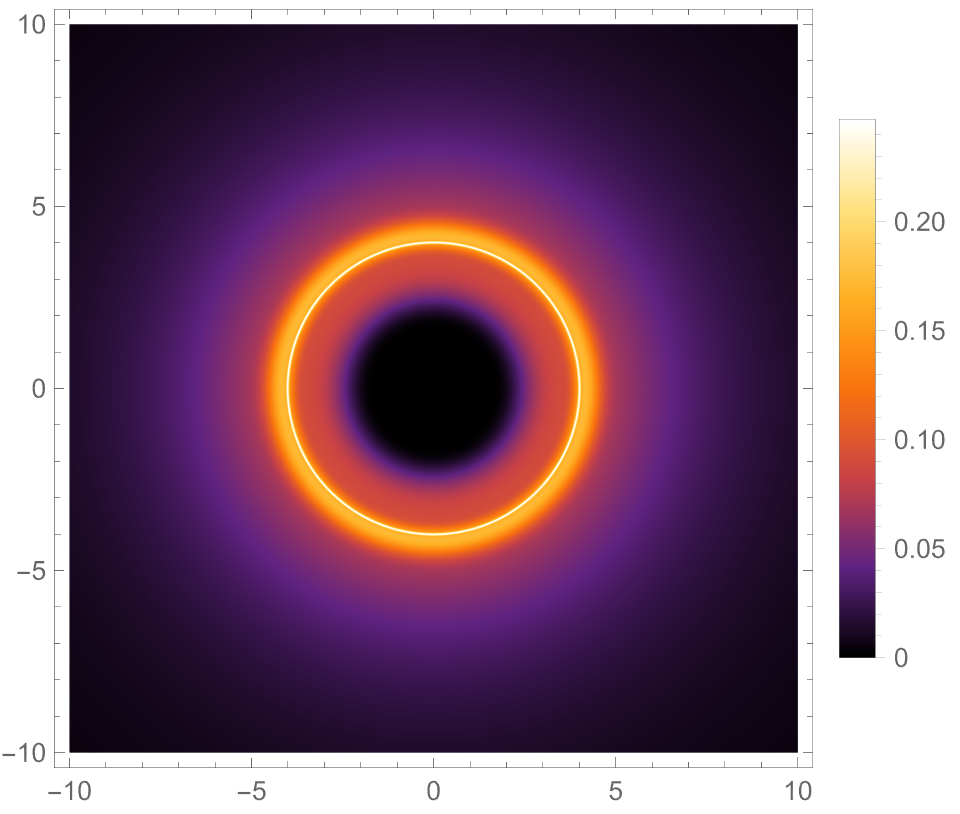}}\hfill
\subfigure[]{\includegraphics[scale=0.25]{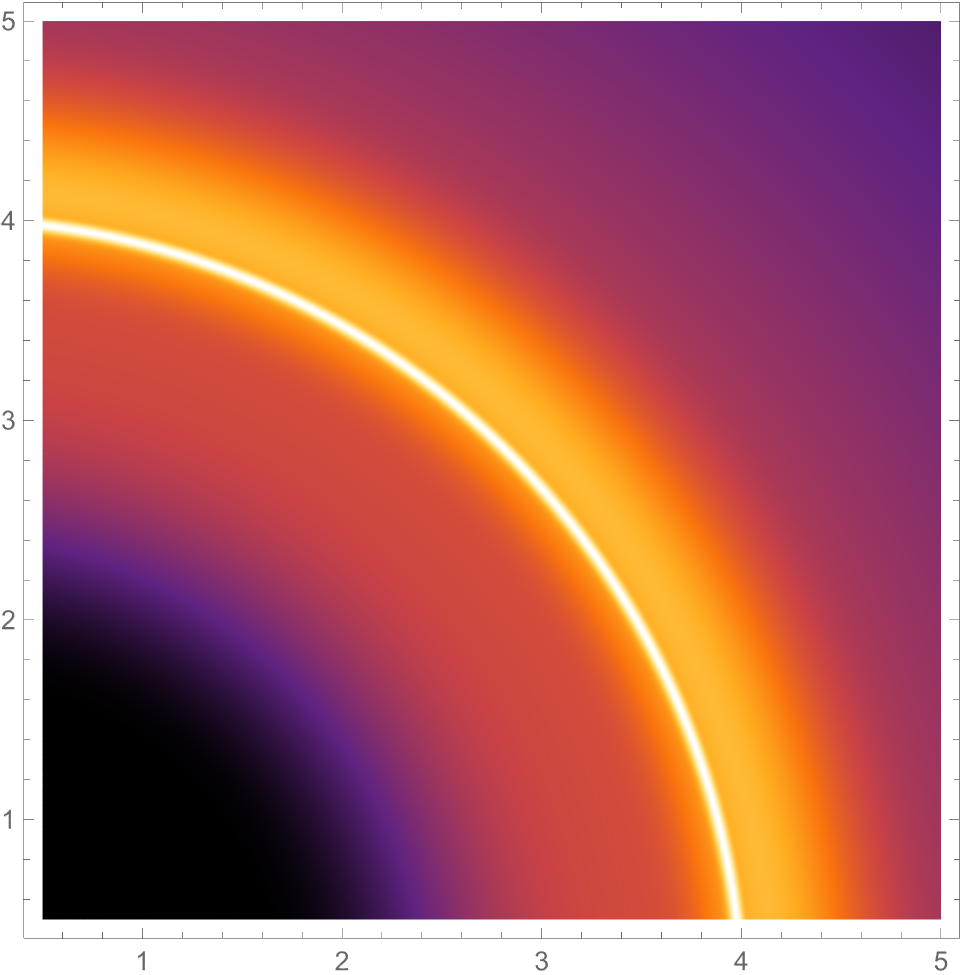}} \\
%\vskip 0.5cm
\subfigure[]{\includegraphics[scale=0.3]{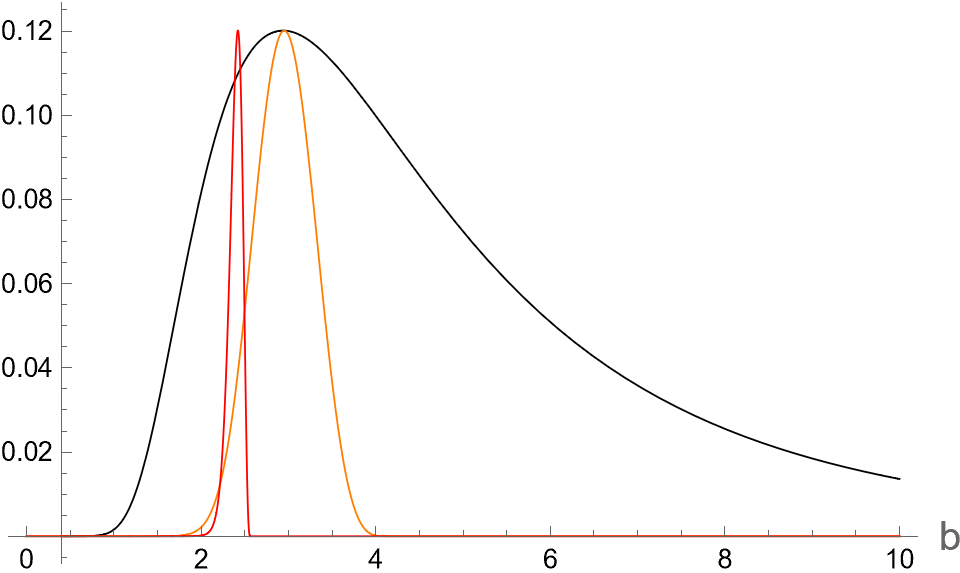}}\hfill
\subfigure[]{\includegraphics[scale=0.3]{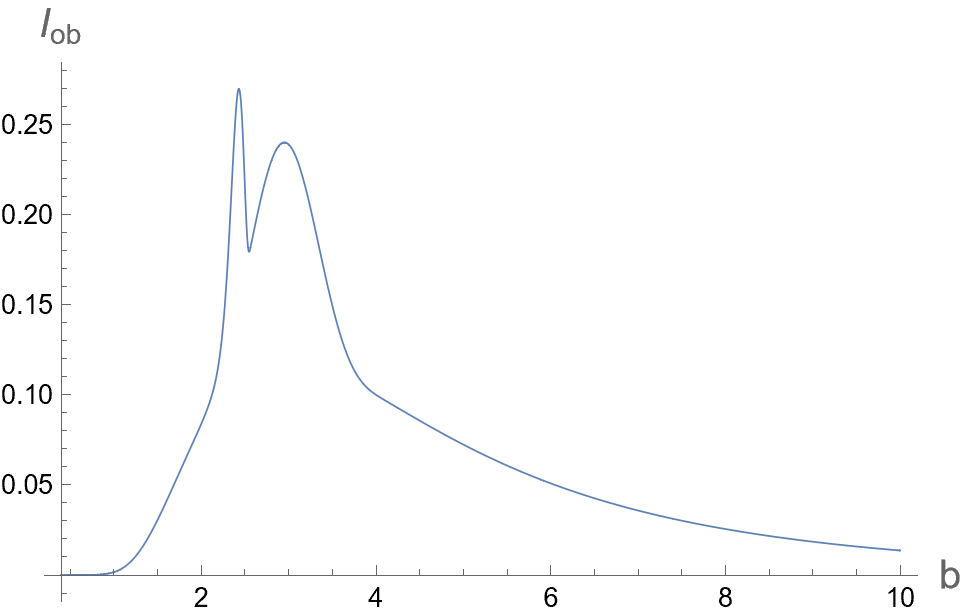}}\hfill
\subfigure[]{\includegraphics[scale=0.31]{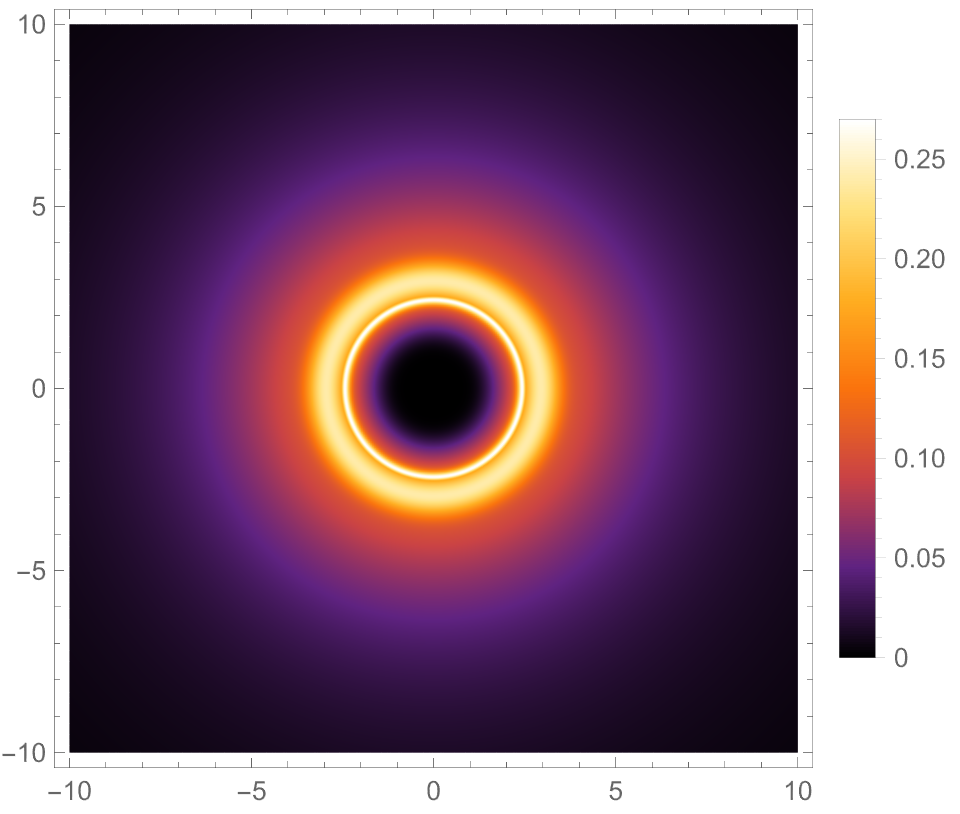}}\hfill
\subfigure[]{\includegraphics[scale=0.25]{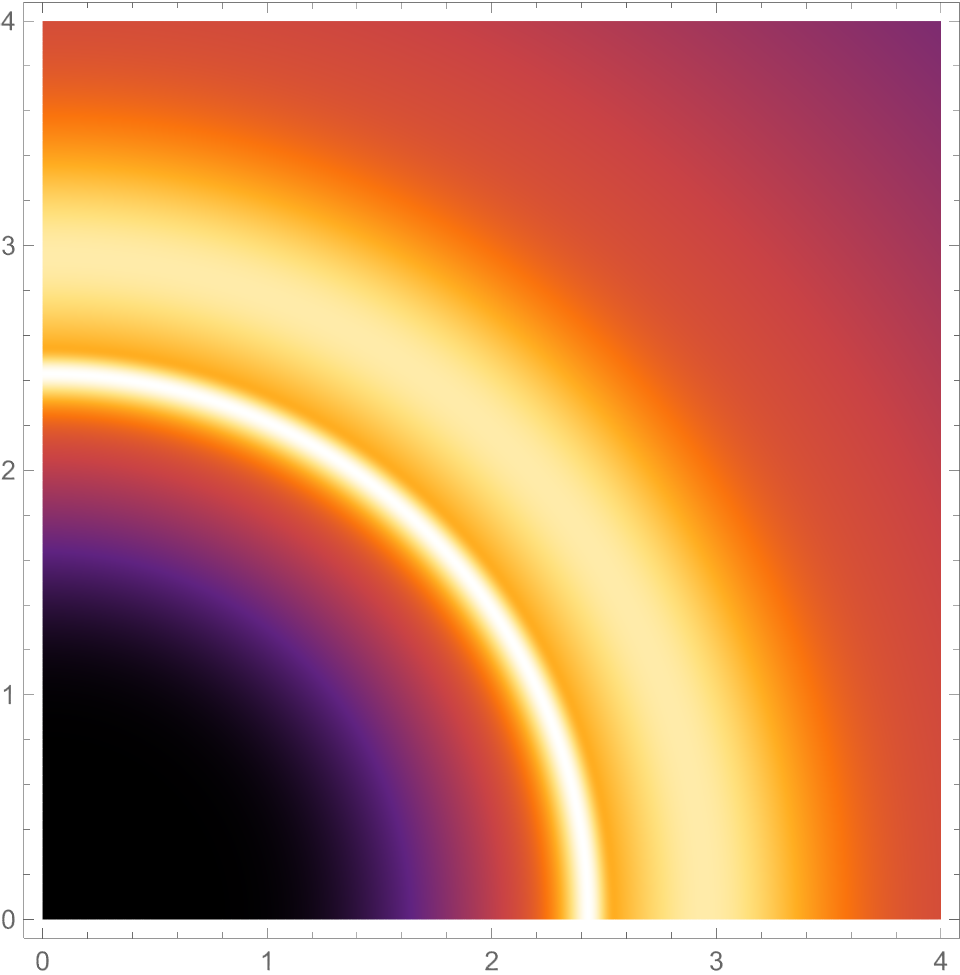}} \\
%\vskip 0.5cm
\subfigure[]{\includegraphics[scale=0.3]{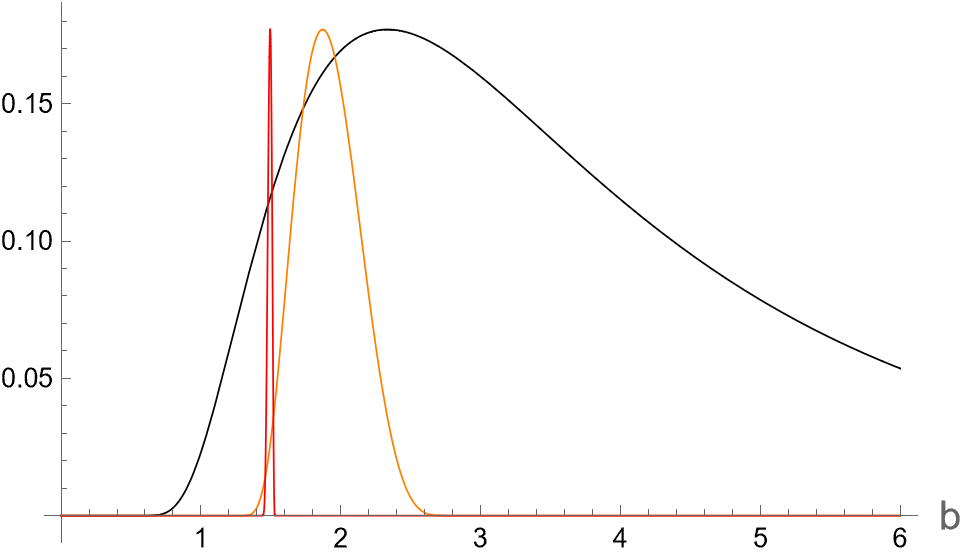}}\hfill
\subfigure[]{\includegraphics[scale=0.3]{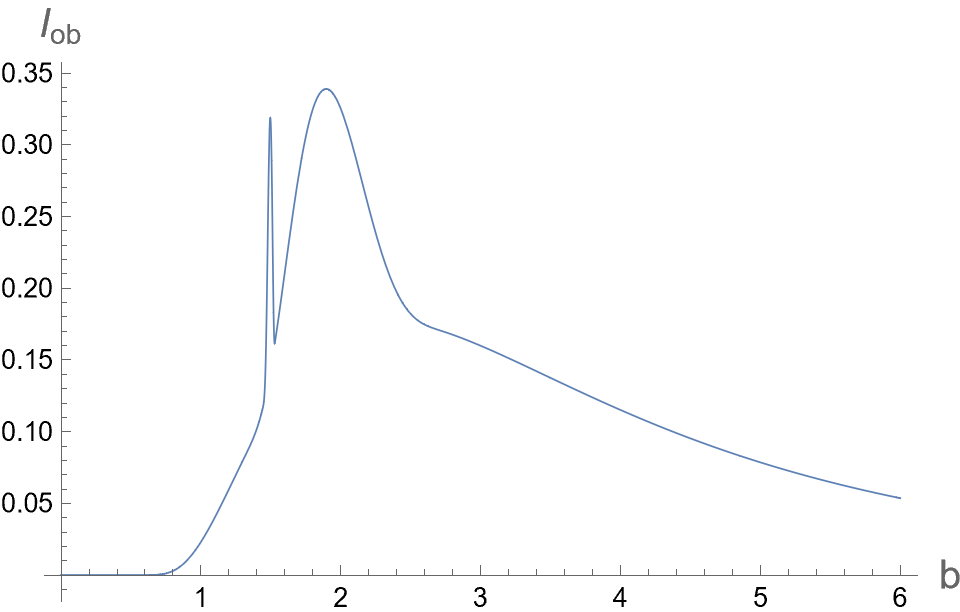}}\hfill
\subfigure[]{\includegraphics[scale=0.31]{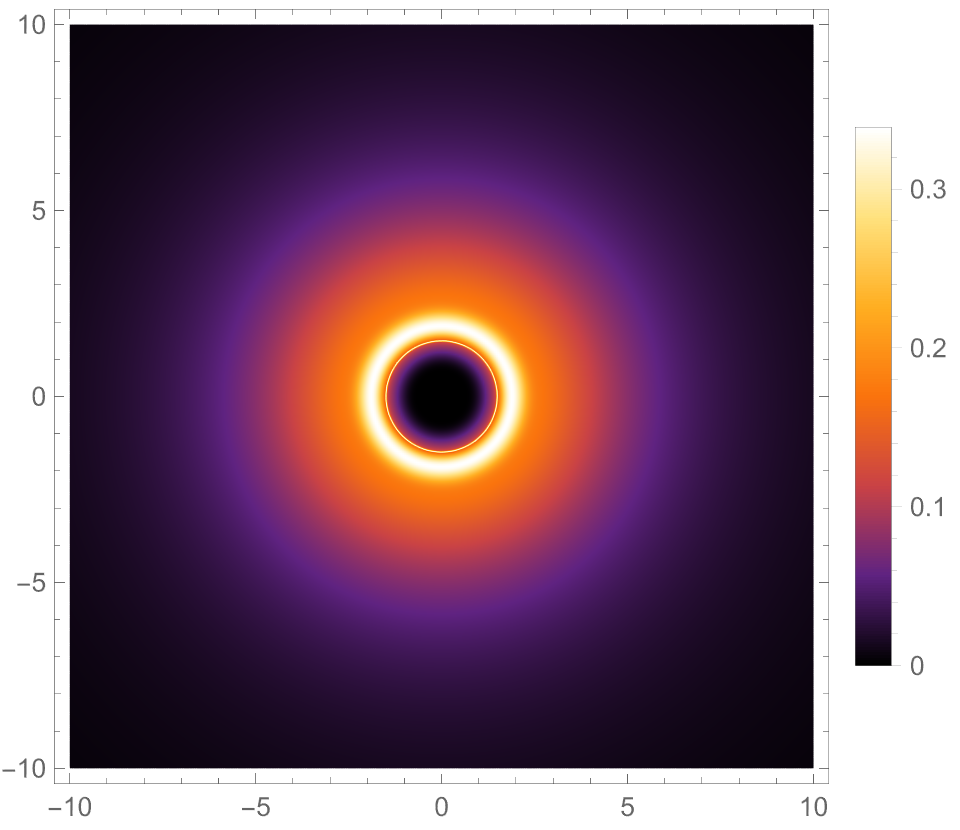}}\hfill
\subfigure[]{\includegraphics[scale=0.25]{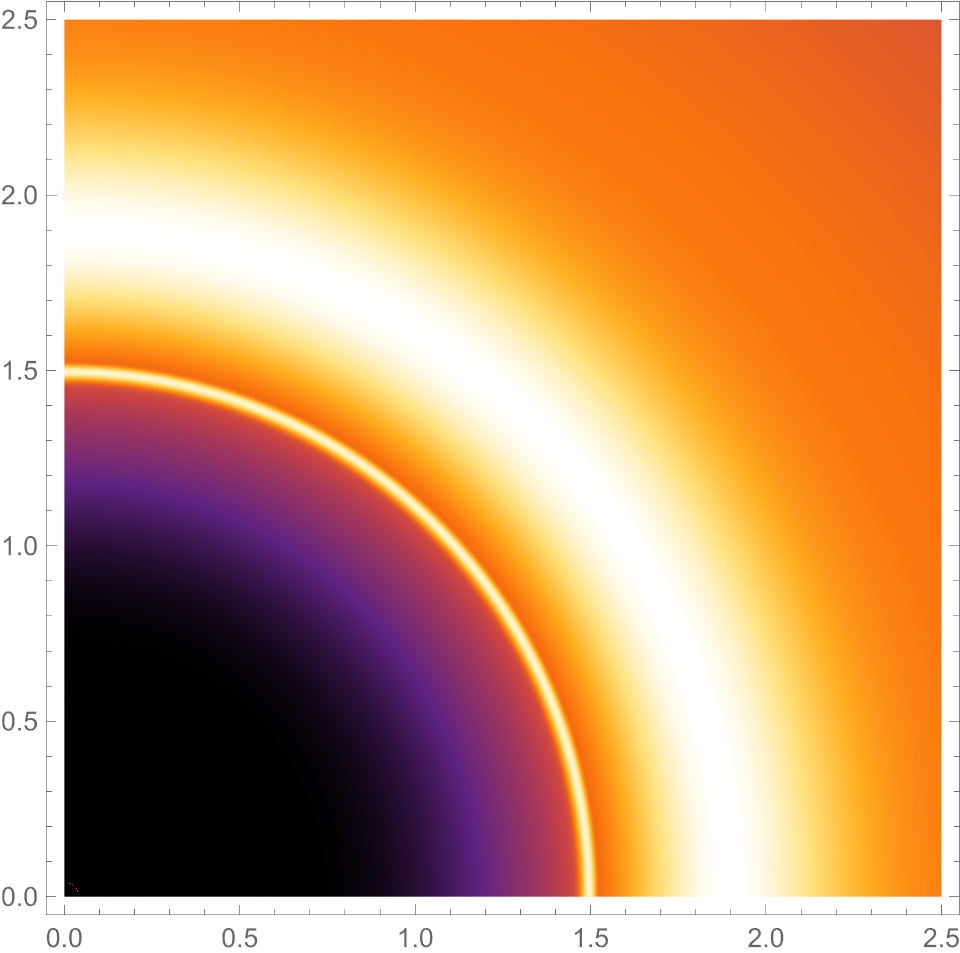}}
\caption{Observational appearance of the \textit{Case III} in the GLM profile for the extremal EMD BHs, i.e., $q_\text{ext}=1.005$ (upper row), $q_\text{ext}=\sqrt{2}$ (second row) and $q=1.9$ (third row) which corresponding to $\alpha  =0.1, 1$ and $\sqrt{3}$, respectively.
Note that we consider near extremal BH in $\alpha =\sqrt{3}$ case.
\textit{First column} represents separately the observed intensities of $m=1, 2$ and $3$ in black, orange and red curves, respectively.
\textit{Second column}, we plot the total observed intensities $I_\text{ob}$ against induced impact parameter $b$.
\textit{Third column} is the density plots of $I_\text{ob}$.
\textit{Fourth column} represents zoom in of the third column.}
\label{fig: GLM image ext case 3}
\end{figure}

In Figs.~\ref{fig: GLM image case 1}, \ref{fig: GLM image case 2} and \ref{fig: GLM image case 3}, we show the observational appearance of the region near the EMD BH for thin accretion disk in \textit{Case I, II} and \textit{III}, respectively.
In these figures, we fix $q=0.9$ with $\alpha =0.1$ (upper row), $1$ (second row) and $\sqrt{3}$ (third row).
The first column distinctly shows the observed intensities arising from the first, second, and third transfer functions, presented in black, orange, and red curves, respectively.
The second column corresponds to the total observed intensities  $I_\text{ob}$ as a function of  impact parameter $b$.
In the third column, we show the density plot of $I_\text{ob}$ in the plane.
The zoom-in of the density plot sector is shown in the fourth column.
Figs.~\ref{fig: GLM image ext case 1}, \ref{fig: GLM image ext case 2} and \ref{fig: GLM image ext case 3} illustrate the images in the large $q$ regime, near or equal to the extremal limit, of EMD BH for \textit{Case I, II} and \textit{III}, respectively.
As the angular Lyapunov exponent $\gamma$ decreases with an increase in $q$ while keeping $\alpha$ constant, the visibility of the photon ring improves with increasing $q$. 
Conversely, when $q$ remains fixed and $\alpha$ is varied, an increase in $\gamma$ results in a reduction in the thickness of the photon ring. This decrease in thickness poses challenges for the observation and identification of the photon ring in the images of BH.

In \textit{Case I}, the emission profile starts near the radius of ISCO and then monotonically decays to zero at infinity. See the red solid curves in Fig.~\ref{fig: GLM profiles}. 
As seen in the Tables \ref{table: alpha 0.1}, \ref{table: alpha 1} and \ref{table: alpha root 3}, the emission region lies outside the photon sphere, namely $r_\text{ISCO} > r_\text{ph}$, for coupling $\alpha =0.1, 1$ and $\sqrt{3}$ with fixed $q=0.9$.
Therefore, the $2$-dimensional shadow image within the photon ring is totally dark.  
With this model of emission profile, the simulation provides the BH's optical appearance, as shown in Fig.~\ref{fig: GLM image case 1}.
The observed intensities $I_\text{ob}$ in the second column of the figure also display three peaks as observed in the first column.
This result suggests a nearly disjoint between the emission from the direct image, lensing ring and photon ring.
The brightness regions depicted in the third column of the figures correspond to, in order, the photon ring, lensing ring, and direct emission emanating from the center. 
The photon ring emission displays an extremely sharp peak with a narrow width, rendering its contribution to the total flux negligible. 
Consequently, in the fourth column, we provide an enlarged view to examine the photon ring emission more closely.
With increasing $\alpha$, we find that the dilaton hair suppressed a flux originating from the lensing ring and photon ring regions in the total observed flux as mention below Eq.~\eqref{flux}.

Let us consider the large $q$ cases.
For extremal BH of $\alpha=0.1$ with $q_\text{ext}=1.005$, we find the first peak of $I_\text{ob}$
%that the maximum peak of $\Tilde{I}_\text{ob}$ appears from the superposition of direct emission and lensing ring intensities 
near $b\sim 4.78$ while photon ring $b_c\sim 2$ is outside the region of disk emission, see figures (a) and (b) in Fig~\ref{fig: GLM image ext case 1}. 
Therefore, the region within the shadow is totally dark.
Interestingly, $r_+\sim r_\text{ph}\sim r_\text{ISCO}\sim 2$ for extremal GMGHS BH ($\alpha =1$ with $q_\text{ext}=\sqrt{2}$) and the transfer function begins at $b\sim 0.39$ which corresponds to radius $r\sim 2$, see in figure (f) of Fig.~\ref{fig:Transfer fn}.
Therefore, the direct, lensing and photon rings are within the emitted accretion region.  
As the boundary of the shadow region is defined by the photon ring, $b_c\sim 2$ in this case, the illuminating region can surprisingly appear inside the shadow region in the image where $0.39<b<2$.
%\fixme{so the bright ring starts from the event horizon radius to some extent.}
Additionally, the slope of $r_1(b)$ is smaller than $1$, the direct image of the disk is not only a redshifted source,  but the direct emission's width on the equatorial intersection area is also expanded as transferred to the image due to its demagnification less than 1.
In this case, direct emission, lensing ring and photon ring intensities are then summed up, such that the light intensity has the maximum peak near $b\sim 2.42$ with a narrower width than another lower peak around $b\sim 2.85$ as seen in figures (e) and (f) in Fig~\ref{fig: GLM image ext case 1}.  
Collectively, the bright ring around the BH shadow in this case appears substantially clearer as shown in figures (g) and (h) in Fig~\ref{fig: GLM image ext case 1}, rather than appearing as a thin ring that is hard to observe astronomically.  
%In contrast, the BH shadow zone appears as a small dark circle. This is shown in Fig.~\ref{fig: Ext BH image model 1}(g) and (h).
Moreover, we also exhibit the BH image representing the extremal case with $\sqrt{3}$. However, we rather consider the near extremal case with $q=1.9$ rather than extremal BH with $q_\text{ext}=2$ since photons have scattered away from extremal BH due to the infinity barrier potential.
In this case, the inner edge of the accretion disk starts at $r_\text{ISCO}\sim 4.22$, so the observed intensity of $m=1$ begins near $b\sim 1.28$ as shown by black curve in figure (i) in Fig~\ref{fig: GLM image ext case 1}.
Consequently, the direct image can manifest within the shadow region since the photon ring becomes occurs at $b_c\sim 1.46$ in the image of BH.
Figure (j) in Fig~\ref{fig: GLM image ext case 1} shows that the emission, lensing ring, and photon ring are superposition in such a way that the observed intensity has two peaks at $b\sim 1.49$ and $b\sim 1.87$ correspond to the photon ring and lensing ring region, respectively.
The bright ring around the BH, in this case, is clearer than the extremal BH with $\alpha =0.1$ case due to its larger width of photon ring intensity as seen in figures (k) and (l) in Fig~\ref{fig: GLM image ext case 1}.
%while the transfer function begins to have values for $b\gtrsim 0.63$ which map to radius $r\gtrsim 3.87$. 

In \textit{Case II}, the emission profile starts near the radius of photon sphere, rather than ISCO as in \textit{Case I}, and then decays monotonically to zero at infinity. 
Recall that, as discussed in section~\ref{section 3}, $r_\text{ph}$ is smaller as $q$ increases at $0<\alpha\leq 1$, while it increases as $q$ increases at $\alpha>1$. 
Consequently, the peak of $I_\text{em}$ shifts to the left in the profile when $q$ is larger in the case of $0<\alpha \leq 1$, as shown in the green and dashed-green curves in the left and middle panels of Fig.~\ref{fig: GLM profiles}.
This behavior is similar to the case of RN BH $(\alpha =0)$ \cite{Gao:2023mjb}. 
Conversely, $I_\text{em}$ shifts to the right in the profile when $q$ is larger in the case of $\alpha >1$, as shown in the right panel of Fig.~\ref{fig: GLM profiles}.
On the other hand, for increasing $\alpha$  with fixed $q$, we observe a rightward shift in the peak of $I_\text{em}$, as shown in the figure.

In contrast to the \textit{Case I}, the EMD BH image emerging from the \textit{Case II} emission profile renders the peak of direct emission intensities located within the peaks of lensing ring and photon ring. 
This is shown in the left column of Fig~\ref{fig: GLM image case 2}, where we have used $q=0.9$ for all panels.
Moreover, the lensing ring and photon ring overlap with the direct emission in a manner that results in two closely positioned peaks of observation intensities $I_\text{ob}$.  
In other words, it is difficult to distinguish these two peaks in the BH image, as shown in the second column of Fig~\ref{fig: GLM image case 2}. 
As illustrated in the third and fourth columns in Fig~\ref{fig: GLM image case 2}, the maximum peak corresponds to the position of the photon ring.

Comparing with $q=0.9$ cases as discussed above, the extremal dilatonic BH with \textit{Case II}  emission profile provides that two closely positioned peaks turn out to be apart from each other, while the direct emission approaches these two peaks as shown in the first column of Fig~\ref{fig: GLM image ext case 2}. 
%In other words, these two peaks become further apart as $q$ increases. Importantly, the extremal EMD BH gives the upper bound of the displacement between these two peaks.
For the case of $\alpha =1$ and $\sqrt{3}$, the result indicates that direct emission, lensing, and photon rings regions of emission intensities are degenerate.
The optical appearances of the region near the extremal GMGHS BH ($\alpha=1$) surrounding with accretion disk in three cases of GLM profile are similar to those depicted in figures (g) and (h) of Figs.~\ref{fig: GLM image ext case 1}, \ref{fig: GLM image ext case 2} and \ref{fig: GLM image ext case 3}. 
This results from the fact that $r_\text{ph}=r_\text{ISCO}=r_+$ in the case of extremal GMGHS BH, which can be seen by comparing $I_\text{em}$  (dashed curves) for each case in the middle panel in Fig~\ref{fig: GLM profiles}. 
%However, the value of observed intensity at the maximum peak of accretion Model 2 larger than Model 1.
For near extremal BH with $q=1.9$ and $\alpha =\sqrt{3}$, the total intensities on the BH image, contributed by the direct, lensing ring and photon ring intensities, have the maximum peak at $b\sim 1.48$, corresponding the peak location of the photon ring contribution, with narrower width than another lower peak around $b\sim 1.67$, corresponding to the peak location of the lensing ring contribution, as shown in the panel (i) and (j) of Fig.~\ref{fig: GLM image ext case 2}. 

Lastly, we investigate the effect of the accretion disk in \textit{Case III} on the image of charged dilaton BH. 
In this model, the emission intensities start just outside the horizon and decrease to zero at infinity.
In the cases of $\alpha =0.1, 1$ and $\sqrt{3}$ with $q=0.9$, the maximum peaks of the direct image, lensing ring, and photon ring are all located in virtually the same region.  
As a result of the superposition of intensities, two closely positioned peaks emerge, corresponding to the photon ring (narrower) and the lensing ring, as shown in Fig.~\ref{fig: GLM image case 3}. 
Moreover, the peaks of the photon ring and lensing ring shift to the right as $\alpha$ increases due to the effect of dilaton hair. Compared with the results obtained using emission profiles \textit{Case I} and \textit{II}, the \textit{Case III} gives rise to a clearer light ring on the black hole image. 
In other words, the location of the photon ring around the BH surrounded by the accretion disk \textit{Case III} can be specified more easily than with \textit{Case I} and \textit{II}, as shown in the third and fourth rows of Fig.~\ref{fig: GLM image case 3}.

When $q$ is increased while $\alpha$ is held fixed, the widths of the three intensities broaden, and their peaks move further apart. As a result, the photon ring and lensing ring both shift closer to the BH event horizon. This can be observed by comparing the first column of Fig.~\ref{fig: GLM image ext case 3}, which represents the extremal limit, with the first column of Fig.~\ref{fig: GLM image case 3}, which represents the case with $q=0.9$.
For $\alpha=\sqrt{3}$ with near extremal charge, the direct image contributes less to the maximum peak of observed intensities than for smaller $q$ values. 
This is due to the shift of the direct emission peak apart from the peaks of lensing and photon ring, as seen in (i) and (j) of Fig.~\ref{fig: GLM image ext case 3}.
Therefore, the maximum peak of $I_\text{ob}$ occur in the lensing ring region rather than the photon ring for near extremal charge limit of $\alpha >1$ cases.

\section{Conclusions} \label{section 6}

%The exploration of string theory as a viable quantum theory of gravity holds significant interest, particularly when examining the effects of dilaton hair in scenarios involving strong field gravity. 
In this work, we study the effects of electric charge and dilaton hairs on the null geodesics around a static and spherically symmetric BH in the EMD gravity. The shadows and light rings express remarkable features of spacetime around the BH when the dilaton coupling $\alpha > 1$. Then, the influences of electric charge and dilaton hair on the optical appearance of BHs illuminated by GLM profile of thin accretion disks are explored. 

For arbitrary values of $\alpha$, the results show that the geometric characteristics of dilatonic BHs crucially depend on the ranges of $\alpha$ values, i.e., $0<\alpha <1$, $\alpha =1$ and $\alpha >1$.
For $0<\alpha <1$, the horizon radius $r_+$ becomes smaller when parameter $q$ increases.
However, $r_+$ of the dilatonic BH with $\alpha =1$, so-called the GMGHS BH, does not depend on $q$, namely, $r_+=2$ in a similar way to the Schwarzschild BH.
For $\alpha >1$, $r_+$ increases when $q$ is larger.

Next, we consider the motion of photons with different impact parameters $b$ in the background of dilatonic BH and divide them into the direct emission, lensing ring, and photon ring regions following the GHW classification.
With $\alpha$ being held fixed, it is evident that the lensing ring and photon ring regions in a BH image have broader bands of $b$ as $q$ increases. 
Conversely, these two regions become narrower when $\alpha$ is increased while keeping $q$ fixed.
We include an analytical study that examines the dependence of the width of the photon ring region within GHW classification on $q$ and $\alpha$.
The behaviors of the ring's width are found to be suppressed by the angular Lyapunov exponent.
For a charged dilaton black hole, we observe that $\gamma$ decreases as $q$ increases while $\alpha$ is held constant. Consequently, with smaller values of $q$, the width of the photon ring region are more narrower compared to the cases with larger $q$. 
Similarly, when $\alpha$ increases with fixed $q$, the results show that $\gamma$ becomes larger, leading to a more narrower photon ring width.

%To enhance the clarity of the second sentence for the reader, we include an analytical study that examines the dependence of the observed flux and width of the light rings on $q$ and $\alpha$. This is achieved by linearizing the equation of motion that describes the geodesic in the radius $r$ around the critical radius $r_\text{ph}$. Thus, the behaviors of the flux and ring's width are found to be suppressed by $e^{-\gamma}$ and $e^{-m\gamma}$, respectively.
%Here, $m$ is the number of half-orbit and $\gamma$ is the angular dependent Lyapunov exponent.
%For a charged dilaton black hole, we observe that $\gamma$ decreases as $q$ increases while $\alpha$ is held constant. Consequently, with smaller values of $q$, the observed flux and the width of the light ring are more magnified compared to the cases with larger $q$. Similarly, when $\alpha$ increases with fixed $q$, the results show that $\gamma$ becomes larger, leading to a more magnified observed flux and light ring width.

Depending on the coupling $\alpha$, we find that the null geodesics in this spacetime show different behaviors as follows. 
For $0<\alpha < 1$, the photon critical orbits 
exist as long as $q \leq q_\text{ext}$.
Remarkably, in the situation with $\alpha=1$ and $q=q_\text{ext}$, the photon ring can exist without the radial effective potential of photons $V_\text{eff}(r)$ possessing the maximum point as usual. Although the effective potential, in this case, is of repulsive type ($V_\text{eff}'(r) < 0$) outside the event horizon and going to the infinite value as $r\to 0$ rather than has the maximum just outside the event horizon, the photon ring region in the BH image are found to exist due to the fact that the photon trajectories can still satisfy the criterion $\displaystyle n > \frac{5}{4}$ according to the GHW classification. 
On the contrary, there is no photon ring region for extremal BH when $\alpha >1$, unlike the GMGHS BH, since $V_\text{eff}(r)$ is of repulsive type and positively diverge at the event horizon of the BH. 
This is further evidence supporting the statement that these BHs behave as an elementary particle as proposed in \cite{holzhey1992black}.

To investigate the shadow cast by hairy BH in the EMD gravity, we have derived celestial coordinates $(X, Y)$ based on null geodesics, leading to the apparent shadow radius being equal to the critical value of the impact parameter $R_s=b_c$.
In the presence of dilaton hair, the dark region enclosed by the critical curve is larger when compared to the RN BH at the same value of $q$.
On the other hand, the shadow region decreases when we increase $q$ and fix dilaton coupling $\alpha$ \cite{PhysRevD.105.124009}.
Intriguingly, for $\alpha >1$, we have identified a curve in the $q-\alpha$ plane, denoted as $q^*(\alpha)$, at which the apparent shadow radius $R_s$ or the critical value of the impact parameter $b_c$ coincides with the photon sphere radius $r_\text{ph}$. Beyond this curve, $b_c$ becomes smaller than $r_\text{ph}$.
Furthermore, as $q$ approaches the extremal limit, we have observed that the BH shadow and photon sphere become absent in the case of $\alpha >1$, whereas the radius of the shadow and the photon sphere maintain a non-zero minimum value for $0<\alpha <1$ and $\alpha =1$ cases.

To clearly illustrate the comparison between $R_s$ and $r_\text{ph}$, we have plotted a graph depicting the $R_s/r_\text{ph}$ as a function of $q$ for different values of $\alpha$.
It is important to note that the dilatonic BH solution with nonzero $\alpha$ has been thought of as in the middle between two extremes, i.e. the RN BH $(\alpha = 0)$ and Schwarzschild BH $(\alpha \rightarrow \infty)$ \cite{Goto:2018iay}.  Interestingly, our results show that $R_s/r_\text{ph}$ as a function of $q$ in the case of the RN BH is different from that in the case of the dilatonic BH. Namely, we find that $R_s/r_\text{ph}$ is a decreasing function of $q$ for charged dilaton BHs, while it becomes an increasing function of $q$ for the RN BHs.
As a consequence, one can potentially use the ratio of shadow radius to photon sphere radius $R_s/r_\text{ph}$ to distinguish between charged dilaton BH and RN BH through the astronomical observation.

Finally, our research has provided a comprehensive examination of the images produced by dilatonic BHs when subjected to three distinct GLM profiles of thin accretion disks.
Notably, both dilaton hair and charge parameters play a significant role in the characteristics of emission profiles of the thin accretion disk, leading to shifts in peak positions and variations in their intensities.
The transfer functions in the case of non-extremal BHs illustrate that the demagnification factors of direct emission are almost equal to one while the lensing ring and photon ring are noticeably demagnified.
We observe that the lensing ring and photon ring in the BH image become clearer as the parameter $q$ increases. 
This phenomenon is attributed to the corresponding decrease in demagnification factors.
Moreover, in certain regions of $b$ of extremal BH, the demagnification factors associated with the second and third transfer functions become smaller than that of the first transfer function.
With fixing $q$, the demagnification factors of the second and third transfer functions increase as the value of $\alpha$ becomes larger. 
Consequently, this indicates that the presence of dilaton hair results in a suppression of the observed flux originating from the lensing ring and photon ring.

Our findings on the characteristics of shadows and light rings in the EMD BH image could potentially contribute valuable insights to the forthcoming examinations of the no-hair theorem and offer potential indications of string theory through optical observations of BHs.

\section*{Acknowledgement}
We are grateful to Supakchai Ponglertsakul, Lunchakorn Tannukij, Ratchaphat Nakarachinda and Chawitt Sakkawattana for helpful discussion. The research received funding support from Thailand National Science, Research and Innovation Fund (NSRF) via the Program Management Unit for Human Resources \& Institutional Development, Research and Innovation (PMU-B) under Grant No. B05F650021. WH is supported by the National Science and Technology Development Agency under the Junior Science Talent Project (JSTP) scholarship, grant number SCA-CO-2565-16992-TH.

\appendix

\section{The Innermost Stable Circular Orbit (ISCO)}\label{App A}
\begin{figure}[H]
    \centering
    \includegraphics[width = 8.cm]{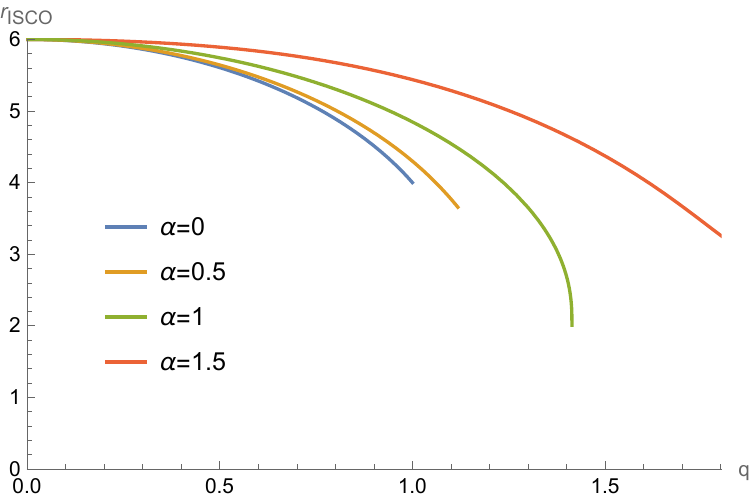}
    \caption{Profiles of $r_\text{ISCO}$ as the function of $q$ with $\alpha =0, 0.5, 1$ and $1.5$.} 
    \label{fig: risco vs q}
\end{figure}

In this appendix, we will investigate the behavior of the innermost stable circular orbit radius $r_\text{ISCO}$ for massive particles in the EMD BH spacetime. Similar to the calculation of null geodesic in section \ref{section 3}, one can obtain the orbit equation for massive particles in the $r-\phi$ plane as
\begin{eqnarray}
    \left(\frac{dr}{d\phi}\right)^2 = \mathcal{V}_\text{eff}.
\end{eqnarray}
Here, we introduce an effective potential of massive particle as $\mathcal{V}_\text{eff}$ where 
\begin{eqnarray}
    \mathcal{V}_\text{eff} = R^4(r)\left(\frac{E^2}{L^2}-\frac{g(r)}{L^2}-\frac{g(r)}{R^2(r)}\right).
\end{eqnarray}
The quantities $E$ and $L$ are the energy and angular momentum of massive particle, respectively.
The $r_\text{ISCO}$ is determined by three conditions, namely
\begin{eqnarray}
    \mathcal{V}_\text{eff}=0, \ \ \ \frac{d\mathcal{V}_\text{eff}}{dr}=0, \ \ \ \frac{d^2\mathcal{V}_\text{eff}}{dr^2}=0.
\end{eqnarray}
Solving these equations, $r_\text{ISCO}$ satisfies the following relation
\begin{eqnarray}
    \left[4g'(r)R(r)R'(r)-\frac{2g(r)g''(r)R(r)R'(r)}{g'(r)}+2g(r)R(r)R''(r)-6g(r)R'^2(r)\right]_{r=r_\text{ISCO}}=0. \label{risco} \nonumber \\
\end{eqnarray}
Consequently, we obtain $r_\text{ISCO}$ as a function of $q$ and $\alpha$. Note that Fig~\ref{fig: risco vs q} shows the dependence of $r_\text{ISCO}$ on $q$ at different values of $\alpha$.
Additionally, we find that in the case of extremal BHs when $\alpha >1$, there does not exist ISCO due to the absence of a real solution satisfying Eq.~\eqref{risco}.

%There is no solution for $r_{\text{isco}}$ of extremal black hole with $\alpha=\sqrt{3}$. This is due to the fact that $r_{\text{isco}}$ is real only  for $q$.

\section{Null Energy Condition and Focusing Theorem} \label{Appendix B}

In the background of a charged dilaton black hole with \(\alpha > 1\) and \(q \geq q^*\), the black hole's shadow radius is smaller than the photon sphere radius. This raises the question of whether the focusing theorem for null geodesics might be violated, suggesting a potential violation of the null energy condition (NEC) within this spacetime. In this appendix, we analyze the congruence of null geodesics in the black hole solution with arbitrary \(\alpha\) in the Einstein-Maxwell-Dilaton (EMD) theory of gravity.

The NEC is one of the important energy conditions in general relativity for ensuring the physical viability of solutions.
For the null tangent vector $\displaystyle k^\mu =\frac{dx^\mu}{d\lambda}$, the NEC indicates that the energy-momentum tensor satisfy
\begin{eqnarray}
    T_{\mu \nu}k^\mu k^\nu \geq 0. \label{NEC}
\end{eqnarray}
Assuming the Einstein field equations, this translates to a condition on the Ricci tensor $R_{\mu \nu}$ associated with $k^\mu$, specifically $R_{\mu \nu}k^\mu k^\nu \geq 0$.
According to the focusing theorem for null geodesics, the Raychaudhuri equation implies that matter obeying the null energy condition (NEC) will never cause the geodesics to diverge when a congruence of null geodesics is hypersurface orthogonal and the NEC holds. In other words, under these conditions, the expansion of the null geodesics will always decrease, preventing the geodesics from being pushed apart~\cite{Poisson:2009pwt}.

In the following, we analyze the null energy condition (NEC) associated with photons moving in the spacetime of a dilaton black hole. The Einstein field equation corresponding to the action of Einstein-Maxwell-Dilaton (EMD) gravity in Eq.~\eqref{action} is given by
\begin{eqnarray}
    R_{\mu \nu}-\frac{1}{2}Rg_{\mu \nu}=2 T_{\mu \nu} =2\left(T_{\mu \nu}^\Phi +T_{\mu \nu}^\text{EM} \right),
\end{eqnarray}
where the energy momentum tensor of dilaton field and Maxwell field are respectively given by
\begin{eqnarray}
    T_{\mu \nu}^\Phi &=& \partial_\mu \Phi \partial_\nu \Phi -\frac{1}{2}g_{\mu \nu}\partial_\rho \Phi \partial^\rho \Phi,
\end{eqnarray}
and
\begin{eqnarray}
    T_{\mu \nu}^\text{EM}&=&f(\Phi)\left( g^{\rho \sigma}F_{\mu \rho}F_{\nu \sigma}-\frac{1}{4}g_{\mu \nu}F_{\rho \sigma}F^{\rho \sigma} \right).
\end{eqnarray}
A null tangent vector associated to the photon trajectory in the background of spherical symmetric BH solutions is given by
\begin{eqnarray}
    k^\mu =( k^t, k^r, 0, k^\phi )=( \Dot{t}, \Dot{r}, 0, \Dot{\phi} ),
\end{eqnarray}
where $\Dot{t}$, $\Dot{\phi}$ and $\Dot{r}$ are given in Eqs.\eqref{E} \eqref{L} and \eqref{Hamiltonian}.
Since $k^\mu$ is the null vector, therefore 
\begin{eqnarray}
k_\mu k^\mu =-g(r)k^tk^t+\frac{1}{g(r)}k^rk^r+R^2(r)k^\phi k^\phi =0. \label{prop}
\end{eqnarray}
Consequently, the contribution of dilaton field to the Eq.~\eqref{NEC} is
\begin{eqnarray}
    T_{\mu \nu}^\Phi k^\mu k^\nu =k^\mu k^\nu \partial_\mu \Phi \partial_\nu \Phi -\frac{1}{2}g_{\mu \nu}k^\mu k^\nu \partial_\rho \Phi \partial^\rho \Phi .
\end{eqnarray}
With the relation in Eq.\eqref{prop}, the second term in the above equation vanishes, the result becomes
\begin{eqnarray}
    T_{\mu \nu}^\Phi k^\mu k^\nu = (k^r\partial_r\Phi )^2, \label{NEC dilaton}
\end{eqnarray}
where $k^t\partial_t\Phi =0$ and $k^\phi \partial_\phi \Phi =0$ since the profile of dilaton $\Phi$ in Eq.\eqref{field profile} does not depend on $t$ and $\phi$.
For Maxwell field, we have
\begin{eqnarray}
    T_{\mu \nu}^\text{EM}k^\mu k^\nu &=& f(\Phi)\left( g^{\rho \sigma}F_{\mu \rho}F_{\nu \sigma} k^\mu k^\nu-\frac{1}{4}g_{\mu \nu}k^\mu k^\nu F_{\rho \sigma}F^{\rho \sigma} \right).
\end{eqnarray}
By using the fact that $F_{tr}=-F_{rt}$ and Eq.\eqref{prop}, we obtain
\begin{eqnarray}
    T_{\mu \nu}^\text{EM}k^\mu k^\nu =f(\Phi)\left[ F_{tr}R(r)k^\phi \right]^2. \label{NEC EM}
\end{eqnarray}
Combining Eqs.\eqref{NEC dilaton} and \eqref{NEC EM}, we obtain
\begin{eqnarray}
    T_{\mu \nu}k^\mu k^\nu = \left(T_{\mu \nu}^\Phi+T_{\mu \nu}^\text{EM}\right)k^\mu k^\nu =(k^r\partial_r\Phi )^2+f(\Phi)\left[ F_{tr}R(r)k^\phi \right]^2 \geq 0.
\end{eqnarray}
With the coupling function $f(\Phi)\geq 0$, we find that
the above relation is always true for arbitrary coupling constant $\alpha$, when $q$ does not exceed the extremal limit. 
This means that the NEC always holds for the photon trajectories outside the charged dilaton black hole. Thus, according to the focusing theorem, the null geodesics will converge during the evolution of the congruence.

\bibliography{ref}

\begin{thebibliography}{127}%
\makeatletter
\providecommand \@ifxundefined [1]{%
 \@ifx{#1\undefined}
}%
\providecommand \@ifnum [1]{%
 \ifnum #1\expandafter \@firstoftwo
 \else \expandafter \@secondoftwo
 \fi
}%
\providecommand \@ifx [1]{%
 \ifx #1\expandafter \@firstoftwo
 \else \expandafter \@secondoftwo
 \fi
}%
\providecommand \natexlab [1]{#1}%
\providecommand \enquote  [1]{``#1''}%
\providecommand \bibnamefont  [1]{#1}%
\providecommand \bibfnamefont [1]{#1}%
\providecommand \citenamefont [1]{#1}%
\providecommand \href@noop [0]{\@secondoftwo}%
\providecommand \href [0]{\begingroup \@sanitize@url \@href}%
\providecommand \@href[1]{\@@startlink{#1}\@@href}%
\providecommand \@@href[1]{\endgroup#1\@@endlink}%
\providecommand \@sanitize@url [0]{\catcode `\\12\catcode `\$12\catcode
  `\&12\catcode `\#12\catcode `\^12\catcode `\_12\catcode `\%12\relax}%
\providecommand \@@startlink[1]{}%
\providecommand \@@endlink[0]{}%
\providecommand \url  [0]{\begingroup\@sanitize@url \@url }%
\providecommand \@url [1]{\endgroup\@href {#1}{\urlprefix }}%
\providecommand \urlprefix  [0]{URL }%
\providecommand \Eprint [0]{\href }%
\providecommand \doibase [0]{http://dx.doi.org/}%
\providecommand \selectlanguage [0]{\@gobble}%
\providecommand \bibinfo  [0]{\@secondoftwo}%
\providecommand \bibfield  [0]{\@secondoftwo}%
\providecommand \translation [1]{[#1]}%
\providecommand \BibitemOpen [0]{}%
\providecommand \bibitemStop [0]{}%
\providecommand \bibitemNoStop [0]{.\EOS\space}%
\providecommand \EOS [0]{\spacefactor3000\relax}%
\providecommand \BibitemShut  [1]{\csname bibitem#1\endcsname}%
\let\auto@bib@innerbib\@empty
%</preamble>
\bibitem [{\citenamefont {Abbott}\ \emph {et~al.}(2016)\citenamefont {Abbott}
  \emph {et~al.}}]{LIGOScientific:2016aoc}%
  \BibitemOpen
  \bibfield  {author} {\bibinfo {author} {\bibfnamefont {B.~P.}\ \bibnamefont
  {Abbott}} \emph {et~al.} (\bibinfo {collaboration} {LIGO Scientific,
  Virgo}),\ }\href {\doibase 10.1103/PhysRevLett.116.061102} {\bibfield
  {journal} {\bibinfo  {journal} {Phys. Rev. Lett.}\ }\textbf {\bibinfo
  {volume} {116}},\ \bibinfo {pages} {061102} (\bibinfo {year} {2016})},\
  \Eprint {http://arxiv.org/abs/1602.03837} {arXiv:1602.03837 [gr-qc]}
  \BibitemShut {NoStop}%
\bibitem [{\citenamefont {Akiyama}\ \emph
  {et~al.}(2019{\natexlab{a}})\citenamefont {Akiyama} \emph {et~al.}}]{ETH1}%
  \BibitemOpen
  \bibfield  {author} {\bibinfo {author} {\bibfnamefont {K.}~\bibnamefont
  {Akiyama}} \emph {et~al.} (\bibinfo {collaboration} {Event Horizon
  Telescope}),\ }\href {\doibase 10.3847/2041-8213/ab0ec7} {\bibfield
  {journal} {\bibinfo  {journal} {Astrophys. J. Lett.}\ }\textbf {\bibinfo
  {volume} {875}},\ \bibinfo {pages} {L1} (\bibinfo {year}
  {2019}{\natexlab{a}})},\ \Eprint {http://arxiv.org/abs/1906.11238}
  {arXiv:1906.11238 [astro-ph.GA]} \BibitemShut {NoStop}%
\bibitem [{\citenamefont {Akiyama}\ \emph
  {et~al.}(2019{\natexlab{b}})\citenamefont {Akiyama} \emph {et~al.}}]{ETH2}%
  \BibitemOpen
  \bibfield  {author} {\bibinfo {author} {\bibfnamefont {K.}~\bibnamefont
  {Akiyama}} \emph {et~al.} (\bibinfo {collaboration} {Event Horizon
  Telescope}),\ }\href {\doibase 10.3847/2041-8213/ab0c96} {\bibfield
  {journal} {\bibinfo  {journal} {Astrophys. J. Lett.}\ }\textbf {\bibinfo
  {volume} {875}},\ \bibinfo {pages} {L2} (\bibinfo {year}
  {2019}{\natexlab{b}})},\ \Eprint {http://arxiv.org/abs/1906.11239}
  {arXiv:1906.11239 [astro-ph.IM]} \BibitemShut {NoStop}%
\bibitem [{\citenamefont {Akiyama}\ \emph
  {et~al.}(2019{\natexlab{c}})\citenamefont {Akiyama} \emph {et~al.}}]{ETH3}%
  \BibitemOpen
  \bibfield  {author} {\bibinfo {author} {\bibfnamefont {K.}~\bibnamefont
  {Akiyama}} \emph {et~al.} (\bibinfo {collaboration} {Event Horizon
  Telescope}),\ }\href {\doibase 10.3847/2041-8213/ab0c57} {\bibfield
  {journal} {\bibinfo  {journal} {Astrophys. J. Lett.}\ }\textbf {\bibinfo
  {volume} {875}},\ \bibinfo {pages} {L3} (\bibinfo {year}
  {2019}{\natexlab{c}})},\ \Eprint {http://arxiv.org/abs/1906.11240}
  {arXiv:1906.11240 [astro-ph.GA]} \BibitemShut {NoStop}%
\bibitem [{\citenamefont {Akiyama}\ \emph
  {et~al.}(2019{\natexlab{d}})\citenamefont {Akiyama} \emph {et~al.}}]{ETH4}%
  \BibitemOpen
  \bibfield  {author} {\bibinfo {author} {\bibfnamefont {K.}~\bibnamefont
  {Akiyama}} \emph {et~al.} (\bibinfo {collaboration} {Event Horizon
  Telescope}),\ }\href {\doibase 10.3847/2041-8213/ab0e85} {\bibfield
  {journal} {\bibinfo  {journal} {Astrophys. J. Lett.}\ }\textbf {\bibinfo
  {volume} {875}},\ \bibinfo {pages} {L4} (\bibinfo {year}
  {2019}{\natexlab{d}})},\ \Eprint {http://arxiv.org/abs/1906.11241}
  {arXiv:1906.11241 [astro-ph.GA]} \BibitemShut {NoStop}%
\bibitem [{\citenamefont {Akiyama}\ \emph
  {et~al.}(2019{\natexlab{e}})\citenamefont {Akiyama} \emph {et~al.}}]{ETH5}%
  \BibitemOpen
  \bibfield  {author} {\bibinfo {author} {\bibfnamefont {K.}~\bibnamefont
  {Akiyama}} \emph {et~al.} (\bibinfo {collaboration} {Event Horizon
  Telescope}),\ }\href {\doibase 10.3847/2041-8213/ab0f43} {\bibfield
  {journal} {\bibinfo  {journal} {Astrophys. J. Lett.}\ }\textbf {\bibinfo
  {volume} {875}},\ \bibinfo {pages} {L5} (\bibinfo {year}
  {2019}{\natexlab{e}})},\ \Eprint {http://arxiv.org/abs/1906.11242}
  {arXiv:1906.11242 [astro-ph.GA]} \BibitemShut {NoStop}%
\bibitem [{\citenamefont {Akiyama}\ \emph
  {et~al.}(2019{\natexlab{f}})\citenamefont {Akiyama} \emph {et~al.}}]{ETH6}%
  \BibitemOpen
  \bibfield  {author} {\bibinfo {author} {\bibfnamefont {K.}~\bibnamefont
  {Akiyama}} \emph {et~al.} (\bibinfo {collaboration} {Event Horizon
  Telescope}),\ }\href {\doibase 10.3847/2041-8213/ab1141} {\bibfield
  {journal} {\bibinfo  {journal} {Astrophys. J. Lett.}\ }\textbf {\bibinfo
  {volume} {875}},\ \bibinfo {pages} {L6} (\bibinfo {year}
  {2019}{\natexlab{f}})},\ \Eprint {http://arxiv.org/abs/1906.11243}
  {arXiv:1906.11243 [astro-ph.GA]} \BibitemShut {NoStop}%
\bibitem [{\citenamefont {Akiyama}\ \emph
  {et~al.}(2022{\natexlab{a}})\citenamefont {Akiyama} \emph
  {et~al.}}]{EventHorizonTelescope:2022wkp}%
  \BibitemOpen
  \bibfield  {author} {\bibinfo {author} {\bibfnamefont {K.}~\bibnamefont
  {Akiyama}} \emph {et~al.} (\bibinfo {collaboration} {Event Horizon
  Telescope}),\ }\href {\doibase 10.3847/2041-8213/ac6674} {\bibfield
  {journal} {\bibinfo  {journal} {Astrophys. J. Lett.}\ }\textbf {\bibinfo
  {volume} {930}},\ \bibinfo {pages} {L12} (\bibinfo {year}
  {2022}{\natexlab{a}})}\BibitemShut {NoStop}%
\bibitem [{\citenamefont {Akiyama}\ \emph
  {et~al.}(2022{\natexlab{b}})\citenamefont {Akiyama} \emph
  {et~al.}}]{EventHorizonTelescope:2022apq}%
  \BibitemOpen
  \bibfield  {author} {\bibinfo {author} {\bibfnamefont {K.}~\bibnamefont
  {Akiyama}} \emph {et~al.} (\bibinfo {collaboration} {Event Horizon
  Telescope}),\ }\href {\doibase 10.3847/2041-8213/ac6675} {\bibfield
  {journal} {\bibinfo  {journal} {Astrophys. J. Lett.}\ }\textbf {\bibinfo
  {volume} {930}},\ \bibinfo {pages} {L13} (\bibinfo {year}
  {2022}{\natexlab{b}})}\BibitemShut {NoStop}%
\bibitem [{\citenamefont {Akiyama}\ \emph
  {et~al.}(2022{\natexlab{c}})\citenamefont {Akiyama} \emph
  {et~al.}}]{EventHorizonTelescope:2022wok}%
  \BibitemOpen
  \bibfield  {author} {\bibinfo {author} {\bibfnamefont {K.}~\bibnamefont
  {Akiyama}} \emph {et~al.} (\bibinfo {collaboration} {Event Horizon
  Telescope}),\ }\href {\doibase 10.3847/2041-8213/ac6429} {\bibfield
  {journal} {\bibinfo  {journal} {Astrophys. J. Lett.}\ }\textbf {\bibinfo
  {volume} {930}},\ \bibinfo {pages} {L14} (\bibinfo {year}
  {2022}{\natexlab{c}})}\BibitemShut {NoStop}%
\bibitem [{\citenamefont {Akiyama}\ \emph
  {et~al.}(2022{\natexlab{d}})\citenamefont {Akiyama} \emph
  {et~al.}}]{EventHorizonTelescope:2022exc}%
  \BibitemOpen
  \bibfield  {author} {\bibinfo {author} {\bibfnamefont {K.}~\bibnamefont
  {Akiyama}} \emph {et~al.} (\bibinfo {collaboration} {Event Horizon
  Telescope}),\ }\href {\doibase 10.3847/2041-8213/ac6736} {\bibfield
  {journal} {\bibinfo  {journal} {Astrophys. J. Lett.}\ }\textbf {\bibinfo
  {volume} {930}},\ \bibinfo {pages} {L15} (\bibinfo {year}
  {2022}{\natexlab{d}})}\BibitemShut {NoStop}%
\bibitem [{\citenamefont {Akiyama}\ \emph
  {et~al.}(2022{\natexlab{e}})\citenamefont {Akiyama} \emph
  {et~al.}}]{EventHorizonTelescope:2022urf}%
  \BibitemOpen
  \bibfield  {author} {\bibinfo {author} {\bibfnamefont {K.}~\bibnamefont
  {Akiyama}} \emph {et~al.} (\bibinfo {collaboration} {Event Horizon
  Telescope}),\ }\href {\doibase 10.3847/2041-8213/ac6672} {\bibfield
  {journal} {\bibinfo  {journal} {Astrophys. J. Lett.}\ }\textbf {\bibinfo
  {volume} {930}},\ \bibinfo {pages} {L16} (\bibinfo {year}
  {2022}{\natexlab{e}})}\BibitemShut {NoStop}%
\bibitem [{\citenamefont {Akiyama}\ \emph
  {et~al.}(2022{\natexlab{f}})\citenamefont {Akiyama} \emph
  {et~al.}}]{EventHorizonTelescope:2022xqj}%
  \BibitemOpen
  \bibfield  {author} {\bibinfo {author} {\bibfnamefont {K.}~\bibnamefont
  {Akiyama}} \emph {et~al.} (\bibinfo {collaboration} {Event Horizon
  Telescope}),\ }\href {\doibase 10.3847/2041-8213/ac6756} {\bibfield
  {journal} {\bibinfo  {journal} {Astrophys. J. Lett.}\ }\textbf {\bibinfo
  {volume} {930}},\ \bibinfo {pages} {L17} (\bibinfo {year}
  {2022}{\natexlab{f}})}\BibitemShut {NoStop}%
\bibitem [{\citenamefont {{Darwin}}(1959)}]{1959RSPSA.249180D}%
  \BibitemOpen
  \bibfield  {author} {\bibinfo {author} {\bibfnamefont {C.}~\bibnamefont
  {{Darwin}}},\ }\href {\doibase 10.1098/rspa.1959.0015} {\bibfield  {journal}
  {\bibinfo  {journal} {Proceedings of the Royal Society of London Series A}\
  }\textbf {\bibinfo {volume} {249}},\ \bibinfo {pages} {180} (\bibinfo {year}
  {1959})}\BibitemShut {NoStop}%
\bibitem [{\citenamefont {Synge}(1966)}]{Synge:1966okc}%
  \BibitemOpen
  \bibfield  {author} {\bibinfo {author} {\bibfnamefont {J.~L.}\ \bibnamefont
  {Synge}},\ }\href {\doibase 10.1093/mnras/131.3.463} {\bibfield  {journal}
  {\bibinfo  {journal} {Mon. Not. Roy. Astron. Soc.}\ }\textbf {\bibinfo
  {volume} {131}},\ \bibinfo {pages} {463} (\bibinfo {year}
  {1966})}\BibitemShut {NoStop}%
\bibitem [{\citenamefont {Ohanian}(1987)}]{ohanian1987black}%
  \BibitemOpen
  \bibfield  {author} {\bibinfo {author} {\bibfnamefont {H.~C.}\ \bibnamefont
  {Ohanian}},\ }\href@noop {} {\bibfield  {journal} {\bibinfo  {journal}
  {American Journal of Physics}\ }\textbf {\bibinfo {volume} {55}},\ \bibinfo
  {pages} {428} (\bibinfo {year} {1987})}\BibitemShut {NoStop}%
\bibitem [{\citenamefont {Zakharov}\ \emph {et~al.}(2005)\citenamefont
  {Zakharov}, \citenamefont {De~Paolis}, \citenamefont {Ingrosso},\ and\
  \citenamefont {Nucita}}]{Zakharov:2005ek}%
  \BibitemOpen
  \bibfield  {author} {\bibinfo {author} {\bibfnamefont {A.~F.}\ \bibnamefont
  {Zakharov}}, \bibinfo {author} {\bibfnamefont {F.}~\bibnamefont {De~Paolis}},
  \bibinfo {author} {\bibfnamefont {G.}~\bibnamefont {Ingrosso}}, \ and\
  \bibinfo {author} {\bibfnamefont {A.~A.}\ \bibnamefont {Nucita}},\ }\href
  {\doibase 10.1051/0004-6361:20053432} {\bibfield  {journal} {\bibinfo
  {journal} {Astron. Astrophys.}\ }\textbf {\bibinfo {volume} {442}},\ \bibinfo
  {pages} {795} (\bibinfo {year} {2005})},\ \Eprint
  {http://arxiv.org/abs/astro-ph/0505286} {arXiv:astro-ph/0505286} \BibitemShut
  {NoStop}%
\bibitem [{\citenamefont {Zakharov}(2014)}]{Zakharov:2014lqa}%
  \BibitemOpen
  \bibfield  {author} {\bibinfo {author} {\bibfnamefont {A.~F.}\ \bibnamefont
  {Zakharov}},\ }\href {\doibase 10.1103/PhysRevD.90.062007} {\bibfield
  {journal} {\bibinfo  {journal} {Phys. Rev. D}\ }\textbf {\bibinfo {volume}
  {90}},\ \bibinfo {pages} {062007} (\bibinfo {year} {2014})},\ \Eprint
  {http://arxiv.org/abs/1407.7457} {arXiv:1407.7457 [gr-qc]} \BibitemShut
  {NoStop}%
\bibitem [{\citenamefont {Bardeen}(1970)}]{bardeen1970kerr}%
  \BibitemOpen
  \bibfield  {author} {\bibinfo {author} {\bibfnamefont {J.~M.}\ \bibnamefont
  {Bardeen}},\ }\href@noop {} {\bibfield  {journal} {\bibinfo  {journal}
  {Nature}\ }\textbf {\bibinfo {volume} {226}},\ \bibinfo {pages} {64}
  (\bibinfo {year} {1970})}\BibitemShut {NoStop}%
\bibitem [{\citenamefont {Bardeen}(1973)}]{Bardeen:1973tla}%
  \BibitemOpen
  \bibfield  {author} {\bibinfo {author} {\bibfnamefont {J.~M.}\ \bibnamefont
  {Bardeen}},\ }\href@noop {} {\bibfield  {journal} {\bibinfo  {journal}
  {Proceedings, Ecole d'Et\'e de Physique Th\'eorique: Les Astres Occlus : Les
  Houches, France, August, 1972, 215-240}\ ,\ \bibinfo {pages} {215}} (\bibinfo
  {year} {1973})}\BibitemShut {NoStop}%
\bibitem [{\citenamefont {Hioki}\ and\ \citenamefont
  {Maeda}(2009)}]{Hioki:2009na}%
  \BibitemOpen
  \bibfield  {author} {\bibinfo {author} {\bibfnamefont {K.}~\bibnamefont
  {Hioki}}\ and\ \bibinfo {author} {\bibfnamefont {K.-i.}\ \bibnamefont
  {Maeda}},\ }\href {\doibase 10.1103/PhysRevD.80.024042} {\bibfield  {journal}
  {\bibinfo  {journal} {Phys. Rev. D}\ }\textbf {\bibinfo {volume} {80}},\
  \bibinfo {pages} {024042} (\bibinfo {year} {2009})},\ \Eprint
  {http://arxiv.org/abs/0904.3575} {arXiv:0904.3575 [astro-ph.HE]} \BibitemShut
  {NoStop}%
\bibitem [{\citenamefont {Takahashi}(2005)}]{Takahashi:2005hy}%
  \BibitemOpen
  \bibfield  {author} {\bibinfo {author} {\bibfnamefont {R.}~\bibnamefont
  {Takahashi}},\ }\href {\doibase 10.1093/pasj/57.2.273} {\bibfield  {journal}
  {\bibinfo  {journal} {Publ. Astron. Soc. Jap.}\ }\textbf {\bibinfo {volume}
  {57}},\ \bibinfo {pages} {273} (\bibinfo {year} {2005})},\ \Eprint
  {http://arxiv.org/abs/astro-ph/0505316} {arXiv:astro-ph/0505316} \BibitemShut
  {NoStop}%
\bibitem [{\citenamefont {Tsukamoto}(2018)}]{Tsukamoto:2017fxq}%
  \BibitemOpen
  \bibfield  {author} {\bibinfo {author} {\bibfnamefont {N.}~\bibnamefont
  {Tsukamoto}},\ }\href {\doibase 10.1103/PhysRevD.97.064021} {\bibfield
  {journal} {\bibinfo  {journal} {Phys. Rev. D}\ }\textbf {\bibinfo {volume}
  {97}},\ \bibinfo {pages} {064021} (\bibinfo {year} {2018})},\ \Eprint
  {http://arxiv.org/abs/1708.07427} {arXiv:1708.07427 [gr-qc]} \BibitemShut
  {NoStop}%
\bibitem [{\citenamefont {Abdujabbarov}\ \emph {et~al.}(2013)\citenamefont
  {Abdujabbarov}, \citenamefont {Atamurotov}, \citenamefont {Kucukakca},
  \citenamefont {Ahmedov},\ and\ \citenamefont {Camci}}]{Abdujabbarov:2012bn}%
  \BibitemOpen
  \bibfield  {author} {\bibinfo {author} {\bibfnamefont {A.}~\bibnamefont
  {Abdujabbarov}}, \bibinfo {author} {\bibfnamefont {F.}~\bibnamefont
  {Atamurotov}}, \bibinfo {author} {\bibfnamefont {Y.}~\bibnamefont
  {Kucukakca}}, \bibinfo {author} {\bibfnamefont {B.}~\bibnamefont {Ahmedov}},
  \ and\ \bibinfo {author} {\bibfnamefont {U.}~\bibnamefont {Camci}},\ }\href
  {\doibase 10.1007/s10509-012-1337-6} {\bibfield  {journal} {\bibinfo
  {journal} {Astrophys. Space Sci.}\ }\textbf {\bibinfo {volume} {344}},\
  \bibinfo {pages} {429} (\bibinfo {year} {2013})},\ \Eprint
  {http://arxiv.org/abs/1212.4949} {arXiv:1212.4949 [physics.gen-ph]}
  \BibitemShut {NoStop}%
\bibitem [{\citenamefont {Mandal}\ \emph {et~al.}(2023)\citenamefont {Mandal},
  \citenamefont {Upadhyay}, \citenamefont {Myrzakulov},\ and\ \citenamefont
  {Yergaliyeva}}]{Mandal:2022oma}%
  \BibitemOpen
  \bibfield  {author} {\bibinfo {author} {\bibfnamefont {S.}~\bibnamefont
  {Mandal}}, \bibinfo {author} {\bibfnamefont {S.}~\bibnamefont {Upadhyay}},
  \bibinfo {author} {\bibfnamefont {Y.}~\bibnamefont {Myrzakulov}}, \ and\
  \bibinfo {author} {\bibfnamefont {G.}~\bibnamefont {Yergaliyeva}},\ }\href
  {\doibase 10.1142/S0217751X23500471} {\bibfield  {journal} {\bibinfo
  {journal} {Int. J. Mod. Phys. A}\ }\textbf {\bibinfo {volume} {38}},\
  \bibinfo {pages} {2350047} (\bibinfo {year} {2023})},\ \Eprint
  {http://arxiv.org/abs/2207.10085} {arXiv:2207.10085 [gr-qc]} \BibitemShut
  {NoStop}%
\bibitem [{\citenamefont {Perlick}\ \emph {et~al.}(2018)\citenamefont
  {Perlick}, \citenamefont {Tsupko},\ and\ \citenamefont
  {Bisnovatyi-Kogan}}]{PhysRevD.97.104062}%
  \BibitemOpen
  \bibfield  {author} {\bibinfo {author} {\bibfnamefont {V.}~\bibnamefont
  {Perlick}}, \bibinfo {author} {\bibfnamefont {O.~Y.}\ \bibnamefont {Tsupko}},
  \ and\ \bibinfo {author} {\bibfnamefont {G.~S.}\ \bibnamefont
  {Bisnovatyi-Kogan}},\ }\href {\doibase 10.1103/PhysRevD.97.104062} {\bibfield
   {journal} {\bibinfo  {journal} {Phys. Rev. D}\ }\textbf {\bibinfo {volume}
  {97}},\ \bibinfo {pages} {104062} (\bibinfo {year} {2018})}\BibitemShut
  {NoStop}%
\bibitem [{\citenamefont {Khodadi}\ \emph {et~al.}(2020)\citenamefont
  {Khodadi}, \citenamefont {Allahyari}, \citenamefont {Vagnozzi},\ and\
  \citenamefont {Mota}}]{Khodadi:2020jij}%
  \BibitemOpen
  \bibfield  {author} {\bibinfo {author} {\bibfnamefont {M.}~\bibnamefont
  {Khodadi}}, \bibinfo {author} {\bibfnamefont {A.}~\bibnamefont {Allahyari}},
  \bibinfo {author} {\bibfnamefont {S.}~\bibnamefont {Vagnozzi}}, \ and\
  \bibinfo {author} {\bibfnamefont {D.~F.}\ \bibnamefont {Mota}},\ }\href
  {\doibase 10.1088/1475-7516/2020/09/026} {\bibfield  {journal} {\bibinfo
  {journal} {JCAP}\ }\textbf {\bibinfo {volume} {09}},\ \bibinfo {pages} {026}
  (\bibinfo {year} {2020})},\ \Eprint {http://arxiv.org/abs/2005.05992}
  {arXiv:2005.05992 [gr-qc]} \BibitemShut {NoStop}%
\bibitem [{\citenamefont {Wei}\ and\ \citenamefont {Liu}(2013)}]{Wei:2013kza}%
  \BibitemOpen
  \bibfield  {author} {\bibinfo {author} {\bibfnamefont {S.-W.}\ \bibnamefont
  {Wei}}\ and\ \bibinfo {author} {\bibfnamefont {Y.-X.}\ \bibnamefont {Liu}},\
  }\href {\doibase 10.1088/1475-7516/2013/11/063} {\bibfield  {journal}
  {\bibinfo  {journal} {JCAP}\ }\textbf {\bibinfo {volume} {11}},\ \bibinfo
  {pages} {063} (\bibinfo {year} {2013})},\ \Eprint
  {http://arxiv.org/abs/1311.4251} {arXiv:1311.4251 [gr-qc]} \BibitemShut
  {NoStop}%
\bibitem [{\citenamefont {Das}\ \emph {et~al.}(2020)\citenamefont {Das},
  \citenamefont {Saha},\ and\ \citenamefont {Gangopadhyay}}]{Das:2019sty}%
  \BibitemOpen
  \bibfield  {author} {\bibinfo {author} {\bibfnamefont {A.}~\bibnamefont
  {Das}}, \bibinfo {author} {\bibfnamefont {A.}~\bibnamefont {Saha}}, \ and\
  \bibinfo {author} {\bibfnamefont {S.}~\bibnamefont {Gangopadhyay}},\ }\href
  {\doibase 10.1140/epjc/s10052-020-7726-z} {\bibfield  {journal} {\bibinfo
  {journal} {Eur. Phys. J. C}\ }\textbf {\bibinfo {volume} {80}},\ \bibinfo
  {pages} {180} (\bibinfo {year} {2020})},\ \Eprint
  {http://arxiv.org/abs/1909.01988} {arXiv:1909.01988 [gr-qc]} \BibitemShut
  {NoStop}%
\bibitem [{\citenamefont {Papnoi}\ and\ \citenamefont
  {Atamurotov}(2022)}]{Papnoi:2021rvw}%
  \BibitemOpen
  \bibfield  {author} {\bibinfo {author} {\bibfnamefont {U.}~\bibnamefont
  {Papnoi}}\ and\ \bibinfo {author} {\bibfnamefont {F.}~\bibnamefont
  {Atamurotov}},\ }\href {\doibase 10.1016/j.dark.2021.100916} {\bibfield
  {journal} {\bibinfo  {journal} {Phys. Dark Univ.}\ }\textbf {\bibinfo
  {volume} {35}},\ \bibinfo {pages} {100916} (\bibinfo {year} {2022})},\
  \Eprint {http://arxiv.org/abs/2111.15523} {arXiv:2111.15523 [gr-qc]}
  \BibitemShut {NoStop}%
\bibitem [{\citenamefont {Panpanich}\ \emph {et~al.}(2019)\citenamefont
  {Panpanich}, \citenamefont {Ponglertsakul},\ and\ \citenamefont
  {Tannukij}}]{Panpanich:2019mll}%
  \BibitemOpen
  \bibfield  {author} {\bibinfo {author} {\bibfnamefont {S.}~\bibnamefont
  {Panpanich}}, \bibinfo {author} {\bibfnamefont {S.}~\bibnamefont
  {Ponglertsakul}}, \ and\ \bibinfo {author} {\bibfnamefont {L.}~\bibnamefont
  {Tannukij}},\ }\href {\doibase 10.1103/PhysRevD.100.044031} {\bibfield
  {journal} {\bibinfo  {journal} {Phys. Rev. D}\ }\textbf {\bibinfo {volume}
  {100}},\ \bibinfo {pages} {044031} (\bibinfo {year} {2019})},\ \Eprint
  {http://arxiv.org/abs/1904.02915} {arXiv:1904.02915 [gr-qc]} \BibitemShut
  {NoStop}%
\bibitem [{\citenamefont {Hendi}\ \emph {et~al.}(2023)\citenamefont {Hendi},
  \citenamefont {Jafarzade},\ and\ \citenamefont
  {Eslam~Panah}}]{Hendi:2022qgi}%
  \BibitemOpen
  \bibfield  {author} {\bibinfo {author} {\bibfnamefont {S.~H.}\ \bibnamefont
  {Hendi}}, \bibinfo {author} {\bibfnamefont {K.}~\bibnamefont {Jafarzade}}, \
  and\ \bibinfo {author} {\bibfnamefont {B.}~\bibnamefont {Eslam~Panah}},\
  }\href {\doibase 10.1088/1475-7516/2023/02/022} {\bibfield  {journal}
  {\bibinfo  {journal} {JCAP}\ }\textbf {\bibinfo {volume} {02}},\ \bibinfo
  {pages} {022} (\bibinfo {year} {2023})},\ \Eprint
  {http://arxiv.org/abs/2206.05132} {arXiv:2206.05132 [gr-qc]} \BibitemShut
  {NoStop}%
\bibitem [{\citenamefont {Khodadi}\ and\ \citenamefont
  {Saridakis}(2021)}]{Khodadi:2020gns}%
  \BibitemOpen
  \bibfield  {author} {\bibinfo {author} {\bibfnamefont {M.}~\bibnamefont
  {Khodadi}}\ and\ \bibinfo {author} {\bibfnamefont {E.~N.}\ \bibnamefont
  {Saridakis}},\ }\href {\doibase 10.1016/j.dark.2021.100835} {\bibfield
  {journal} {\bibinfo  {journal} {Phys. Dark Univ.}\ }\textbf {\bibinfo
  {volume} {32}},\ \bibinfo {pages} {100835} (\bibinfo {year} {2021})},\
  \Eprint {http://arxiv.org/abs/2012.05186} {arXiv:2012.05186 [gr-qc]}
  \BibitemShut {NoStop}%
\bibitem [{\citenamefont {Zhu}\ \emph {et~al.}(2019)\citenamefont {Zhu},
  \citenamefont {Wu}, \citenamefont {Jamil},\ and\ \citenamefont
  {Jusufi}}]{Zhu:2019ura}%
  \BibitemOpen
  \bibfield  {author} {\bibinfo {author} {\bibfnamefont {T.}~\bibnamefont
  {Zhu}}, \bibinfo {author} {\bibfnamefont {Q.}~\bibnamefont {Wu}}, \bibinfo
  {author} {\bibfnamefont {M.}~\bibnamefont {Jamil}}, \ and\ \bibinfo {author}
  {\bibfnamefont {K.}~\bibnamefont {Jusufi}},\ }\href {\doibase
  10.1103/PhysRevD.100.044055} {\bibfield  {journal} {\bibinfo  {journal}
  {Phys. Rev. D}\ }\textbf {\bibinfo {volume} {100}},\ \bibinfo {pages}
  {044055} (\bibinfo {year} {2019})},\ \Eprint
  {http://arxiv.org/abs/1906.05673} {arXiv:1906.05673 [gr-qc]} \BibitemShut
  {NoStop}%
\bibitem [{\citenamefont {Belhaj}\ \emph {et~al.}(2020)\citenamefont {Belhaj},
  \citenamefont {Benali}, \citenamefont {El~Balali}, \citenamefont
  {El~Moumni},\ and\ \citenamefont {Ennadifi}}]{Belhaj:2020rdb}%
  \BibitemOpen
  \bibfield  {author} {\bibinfo {author} {\bibfnamefont {A.}~\bibnamefont
  {Belhaj}}, \bibinfo {author} {\bibfnamefont {M.}~\bibnamefont {Benali}},
  \bibinfo {author} {\bibfnamefont {A.}~\bibnamefont {El~Balali}}, \bibinfo
  {author} {\bibfnamefont {H.}~\bibnamefont {El~Moumni}}, \ and\ \bibinfo
  {author} {\bibfnamefont {S.~E.}\ \bibnamefont {Ennadifi}},\ }\href {\doibase
  10.1088/1361-6382/abbaa9} {\bibfield  {journal} {\bibinfo  {journal} {Class.
  Quant. Grav.}\ }\textbf {\bibinfo {volume} {37}},\ \bibinfo {pages} {215004}
  (\bibinfo {year} {2020})},\ \Eprint {http://arxiv.org/abs/2006.01078}
  {arXiv:2006.01078 [gr-qc]} \BibitemShut {NoStop}%
\bibitem [{\citenamefont {Hod}(2013)}]{Hod:2013jhd}%
  \BibitemOpen
  \bibfield  {author} {\bibinfo {author} {\bibfnamefont {S.}~\bibnamefont
  {Hod}},\ }\href {\doibase 10.1016/j.physletb.2013.10.047} {\bibfield
  {journal} {\bibinfo  {journal} {Phys. Lett. B}\ }\textbf {\bibinfo {volume}
  {727}},\ \bibinfo {pages} {345} (\bibinfo {year} {2013})},\ \Eprint
  {http://arxiv.org/abs/1701.06587} {arXiv:1701.06587 [gr-qc]} \BibitemShut
  {NoStop}%
\bibitem [{\citenamefont {Cvetic}\ \emph {et~al.}(2016)\citenamefont {Cvetic},
  \citenamefont {Gibbons},\ and\ \citenamefont {Pope}}]{Cvetic:2016bxi}%
  \BibitemOpen
  \bibfield  {author} {\bibinfo {author} {\bibfnamefont {M.}~\bibnamefont
  {Cvetic}}, \bibinfo {author} {\bibfnamefont {G.~W.}\ \bibnamefont {Gibbons}},
  \ and\ \bibinfo {author} {\bibfnamefont {C.~N.}\ \bibnamefont {Pope}},\
  }\href {\doibase 10.1103/PhysRevD.94.106005} {\bibfield  {journal} {\bibinfo
  {journal} {Phys. Rev. D}\ }\textbf {\bibinfo {volume} {94}},\ \bibinfo
  {pages} {106005} (\bibinfo {year} {2016})},\ \Eprint
  {http://arxiv.org/abs/1608.02202} {arXiv:1608.02202 [gr-qc]} \BibitemShut
  {NoStop}%
\bibitem [{\citenamefont {Lu}\ and\ \citenamefont {Lyu}(2020)}]{Lu:2019zxb}%
  \BibitemOpen
  \bibfield  {author} {\bibinfo {author} {\bibfnamefont {H.}~\bibnamefont
  {Lu}}\ and\ \bibinfo {author} {\bibfnamefont {H.-D.}\ \bibnamefont {Lyu}},\
  }\href {\doibase 10.1103/PhysRevD.101.044059} {\bibfield  {journal} {\bibinfo
   {journal} {Phys. Rev. D}\ }\textbf {\bibinfo {volume} {101}},\ \bibinfo
  {pages} {044059} (\bibinfo {year} {2020})},\ \Eprint
  {http://arxiv.org/abs/1911.02019} {arXiv:1911.02019 [gr-qc]} \BibitemShut
  {NoStop}%
\bibitem [{\citenamefont {Ma}\ and\ \citenamefont {Lu}(2020)}]{Ma:2019ybz}%
  \BibitemOpen
  \bibfield  {author} {\bibinfo {author} {\bibfnamefont {L.}~\bibnamefont
  {Ma}}\ and\ \bibinfo {author} {\bibfnamefont {H.}~\bibnamefont {Lu}},\ }\href
  {\doibase 10.1016/j.physletb.2020.135535} {\bibfield  {journal} {\bibinfo
  {journal} {Phys. Lett. B}\ }\textbf {\bibinfo {volume} {807}},\ \bibinfo
  {pages} {135535} (\bibinfo {year} {2020})},\ \Eprint
  {http://arxiv.org/abs/1912.05569} {arXiv:1912.05569 [gr-qc]} \BibitemShut
  {NoStop}%
\bibitem [{\citenamefont {Chakraborty}(2021)}]{Chakraborty:2021dmu}%
  \BibitemOpen
  \bibfield  {author} {\bibinfo {author} {\bibfnamefont {S.}~\bibnamefont
  {Chakraborty}},\ }\href {\doibase 10.3390/galaxies9040096} {\bibfield
  {journal} {\bibinfo  {journal} {Galaxies}\ }\textbf {\bibinfo {volume} {9}},\
  \bibinfo {pages} {96} (\bibinfo {year} {2021})},\ \Eprint
  {http://arxiv.org/abs/2111.04912} {arXiv:2111.04912 [gr-qc]} \BibitemShut
  {NoStop}%
\bibitem [{\citenamefont {Abramowicz}\ and\ \citenamefont
  {Fragile}(2013)}]{Abramowicz:2011xu}%
  \BibitemOpen
  \bibfield  {author} {\bibinfo {author} {\bibfnamefont {M.~A.}\ \bibnamefont
  {Abramowicz}}\ and\ \bibinfo {author} {\bibfnamefont {P.~C.}\ \bibnamefont
  {Fragile}},\ }\href {\doibase 10.12942/lrr-2013-1} {\bibfield  {journal}
  {\bibinfo  {journal} {Living Rev. Rel.}\ }\textbf {\bibinfo {volume} {16}},\
  \bibinfo {pages} {1} (\bibinfo {year} {2013})},\ \Eprint
  {http://arxiv.org/abs/1104.5499} {arXiv:1104.5499 [astro-ph.HE]} \BibitemShut
  {NoStop}%
\bibitem [{\citenamefont {Luminet}(1979)}]{luminet1979image}%
  \BibitemOpen
  \bibfield  {author} {\bibinfo {author} {\bibfnamefont {J.-P.}\ \bibnamefont
  {Luminet}},\ }\href@noop {} {\bibfield  {journal} {\bibinfo  {journal}
  {Astronomy and Astrophysics, vol. 75, no. 1-2, May 1979, p. 228-235.}\
  }\textbf {\bibinfo {volume} {75}},\ \bibinfo {pages} {228} (\bibinfo {year}
  {1979})}\BibitemShut {NoStop}%
\bibitem [{\citenamefont {Falcke}\ \emph {et~al.}(2000)\citenamefont {Falcke},
  \citenamefont {Melia},\ and\ \citenamefont {Agol}}]{Falcke:1999pj}%
  \BibitemOpen
  \bibfield  {author} {\bibinfo {author} {\bibfnamefont {H.}~\bibnamefont
  {Falcke}}, \bibinfo {author} {\bibfnamefont {F.}~\bibnamefont {Melia}}, \
  and\ \bibinfo {author} {\bibfnamefont {E.}~\bibnamefont {Agol}},\ }\href
  {\doibase 10.1086/312423} {\bibfield  {journal} {\bibinfo  {journal}
  {Astrophys. J. Lett.}\ }\textbf {\bibinfo {volume} {528}},\ \bibinfo {pages}
  {L13} (\bibinfo {year} {2000})},\ \Eprint
  {http://arxiv.org/abs/astro-ph/9912263} {arXiv:astro-ph/9912263} \BibitemShut
  {NoStop}%
\bibitem [{\citenamefont {Narayan}\ \emph {et~al.}(2019)\citenamefont
  {Narayan}, \citenamefont {Johnson},\ and\ \citenamefont
  {Gammie}}]{Narayan2019xty}%
  \BibitemOpen
  \bibfield  {author} {\bibinfo {author} {\bibfnamefont {R.}~\bibnamefont
  {Narayan}}, \bibinfo {author} {\bibfnamefont {M.~D.}\ \bibnamefont
  {Johnson}}, \ and\ \bibinfo {author} {\bibfnamefont {C.~F.}\ \bibnamefont
  {Gammie}},\ }\href {\doibase 10.3847/2041-8213/ab518c} {\bibfield  {journal}
  {\bibinfo  {journal} {The Astrophysical Journal}\ }\textbf {\bibinfo {volume}
  {885}},\ \bibinfo {pages} {L33} (\bibinfo {year} {2019})}\BibitemShut
  {NoStop}%
\bibitem [{\citenamefont {Heydari-Fard}(2022)}]{Heydari-Fard:2022jdu}%
  \BibitemOpen
  \bibfield  {author} {\bibinfo {author} {\bibfnamefont {M.}~\bibnamefont
  {Heydari-Fard}},\ }\href@noop {} {\  (\bibinfo {year} {2022})},\ \Eprint
  {http://arxiv.org/abs/2209.09103} {arXiv:2209.09103 [gr-qc]} \BibitemShut
  {NoStop}%
\bibitem [{\citenamefont {Heydari-Fard}\ \emph {et~al.}(2023)\citenamefont
  {Heydari-Fard}, \citenamefont {Heydari-Fard},\ and\ \citenamefont
  {Riazi}}]{Heydari-Fard:2023ent}%
  \BibitemOpen
  \bibfield  {author} {\bibinfo {author} {\bibfnamefont {M.}~\bibnamefont
  {Heydari-Fard}}, \bibinfo {author} {\bibfnamefont {M.}~\bibnamefont
  {Heydari-Fard}}, \ and\ \bibinfo {author} {\bibfnamefont {N.}~\bibnamefont
  {Riazi}},\ }\href@noop {} {\  (\bibinfo {year} {2023})},\ \Eprint
  {http://arxiv.org/abs/2307.01529} {arXiv:2307.01529 [gr-qc]} \BibitemShut
  {NoStop}%
\bibitem [{\citenamefont {Gralla}\ \emph {et~al.}(2019)\citenamefont {Gralla},
  \citenamefont {Holz},\ and\ \citenamefont {Wald}}]{Gralla:2019xty}%
  \BibitemOpen
  \bibfield  {author} {\bibinfo {author} {\bibfnamefont {S.~E.}\ \bibnamefont
  {Gralla}}, \bibinfo {author} {\bibfnamefont {D.~E.}\ \bibnamefont {Holz}}, \
  and\ \bibinfo {author} {\bibfnamefont {R.~M.}\ \bibnamefont {Wald}},\ }\href
  {\doibase 10.1103/PhysRevD.100.024018} {\bibfield  {journal} {\bibinfo
  {journal} {Phys. Rev. D}\ }\textbf {\bibinfo {volume} {100}},\ \bibinfo
  {pages} {024018} (\bibinfo {year} {2019})},\ \Eprint
  {http://arxiv.org/abs/1906.00873} {arXiv:1906.00873 [astro-ph.HE]}
  \BibitemShut {NoStop}%
\bibitem [{\citenamefont {Zeng}\ \emph {et~al.}(2020)\citenamefont {Zeng},
  \citenamefont {Zhang},\ and\ \citenamefont {Zhang}}]{Zeng2020arx}%
  \BibitemOpen
  \bibfield  {author} {\bibinfo {author} {\bibfnamefont {X.-X.}\ \bibnamefont
  {Zeng}}, \bibinfo {author} {\bibfnamefont {H.-Q.}\ \bibnamefont {Zhang}}, \
  and\ \bibinfo {author} {\bibfnamefont {H.}~\bibnamefont {Zhang}},\ }\href
  {\doibase 10.1140/epjc/s10052-020-08449-y} {\bibfield  {journal} {\bibinfo
  {journal} {The European Physical Journal C}\ }\textbf {\bibinfo {volume}
  {80}} (\bibinfo {year} {2020}),\ 10.1140/epjc/s10052-020-08449-y}\BibitemShut
  {NoStop}%
\bibitem [{\citenamefont {Zeng}\ \emph {et~al.}(2023)\citenamefont {Zeng},
  \citenamefont {Aslam},\ and\ \citenamefont {Saleem}}]{Zeng:2022fdm}%
  \BibitemOpen
  \bibfield  {author} {\bibinfo {author} {\bibfnamefont {X.-X.}\ \bibnamefont
  {Zeng}}, \bibinfo {author} {\bibfnamefont {M.~I.}\ \bibnamefont {Aslam}}, \
  and\ \bibinfo {author} {\bibfnamefont {R.}~\bibnamefont {Saleem}},\ }\href
  {\doibase 10.1140/epjc/s10052-023-11274-8} {\bibfield  {journal} {\bibinfo
  {journal} {Eur. Phys. J. C}\ }\textbf {\bibinfo {volume} {83}},\ \bibinfo
  {pages} {129} (\bibinfo {year} {2023})},\ \Eprint
  {http://arxiv.org/abs/2208.06246} {arXiv:2208.06246 [gr-qc]} \BibitemShut
  {NoStop}%
\bibitem [{\citenamefont {Zeng}\ and\ \citenamefont
  {Zhang}(2020)}]{Zeng2020qun}%
  \BibitemOpen
  \bibfield  {author} {\bibinfo {author} {\bibfnamefont {X.-X.}\ \bibnamefont
  {Zeng}}\ and\ \bibinfo {author} {\bibfnamefont {H.-Q.}\ \bibnamefont
  {Zhang}},\ }\href {\doibase 10.1140/epjc/s10052-020-08656-7} {\bibfield
  {journal} {\bibinfo  {journal} {The European Physical Journal C}\ }\textbf
  {\bibinfo {volume} {80}} (\bibinfo {year} {2020}),\
  10.1140/epjc/s10052-020-08656-7}\BibitemShut {NoStop}%
\bibitem [{\citenamefont {Zeng}\ \emph {et~al.}(2022)\citenamefont {Zeng},
  \citenamefont {Li},\ and\ \citenamefont {He}}]{Zeng:2021dlj}%
  \BibitemOpen
  \bibfield  {author} {\bibinfo {author} {\bibfnamefont {X.-X.}\ \bibnamefont
  {Zeng}}, \bibinfo {author} {\bibfnamefont {G.-P.}\ \bibnamefont {Li}}, \ and\
  \bibinfo {author} {\bibfnamefont {K.-J.}\ \bibnamefont {He}},\ }\href
  {\doibase 10.1016/j.nuclphysb.2021.115639} {\bibfield  {journal} {\bibinfo
  {journal} {Nucl. Phys. B}\ }\textbf {\bibinfo {volume} {974}},\ \bibinfo
  {pages} {115639} (\bibinfo {year} {2022})},\ \Eprint
  {http://arxiv.org/abs/2106.14478} {arXiv:2106.14478 [hep-th]} \BibitemShut
  {NoStop}%
\bibitem [{\citenamefont {Wang}\ \emph {et~al.}(2022)\citenamefont {Wang},
  \citenamefont {Lin},\ and\ \citenamefont {Wei}}]{Wang:2022yvi}%
  \BibitemOpen
  \bibfield  {author} {\bibinfo {author} {\bibfnamefont {H.-M.}\ \bibnamefont
  {Wang}}, \bibinfo {author} {\bibfnamefont {Z.-C.}\ \bibnamefont {Lin}}, \
  and\ \bibinfo {author} {\bibfnamefont {S.-W.}\ \bibnamefont {Wei}},\ }\href
  {\doibase 10.1016/j.nuclphysb.2022.116026} {\bibfield  {journal} {\bibinfo
  {journal} {Nucl. Phys. B}\ }\textbf {\bibinfo {volume} {985}},\ \bibinfo
  {pages} {116026} (\bibinfo {year} {2022})},\ \Eprint
  {http://arxiv.org/abs/2205.13174} {arXiv:2205.13174 [gr-qc]} \BibitemShut
  {NoStop}%
\bibitem [{\citenamefont {Gan}\ \emph {et~al.}(2021{\natexlab{a}})\citenamefont
  {Gan}, \citenamefont {Wang}, \citenamefont {Wu},\ and\ \citenamefont
  {Yang}}]{PhysRevD.104.044049}%
  \BibitemOpen
  \bibfield  {author} {\bibinfo {author} {\bibfnamefont {Q.}~\bibnamefont
  {Gan}}, \bibinfo {author} {\bibfnamefont {P.}~\bibnamefont {Wang}}, \bibinfo
  {author} {\bibfnamefont {H.}~\bibnamefont {Wu}}, \ and\ \bibinfo {author}
  {\bibfnamefont {H.}~\bibnamefont {Yang}},\ }\href {\doibase
  10.1103/PhysRevD.104.044049} {\bibfield  {journal} {\bibinfo  {journal}
  {Phys. Rev. D}\ }\textbf {\bibinfo {volume} {104}},\ \bibinfo {pages}
  {044049} (\bibinfo {year} {2021}{\natexlab{a}})}\BibitemShut {NoStop}%
\bibitem [{\citenamefont {Gan}\ \emph {et~al.}(2021{\natexlab{b}})\citenamefont
  {Gan}, \citenamefont {Wang}, \citenamefont {Wu},\ and\ \citenamefont
  {Yang}}]{Gan:2021pwu}%
  \BibitemOpen
  \bibfield  {author} {\bibinfo {author} {\bibfnamefont {Q.}~\bibnamefont
  {Gan}}, \bibinfo {author} {\bibfnamefont {P.}~\bibnamefont {Wang}}, \bibinfo
  {author} {\bibfnamefont {H.}~\bibnamefont {Wu}}, \ and\ \bibinfo {author}
  {\bibfnamefont {H.}~\bibnamefont {Yang}},\ }\href {\doibase
  10.1103/PhysRevD.104.024003} {\bibfield  {journal} {\bibinfo  {journal}
  {Phys. Rev. D}\ }\textbf {\bibinfo {volume} {104}},\ \bibinfo {pages}
  {024003} (\bibinfo {year} {2021}{\natexlab{b}})},\ \Eprint
  {http://arxiv.org/abs/2104.08703} {arXiv:2104.08703 [gr-qc]} \BibitemShut
  {NoStop}%
\bibitem [{\citenamefont {Uniyal}\ \emph {et~al.}(2023)\citenamefont {Uniyal},
  \citenamefont {Pantig},\ and\ \citenamefont {\"Ovg\"un}}]{Uniyal:2022vdu}%
  \BibitemOpen
  \bibfield  {author} {\bibinfo {author} {\bibfnamefont {A.}~\bibnamefont
  {Uniyal}}, \bibinfo {author} {\bibfnamefont {R.~C.}\ \bibnamefont {Pantig}},
  \ and\ \bibinfo {author} {\bibfnamefont {A.}~\bibnamefont {\"Ovg\"un}},\
  }\href {\doibase 10.1016/j.dark.2023.101178} {\bibfield  {journal} {\bibinfo
  {journal} {Phys. Dark Univ.}\ }\textbf {\bibinfo {volume} {40}},\ \bibinfo
  {pages} {101178} (\bibinfo {year} {2023})},\ \Eprint
  {http://arxiv.org/abs/2205.11072} {arXiv:2205.11072 [gr-qc]} \BibitemShut
  {NoStop}%
\bibitem [{\citenamefont {Wang}\ \emph {et~al.}(2023)\citenamefont {Wang},
  \citenamefont {Kuang}, \citenamefont {Meng}, \citenamefont {Wang},\ and\
  \citenamefont {Wu}}]{Wang:2023vcv}%
  \BibitemOpen
  \bibfield  {author} {\bibinfo {author} {\bibfnamefont {X.-J.}\ \bibnamefont
  {Wang}}, \bibinfo {author} {\bibfnamefont {X.-M.}\ \bibnamefont {Kuang}},
  \bibinfo {author} {\bibfnamefont {Y.}~\bibnamefont {Meng}}, \bibinfo {author}
  {\bibfnamefont {B.}~\bibnamefont {Wang}}, \ and\ \bibinfo {author}
  {\bibfnamefont {J.-P.}\ \bibnamefont {Wu}},\ }\href {\doibase
  10.1103/PhysRevD.107.124052} {\bibfield  {journal} {\bibinfo  {journal}
  {Phys. Rev. D}\ }\textbf {\bibinfo {volume} {107}},\ \bibinfo {pages}
  {124052} (\bibinfo {year} {2023})},\ \Eprint
  {http://arxiv.org/abs/2304.10015} {arXiv:2304.10015 [gr-qc]} \BibitemShut
  {NoStop}%
\bibitem [{\citenamefont {Peng}\ \emph
  {et~al.}(2021{\natexlab{a}})\citenamefont {Peng}, \citenamefont {Guo},\ and\
  \citenamefont {Feng}}]{Peng:2020wun}%
  \BibitemOpen
  \bibfield  {author} {\bibinfo {author} {\bibfnamefont {J.}~\bibnamefont
  {Peng}}, \bibinfo {author} {\bibfnamefont {M.}~\bibnamefont {Guo}}, \ and\
  \bibinfo {author} {\bibfnamefont {X.-H.}\ \bibnamefont {Feng}},\ }\href
  {\doibase 10.1088/1674-1137/ac06bb} {\bibfield  {journal} {\bibinfo
  {journal} {Chin. Phys. C}\ }\textbf {\bibinfo {volume} {45}},\ \bibinfo
  {pages} {085103} (\bibinfo {year} {2021}{\natexlab{a}})},\ \Eprint
  {http://arxiv.org/abs/2008.00657} {arXiv:2008.00657 [gr-qc]} \BibitemShut
  {NoStop}%
\bibitem [{\citenamefont {Peng}\ \emph
  {et~al.}(2021{\natexlab{b}})\citenamefont {Peng}, \citenamefont {Guo},\ and\
  \citenamefont {Feng}}]{Peng:2021osd}%
  \BibitemOpen
  \bibfield  {author} {\bibinfo {author} {\bibfnamefont {J.}~\bibnamefont
  {Peng}}, \bibinfo {author} {\bibfnamefont {M.}~\bibnamefont {Guo}}, \ and\
  \bibinfo {author} {\bibfnamefont {X.-H.}\ \bibnamefont {Feng}},\ }\href
  {\doibase 10.1103/PhysRevD.104.124010} {\bibfield  {journal} {\bibinfo
  {journal} {Phys. Rev. D}\ }\textbf {\bibinfo {volume} {104}},\ \bibinfo
  {pages} {124010} (\bibinfo {year} {2021}{\natexlab{b}})},\ \Eprint
  {http://arxiv.org/abs/2102.05488} {arXiv:2102.05488 [gr-qc]} \BibitemShut
  {NoStop}%
\bibitem [{\citenamefont {Guerrero}\ \emph
  {et~al.}(2022{\natexlab{a}})\citenamefont {Guerrero}, \citenamefont {Olmo},
  \citenamefont {Rubiera-Garcia},\ and\ \citenamefont
  {G\'omez}}]{PhysRevD.105.084057}%
  \BibitemOpen
  \bibfield  {author} {\bibinfo {author} {\bibfnamefont {M.}~\bibnamefont
  {Guerrero}}, \bibinfo {author} {\bibfnamefont {G.~J.}\ \bibnamefont {Olmo}},
  \bibinfo {author} {\bibfnamefont {D.}~\bibnamefont {Rubiera-Garcia}}, \ and\
  \bibinfo {author} {\bibfnamefont {D.~S.-C.}\ \bibnamefont {G\'omez}},\ }\href
  {\doibase 10.1103/PhysRevD.105.084057} {\bibfield  {journal} {\bibinfo
  {journal} {Phys. Rev. D}\ }\textbf {\bibinfo {volume} {105}},\ \bibinfo
  {pages} {084057} (\bibinfo {year} {2022}{\natexlab{a}})}\BibitemShut
  {NoStop}%
\bibitem [{\citenamefont {Guerrero}\ \emph
  {et~al.}(2022{\natexlab{b}})\citenamefont {Guerrero}, \citenamefont {Olmo},
  \citenamefont {Rubiera-Garcia},\ and\ \citenamefont
  {S\'aez-Chill\'on~G\'omez}}]{Guerrero:2022msp}%
  \BibitemOpen
  \bibfield  {author} {\bibinfo {author} {\bibfnamefont {M.}~\bibnamefont
  {Guerrero}}, \bibinfo {author} {\bibfnamefont {G.~J.}\ \bibnamefont {Olmo}},
  \bibinfo {author} {\bibfnamefont {D.}~\bibnamefont {Rubiera-Garcia}}, \ and\
  \bibinfo {author} {\bibfnamefont {D.}~\bibnamefont
  {S\'aez-Chill\'on~G\'omez}},\ }\href {\doibase 10.1103/PhysRevD.106.044070}
  {\bibfield  {journal} {\bibinfo  {journal} {Phys. Rev. D}\ }\textbf {\bibinfo
  {volume} {106}},\ \bibinfo {pages} {044070} (\bibinfo {year}
  {2022}{\natexlab{b}})},\ \Eprint {http://arxiv.org/abs/2205.12147}
  {arXiv:2205.12147 [gr-qc]} \BibitemShut {NoStop}%
\bibitem [{\citenamefont {Huang}\ \emph {et~al.}(2023)\citenamefont {Huang},
  \citenamefont {Kunz}, \citenamefont {Yang},\ and\ \citenamefont
  {Zhang}}]{Huang:2023yqd}%
  \BibitemOpen
  \bibfield  {author} {\bibinfo {author} {\bibfnamefont {H.}~\bibnamefont
  {Huang}}, \bibinfo {author} {\bibfnamefont {J.}~\bibnamefont {Kunz}},
  \bibinfo {author} {\bibfnamefont {J.}~\bibnamefont {Yang}}, \ and\ \bibinfo
  {author} {\bibfnamefont {C.}~\bibnamefont {Zhang}},\ }\href {\doibase
  10.1103/PhysRevD.107.104060} {\bibfield  {journal} {\bibinfo  {journal}
  {Phys. Rev. D}\ }\textbf {\bibinfo {volume} {107}},\ \bibinfo {pages}
  {104060} (\bibinfo {year} {2023})},\ \Eprint
  {http://arxiv.org/abs/2303.11885} {arXiv:2303.11885 [gr-qc]} \BibitemShut
  {NoStop}%
\bibitem [{\citenamefont {Hawking}(1976)}]{PhysRevD.14.2460}%
  \BibitemOpen
  \bibfield  {author} {\bibinfo {author} {\bibfnamefont {S.~W.}\ \bibnamefont
  {Hawking}},\ }\href {\doibase 10.1103/PhysRevD.14.2460} {\bibfield  {journal}
  {\bibinfo  {journal} {Phys. Rev. D}\ }\textbf {\bibinfo {volume} {14}},\
  \bibinfo {pages} {2460} (\bibinfo {year} {1976})}\BibitemShut {NoStop}%
\bibitem [{\citenamefont {Page}(1993)}]{Page:1993wv}%
  \BibitemOpen
  \bibfield  {author} {\bibinfo {author} {\bibfnamefont {D.~N.}\ \bibnamefont
  {Page}},\ }\href {\doibase 10.1103/PhysRevLett.71.3743} {\bibfield  {journal}
  {\bibinfo  {journal} {Phys. Rev. Lett.}\ }\textbf {\bibinfo {volume} {71}},\
  \bibinfo {pages} {3743} (\bibinfo {year} {1993})},\ \Eprint
  {http://arxiv.org/abs/hep-th/9306083} {arXiv:hep-th/9306083} \BibitemShut
  {NoStop}%
\bibitem [{\citenamefont {Page}(2013)}]{Page:2013dx}%
  \BibitemOpen
  \bibfield  {author} {\bibinfo {author} {\bibfnamefont {D.~N.}\ \bibnamefont
  {Page}},\ }\href {\doibase 10.1088/1475-7516/2013/09/028} {\bibfield
  {journal} {\bibinfo  {journal} {JCAP}\ }\textbf {\bibinfo {volume} {09}},\
  \bibinfo {pages} {028} (\bibinfo {year} {2013})},\ \Eprint
  {http://arxiv.org/abs/1301.4995} {arXiv:1301.4995 [hep-th]} \BibitemShut
  {NoStop}%
\bibitem [{\citenamefont {Mathur}(2009)}]{Mathur:2009hf}%
  \BibitemOpen
  \bibfield  {author} {\bibinfo {author} {\bibfnamefont {S.~D.}\ \bibnamefont
  {Mathur}},\ }\href {\doibase 10.1088/0264-9381/26/22/224001} {\bibfield
  {journal} {\bibinfo  {journal} {Class. Quant. Grav.}\ }\textbf {\bibinfo
  {volume} {26}},\ \bibinfo {pages} {224001} (\bibinfo {year} {2009})},\
  \Eprint {http://arxiv.org/abs/0909.1038} {arXiv:0909.1038 [hep-th]}
  \BibitemShut {NoStop}%
\bibitem [{\citenamefont {Harlow}(2016)}]{RevModPhys.88.015002}%
  \BibitemOpen
  \bibfield  {author} {\bibinfo {author} {\bibfnamefont {D.}~\bibnamefont
  {Harlow}},\ }\href {\doibase 10.1103/RevModPhys.88.015002} {\bibfield
  {journal} {\bibinfo  {journal} {Rev. Mod. Phys.}\ }\textbf {\bibinfo {volume}
  {88}},\ \bibinfo {pages} {015002} (\bibinfo {year} {2016})}\BibitemShut
  {NoStop}%
\bibitem [{\citenamefont {Raju}(2022)}]{Raju:2020smc}%
  \BibitemOpen
  \bibfield  {author} {\bibinfo {author} {\bibfnamefont {S.}~\bibnamefont
  {Raju}},\ }\href {\doibase 10.1016/j.physrep.2021.10.001} {\bibfield
  {journal} {\bibinfo  {journal} {Phys. Rept.}\ }\textbf {\bibinfo {volume}
  {943}},\ \bibinfo {pages} {1} (\bibinfo {year} {2022})},\ \Eprint
  {http://arxiv.org/abs/2012.05770} {arXiv:2012.05770 [hep-th]} \BibitemShut
  {NoStop}%
\bibitem [{\citenamefont {Almheiri}\ \emph {et~al.}(2021)\citenamefont
  {Almheiri}, \citenamefont {Hartman}, \citenamefont {Maldacena}, \citenamefont
  {Shaghoulian},\ and\ \citenamefont {Tajdini}}]{RevModPhys.93.035002}%
  \BibitemOpen
  \bibfield  {author} {\bibinfo {author} {\bibfnamefont {A.}~\bibnamefont
  {Almheiri}}, \bibinfo {author} {\bibfnamefont {T.}~\bibnamefont {Hartman}},
  \bibinfo {author} {\bibfnamefont {J.}~\bibnamefont {Maldacena}}, \bibinfo
  {author} {\bibfnamefont {E.}~\bibnamefont {Shaghoulian}}, \ and\ \bibinfo
  {author} {\bibfnamefont {A.}~\bibnamefont {Tajdini}},\ }\href {\doibase
  10.1103/RevModPhys.93.035002} {\bibfield  {journal} {\bibinfo  {journal}
  {Rev. Mod. Phys.}\ }\textbf {\bibinfo {volume} {93}},\ \bibinfo {pages}
  {035002} (\bibinfo {year} {2021})}\BibitemShut {NoStop}%
\bibitem [{\citenamefont {Gibbons}\ and\ \citenamefont
  {Maeda}(1988)}]{Gibbons:1987ps}%
  \BibitemOpen
  \bibfield  {author} {\bibinfo {author} {\bibfnamefont {G.~W.}\ \bibnamefont
  {Gibbons}}\ and\ \bibinfo {author} {\bibfnamefont {K.-i.}\ \bibnamefont
  {Maeda}},\ }\href {\doibase 10.1016/0550-3213(88)90006-5} {\bibfield
  {journal} {\bibinfo  {journal} {Nucl. Phys. B}\ }\textbf {\bibinfo {volume}
  {298}},\ \bibinfo {pages} {741} (\bibinfo {year} {1988})}\BibitemShut
  {NoStop}%
\bibitem [{\citenamefont {Garfinkle}\ \emph {et~al.}(1991)\citenamefont
  {Garfinkle}, \citenamefont {Horowitz},\ and\ \citenamefont
  {Strominger}}]{Garfinkle:1990qj}%
  \BibitemOpen
  \bibfield  {author} {\bibinfo {author} {\bibfnamefont {D.}~\bibnamefont
  {Garfinkle}}, \bibinfo {author} {\bibfnamefont {G.~T.}\ \bibnamefont
  {Horowitz}}, \ and\ \bibinfo {author} {\bibfnamefont {A.}~\bibnamefont
  {Strominger}},\ }\href {\doibase 10.1103/PhysRevD.43.3140} {\bibfield
  {journal} {\bibinfo  {journal} {Phys. Rev. D}\ }\textbf {\bibinfo {volume}
  {43}},\ \bibinfo {pages} {3140} (\bibinfo {year} {1991})},\ \bibinfo {note}
  {[Erratum: Phys.Rev.D 45, 3888 (1992)]}\BibitemShut {NoStop}%
\bibitem [{\citenamefont {Misner}\ \emph {et~al.}(1973)\citenamefont {Misner},
  \citenamefont {Thorne},\ and\ \citenamefont {Wheeler}}]{gravitation}%
  \BibitemOpen
  \bibfield  {author} {\bibinfo {author} {\bibfnamefont {C.~W.}\ \bibnamefont
  {Misner}}, \bibinfo {author} {\bibfnamefont {K.~S.}\ \bibnamefont {Thorne}},
  \ and\ \bibinfo {author} {\bibfnamefont {J.~A.}\ \bibnamefont {Wheeler}},\
  }\href {http://example.com/history_of_time.pdf} {\emph {\bibinfo {title}
  {Gravitation}}}\ (\bibinfo  {publisher} {Princeton University Press},\
  \bibinfo {year} {1973})\BibitemShut {NoStop}%
\bibitem [{\citenamefont {Doneva}\ and\ \citenamefont
  {Yazadjiev}(2018)}]{PhysRevLett.120.131103}%
  \BibitemOpen
  \bibfield  {author} {\bibinfo {author} {\bibfnamefont {D.~D.}\ \bibnamefont
  {Doneva}}\ and\ \bibinfo {author} {\bibfnamefont {S.~S.}\ \bibnamefont
  {Yazadjiev}},\ }\href {\doibase 10.1103/PhysRevLett.120.131103} {\bibfield
  {journal} {\bibinfo  {journal} {Phys. Rev. Lett.}\ }\textbf {\bibinfo
  {volume} {120}},\ \bibinfo {pages} {131103} (\bibinfo {year}
  {2018})}\BibitemShut {NoStop}%
\bibitem [{\citenamefont {Silva}\ \emph {et~al.}(2018)\citenamefont {Silva},
  \citenamefont {Sakstein}, \citenamefont {Gualtieri}, \citenamefont
  {Sotiriou},\ and\ \citenamefont {Berti}}]{PhysRevLett.120.131104}%
  \BibitemOpen
  \bibfield  {author} {\bibinfo {author} {\bibfnamefont {H.~O.}\ \bibnamefont
  {Silva}}, \bibinfo {author} {\bibfnamefont {J.}~\bibnamefont {Sakstein}},
  \bibinfo {author} {\bibfnamefont {L.}~\bibnamefont {Gualtieri}}, \bibinfo
  {author} {\bibfnamefont {T.~P.}\ \bibnamefont {Sotiriou}}, \ and\ \bibinfo
  {author} {\bibfnamefont {E.}~\bibnamefont {Berti}},\ }\href {\doibase
  10.1103/PhysRevLett.120.131104} {\bibfield  {journal} {\bibinfo  {journal}
  {Phys. Rev. Lett.}\ }\textbf {\bibinfo {volume} {120}},\ \bibinfo {pages}
  {131104} (\bibinfo {year} {2018})}\BibitemShut {NoStop}%
\bibitem [{\citenamefont {Herdeiro}\ \emph {et~al.}(2018)\citenamefont
  {Herdeiro}, \citenamefont {Radu}, \citenamefont {Sanchis-Gual},\ and\
  \citenamefont {Font}}]{PhysRevLett.121.101102}%
  \BibitemOpen
  \bibfield  {author} {\bibinfo {author} {\bibfnamefont {C.~A.~R.}\
  \bibnamefont {Herdeiro}}, \bibinfo {author} {\bibfnamefont {E.}~\bibnamefont
  {Radu}}, \bibinfo {author} {\bibfnamefont {N.}~\bibnamefont {Sanchis-Gual}},
  \ and\ \bibinfo {author} {\bibfnamefont {J.~A.}\ \bibnamefont {Font}},\
  }\href {\doibase 10.1103/PhysRevLett.121.101102} {\bibfield  {journal}
  {\bibinfo  {journal} {Phys. Rev. Lett.}\ }\textbf {\bibinfo {volume} {121}},\
  \bibinfo {pages} {101102} (\bibinfo {year} {2018})}\BibitemShut {NoStop}%
\bibitem [{\citenamefont {Fernandes}\ \emph {et~al.}(2019)\citenamefont
  {Fernandes}, \citenamefont {Herdeiro}, \citenamefont {Pombo}, \citenamefont
  {Radu},\ and\ \citenamefont {Sanchis-Gual}}]{Fernandes:2019rez}%
  \BibitemOpen
  \bibfield  {author} {\bibinfo {author} {\bibfnamefont {P.~G.~S.}\
  \bibnamefont {Fernandes}}, \bibinfo {author} {\bibfnamefont {C.~A.~R.}\
  \bibnamefont {Herdeiro}}, \bibinfo {author} {\bibfnamefont {A.~M.}\
  \bibnamefont {Pombo}}, \bibinfo {author} {\bibfnamefont {E.}~\bibnamefont
  {Radu}}, \ and\ \bibinfo {author} {\bibfnamefont {N.}~\bibnamefont
  {Sanchis-Gual}},\ }\href {\doibase 10.1088/1361-6382/ab23a1} {\bibfield
  {journal} {\bibinfo  {journal} {Class. Quant. Grav.}\ }\textbf {\bibinfo
  {volume} {36}},\ \bibinfo {pages} {134002} (\bibinfo {year} {2019})},\
  \bibinfo {note} {[Erratum: Class.Quant.Grav. 37, 049501 (2020)]},\ \Eprint
  {http://arxiv.org/abs/1902.05079} {arXiv:1902.05079 [gr-qc]} \BibitemShut
  {NoStop}%
\bibitem [{\citenamefont {Guo}\ \emph {et~al.}(2021)\citenamefont {Guo},
  \citenamefont {Wang}, \citenamefont {Wu},\ and\ \citenamefont
  {Yang}}]{Guo:2021zed}%
  \BibitemOpen
  \bibfield  {author} {\bibinfo {author} {\bibfnamefont {G.}~\bibnamefont
  {Guo}}, \bibinfo {author} {\bibfnamefont {P.}~\bibnamefont {Wang}}, \bibinfo
  {author} {\bibfnamefont {H.}~\bibnamefont {Wu}}, \ and\ \bibinfo {author}
  {\bibfnamefont {H.}~\bibnamefont {Yang}},\ }\href {\doibase
  10.1140/epjc/s10052-021-09614-7} {\bibfield  {journal} {\bibinfo  {journal}
  {Eur. Phys. J. C}\ }\textbf {\bibinfo {volume} {81}},\ \bibinfo {pages} {864}
  (\bibinfo {year} {2021})},\ \Eprint {http://arxiv.org/abs/2102.04015}
  {arXiv:2102.04015 [gr-qc]} \BibitemShut {NoStop}%
\bibitem [{\citenamefont {Promsiri}\ \emph {et~al.}(2023)\citenamefont
  {Promsiri}, \citenamefont {Tangphati}, \citenamefont {Hirunsirisawat},\ and\
  \citenamefont {Ponglertsakul}}]{PhysRevD.108.024015}%
  \BibitemOpen
  \bibfield  {author} {\bibinfo {author} {\bibfnamefont {C.}~\bibnamefont
  {Promsiri}}, \bibinfo {author} {\bibfnamefont {T.}~\bibnamefont {Tangphati}},
  \bibinfo {author} {\bibfnamefont {E.}~\bibnamefont {Hirunsirisawat}}, \ and\
  \bibinfo {author} {\bibfnamefont {S.}~\bibnamefont {Ponglertsakul}},\ }\href
  {\doibase 10.1103/PhysRevD.108.024015} {\bibfield  {journal} {\bibinfo
  {journal} {Phys. Rev. D}\ }\textbf {\bibinfo {volume} {108}},\ \bibinfo
  {pages} {024015} (\bibinfo {year} {2023})}\BibitemShut {NoStop}%
\bibitem [{\citenamefont {Fernando}\ and\ \citenamefont
  {Arnold}(2004)}]{Fernando:2003wc}%
  \BibitemOpen
  \bibfield  {author} {\bibinfo {author} {\bibfnamefont {S.}~\bibnamefont
  {Fernando}}\ and\ \bibinfo {author} {\bibfnamefont {K.}~\bibnamefont
  {Arnold}},\ }\href {\doibase 10.1023/B:GERG.0000035953.31652.88} {\bibfield
  {journal} {\bibinfo  {journal} {Gen. Rel. Grav.}\ }\textbf {\bibinfo {volume}
  {36}},\ \bibinfo {pages} {1805} (\bibinfo {year} {2004})},\ \Eprint
  {http://arxiv.org/abs/hep-th/0312041} {arXiv:hep-th/0312041} \BibitemShut
  {NoStop}%
\bibitem [{\citenamefont {Fernando}(2012)}]{PhysRevD.85.024033}%
  \BibitemOpen
  \bibfield  {author} {\bibinfo {author} {\bibfnamefont {S.}~\bibnamefont
  {Fernando}},\ }\href {\doibase 10.1103/PhysRevD.85.024033} {\bibfield
  {journal} {\bibinfo  {journal} {Phys. Rev. D}\ }\textbf {\bibinfo {volume}
  {85}},\ \bibinfo {pages} {024033} (\bibinfo {year} {2012})}\BibitemShut
  {NoStop}%
\bibitem [{\citenamefont {Heydari-Fard}\ \emph {et~al.}(2022)\citenamefont
  {Heydari-Fard}, \citenamefont {Heydari-Fard},\ and\ \citenamefont
  {Sepangi}}]{PhysRevD.105.124009}%
  \BibitemOpen
  \bibfield  {author} {\bibinfo {author} {\bibfnamefont {M.}~\bibnamefont
  {Heydari-Fard}}, \bibinfo {author} {\bibfnamefont {M.}~\bibnamefont
  {Heydari-Fard}}, \ and\ \bibinfo {author} {\bibfnamefont {H.~R.}\
  \bibnamefont {Sepangi}},\ }\href {\doibase 10.1103/PhysRevD.105.124009}
  {\bibfield  {journal} {\bibinfo  {journal} {Phys. Rev. D}\ }\textbf {\bibinfo
  {volume} {105}},\ \bibinfo {pages} {124009} (\bibinfo {year}
  {2022})}\BibitemShut {NoStop}%
\bibitem [{\citenamefont {Maki}\ and\ \citenamefont
  {Shiraishi}(1994)}]{Maki:1992up}%
  \BibitemOpen
  \bibfield  {author} {\bibinfo {author} {\bibfnamefont {T.}~\bibnamefont
  {Maki}}\ and\ \bibinfo {author} {\bibfnamefont {K.}~\bibnamefont
  {Shiraishi}},\ }\href {\doibase 10.1088/0264-9381/11/1/022} {\bibfield
  {journal} {\bibinfo  {journal} {Class. Quant. Grav.}\ }\textbf {\bibinfo
  {volume} {11}},\ \bibinfo {pages} {227} (\bibinfo {year} {1994})},\ \Eprint
  {http://arxiv.org/abs/1707.05463} {arXiv:1707.05463 [gr-qc]} \BibitemShut
  {NoStop}%
\bibitem [{\citenamefont {Olivares}\ and\ \citenamefont
  {Villanueva}(2013)}]{Olivares:2013jza}%
  \BibitemOpen
  \bibfield  {author} {\bibinfo {author} {\bibfnamefont {M.}~\bibnamefont
  {Olivares}}\ and\ \bibinfo {author} {\bibfnamefont {J.~R.}\ \bibnamefont
  {Villanueva}},\ }\href {\doibase 10.1140/epjc/s10052-013-2659-4} {\bibfield
  {journal} {\bibinfo  {journal} {Eur. Phys. J. C}\ }\textbf {\bibinfo {volume}
  {73}},\ \bibinfo {pages} {2659} (\bibinfo {year} {2013})},\ \Eprint
  {http://arxiv.org/abs/1311.4236} {arXiv:1311.4236 [gr-qc]} \BibitemShut
  {NoStop}%
\bibitem [{\citenamefont {Pradhan}(2015)}]{Pradhan:2012id}%
  \BibitemOpen
  \bibfield  {author} {\bibinfo {author} {\bibfnamefont {P.~P.}\ \bibnamefont
  {Pradhan}},\ }\href {\doibase 10.1142/S0218271815500868} {\bibfield
  {journal} {\bibinfo  {journal} {Int. J. Mod. Phys. D}\ }\textbf {\bibinfo
  {volume} {24}},\ \bibinfo {pages} {1550086} (\bibinfo {year} {2015})},\
  \Eprint {http://arxiv.org/abs/1210.0221} {arXiv:1210.0221 [gr-qc]}
  \BibitemShut {NoStop}%
\bibitem [{\citenamefont {Blaga}(2015)}]{Blaga:2014spa}%
  \BibitemOpen
  \bibfield  {author} {\bibinfo {author} {\bibfnamefont {C.}~\bibnamefont
  {Blaga}},\ }\href {\doibase 10.2298/SAJ1590041B} {\bibfield  {journal}
  {\bibinfo  {journal} {Serb. Astron. J.}\ }\textbf {\bibinfo {volume} {190}},\
  \bibinfo {pages} {41} (\bibinfo {year} {2015})},\ \Eprint
  {http://arxiv.org/abs/1407.1504} {arXiv:1407.1504 [gr-qc]} \BibitemShut
  {NoStop}%
\bibitem [{\citenamefont {Bhadra}(2003)}]{Bhadra:2003zs}%
  \BibitemOpen
  \bibfield  {author} {\bibinfo {author} {\bibfnamefont {A.}~\bibnamefont
  {Bhadra}},\ }\href {\doibase 10.1103/PhysRevD.67.103009} {\bibfield
  {journal} {\bibinfo  {journal} {Phys. Rev. D}\ }\textbf {\bibinfo {volume}
  {67}},\ \bibinfo {pages} {103009} (\bibinfo {year} {2003})},\ \Eprint
  {http://arxiv.org/abs/gr-qc/0306016} {arXiv:gr-qc/0306016} \BibitemShut
  {NoStop}%
\bibitem [{\citenamefont {Mukherjee}\ and\ \citenamefont
  {Majumdar}(2007)}]{Mukherjee:2006ru}%
  \BibitemOpen
  \bibfield  {author} {\bibinfo {author} {\bibfnamefont {N.}~\bibnamefont
  {Mukherjee}}\ and\ \bibinfo {author} {\bibfnamefont {A.~S.}\ \bibnamefont
  {Majumdar}},\ }\href {\doibase 10.1007/s10714-007-0407-5} {\bibfield
  {journal} {\bibinfo  {journal} {Gen. Rel. Grav.}\ }\textbf {\bibinfo {volume}
  {39}},\ \bibinfo {pages} {583} (\bibinfo {year} {2007})},\ \Eprint
  {http://arxiv.org/abs/astro-ph/0605224} {arXiv:astro-ph/0605224} \BibitemShut
  {NoStop}%
\bibitem [{\citenamefont {Ghosh}\ and\ \citenamefont
  {Sengupta}(2010)}]{Ghosh:2010uw}%
  \BibitemOpen
  \bibfield  {author} {\bibinfo {author} {\bibfnamefont {T.}~\bibnamefont
  {Ghosh}}\ and\ \bibinfo {author} {\bibfnamefont {S.}~\bibnamefont
  {Sengupta}},\ }\href {\doibase 10.1103/PhysRevD.81.044013} {\bibfield
  {journal} {\bibinfo  {journal} {Phys. Rev. D}\ }\textbf {\bibinfo {volume}
  {81}},\ \bibinfo {pages} {044013} (\bibinfo {year} {2010})},\ \Eprint
  {http://arxiv.org/abs/1001.5129} {arXiv:1001.5129 [gr-qc]} \BibitemShut
  {NoStop}%
\bibitem [{\citenamefont {Shiraishi}(1992)}]{Shiraishi:1992np}%
  \BibitemOpen
  \bibfield  {author} {\bibinfo {author} {\bibfnamefont {K.}~\bibnamefont
  {Shiraishi}},\ }\href {\doibase 10.1016/0375-9601(92)90712-U} {\bibfield
  {journal} {\bibinfo  {journal} {Phys. Lett. A}\ }\textbf {\bibinfo {volume}
  {166}},\ \bibinfo {pages} {298} (\bibinfo {year} {1992})},\ \Eprint
  {http://arxiv.org/abs/1511.08543} {arXiv:1511.08543 [gr-qc]} \BibitemShut
  {NoStop}%
\bibitem [{\citenamefont {Horne}\ and\ \citenamefont
  {Horowitz}(1992{\natexlab{a}})}]{Horne:1992zy}%
  \BibitemOpen
  \bibfield  {author} {\bibinfo {author} {\bibfnamefont {J.~H.}\ \bibnamefont
  {Horne}}\ and\ \bibinfo {author} {\bibfnamefont {G.~T.}\ \bibnamefont
  {Horowitz}},\ }\href {\doibase 10.1103/PhysRevD.46.1340} {\bibfield
  {journal} {\bibinfo  {journal} {Phys. Rev. D}\ }\textbf {\bibinfo {volume}
  {46}},\ \bibinfo {pages} {1340} (\bibinfo {year} {1992}{\natexlab{a}})},\
  \Eprint {http://arxiv.org/abs/hep-th/9203083} {arXiv:hep-th/9203083}
  \BibitemShut {NoStop}%
\bibitem [{\citenamefont {Sen}(1992)}]{Sen:1992ua}%
  \BibitemOpen
  \bibfield  {author} {\bibinfo {author} {\bibfnamefont {A.}~\bibnamefont
  {Sen}},\ }\href {\doibase 10.1103/PhysRevLett.69.1006} {\bibfield  {journal}
  {\bibinfo  {journal} {Phys. Rev. Lett.}\ }\textbf {\bibinfo {volume} {69}},\
  \bibinfo {pages} {1006} (\bibinfo {year} {1992})},\ \Eprint
  {http://arxiv.org/abs/hep-th/9204046} {arXiv:hep-th/9204046} \BibitemShut
  {NoStop}%
\bibitem [{\citenamefont {Younsi}\ \emph {et~al.}(2016)\citenamefont {Younsi},
  \citenamefont {Zhidenko}, \citenamefont {Rezzolla}, \citenamefont
  {Konoplya},\ and\ \citenamefont {Mizuno}}]{Younsi:2016azx}%
  \BibitemOpen
  \bibfield  {author} {\bibinfo {author} {\bibfnamefont {Z.}~\bibnamefont
  {Younsi}}, \bibinfo {author} {\bibfnamefont {A.}~\bibnamefont {Zhidenko}},
  \bibinfo {author} {\bibfnamefont {L.}~\bibnamefont {Rezzolla}}, \bibinfo
  {author} {\bibfnamefont {R.}~\bibnamefont {Konoplya}}, \ and\ \bibinfo
  {author} {\bibfnamefont {Y.}~\bibnamefont {Mizuno}},\ }\href {\doibase
  10.1103/PhysRevD.94.084025} {\bibfield  {journal} {\bibinfo  {journal} {Phys.
  Rev. D}\ }\textbf {\bibinfo {volume} {94}},\ \bibinfo {pages} {084025}
  (\bibinfo {year} {2016})},\ \Eprint {http://arxiv.org/abs/1607.05767}
  {arXiv:1607.05767 [gr-qc]} \BibitemShut {NoStop}%
\bibitem [{\citenamefont {Konoplya}\ and\ \citenamefont
  {Zhidenko}(2021)}]{Konoplya:2021slg}%
  \BibitemOpen
  \bibfield  {author} {\bibinfo {author} {\bibfnamefont {R.~A.}\ \bibnamefont
  {Konoplya}}\ and\ \bibinfo {author} {\bibfnamefont {A.}~\bibnamefont
  {Zhidenko}},\ }\href {\doibase 10.1103/PhysRevD.103.104033} {\bibfield
  {journal} {\bibinfo  {journal} {Phys. Rev. D}\ }\textbf {\bibinfo {volume}
  {103}},\ \bibinfo {pages} {104033} (\bibinfo {year} {2021})},\ \Eprint
  {http://arxiv.org/abs/2103.03855} {arXiv:2103.03855 [gr-qc]} \BibitemShut
  {NoStop}%
\bibitem [{\citenamefont {Bad\'\i{}a}\ and\ \citenamefont
  {Eiroa}(2023)}]{Badia:2022phg}%
  \BibitemOpen
  \bibfield  {author} {\bibinfo {author} {\bibfnamefont {J.}~\bibnamefont
  {Bad\'\i{}a}}\ and\ \bibinfo {author} {\bibfnamefont {E.~F.}\ \bibnamefont
  {Eiroa}},\ }\href {\doibase 10.1103/PhysRevD.107.124028} {\bibfield
  {journal} {\bibinfo  {journal} {Phys. Rev. D}\ }\textbf {\bibinfo {volume}
  {107}},\ \bibinfo {pages} {124028} (\bibinfo {year} {2023})},\ \Eprint
  {http://arxiv.org/abs/2210.03081} {arXiv:2210.03081 [gr-qc]} \BibitemShut
  {NoStop}%
\bibitem [{\citenamefont {Heydari-Fard}\ \emph {et~al.}(2020)\citenamefont
  {Heydari-Fard}, \citenamefont {Heydari-Fard},\ and\ \citenamefont
  {Sepangi}}]{Heydari-Fard:2020ugv}%
  \BibitemOpen
  \bibfield  {author} {\bibinfo {author} {\bibfnamefont {M.}~\bibnamefont
  {Heydari-Fard}}, \bibinfo {author} {\bibfnamefont {M.}~\bibnamefont
  {Heydari-Fard}}, \ and\ \bibinfo {author} {\bibfnamefont {H.~R.}\
  \bibnamefont {Sepangi}},\ }\href {\doibase 10.1140/epjc/s10052-020-7911-0}
  {\bibfield  {journal} {\bibinfo  {journal} {Eur. Phys. J. C}\ }\textbf
  {\bibinfo {volume} {80}},\ \bibinfo {pages} {351} (\bibinfo {year} {2020})},\
  \Eprint {http://arxiv.org/abs/2004.05552} {arXiv:2004.05552 [gr-qc]}
  \BibitemShut {NoStop}%
\bibitem [{\citenamefont {Holzhey}\ and\ \citenamefont
  {Wilczek}(1992)}]{holzhey1992black}%
  \BibitemOpen
  \bibfield  {author} {\bibinfo {author} {\bibfnamefont {C.~F.}\ \bibnamefont
  {Holzhey}}\ and\ \bibinfo {author} {\bibfnamefont {F.}~\bibnamefont
  {Wilczek}},\ }\href@noop {} {\bibfield  {journal} {\bibinfo  {journal}
  {Nuclear Physics B}\ }\textbf {\bibinfo {volume} {380}},\ \bibinfo {pages}
  {447} (\bibinfo {year} {1992})}\BibitemShut {NoStop}%
\bibitem [{\citenamefont {Gralla}\ \emph {et~al.}(2020)\citenamefont {Gralla},
  \citenamefont {Lupsasca},\ and\ \citenamefont
  {Marrone}}]{PhysRevD.102.124004}%
  \BibitemOpen
  \bibfield  {author} {\bibinfo {author} {\bibfnamefont {S.~E.}\ \bibnamefont
  {Gralla}}, \bibinfo {author} {\bibfnamefont {A.}~\bibnamefont {Lupsasca}}, \
  and\ \bibinfo {author} {\bibfnamefont {D.~P.}\ \bibnamefont {Marrone}},\
  }\href {\doibase 10.1103/PhysRevD.102.124004} {\bibfield  {journal} {\bibinfo
   {journal} {Phys. Rev. D}\ }\textbf {\bibinfo {volume} {102}},\ \bibinfo
  {pages} {124004} (\bibinfo {year} {2020})}\BibitemShut {NoStop}%
\bibitem [{\citenamefont {Horne}\ and\ \citenamefont
  {Horowitz}(1992{\natexlab{b}})}]{PhysRevD.46.1340}%
  \BibitemOpen
  \bibfield  {author} {\bibinfo {author} {\bibfnamefont {J.~H.}\ \bibnamefont
  {Horne}}\ and\ \bibinfo {author} {\bibfnamefont {G.~T.}\ \bibnamefont
  {Horowitz}},\ }\href {\doibase 10.1103/PhysRevD.46.1340} {\bibfield
  {journal} {\bibinfo  {journal} {Phys. Rev. D}\ }\textbf {\bibinfo {volume}
  {46}},\ \bibinfo {pages} {1340} (\bibinfo {year}
  {1992}{\natexlab{b}})}\BibitemShut {NoStop}%
\bibitem [{\citenamefont {Astefanesei}\ \emph {et~al.}(2019)\citenamefont
  {Astefanesei}, \citenamefont {Herdeiro}, \citenamefont {Pombo},\ and\
  \citenamefont {Radu}}]{Astefanesei:2019pfq}%
  \BibitemOpen
  \bibfield  {author} {\bibinfo {author} {\bibfnamefont {D.}~\bibnamefont
  {Astefanesei}}, \bibinfo {author} {\bibfnamefont {C.}~\bibnamefont
  {Herdeiro}}, \bibinfo {author} {\bibfnamefont {A.}~\bibnamefont {Pombo}}, \
  and\ \bibinfo {author} {\bibfnamefont {E.}~\bibnamefont {Radu}},\ }\href
  {\doibase 10.1007/JHEP10(2019)078} {\bibfield  {journal} {\bibinfo  {journal}
  {JHEP}\ }\textbf {\bibinfo {volume} {10}},\ \bibinfo {pages} {078} (\bibinfo
  {year} {2019})},\ \Eprint {http://arxiv.org/abs/1905.08304} {arXiv:1905.08304
  [hep-th]} \BibitemShut {NoStop}%
\bibitem [{\citenamefont {Khalil}\ \emph {et~al.}(2018)\citenamefont {Khalil},
  \citenamefont {Sennett}, \citenamefont {Steinhoff}, \citenamefont {Vines},\
  and\ \citenamefont {Buonanno}}]{Khalil:2018aaj}%
  \BibitemOpen
  \bibfield  {author} {\bibinfo {author} {\bibfnamefont {M.}~\bibnamefont
  {Khalil}}, \bibinfo {author} {\bibfnamefont {N.}~\bibnamefont {Sennett}},
  \bibinfo {author} {\bibfnamefont {J.}~\bibnamefont {Steinhoff}}, \bibinfo
  {author} {\bibfnamefont {J.}~\bibnamefont {Vines}}, \ and\ \bibinfo {author}
  {\bibfnamefont {A.}~\bibnamefont {Buonanno}},\ }\href {\doibase
  10.1103/PhysRevD.98.104010} {\bibfield  {journal} {\bibinfo  {journal} {Phys.
  Rev. D}\ }\textbf {\bibinfo {volume} {98}},\ \bibinfo {pages} {104010}
  (\bibinfo {year} {2018})},\ \Eprint {http://arxiv.org/abs/1809.03109}
  {arXiv:1809.03109 [gr-qc]} \BibitemShut {NoStop}%
\bibitem [{\citenamefont {Chandrasekhar}(1985)}]{Chandrasekhar:1985kt}%
  \BibitemOpen
  \bibfield  {author} {\bibinfo {author} {\bibfnamefont {S.}~\bibnamefont
  {Chandrasekhar}},\ }\href@noop {} {\emph {\bibinfo {title} {{The mathematical
  theory of black holes}}}}\ (\bibinfo {year} {1985})\BibitemShut {NoStop}%
\bibitem [{\citenamefont {Carter}(1968)}]{PhysRev.174.1559}%
  \BibitemOpen
  \bibfield  {author} {\bibinfo {author} {\bibfnamefont {B.}~\bibnamefont
  {Carter}},\ }\href {\doibase 10.1103/PhysRev.174.1559} {\bibfield  {journal}
  {\bibinfo  {journal} {Phys. Rev.}\ }\textbf {\bibinfo {volume} {174}},\
  \bibinfo {pages} {1559} (\bibinfo {year} {1968})}\BibitemShut {NoStop}%
\bibitem [{\citenamefont {Johnson}\ \emph {et~al.}(2020)\citenamefont {Johnson}
  \emph {et~al.}}]{Johnson:2019ljv}%
  \BibitemOpen
  \bibfield  {author} {\bibinfo {author} {\bibfnamefont {M.~D.}\ \bibnamefont
  {Johnson}} \emph {et~al.},\ }\href {\doibase 10.1126/sciadv.aaz1310}
  {\bibfield  {journal} {\bibinfo  {journal} {Sci. Adv.}\ }\textbf {\bibinfo
  {volume} {6}},\ \bibinfo {pages} {eaaz1310} (\bibinfo {year} {2020})},\
  \Eprint {http://arxiv.org/abs/1907.04329} {arXiv:1907.04329 [astro-ph.IM]}
  \BibitemShut {NoStop}%
\bibitem [{\citenamefont {Broderick}\ \emph {et~al.}(2023)\citenamefont
  {Broderick}, \citenamefont {Salehi},\ and\ \citenamefont
  {Georgiev}}]{Broderick:2023jfl}%
  \BibitemOpen
  \bibfield  {author} {\bibinfo {author} {\bibfnamefont {A.~E.}\ \bibnamefont
  {Broderick}}, \bibinfo {author} {\bibfnamefont {K.}~\bibnamefont {Salehi}}, \
  and\ \bibinfo {author} {\bibfnamefont {B.}~\bibnamefont {Georgiev}},\ }\href
  {\doibase 10.3847/1538-4357/acf9f6} {\bibfield  {journal} {\bibinfo
  {journal} {Astrophys. J.}\ }\textbf {\bibinfo {volume} {958}},\ \bibinfo
  {pages} {114} (\bibinfo {year} {2023})},\ \Eprint
  {http://arxiv.org/abs/2307.15120} {arXiv:2307.15120 [astro-ph.HE]}
  \BibitemShut {NoStop}%
\bibitem [{\citenamefont {{Teo}}(2003)}]{2003GReGr..35.1909T}%
  \BibitemOpen
  \bibfield  {author} {\bibinfo {author} {\bibfnamefont {E.}~\bibnamefont
  {{Teo}}},\ }\href {\doibase 10.1023/A:1026286607562} {\bibfield  {journal}
  {\bibinfo  {journal} {General Relativity and Gravitation}\ }\textbf {\bibinfo
  {volume} {35}},\ \bibinfo {pages} {1909} (\bibinfo {year}
  {2003})}\BibitemShut {NoStop}%
\bibitem [{\citenamefont {Gralla}\ and\ \citenamefont
  {Lupsasca}(2020)}]{Gralla:2019drh}%
  \BibitemOpen
  \bibfield  {author} {\bibinfo {author} {\bibfnamefont {S.~E.}\ \bibnamefont
  {Gralla}}\ and\ \bibinfo {author} {\bibfnamefont {A.}~\bibnamefont
  {Lupsasca}},\ }\href {\doibase 10.1103/PhysRevD.101.044031} {\bibfield
  {journal} {\bibinfo  {journal} {Phys. Rev. D}\ }\textbf {\bibinfo {volume}
  {101}},\ \bibinfo {pages} {044031} (\bibinfo {year} {2020})},\ \Eprint
  {http://arxiv.org/abs/1910.12873} {arXiv:1910.12873 [gr-qc]} \BibitemShut
  {NoStop}%
\bibitem [{\citenamefont {Bisnovatyi-Kogan}\ and\ \citenamefont
  {Tsupko}(2022)}]{Bisnovatyi-Kogan:2022ujt}%
  \BibitemOpen
  \bibfield  {author} {\bibinfo {author} {\bibfnamefont {G.~S.}\ \bibnamefont
  {Bisnovatyi-Kogan}}\ and\ \bibinfo {author} {\bibfnamefont {O.~Y.}\
  \bibnamefont {Tsupko}},\ }\href {\doibase 10.1103/PhysRevD.105.064040}
  {\bibfield  {journal} {\bibinfo  {journal} {Phys. Rev. D}\ }\textbf {\bibinfo
  {volume} {105}},\ \bibinfo {pages} {064040} (\bibinfo {year} {2022})},\
  \Eprint {http://arxiv.org/abs/2201.01716} {arXiv:2201.01716 [gr-qc]}
  \BibitemShut {NoStop}%
\bibitem [{\citenamefont {Chang}\ and\ \citenamefont
  {Zhu}(2020{\natexlab{a}})}]{Chang:2020miq}%
  \BibitemOpen
  \bibfield  {author} {\bibinfo {author} {\bibfnamefont {Z.}~\bibnamefont
  {Chang}}\ and\ \bibinfo {author} {\bibfnamefont {Q.-H.}\ \bibnamefont
  {Zhu}},\ }\href {\doibase 10.1103/PhysRevD.101.084029} {\bibfield  {journal}
  {\bibinfo  {journal} {Phys. Rev. D}\ }\textbf {\bibinfo {volume} {101}},\
  \bibinfo {pages} {084029} (\bibinfo {year} {2020}{\natexlab{a}})},\ \Eprint
  {http://arxiv.org/abs/2001.05175} {arXiv:2001.05175 [gr-qc]} \BibitemShut
  {NoStop}%
\bibitem [{\citenamefont {Chang}\ and\ \citenamefont
  {Zhu}(2020{\natexlab{b}})}]{Chang:2020lmg}%
  \BibitemOpen
  \bibfield  {author} {\bibinfo {author} {\bibfnamefont {Z.}~\bibnamefont
  {Chang}}\ and\ \bibinfo {author} {\bibfnamefont {Q.-H.}\ \bibnamefont
  {Zhu}},\ }\href {\doibase 10.1103/PhysRevD.102.044012} {\bibfield  {journal}
  {\bibinfo  {journal} {Phys. Rev. D}\ }\textbf {\bibinfo {volume} {102}},\
  \bibinfo {pages} {044012} (\bibinfo {year} {2020}{\natexlab{b}})},\ \Eprint
  {http://arxiv.org/abs/2006.00685} {arXiv:2006.00685 [gr-qc]} \BibitemShut
  {NoStop}%
\bibitem [{\citenamefont {Vazquez}\ and\ \citenamefont
  {Esteban}(2004)}]{Vazquez:2003zm}%
  \BibitemOpen
  \bibfield  {author} {\bibinfo {author} {\bibfnamefont {S.~E.}\ \bibnamefont
  {Vazquez}}\ and\ \bibinfo {author} {\bibfnamefont {E.~P.}\ \bibnamefont
  {Esteban}},\ }\href {\doibase 10.1393/ncb/i2004-10121-y} {\bibfield
  {journal} {\bibinfo  {journal} {Nuovo Cim. B}\ }\textbf {\bibinfo {volume}
  {119}},\ \bibinfo {pages} {489} (\bibinfo {year} {2004})},\ \Eprint
  {http://arxiv.org/abs/gr-qc/0308023} {arXiv:gr-qc/0308023} \BibitemShut
  {NoStop}%
\bibitem [{\citenamefont {Papnoi}\ \emph {et~al.}(2014)\citenamefont {Papnoi},
  \citenamefont {Atamurotov}, \citenamefont {Ghosh},\ and\ \citenamefont
  {Ahmedov}}]{PhysRevD.90.024073}%
  \BibitemOpen
  \bibfield  {author} {\bibinfo {author} {\bibfnamefont {U.}~\bibnamefont
  {Papnoi}}, \bibinfo {author} {\bibfnamefont {F.}~\bibnamefont {Atamurotov}},
  \bibinfo {author} {\bibfnamefont {S.~G.}\ \bibnamefont {Ghosh}}, \ and\
  \bibinfo {author} {\bibfnamefont {B.}~\bibnamefont {Ahmedov}},\ }\href
  {\doibase 10.1103/PhysRevD.90.024073} {\bibfield  {journal} {\bibinfo
  {journal} {Phys. Rev. D}\ }\textbf {\bibinfo {volume} {90}},\ \bibinfo
  {pages} {024073} (\bibinfo {year} {2014})}\BibitemShut {NoStop}%
\bibitem [{\citenamefont {{Hartle}}(2003)}]{2003gieg.book.H}%
  \BibitemOpen
  \bibfield  {author} {\bibinfo {author} {\bibfnamefont {J.~B.}\ \bibnamefont
  {{Hartle}}},\ }\href@noop {} {\emph {\bibinfo {title} {{Gravity : an
  introduction to Einstein's general relativity}}}}\ (\bibinfo {year}
  {2003})\BibitemShut {NoStop}%
\bibitem [{\citenamefont {Gott}\ \emph {et~al.}(2019)\citenamefont {Gott},
  \citenamefont {Ayzenberg}, \citenamefont {Yunes},\ and\ \citenamefont
  {Lohfink}}]{Gott:2018ocn}%
  \BibitemOpen
  \bibfield  {author} {\bibinfo {author} {\bibfnamefont {H.}~\bibnamefont
  {Gott}}, \bibinfo {author} {\bibfnamefont {D.}~\bibnamefont {Ayzenberg}},
  \bibinfo {author} {\bibfnamefont {N.}~\bibnamefont {Yunes}}, \ and\ \bibinfo
  {author} {\bibfnamefont {A.}~\bibnamefont {Lohfink}},\ }\href {\doibase
  10.1088/1361-6382/ab01b0} {\bibfield  {journal} {\bibinfo  {journal} {Class.
  Quant. Grav.}\ }\textbf {\bibinfo {volume} {36}},\ \bibinfo {pages} {055007}
  (\bibinfo {year} {2019})},\ \Eprint {http://arxiv.org/abs/1808.05703}
  {arXiv:1808.05703 [gr-qc]} \BibitemShut {NoStop}%
\bibitem [{\citenamefont {Perlick}\ and\ \citenamefont
  {Tsupko}(2022)}]{Perlick:2021aok}%
  \BibitemOpen
  \bibfield  {author} {\bibinfo {author} {\bibfnamefont {V.}~\bibnamefont
  {Perlick}}\ and\ \bibinfo {author} {\bibfnamefont {O.~Y.}\ \bibnamefont
  {Tsupko}},\ }\href {\doibase 10.1016/j.physrep.2021.10.004} {\bibfield
  {journal} {\bibinfo  {journal} {Phys. Rept.}\ }\textbf {\bibinfo {volume}
  {947}},\ \bibinfo {pages} {1} (\bibinfo {year} {2022})},\ \Eprint
  {http://arxiv.org/abs/2105.07101} {arXiv:2105.07101 [gr-qc]} \BibitemShut
  {NoStop}%
\bibitem [{\citenamefont {Cardoso}\ and\ \citenamefont
  {Pani}(2019)}]{Cardoso:2019rvt}%
  \BibitemOpen
  \bibfield  {author} {\bibinfo {author} {\bibfnamefont {V.}~\bibnamefont
  {Cardoso}}\ and\ \bibinfo {author} {\bibfnamefont {P.}~\bibnamefont {Pani}},\
  }\href {\doibase 10.1007/s41114-019-0020-4} {\bibfield  {journal} {\bibinfo
  {journal} {Living Rev. Rel.}\ }\textbf {\bibinfo {volume} {22}},\ \bibinfo
  {pages} {4} (\bibinfo {year} {2019})},\ \Eprint
  {http://arxiv.org/abs/1904.05363} {arXiv:1904.05363 [gr-qc]} \BibitemShut
  {NoStop}%
\bibitem [{\citenamefont {Claudel}\ \emph {et~al.}(2001)\citenamefont
  {Claudel}, \citenamefont {Virbhadra},\ and\ \citenamefont
  {Ellis}}]{Claudel:2000yi}%
  \BibitemOpen
  \bibfield  {author} {\bibinfo {author} {\bibfnamefont {C.-M.}\ \bibnamefont
  {Claudel}}, \bibinfo {author} {\bibfnamefont {K.~S.}\ \bibnamefont
  {Virbhadra}}, \ and\ \bibinfo {author} {\bibfnamefont {G.~F.~R.}\
  \bibnamefont {Ellis}},\ }\href {\doibase 10.1063/1.1308507} {\bibfield
  {journal} {\bibinfo  {journal} {J. Math. Phys.}\ }\textbf {\bibinfo {volume}
  {42}},\ \bibinfo {pages} {818} (\bibinfo {year} {2001})},\ \Eprint
  {http://arxiv.org/abs/gr-qc/0005050} {arXiv:gr-qc/0005050} \BibitemShut
  {NoStop}%
\bibitem [{\citenamefont {Cunha}\ \emph {et~al.}(2017)\citenamefont {Cunha},
  \citenamefont {Berti},\ and\ \citenamefont {Herdeiro}}]{Cunha:2017qtt}%
  \BibitemOpen
  \bibfield  {author} {\bibinfo {author} {\bibfnamefont {P.~V.~P.}\
  \bibnamefont {Cunha}}, \bibinfo {author} {\bibfnamefont {E.}~\bibnamefont
  {Berti}}, \ and\ \bibinfo {author} {\bibfnamefont {C.~A.~R.}\ \bibnamefont
  {Herdeiro}},\ }\href {\doibase 10.1103/PhysRevLett.119.251102} {\bibfield
  {journal} {\bibinfo  {journal} {Phys. Rev. Lett.}\ }\textbf {\bibinfo
  {volume} {119}},\ \bibinfo {pages} {251102} (\bibinfo {year} {2017})},\
  \Eprint {http://arxiv.org/abs/1708.04211} {arXiv:1708.04211 [gr-qc]}
  \BibitemShut {NoStop}%
\bibitem [{\citenamefont {Cunha}\ and\ \citenamefont
  {Herdeiro}(2018)}]{Cunha:2018acu}%
  \BibitemOpen
  \bibfield  {author} {\bibinfo {author} {\bibfnamefont {P.~V.~P.}\
  \bibnamefont {Cunha}}\ and\ \bibinfo {author} {\bibfnamefont {C.~A.~R.}\
  \bibnamefont {Herdeiro}},\ }\href {\doibase 10.1007/s10714-018-2361-9}
  {\bibfield  {journal} {\bibinfo  {journal} {Gen. Rel. Grav.}\ }\textbf
  {\bibinfo {volume} {50}},\ \bibinfo {pages} {42} (\bibinfo {year} {2018})},\
  \Eprint {http://arxiv.org/abs/1801.00860} {arXiv:1801.00860 [gr-qc]}
  \BibitemShut {NoStop}%
\bibitem [{\citenamefont {Wei}(2020)}]{PhysRevD.102.064039}%
  \BibitemOpen
  \bibfield  {author} {\bibinfo {author} {\bibfnamefont {S.-W.}\ \bibnamefont
  {Wei}},\ }\href {\doibase 10.1103/PhysRevD.102.064039} {\bibfield  {journal}
  {\bibinfo  {journal} {Phys. Rev. D}\ }\textbf {\bibinfo {volume} {102}},\
  \bibinfo {pages} {064039} (\bibinfo {year} {2020})}\BibitemShut {NoStop}%
\bibitem [{\citenamefont {Vincent}\ \emph {et~al.}(2022)\citenamefont
  {Vincent}, \citenamefont {Gralla}, \citenamefont {Lupsasca},\ and\
  \citenamefont {Wielgus}}]{Vincent:2022fwj}%
  \BibitemOpen
  \bibfield  {author} {\bibinfo {author} {\bibfnamefont {F.~H.}\ \bibnamefont
  {Vincent}}, \bibinfo {author} {\bibfnamefont {S.~E.}\ \bibnamefont {Gralla}},
  \bibinfo {author} {\bibfnamefont {A.}~\bibnamefont {Lupsasca}}, \ and\
  \bibinfo {author} {\bibfnamefont {M.}~\bibnamefont {Wielgus}},\ }\href
  {\doibase 10.1051/0004-6361/202244339} {\bibfield  {journal} {\bibinfo
  {journal} {Astron. Astrophys.}\ }\textbf {\bibinfo {volume} {667}},\ \bibinfo
  {pages} {A170} (\bibinfo {year} {2022})},\ \Eprint
  {http://arxiv.org/abs/2206.12066} {arXiv:2206.12066 [astro-ph.HE]}
  \BibitemShut {NoStop}%
\bibitem [{\citenamefont {De~Martino}\ \emph {et~al.}(2023)\citenamefont
  {De~Martino}, \citenamefont {Della~Monica},\ and\ \citenamefont
  {Rubiera-Garcia}}]{DeMartino:2023ovj}%
  \BibitemOpen
  \bibfield  {author} {\bibinfo {author} {\bibfnamefont {I.}~\bibnamefont
  {De~Martino}}, \bibinfo {author} {\bibfnamefont {R.}~\bibnamefont
  {Della~Monica}}, \ and\ \bibinfo {author} {\bibfnamefont {D.}~\bibnamefont
  {Rubiera-Garcia}},\ }\href {\doibase 10.1103/PhysRevD.108.124054} {\bibfield
  {journal} {\bibinfo  {journal} {Phys. Rev. D}\ }\textbf {\bibinfo {volume}
  {108}},\ \bibinfo {pages} {124054} (\bibinfo {year} {2023})},\ \Eprint
  {http://arxiv.org/abs/2310.11039} {arXiv:2310.11039 [gr-qc]} \BibitemShut
  {NoStop}%
\bibitem [{\citenamefont {Gao}\ \emph {et~al.}(2023)\citenamefont {Gao},
  \citenamefont {Sui}, \citenamefont {Zeng}, \citenamefont {An},\ and\
  \citenamefont {Hu}}]{Gao:2023mjb}%
  \BibitemOpen
  \bibfield  {author} {\bibinfo {author} {\bibfnamefont {X.-J.}\ \bibnamefont
  {Gao}}, \bibinfo {author} {\bibfnamefont {T.-T.}\ \bibnamefont {Sui}},
  \bibinfo {author} {\bibfnamefont {X.-X.}\ \bibnamefont {Zeng}}, \bibinfo
  {author} {\bibfnamefont {Y.-S.}\ \bibnamefont {An}}, \ and\ \bibinfo {author}
  {\bibfnamefont {Y.-P.}\ \bibnamefont {Hu}},\ }\href {\doibase
  10.1140/epjc/s10052-023-12231-1} {\bibfield  {journal} {\bibinfo  {journal}
  {Eur. Phys. J. C}\ }\textbf {\bibinfo {volume} {83}},\ \bibinfo {pages}
  {1052} (\bibinfo {year} {2023})},\ \Eprint {http://arxiv.org/abs/2311.11780}
  {arXiv:2311.11780 [gr-qc]} \BibitemShut {NoStop}%
\bibitem [{\citenamefont {Macedo}\ \emph {et~al.}(2024)\citenamefont {Macedo},
  \citenamefont {Rosa},\ and\ \citenamefont
  {Rubiera-Garcia}}]{macedo2024optical}%
  \BibitemOpen
  \bibfield  {author} {\bibinfo {author} {\bibfnamefont {C.~F.}\ \bibnamefont
  {Macedo}}, \bibinfo {author} {\bibfnamefont {J.~L.}\ \bibnamefont {Rosa}}, \
  and\ \bibinfo {author} {\bibfnamefont {D.}~\bibnamefont {Rubiera-Garcia}},\
  }\href@noop {} {\bibfield  {journal} {\bibinfo  {journal} {arXiv preprint
  arXiv:2402.13047}\ } (\bibinfo {year} {2024})}\BibitemShut {NoStop}%
\bibitem [{\citenamefont {Olmo}\ \emph {et~al.}(2023)\citenamefont {Olmo},
  \citenamefont {Rosa}, \citenamefont {Rubiera-Garcia},\ and\ \citenamefont
  {Saez-Chillon~Gomez}}]{Olmo:2023lil}%
  \BibitemOpen
  \bibfield  {author} {\bibinfo {author} {\bibfnamefont {G.~J.}\ \bibnamefont
  {Olmo}}, \bibinfo {author} {\bibfnamefont {J.~L.}\ \bibnamefont {Rosa}},
  \bibinfo {author} {\bibfnamefont {D.}~\bibnamefont {Rubiera-Garcia}}, \ and\
  \bibinfo {author} {\bibfnamefont {D.}~\bibnamefont {Saez-Chillon~Gomez}},\
  }\href {\doibase 10.1088/1361-6382/aceacd} {\bibfield  {journal} {\bibinfo
  {journal} {Class. Quant. Grav.}\ }\textbf {\bibinfo {volume} {40}},\ \bibinfo
  {pages} {174002} (\bibinfo {year} {2023})},\ \Eprint
  {http://arxiv.org/abs/2302.12064} {arXiv:2302.12064 [gr-qc]} \BibitemShut
  {NoStop}%
\bibitem [{\citenamefont {Rosa}(2023)}]{Rosa:2023hfm}%
  \BibitemOpen
  \bibfield  {author} {\bibinfo {author} {\bibfnamefont {J.~a.~L.}\
  \bibnamefont {Rosa}},\ }\href {\doibase 10.1103/PhysRevD.107.084048}
  {\bibfield  {journal} {\bibinfo  {journal} {Phys. Rev. D}\ }\textbf {\bibinfo
  {volume} {107}},\ \bibinfo {pages} {084048} (\bibinfo {year} {2023})},\
  \Eprint {http://arxiv.org/abs/2302.11915} {arXiv:2302.11915 [gr-qc]}
  \BibitemShut {NoStop}%
\bibitem [{\citenamefont {Rosa}\ \emph {et~al.}(2023)\citenamefont {Rosa},
  \citenamefont {Macedo},\ and\ \citenamefont {Rubiera-Garcia}}]{Rosa:2023qcv}%
  \BibitemOpen
  \bibfield  {author} {\bibinfo {author} {\bibfnamefont {J.~a.~L.}\
  \bibnamefont {Rosa}}, \bibinfo {author} {\bibfnamefont {C.~F.~B.}\
  \bibnamefont {Macedo}}, \ and\ \bibinfo {author} {\bibfnamefont
  {D.}~\bibnamefont {Rubiera-Garcia}},\ }\href {\doibase
  10.1103/PhysRevD.108.044021} {\bibfield  {journal} {\bibinfo  {journal}
  {Phys. Rev. D}\ }\textbf {\bibinfo {volume} {108}},\ \bibinfo {pages}
  {044021} (\bibinfo {year} {2023})},\ \Eprint
  {http://arxiv.org/abs/2303.17296} {arXiv:2303.17296 [gr-qc]} \BibitemShut
  {NoStop}%
\bibitem [{\citenamefont {Goto}\ \emph {et~al.}(2019)\citenamefont {Goto},
  \citenamefont {Marrochio}, \citenamefont {Myers}, \citenamefont {Queimada},\
  and\ \citenamefont {Yoshida}}]{Goto:2018iay}%
  \BibitemOpen
  \bibfield  {author} {\bibinfo {author} {\bibfnamefont {K.}~\bibnamefont
  {Goto}}, \bibinfo {author} {\bibfnamefont {H.}~\bibnamefont {Marrochio}},
  \bibinfo {author} {\bibfnamefont {R.~C.}\ \bibnamefont {Myers}}, \bibinfo
  {author} {\bibfnamefont {L.}~\bibnamefont {Queimada}}, \ and\ \bibinfo
  {author} {\bibfnamefont {B.}~\bibnamefont {Yoshida}},\ }\href {\doibase
  10.1007/JHEP02(2019)160} {\bibfield  {journal} {\bibinfo  {journal} {JHEP}\
  }\textbf {\bibinfo {volume} {02}},\ \bibinfo {pages} {160} (\bibinfo {year}
  {2019})},\ \Eprint {http://arxiv.org/abs/1901.00014} {arXiv:1901.00014
  [hep-th]} \BibitemShut {NoStop}%
\bibitem [{\citenamefont {Poisson}(2009)}]{Poisson:2009pwt}%
  \BibitemOpen
  \bibfield  {author} {\bibinfo {author} {\bibfnamefont {E.}~\bibnamefont
  {Poisson}},\ }\href {\doibase 10.1017/CBO9780511606601} {\emph {\bibinfo
  {title} {{A Relativist's Toolkit: The Mathematics of Black-Hole
  Mechanics}}}}\ (\bibinfo  {publisher} {Cambridge University Press},\ \bibinfo
  {year} {2009})\BibitemShut {NoStop}%
\end{thebibliography}%
\end{document}